\documentclass[12pt, a4paper]{article}
\usepackage[T2A]{fontenc}
\usepackage[utf8]{inputenc}
\usepackage{cite}
\usepackage{mathtext}
\usepackage{amssymb,amsthm,amsmath}
\usepackage[mathscr]{eucal}
\usepackage{colortbl}
\usepackage{array}
\usepackage[english]{babel}
\usepackage[title,titletoc]{appendix}
\usepackage{graphicx}
\usepackage{graphics}
\usepackage{float}

\begin{document}
\author{\bf Yu.A.\,Markov$\!\,$\thanks{e-mail:markov@icc.ru}\,,\,
M.A.\,Markova$\!\,$\thanks{e-mail:markova@icc.ru}\,,\,
N.Yu.\,Markov$\!\:$\thanks{e-mail:karvedys1398@mail.ru}}

\title{Hamiltonian formalism for Fermi excitations in a plasma with a non-Abelian interaction}
%
%
\date{\it\normalsize
\begin{itemize}
\item[]Matrosov Institute for System Dynamics and Control Theory, Siberian Branch, Russian Academy of Sciences, Irkutsk, 664033 Russia
\vspace{-0.3cm}
%
\end{itemize}}
%
\thispagestyle{empty}
\maketitle{}


\def\theequation{\arabic{section}.\arabic{equation}}
{
\[
\mbox{\bf Abstract}
\]

The Hamiltonian theory for the collective longitudinally polarized colorless gluon excitations (plasmons) and for collective quark-antiquark excitations with abnormal relation between chirality and helicity (plasminos) in a high-temperature quark-gluon plasma (QGP) is developed. For this purpose, Zakharov's forma\-lism for constructing the wave theory in nonlinear media with dispersion is used. A generalization of the Poisson superbracket involving both commuting and anticommuting variables to the case of a continuous medium is  performed and the corresponding Hamilton equations are presented. The canonical transformations including simultaneously both bosonic and fermionic degrees of freedom of the collective excitations in QGP are discussed and a complete system of the canonicity conditions for these transformations is written out. An explicit form of the effective fourth-order Hamiltonians describing the elastic scattering of plasmino off plasmino and plasmino off plasmon is found and the Boltzmann type kinetic equations describing the processes of elastic scattering are obtained. A detailed comparison of the effective amplitudes defined within the (pseudo)classical Hamiltonian theory, with the corresponding matrix elements calcu\-la\-ted early in the framework of high-temperature quantum chromodynamics in the so-called hard thermal loop approximation is performed. This enables one to obtain, in particular, an explicit form of the vertex and coefficient functions in the effective amplitudes and in the canonical transformations, correspondingly, and also to define the validity of a purely pseudoclassical approach in the Hamiltonian description of the dynamics of a quark-gluon plasma. The problem of determining the higher-order coefficient functions in the canonical transformations of fermionic and bosonic normal variables is considered. With the help of the coefficient functions obtained, the totally symmetric effective amplitudes of the elastic scattering of plasmino off plasmon and plasmino off plasmino are written out.

}



\newpage
\section{Introduction}
\setcounter{equation}{0}

When one studies wave phenomena in various nonlinear continuous media their similarity stands out. That is why it was rather important to construct a general wave theory in nonlinear media which would consider all wave processes from a single point of view without considering the specifics of the medium. Classical mechanics formulated in the language of canonical variables can serve as an example of such a theory. From the fundamental point of view the formulation of a general theory of waves in nonlinear continuous media is no more than a consistent generalization of the ideas of classical Hamiltonian mechanics to the case of an infinite number of degrees of freedom. This generalization was systematically developed by Zakharov \cite{zakharov_1971, zakharov_1974} and presented in detail with  many examples of concrete physical systems in comprehensive reviews \cite{zakharov_1985, zakharov_1997} and in the monograph \cite{zakharov_book_1992} (see also Krasitskii \cite{krasitskii_1990}). The approach suggested by Zakharov has a particular value in  the fact that it gives universal algorithm of the construction of a general wave theory in a wide range of nonlinear media with dispersion.\\
\indent In our early paper \cite{markov_2020} a step forward to the construction of the classical Hamiltonian formalism for the description of the nonlinear interaction processes for soft collective excitations in a hot weakly coupled quark-gluon plasma (QGP) was taken. The construction was based on the fact that equations describing a collisionless high-temperature plasma in the hard thermal loop approximation have the Hamiltonian structure that had been determined in the papers by Nair \cite{nair_1993, nair_1994}, Blaizot and Iancu \cite{blaizot_1994(1), blaizot_2002}. This allowed us, at least for weakly excited states, to develop an independent approach to the derivation of the kinetic equation for soft longitudinally polarized gluonic plasma excitations. In the paper \cite{markov_2020} the simplest case of interactions, namely the interaction of collective longitudinal-polarized colorless gluon excitations (plasmons) in pure gluon plasma was considered. Within the framework of a general Hamiltonian approach to the construction of the wave theory in nonlinear media with dispersion \cite{zakharov_1971, zakharov_1974, zakharov_1985, zakharov_book_1992, zakharov_1997, krasitskii_1990} we have defined in the explicit form a special canonical transformation up to cubic terms in the plasmon creation and annihilation operators  $\hat{a}^{\dagger\ \!\!a}_{{\bf k}}$ and $\hat{a}^{\phantom{\dagger}\! a}_{{\bf k}}$. A system of the canonicity conditions connecting among themselves the highest and lowest coefficient functions in the integrands of the expansion terms of the canonical transformations was defined. By virtue of three-wave non-decay nature of the dispersion relation for the colorless plasmons, the canonical transformations enabled us to eliminate the third-order Hamiltonian $H_{3}$ in powers of the quasiparticle operators 
$\hat{a}^{\dagger\ \!\!a}_{{\bf k}}$ and $\hat{a}^{\phantom{\dagger}\! a}_{{\bf k}}$. Excluding ``nonessential'' (in the terminology of V.E. Zakharov \cite{zakharov_1974}) interaction Hamiltonian $H_{3}$ gave us a possibility to obtain a new effective fourth-order Hamiltonian $\widetilde{H}_{4}$, the integrand of which contains the gauge-invariant scattering amplitude defining the simplest elastic scattering process of two plasmons off each other. A graphic interpretation of all of the terms in the effective scattering amplitude was presented.\\
\indent The developed Hamiltonian approach was used for the construction of a Boltzmann-type kinetic equation, which describes a change of the number density $N^{\hspace{0.03cm}l}_{\bf k}$ of colorless plasmons in a weakly inhomogeneous and weakly non-stationary gluon plasma by virtue of the elastic scattering process of plasmon off plasmon and of the so-called nonlinear Landau damping. When constructing the desired kinetic equation, a self-consistent system of two integro-differential equations for correlation functions of the second and fourth orders in the new quasiparticle operators $\hat{c}^{\dagger\ \!\!a}_{{\bf k}}$ and $\hat{c}^{\phantom{\dagger}\! a}_{{\bf k}}$ was defined. In order to close the system, the sixth-order correlation function was expressed in the terms of two-point correlators and an approximate solution of the equation for the fourth-order correlation function was obtained. This solution enabled us to reduce the system to a single kinetic equation for the two-point correlation function. We have also compared in detail the effective amplitude of plasmon-plasmon scattering derived within the classical Hamiltonian theory and the correspondent matrix element derived early in the framework of high-temperature quantum chromodynamics \cite{markov_2002}. This allowed us, in particular, to define an explicit form of the vertex functions in  the effective scattering amplitude, an explicit form of the third- and fourth-order coefficient functions in the canonical transformation of the operator $\hat{a}_{\bf k}$ and the limits of validity of a pure classical approach in the construction of the Hamiltonian formalism for a hot gluon plasma. To compare the effective amplitude and the matrix element, we needed an approximation of the effective gluon propagator at the plasmon pole in a spirit of the works by Weldon \cite{weldon_1998} and Blaizot and Iancu \cite{blaizot_1994(1)}.\\
\indent In the present paper, we enlarge the analysis carried out in \cite{markov_2020} to the fermion sector of soft plasma waves. It is a commonly known fact that the hot QCD plasma which includes massless quarks and antiquarks possesses also the collective fermion (quark-antiquark) excitations. These soft fermion modes obey a dispersion relation, which has two independent branches \cite{klimov_1981, klimov_1982, weldon_1982}.  The first branch describes normal particle mode having a relation between chirality and helicity as it takes place in the usual quantum field theory, or in other words in a system at zero temperature. The second branch is purely collective excitation in which the standard relation between chirality and helicity is replaced by the relation with opposite (abnormal) sign \cite{pisarski_1989}.  By analogy with the plasmon (longitudinal) mode of gluons having a purely collective character, this abnormal mode was called a plasmino \cite{braaten_1992}. In the works by Weldon \cite{weldon_1989(1), weldon_1989(2)} a physical particle-hole interpretation for the splitting of the fermion spectrum at high temperature has been given. This interpretation was based on an analysis of the quantum numbers of modes, and on an analogy with BCS theory of superconductivity. A deeper study of the nature of fermionic excitations for the case of the Yukawa and QED theories at finite temperature and with zero chemical potential has been given by Baym, Blaizot and Svetitsky \cite{baym_1992} making use of numerical analysis of the one-loop electron propagator. The case of a cold ultrarelativistic plasma with a large chemical potential was considered by Blaizot and Ollitrault \cite{blaizot_1993}. The latter authors have explicitly constructed the quantum states associated with the two branches, which made it possible to clarify more clearly the collective nature of the long wavelength Fermi-excitations.\\
\indent It should also be noted that because of the processes of nonlinear interaction of soft collective QGP plasma waves (quasiparticles) among themselves, such quasiparticles do not have sharp energy shells, but instead behaves like resonances with the widths proportional to the corresponding damping rates $\gamma\sim g^{2}\hspace{0.02cm}T$, where $g$ is the QCD coupling constant and $T$ is a temperature of the system under consideration. This circumstance was first pointed out by Braaten and Pisarski in \cite{braaten_1992_2} when calculating the quark damping rate in hot QCD within the framework of the hard thermal loop effective theory. Within the limits of the construction of plasmino and plasmon kinetic equations, this problem was also discussed in \cite{markov_2001}. The latter was based on the Blaizot-Iancu equations, which are a local formulation of the hard thermal loop equations of motion for soft fluctuating quark and gluon fields and their induced sources. Thus, if the coupling constant is small enough, then the width $\gamma\sim g^{2}\hspace{0.02cm}T$ is 'small' compared to the plasma frequency $\omega\sim g\hspace{0.02cm}T$ and therefore the (parametric) lifetime of the plasmino is sufficiently large to consider the plasmino as a real (quasi)particle.\\
\indent Finally, it is worth mentioning that significant efforts have been made in recent years to push the hard thermal loop calculation and to obtain the HTL effective Lagrangian at next-to-leading order, at least for the high temperature quantum electrodynamics. The complete leading power corrections in the $1/T$ for massless QED, including the fermion sector, were calculated in \cite{carignano_2018} and the weak coupling corrections at two loops to the photon self-energy for the case of the soft external momentum were worked out in \cite{carignano_2020, gorda_2022}. All calculations were performed within the Keldysh representation of the real-time formalism utilizing dimensional regularization. As an application, the next-to-leading order corrections to the HTL dispersion relations for the  transverse and longitudinal modes, were calculated. In due time Smilga \cite{smilga_1997} raised the question of the stability of the plasmon and plasmino modes of oscillations with respect to higher-order corrections to the one-loop approximation for the HTL propagators. Based on heuristic considerations, the suggestion was made that these purely collective modes should disappear at all at some value of the momentum $|{\bf k}|\sim gT$, i.e. the plasmon and plasmino dispersion curves cross the line $\omega = |{\bf k}|$ and, as a consequence, at this point the polarization and mass operators acquire the imaginary part due to {\it Landau damping}. Now, in light of the new results \cite{carignano_2018, carignano_2020, gorda_2022}, it is possible to verify this statement in a rigorous way.
\\
\indent Our main purpose is to construct the (pseudo)classical Hamiltonian formalism for a complete and self-consistent description of the nonlinear scattering processes of soft collective excitations of both bosonic and fermionic types in a hot plasma with non-Abelian interaction. To achieve the stated aim, as the guiding principle we will use again a general Hamiltonian approach to the description of wave processes in nonlinear media with nondecay dispersion laws developed by Zakharov. We generalize the pseudoclassical Hamiltonian formalism (see discussion just below) to the case of the systems with distributed parameters for the description of the dynamics of collective excitations obeying Fermi statistics and construct specific canonical transformations that take into account simultaneously both boson and fermion degrees of freedom of the system. It enables us to eliminate the third-order terms in the normal boson and fermion variables from the initial interaction Hamiltonian and thus to define a new effective fourth-order Hamiltonians. Here, we restrict ourselves mainly to research of the processes connected with the interaction of the soft  boson and fermion pure collective excitations (plasmons and plasminos) among themselves without exchange of energy with hard thermal (or external) partons: quarks, antiquarks and gluons. We compare in detail the effective amplitudes of plasmino-plasmino and of plasmino-plasmon interactions obtained within the pseudoclassical Hamiltonian theory and the relevant matrix elements derived early by us within the framework of high-temperature quantum chromodynamics \cite{markov_2006}. It will give us the opportunity to obtain an explicit form of vertex functions in the effective amplitudes, to determine the limits of validity of the pure pseudoclassical Hamiltonian approach, and to construct a self-consistent system of kinetic equations of Boltzmann type describing the scattering processes of plasmino off plasmino or plasmon.\\
\indent It should be noted that the derivation of the Boltzmann equation describing an evolution of the number density of fermion quasiparticles in a quark-gluon plasma was also considered by Ni\'egawa \cite{niegawa_2002}. Within the framework of the Keldysh-Schwinger formalism the kinetic equations for normal and abnormal fermionic modes here, were defined from the requirement of the absence of the large contributions due to pinch singularities of the perturbative scheme proposed by Ni\'egawa. The formalism suggested by the author is quite rigorous and possesses great generality at least for weakly non-stationary and weakly inhomogeneous systems near equili\-bri\-um. Unfortunately, in this paper concrete expressions for collision terms were not given.\\
\indent In the present paper, as opposed to \cite{markov_2020}, we do not use an operator formalism in constructing the Hamiltonian theory of nonlinear wave processes in the quark-gluon plasma. Instead of the quasiparticle creation and annihilation operators for plasmons $\bigl(\hat{a}^{\phantom{\dagger}\!\!a}_{{\bf k}},\, \hat{a}^{\dagger\ \!\!a}_{{\bf k}}\bigr)$ and for plasminos $\bigl(\hat{b}_{\bf q}^{\,i},\,\hat{b}_{\bf q}^{\,\dagger\hspace{0.03cm}i}\bigr)$ obeying the usual (anti)commutation relations for Bose and Fermi operators: 
\begin{align}
&\Bigl[\hat{a}^{\phantom{\dagger}\!\! a}_{{\bf k}},\,\hat{a}^{\phantom{\dagger}\!\! b}_{{\bf k}^{\prime}}\Bigr]
=
\Bigl[\hat{a}^{\dagger\ \!\! a}_{{\bf k}},\,\hat{a}^{\dagger\ \!\!b}_{{\bf k}^{\prime}}\Bigr] = 0,
\qquad
\Bigl[\hat{a}^{\phantom{\dagger}\!\! a}_{{\bf k}},\,\hat{a}^{\dagger\ \!\!b}_{{\bf k}^{\prime}}\Bigr]
=
\delta^{\phantom{\dagger}\!\! a\hspace{0.02cm}b}\ \!\! (2\pi)^{3}\hspace{0.03cm}\delta({\bf k} - {\bf k}^{\prime}),
\notag\\[1.5ex]
&\Bigl\{\hat{b}^{\phantom{\dagger}\!\! i}_{{\bf q}^{\phantom\prime}},\,\hat{b}^{\phantom{\dagger}\!\! j}_{{\bf q}^{\prime}}\Bigr\} 
=
\Bigl\{\hat{b}^{\,\dagger\hspace{0.03cm} i}_{{\bf q}^{\phantom\prime}},\,\hat{b}^{\,\dagger\hspace{0.03cm} j}_{{\bf q}^{\prime}}\Bigr\} = 0, 
\quad\;
\Bigl\{\hat{b}^{\,\phantom{\dagger}\!\!\! i}_{{\bf q}^{\phantom\prime}},\,\hat{b}^{\,\dagger\, j}_{{\bf q}^{\prime}}\Bigr\}
=
\delta^{\phantom{\dag}\!\! i\hspace{0.02cm}j}\ \!\! (2\pi)^{3}\hspace{0.03cm} \delta({\bf q} - {\bf q}^{\prime}),
\notag
\end{align} 
for the description of boson degree of freedom of QGP we use the usual functions, in our case  we employ the complex normal variables $\bigl(a^{\phantom{\ast}\!\!a}_{{\bf k}},\,a^{\ast\ \!\!a}_{{\bf k}}\bigr)$, and for the description of fermion degree of freedom we use the normal variables $\bigl(b_{\bf q}^{\,i},\, b_{\bf q}^{\,\ast\hspace{0.03cm}i}\bigr)$ taking values in a Grassmann algebra. These functions satisfy canonical super-Poisson bracket relations, as they are defined by Eqs.\,(\ref{eq:2i}) and (\ref{eq:2o}). The Heisenberg equations in this case are replaced by the Hamilton equations, the right-hand side of which is defined by the corresponding Poisson superbracket. The Poisson superbracket represents a generalization of the usual Poisson bracket. It was proposed for the first time by Fradkin \cite{fradkin_1973} within the framework of quantum field theory with an arbitrary Lagrangian $L(\varphi_{i}, \psi)$ of interacting Bose and Fermi fields and, independently, by Casalbuoni \cite{casalbuoni_1976_1, casalbuoni_1976_2},  Berezin and Marinov \cite{berezin_1977} for the description of classical Hamiltonian mechanics including anticommuting variables. A systematic and more modern introduction to the concept of the Poisson superbracket can be found in the monographs by Gitman and Tyutin \cite{gitman_book_1990}, and by Henneaux and  Teitelboim \cite{henneaux_book_1995}.\\
\indent In addition, as compared to our previous paper \cite{markov_2020}, here a new element is analyses of excluding the so-called ``nonresonant'' (in the terminology of V.P. Krasitskii \cite{krasitskii_1990}) fourth-order terms in the normal variables from the effective fourth-order Hamiltonians ${\mathcal H}^{(4)}_{qq\rightarrow qq}$ and 
${\mathcal H}^{(4)}_{qg\rightarrow qg}$ describing the elastic scattering processes of plasmino off plasmino or plasmon, respectively. Krasitskii \cite{krasitskii_1990} first pointed out the importance of this problem. He has performed detailed calculations for the case of the simplest boson system with one type of wave without
polarization. This rather nontrivial point has not been completely taken into account in the works by Zakharov with co-workers \cite{zakharov_1974, zakharov_1985, zakharov_book_1992, zakharov_1997}. In fact, in obtaining any effective fourth-order Hamiltonian as a result of canonical transformations not only the nondecay cubic terms, but also the nonresonant terms of fourth order in the normal variables should be eliminated. Excluding the nonresonant fourth-order terms makes it possible to find an explicit form of the higher-order coefficient functions in the integrands of the expansion terms of canonical transformations, and perhaps even more importantly, to obtain the most general form of the effective fourth-order Hamiltonians by itself.\\  
%
%
\indent The paper is organized as follows. In section \ref{section_2}, the general form of the decomposition into plane waves of the gauge field potential and quark-antiquark wave function is written out and the expectation values of the product of two bosonic and fermionic amplitudes are derived. In the same section a generalization of the Poisson superbracket including anticommuting variables to the case of a continuous media is presented, the corresponding Hamilton equations are defined and the most general structure of the third- and fourth-order interaction Hamiltonians in the normal variables $a^{\phantom{\ast}\!\!a}_{{\bf k}}$ and $b_{\bf q}^{\,i}$ of Fermi and Bose fields of a quark-gluon plasma is written out. In section \ref{section_3}, the canonical transformations including simultaneously both bosonic and fermionic degrees of freedom of the collective excitations in QGP are discussed. Two systems of the canonicity conditions for these transformations based on the Poisson superbracket are written out. The most general structure of the canonical transformations in the form of integro-power series in the new normal Grassmann-graded field variables
$c^{\phantom{\ast}\!\!a}_{{\bf k}}$ and $f_{\bf q}^{\,i}$ up to the terms of third order is presented. In section~\ref{section_4}, making use of the above-mentioned canonical transformations the problem of excluding the “non-essential” third-order Hamiltonian $H^{(3)}$ is considered. The explicit expressions for the coefficient functions in the quadratic part of the canonical transformations, which exclude all of the cubic terms of the new Hamiltonian are obtained. Section \ref{section_5} is devoted to the determination of an explicit form of the effective fourth-order Hamiltonian describing elastic scattering of plasmino off plasmon. An effective amplitude for this scattering process is written out and a simple diagrammatic interpretation of the individual terms in the effective amplitude is given. In section \ref{section_6} an explicit form of the effective fourth-order Hamiltonian, which describes elastic scattering of plasmino off plasmino is defined. A diagrammatic interpretation of the individual terms in the effective amplitude for this scattering process is given.\\
\indent Section \ref{section_7} is concerned with the calculation of fourth- and sixth-order correlation functions in the normal variables of anticommuting Fermi field. The notion of the plasmino number density $n^{-}_{\bf q}$ is introduced. On the basis of Hamilton’s equation of motion with the Poisson superbracket, a differential equation which is obeyed by the fourth-order fermionic correlation function is obtained. In section \ref{section_8} an approximate solution of the equation, accounting for the deviation of the four-point correlation function from the Gaussian approximation for a low nonlinearity level of interacting Fermi-excitations is found. On the basis of this solution, the kinetic equation for the number density of plasminos describing the elastic scattering process of collective Fermi-excitations among themselves with  allowance for the Landau nonlinear damping effect is constructed. Section \ref{section_9} is devoted to determining an explicit form of the plasmino-plasmino-plasmon vertex function in the effective amplitudes. For this purpose, a detailed comparison of the effective amplitude of elastic plasmino-plasmino scattering obtained within the pseudo\-classical Hamiltonian theory with the relevant matrix element derived within the frame\-work of the high-temperature quantum chromodynamics (QCD) is performed. The applicability limits of the pure classical Hamiltonian approach are determined. Section \ref{section_10} is concerned with the calculation of the fourth- and sixth-order correlation functions simultaneously containing normal variables of commuting Bose field and anticommuting Fermi field. The notion of the colorless plasmon number density $N^{\hspace{0.03cm}l}_{\bf k}$ is introduced. Making use of Hamilton’s equation of motion for the bosonic variable $c^{\phantom{\ast}\!\!a}_{{\bf k}}$ and for the fermionic one $f^{\,i}_{\bf q}$, a differential equation to which the fourth-order correlation function containing two fermionic and two bosonic variables obeys, is derived. In section \ref{section_11} an approximate solution of the equation, taking into account the deviation of the four-point correlator from the Gaussian approximation for weak nonlinearity level of interacting Fermi- and Bose-excitations is found. On the basis of this solution for the fourth-order correlation function including Bose and Fermi fields, a self-consistent system of two Boltzmann-type kinetic equations for the number density of plasminos and plasmons describing the processes of elastic scattering of plasmino off plasmon and of plasmon off plasmino with allowance for the Landau nonlinear damping effect for Fermi and Bose plasma excitations is obtained.\\
\indent Section \ref{section_12} is devoted to a detail comparison of the effective amplitude of plasmino-plasmon scattering with the appropriate matrix element calculated within the framework of the high-temperature QCD using the hard thermal loop approximation. As a result of this comparison an explicit form of the vertex functions in the effective amplitude of plasmino-plasmon scattering is defined. In section \ref{section_13} the problem of the construction of the third-order coefficient functions $S^{\, (n)\, i\; i_{1}\, i_{2}\, i_{3}}_{\, {\bf q},\, {\bf q}_{1},\, {\bf q}_{2},\, {\bf q}_{3}}$, $n = 1,\ldots,4$ entering into the canonical transformation of the original fermionic variable $b^{\; i}_{{\bf q}}$ is considered. Based on the requirement of making the so-called non-resonant terms vanish in effective interaction Hamiltonian of the fourth-order in $f^{\; i}_{{\bf q}}$ and $f^{\,\ast\, i}_{{\bf q}}$ the explicit form of these coefficients at $n = 1,\,3$ and $4$ is defined. To determine an explicit form of the coefficient function $S^{(2)\, i\; i_{1}\, i_{2}\, i_{3}}_{\; {\bf q},\, {\bf q}_{1},\, {\bf q}_{2},\, {\bf q}_{3}}$ a system of two functional equations is solved. With the help of the function obtained, a complete effective amplitude of elastic scattering of  plasmino off plasmino satisfying all of the symmetry conditions is written out. In section \ref{section_14} a similar problem of defining the third-order coefficient functions  $J^{\, (n)\, a\, a_{1}\, i_{1}\, i_{2}}_{\ {\bf k},\, {\bf k}_{1},\, {\bf q}_{1},\, {\bf q}_{2}}$ and $R^{\,(n)\, i\, a_{1}\, a_{2}\, i_{1}}_{\, {\bf q},\, {\bf k}_{1},\, {\bf k}_{2},\, {\bf q}_{1}},\, n = 1,\ldots,6$ in the canonical transformations of the normal bosonic $a^{\phantom{\ast}\!\!a}_{{\bf k}}$ and fermionic variables $b^{\; i}_{{\bf q}}$, respectively,  is considered. The explicit form of the coefficient functions with $n = 1,\,3,\ldots,6$ is unambiguously determined. To define an explicit form of the remaining coefficient functions $J^{\, (2)\, a\, a_{1}\, i_{1}\, i_{2}}_{\ {\bf k},\, {\bf k}_{1},\, {\bf q}_{1},\, {\bf q}_{2}}$ and $R^{\,(2)\, i\, a_{1}\, a_{2}\, i_{1}}_{\, {\bf q},\, {\bf k}_{1},\, {\bf k}_{2},\, {\bf q}_{1}}$ a self-consistent system of four functional equations is solved. With the help of the functions obtained, a complete effective amplitude of elastic scattering of plasmino off plasmon satisfying all of the required symmetry conditions with respect to the rearrangement of external lines is written out.\\
\indent In the concluding section \ref{section_15} the key points of our work are specified. This section briefly discusses several interesting issues which are very close to the subject of the present research but have not been touched in the paper. They are concerned in particular with the introduction of antiplasmino branch of fermionic excitations, with an analysis of a possibility of the construction  of the so-called odd Poisson bracket and the corresponding odd Hamiltonian for the system under consideration, and with a discussion of a further generalization of the structure of the interaction Hamiltonians $H^{(3)}$ and $H^{(4)}$, Eqs.\,(\ref{eq:2f}) and (\ref{eq:2g}), and also of the canonical transformations (\ref{eq:3t}) and (\ref{eq:3y}) in case when some of the vertex and coefficient functions can take values in a Grassmann algebra.\\
\indent In Appendix \ref{appendix_A} we give all of the basic expressions for the effective gluon vertex functions and gluon propagator within the framework of the hard thermal loop approximation (HTL). In Appendix \ref{appendix_B} an explicit form of  the spinors $u^{(\pm)}$ and $v^{(\pm)}$ describing polarization properties of normal and abnormal fermionic modes of quark-gluon plasma is written out. The expressions for the HTL-induced vertices between a quark pair and a gluon, between quark pair and two gluons and for a medium modified quark propagator are given. The most important properties of these vertices are listed. In Appendices \ref{appendix_C} and \ref{appendix_D} a complete system of independent relations of the canonicity conditions connecting the lowest and highest coefficient functions in the canonical transformations among themselves is given. In Appendix \ref{appendix_E} an explicit form of some third-order coefficient functions, which enter into the canonical transformations is written out.


\section{Interaction Hamiltonian of plasmons and plasminos}
\setcounter{equation}{0}
\label{section_2}

Let us consider the application of the general Zakharov theory to a specific system, namely to a high-temperature quark-gluon plasma in the semi-classical approximation. The gauge field potentials describing the gluon field in the system are $N_c\times N_c$ matrices in the color space and are defined in terms of $A_{\mu}(x) = A_{\mu}^{a}(x)\, t^{a}$ with $N^{\hspace{0.02cm}2}_c - 1$ Hermitian generators $t^{a}$ of the color $SU(N_c)$ group in the fundamental representation\footnote{\,The color indices $a,\,b,\,c,\,\ldots$ run through values $1,2,\,\ldots\,,N^{2}_{c}-1$, while the vector indices $\mu,\,\nu,\,\lambda,\,\ldots$ run through values $0,1,2,3$. Everywhere in this article, we imply summation over repeated indices and use the system of units with $\hbar = c = 1$.}.\\
\indent It is known that there exist two types of the physical soft gluon fields in an equilibrium hot quark-gluon plasma:  transverse- and longitudinal-polarized ones \cite{kalashnikov_1980}. Let us consider the gauge field potential in the form of the decomposition into plane waves \cite{blaizot_1994(1), hakim_book_2011}
\begin{equation}
\begin{split}
A^{a}_{\mu}(x) = &\int\!\frac{d\hspace{0.02cm}{\bf k}}{(2\pi)^{3}}\!\left(\frac{Z_{l}({\bf k})}
{2\omega^{l}_{{\bf k}}}\right)^{\!\!1/2}\!\!
\left\{\epsilon^{\ \! l}_{\mu}({\bf k})\, a^{\phantom{\ast}\!\!a}_{{\bf k}}\ \!e^{-i\hspace{0.03cm}\omega^{l}_{{\bf k}}\hspace{0.02cm}t\hspace{0.03cm} +\hspace{0.03cm} i\hspace{0.03cm}{\bf k}\hspace{0.02cm}\cdot\hspace{0.02cm} {\bf x}}
+
\epsilon^{\ast\, l}_{\mu}({\bf k})\, a^{\ast\ \!\!a}_{{\bf k}}\ \!e^{\hspace{0.02cm}i\hspace{0.03cm}\omega^{l}_{{\bf k}}\hspace{0.02cm}t\hspace{0.03cm} -\hspace{0.03cm} i\hspace{0.03cm}{\bf k}\hspace{0.02cm}\cdot\hspace{0.02cm} {\bf x}}
\right\}
+\\[1.3ex]
+ 
\sum_{\zeta\hspace{0.02cm} =\hspace{0.02cm} 1,\hspace{0.02cm} 2}\hspace{0.03cm} 
&\int\!\frac{d\hspace{0.02cm}{\bf k}}{(2\pi)^{3}}\!\left(\frac{Z_{t}({\bf k})}
{2\omega^{t}_{{\bf k}}}\right)^{\!\!1/2}\!\!
\left\{\epsilon^{\ \! t}_{\mu}({\bf k}, \zeta)\, a^{\phantom{\ast}\!\!a}_{{\bf k}}(\zeta)\ \!e^{-i\hspace{0.03cm}\omega^{t}_{{\bf k}}\hspace{0.02cm}t\hspace{0.03cm} +\hspace{0.03cm} i\hspace{0.03cm}{\bf k}\hspace{0.02cm}\cdot\hspace{0.02cm} {\bf x}}
+
\epsilon^{\ast\, t}_{\mu}({\bf k},\zeta)\, a^{\ast\ \!\!a}_{{\bf k}}(\zeta)\ \!e^{\hspace{0.02cm}i\hspace{0.03cm}\omega^{t}_{{\bf k}}\hspace{0.02cm}t\hspace{0.03cm} -\hspace{0.03cm} i\hspace{0.03cm}{\bf k}\hspace{0.02cm}\cdot\hspace{0.02cm} {\bf x}}\right\},
\end{split}
\label{eq:2q}
\end{equation}
where $\epsilon^{\ \! l}_{\mu} ({\bf k})$ is the polarization vector of a longitudinal mode  (${\bf k}$ is the wave vector) and $\epsilon^{\ \! t}_{\mu} ({\bf k},\zeta)$ is the polarization vector of the transverse one. The symbol $\zeta = 1, 2$ stands for two possible polarization states and the asterisk $\ast$ denotes the complex conjugation. The factors $Z_{l}({\bf k})$ and  $Z_{t}({\bf k})$ are the residues of the effective gluon propagator at the longitudinal and transverse mode poles, correspondingly. Finally, $\omega^{\ \! l}_{{\bf k}}$ and $\omega^{\ \! t}_{{\bf k}}$ are the dispersion relations of the corresponding modes. We consider the amplitudes for longitudinal $a^{\phantom{\ast}\!\!a}_{{\bf k}}$ and transverse $a^{\phantom{\ast}\!\!a}_{{\bf k}}(\zeta)$ excitations as ordinary (complex) random functions. By virtue of the representation (\ref{eq:2q}) the correlation function of soft-boson excitations $\langle A_{\mu}^{\ast\,a}(k) A_{\nu}^{b}(k^{\prime}) \rangle$ has the following structure \cite{markov_2001, markov_2006}:
\[
\bigl\langle A_{\mu}^{\ast\,a}(k) A_{\nu}^{b}(k^{\prime}) \bigr\rangle
=
\vspace{-0.4cm}
\]
\begin{align}
= 
-\hspace{0.03cm}\delta^{ab}\delta^{(4)}(k - k^{\prime})\frac{1}{(2\pi)^{3}}\,\biggl\{
&\widetilde{Q}_{\mu \nu}(k)\biggl[\hspace{0.02cm}\left(\frac{Z_{l}({\bf k})}{2\hspace{0.03cm}\omega_{{\bf k}}^{l}}\right)\! {\mathcal N}^{\hspace{0.03cm}l}_{\bf k}\hspace{0.04cm} \delta(k^{0} - {\omega}_{\mathbf k}^{l}) 
+
\left(\frac{Z_{l}(-{\bf k})}{2\hspace{0.03cm}\omega_{-{\bf k}}^{l}}\right)\! \bigl(1 + {\mathcal N}^{\hspace{0.03cm}l}_{-{\bf k}}\bigr)\hspace{0.03cm}  \delta(k^{0} + {\omega}_{\mathbf k}^{l})\hspace{0.02cm}\biggr]
\notag\\[1.3ex]
+\,
&P_{\mu \nu}(k)\hspace{0.03cm}\biggl[\left(\frac{Z_{t}({\bf k})}{2\hspace{0.03cm}\omega_{{\bf k}}^{t}}\right)\! {\mathcal N}^{\hspace{0.03cm}t}_{\bf k}\hspace{0.04cm} \delta(k^{0} - {\omega}_{\mathbf k}^{t}) 
+
\left(\frac{Z_{t}(-{\bf k})}{2\hspace{0.03cm}\omega_{-{\bf k}}^{t}}\right)\! \bigl(1 + {\mathcal N}^{\hspace{0.03cm}t}_{-{\bf k}}\bigr)\hspace{0.03cm}  \delta(k^{0} + {\omega}_{\mathbf k}^{t})\hspace{0.02cm}\biggr]\!
\biggr\}.
\notag
\end{align}
Here, an explicit form of the longitudinal $\tilde{Q}_{\mu \nu}(k)$ and transverse $P_{\mu \nu}(k)$ projectors (in the $A_0$\hspace{0.02cm}-\hspace{0.02cm}gauge) is written out in Appendix \ref{appendix_A}. In the expression above we have taken into account that the expectation values of the products of two bosonic amplitudes are
\begin{equation}
\bigl\langle\hspace{0.03cm}a^{\ast\hspace{0.03cm}a}_{{\bf k}}\hspace{0.03cm} a^{\phantom{\ast}\!\!b}_{{\bf k}^{\prime}}\bigr\rangle
=
\delta^{ab}(2\pi)^{3}\hspace{0.03cm}\delta({\bf k} - {\bf k}^{\prime})\hspace{0.05cm}{\mathcal N}^{\hspace{0.03cm}l}_{\bf k},
\qquad
\bigl\langle\hspace{0.03cm}a^{\ast\hspace{0.03cm}a}_{{\bf k}}(\zeta)\, a^{\phantom{\ast}\!\!b}_{{\bf k}^{\prime}}(\zeta^{\hspace{0.03cm}\prime})\bigr\rangle
=
\delta^{ab}\delta_{\zeta\hspace{0.01cm} \zeta^{\prime}}(2\pi)^{3}\hspace{0.03cm}\delta({\bf k} - {\bf k}^{\prime})\hspace{0.05cm}{\mathcal N}^{\hspace{0.03cm}t}_{\bf k},
\label{eq:2w}
\end{equation}
where ${\mathcal N}^{\hspace{0.03cm}l}_{\bf k}$ and ${\mathcal N}^{\hspace{0.03cm}t}_{\bf k}$ are the number densities of  the longitudinal and transverse plasma waves.\\
\indent For simplicity, we confine our analysis only to processes involving longitudinally polarized plasma excitations, which are known as {\it plasmons}. These excitations are a purely collective effect of the medium, which has no analogs in the conventional quantum field theory. The dispersion relation $\omega^{\,l}_{{\bf k}}$ for plasmons satisfies the following dispersion equation \cite{kalashnikov_1980}:
\begin{equation}
	{\rm Re} \,^{\ast}\!\Delta^{\!-1\,l} (\omega, {\bf k}) = 0 ,
	\label{eq:2e}
\end{equation}
where
\[
\,^{\ast}\!\Delta^{\!-1\,l}(\omega, {\bf k}) = k^{2}\bigg(
1 + \frac{3\hspace{0.03cm}\omega^{\hspace{0.01cm}2}_{p\hspace{0.02cm}l}}{{\bf k}^2}
\bigg[1 - F \bigg( \frac{\omega}{\vert {\bf k} \vert} \bigg) \bigg]\bigg),
\;
F (x) \equiv \frac{x}{2} \bigg[ \ln \bigg \vert \frac{1 + x}{1 - x}
\bigg \vert - i \pi
\theta ( 1 - \vert x \vert ) \bigg]
\]
is the inverse HTL-resummed longitudinal propagator,
$\omega^{\hspace{0.01cm}2}_{p\hspace{0.02cm}l} = g^{\hspace{0.02cm}2}\hspace{0.02cm}(2\hspace{0.02cm}N_{c} + N_{f})\hspace{0.03cm}T^{\hspace{0.02cm} 2}/18$ is the plasma frequency squared of the gluon sector of plasma excitations, $T$ is the temperature of the system, $g$ is the strong interaction constant, and $N_{f}$ represents the number of flavors of massless quarks.
The dispersion equation (\ref{eq:2e}) is obtained from the pole of the ``scalar'' longitudinal propagator $\,^{\ast}\!\Delta^{l}(k)$ in (\ref{ap:A9}), which in turn includes all effects of the medium-to-leading order in $g$. The equation (\ref{eq:2e}) is only valid when the external frequency $\omega$ and the momentum ${\bf k}$ are of order $g\hspace{0.02cm}T$, that is, in other words, at the {\it soft} scale \cite{pisarski_1989(1)}. Thus, in conditions when the strong coupling constant $g$ is small we will have $\omega,\,|{\bf k}|\ll T$.\\
\indent As mentioned in the Introduction in an equilibrium hot quark-gluon plasma including massless quarks and antiquarks there exist two types of physical fermion soft fields: the first one is normal excitations with the relation between chirality and helicity at zero temperature and the second type is purely collective excitations with abnormal relation between chirality and helicity. In the subsequent discussion we will designate these two modes of Fermi excitations by the symbols $(+)$ and $(-)$. The decomposition of the collective quark-antiquark field into plane waves is  \cite{hakim_book_2011}
\begin{equation}
\begin{split}
&\psi_{\alpha}^{\hspace{0.02cm}i}(x) \!=\! \sum\limits_{s\hspace{0.02cm} =\hspace{0.02cm} \pm}\!
\sum_{\;\lambda\hspace{0.02cm} =\hspace{0.02cm} \pm\hspace{0.02cm} 1}\hspace{0.01cm} 
\int\!\!\frac{d\hspace{0.02cm}{\bf q}}{(2\pi)^{3}}\!\left(\!\frac{Z_{s}({\bf q})}
{2}\right)^{\!\!1/2}\!\!\!\left[\hspace{0.03cm}b_{\bf q}^{\,i\hspace{0.02cm}(s)}(\lambda)
\hspace{0.03cm}u^{(s)}_{\alpha}(\hat{\bf q}, \lambda)\,
e^{-i\hspace{0.03cm}\omega^{s}_{{\bf q}}\hspace{0.02cm}t\hspace{0.03cm} +
\hspace{0.03cm} i\hspace{0.03cm}{\bf q}\hspace{0.02cm}\cdot\hspace{0.02cm} {\bf x}}
\!+
d_{\bf q}^{\,\ast\hspace{0.03cm}i\hspace{0.02cm}(s)}(\lambda)\hspace{0.03cm}v^{(s)}_{\alpha}(\hat{\bf q}, \lambda)\,
e^{\hspace{0.03cm}i\hspace{0.03cm}\omega^{s}_{{\bf q}}\hspace{0.02cm}t\hspace{0.03cm} -\hspace{0.03cm} i\hspace{0.03cm}{\bf q}\hspace{0.02cm}\cdot\hspace{0.02cm} {\bf x}}\hspace{0.03cm}\right]\!,\\[1.5ex]
&\bar{\psi}_{\alpha}^{\hspace{0.02cm}i}(x) \!=\! \sum\limits_{s\hspace{0.02cm} =\hspace{0.02cm} \pm}\!
\sum_{\;\lambda\hspace{0.02cm} =\hspace{0.02cm} \pm\hspace{0.02cm} 1}\hspace{0.01cm} 
\int\!\!\frac{d\hspace{0.02cm}{\bf q}}{(2\pi)^{3}}\!\left(\!\frac{Z_{s}({\bf q})}
{2}\right)^{\!\!1/2}\!\!\!\left[\hspace{0.03cm}b_{\bf q}^{\,\ast\hspace{0.03cm}i\hspace{0.02cm}(s)}(\lambda)\hspace{0.03cm}
\bar{u}^{(s)}_{\alpha}(\hat{\bf q}, \lambda)\,
e^{\hspace{0.03cm}i\hspace{0.03cm}\omega^{s}_{{\bf q}}\hspace{0.02cm}t\hspace{0.03cm} -\hspace{0.03cm} i\hspace{0.03cm}{\bf q}\hspace{0.02cm}\cdot\hspace{0.02cm} {\bf x}}
\!+
d_{\bf q}^{\,i\hspace{0.02cm}(s)}(\lambda)\hspace{0.03cm}\bar{v}^{(s)}_{\alpha}(\hat{\bf q}, \lambda)\,
e^{-i\hspace{0.03cm}\omega^{s}_{{\bf q}}\hspace{0.02cm}t\hspace{0.03cm} +\hspace{0.03cm} i\hspace{0.03cm}{\bf q}\hspace{0.02cm}\cdot\hspace{0.02cm} {\bf x}}\hspace{0.03cm}\right]\!,
\label{eq:2r}
\end{split}
\end{equation}
where $\alpha = 1, 2, 3, 4$ and the spinors $u^{(s)}_{\alpha}(\hat{\bf q}, \lambda)$ and $v^{(s)}_{\alpha}(\hat{\bf q}, \lambda)$ denote solutions of the free massless Dirac equation
\begin{subequations} 
\label{eq:2t}
\begin{align}
&\bigl(\gamma^{0} \mp \hat{\bf q}\cdot\boldsymbol{\gamma}\bigr)u^{(\pm)}(\hspace{0.03cm}\hat{\bf q}, \lambda) = 0,
\label{eq:2ta}\\[1.5ex]
&\bigl(\gamma^{0} \pm \hat{\bf q}\cdot\boldsymbol{\gamma}\bigr)v^{(\pm)}(-\hat{\bf q}, \lambda) = 0
\label{eq:2tb}
\end{align}
\end{subequations} 
with the helicity $\lambda = \pm\hspace{0.02cm} 1$ and $\hat{\bf q} \equiv {\bf q}/|{\bf q}|$. An explicit form of these spinors, as well as some of their properties, are given in the Appendix \ref{appendix_B}. The indices $i,\, j,\, k,\, \ldots$ designate a color degree of freedom of Fermi excitations and run values from 1 to $N_{c}$. The factors $Z_{+}({\bf q})$ and $Z_{-}({\bf q})$ are the residues of the effective quark propagator at the normal and abnormal mode poles, correspondingly. Finally, $\omega^{+}_{{\bf q}}$ and $\omega^{-}_{{\bf q}}$ are the dispersion relations of the corresponding modes. We will consider the amplitudes for the normal $\bigl(\hspace{0.02cm}b_{\bf q}^{\,i\hspace{0.03cm}(+)}(\lambda),\, d_{\bf q}^{\,i\hspace{0.03cm}(+)}(\lambda)\bigr)$ and abnormal 
$\bigl(\hspace{0.02cm}b_{\bf q}^{\,i\hspace{0.03cm}(-)}(\lambda),\, d_{\bf q}^{\,i\hspace{0.03cm}(-)}(\lambda)\bigr)$ fermion excitations as variables taking its values in a Grassmann algebra. By virtue of the represen\-tation (\ref{eq:2r}) the correlation function of soft fermion excitations $\langle \bar{\psi}_{\alpha}^{\hspace{0.02cm}i}(-q) \psi_{\beta}^{\hspace{0.02cm}j}(q^{\prime}) \rangle$ has the following structure \cite{markov_2001, markov_2006}:
\[
\bigl\langle \bar{\psi}_{\alpha}^{\hspace{0.02cm}i}(-q) \psi_{\beta}^{\hspace{0.02cm}j}(q^{\prime}) \bigr\rangle
=
\vspace{-0.4cm}
\]
\begin{align}
= 
\delta^{ji}\delta^{(4)\!}\hspace{0.02cm}(q - q^{\prime})\hspace{0.03cm}\frac{1}{2\hspace{0.02cm}(2\pi)^{3}}\,\Bigl\{
&(h_{+}({\hat{\bf q}}))_{\beta\alpha}
\Bigl[\hspace{0.03cm}Z_{+}({\mathbf q})\hspace{0.02cm} n_{\mathbf q}^{+}\hspace{0.04cm} 
\delta(q^{0} - {\omega}_{\mathbf q}^{+}) 
+
Z_{-}(-{\bf q})\hspace{0.02cm} \bigl(1 - \bar{n}_{-{\mathbf q}}^{-}\bigr)\hspace{0.03cm}  
\delta(q^{0} + 
{\omega}_{\mathbf q}^{-})\hspace{0.02cm}\Bigr]
\notag\\[1.3ex]
+\,
&(h_{-}({\hat{\bf q}}))_{\beta\alpha}\hspace{0.03cm}
\Bigl[\hspace{0.03cm}Z_{-}({\mathbf q})\hspace{0.02cm} n_{\mathbf q}^{-}\hspace{0.04cm} 
\delta(q^{0} - {\omega}_{\mathbf q}^{-}) 
+
Z_{+}(-{\bf q})\hspace{0.02cm} \bigl(1 - \bar{n}_{-{\mathbf q}}^{+}\bigr)\hspace{0.03cm}  
\delta(q^{0} + {\omega}_{\mathbf q}^{+})\hspace{0.02cm}\Bigr]\!\Bigr\}.
\notag
\end{align}
Here, the matrix functions $h_{\pm}(\hat{\bf q}) = (\gamma^0 \mp \hat{{\bf q}}\cdot\boldsymbol{\gamma})/2$ are the spinor projectors\footnote{\,More precisely, the projectors are the spin matrices \cite{blaizot_2001}
\[
\Lambda_{\pm}(\hat{{\bf q}}) \equiv \gamma^{0}h_{\mp}(\hat{\bf q}) 
=
\frac{1 \pm \gamma^{0}\hat{{\bf q}}\cdot\boldsymbol{\gamma}}{2}, 
\quad 
\Lambda_{+} + \Lambda_{-} = 1, \quad \Lambda^{2}_{\pm} = \Lambda_{\pm},
\quad
\Lambda_{+}\Lambda_{-}  = \Lambda_{-}\Lambda_{+} = 0,
\]
which project onto spinors whose chirality is equal ($\Lambda_{+}$), or opposite ($\Lambda_{-}$), to their helicity.
} 
onto eigenstates of helicity and we have taken into account the expectation values of the products of two fermionic amplitudes to be
\begin{equation}
\begin{split}
&\bigl\langle\hspace{0.02cm} b^{\hspace{0.03cm}\ast\,i\hspace{0.03cm}(+)}_{{\bf q}}(\lambda)
\hspace{0.04cm} b^{\,j\hspace{0.03cm}(+)}_{{\bf q}^{\prime}}(\lambda^{\prime})\bigr\rangle
=
\delta^{\hspace{0.02cm}ij}\delta_{\lambda\lambda^{\prime}}\hspace{0.02cm}(2\pi)^{3}\hspace{0.03cm}
\delta({\bf q} - {\bf q}^{\prime})\hspace{0.03cm}{\mathfrak n}^{+}_{\bf q},
\\[1.7ex]
&\bigl\langle\hspace{0.02cm} b^{\hspace{0.02cm}\ast\,i\hspace{0.03cm}(-)}_{{\bf q}}(\lambda)
\hspace{0.04cm} b^{\,j\hspace{0.03cm}(-)}_{{\bf q}^{\prime}}(\lambda^{\prime})\bigr\rangle
=
\delta^{\hspace{0.02cm}ij}\delta_{\lambda\lambda^{\prime}}\hspace{0.02cm}(2\pi)^{3}\hspace{0.03cm}
\delta({\bf q} - {\bf q}^{\prime})\hspace{0.03cm}{\mathfrak n}^{-}_{\bf q},
\\[1.7ex]
&\bigl\langle\hspace{0.02cm} d^{\;i\hspace{0.03cm}(+)}_{-{\bf q}}(\lambda) 
\hspace{0.04cm} d^{\hspace{0.04cm}\ast\,j\hspace{0.03cm}(+)}_{-{\bf q}^{\prime}}(\lambda^{\prime})\bigr\rangle
=
\delta^{\hspace{0.02cm}ij}\delta_{\lambda\lambda^{\prime}}\hspace{0.02cm}(2\pi)^{3}\hspace{0.03cm}
\delta({\bf q} - {\bf q}^{\prime})\hspace{0.03cm}\bigl(1 - \bar{{\mathfrak n}}^{+}_{-{\bf q}}\bigr),
\\[1.7ex]
&\bigl\langle\hspace{0.02cm} d^{\;i\hspace{0.03cm}(-)}_{-{\bf q}}(\lambda)
\hspace{0.04cm} d^{\hspace{0.04cm}\ast\,j\hspace{0.03cm}(-)}_{-{\bf q}^{\prime}}(\lambda^{\prime})\bigr\rangle
=
\delta^{\hspace{0.02cm}ij}\delta_{\lambda\lambda^{\prime}}\hspace{0.02cm}(2\pi)^{3}\hspace{0.03cm}
\delta({\bf q} - {\bf q}^{\prime})\hspace{0.03cm}\bigl(1 - \bar{{\mathfrak n}}^{-}_{-{\bf q}}\bigr),
\end{split}
\label{eq:2y}
\end{equation}
where ${\mathfrak n}^{\pm}_{\bf q}$ and $\bar{{\mathfrak n}}^{\pm}_{\bf q}$ are the number densities of  the normal $(+)$ and abnormal $(-)$, fermion and antifermion plasma waves, correspondingly.\\
\indent Further, for simplicity we will restrict our consideration only to the processes that include collective fermionic plasma excitations called {\it plasmino}. These excitations are purely collective effect of medium having no analogue in ordinary quantum field theory. In this case $q_{0} = \omega_{\bf q}^{-}$  is a solution of the following dispersion equation \cite{klimov_1981, klimov_1982, weldon_1982}:
\begin{equation}
{\rm Re}\,^{\ast}\!\Delta^{\!-1}_{-}(q^0,{\bf q}) = 0,
\label{eq:2u}
\end{equation}
where
\[
\,^{\ast}\!\Delta^{\!-1}_{-}(q^0,{\bf q}) = q^0 + |{\bf q}| + \frac{\omega_0^2}{|{\bf q}|}
\biggl[1 - \biggl(1 + \frac{|{\bf q}|}{q^0}\biggr)F\biggl(\frac{q^0}{\vert {\bf q}\vert}\bigg)\bigg],
\]
and $\omega_{0}^{2} = g^{2}C_FT^2/8$ is the plasma frequency squared of the quark-antiquark sector of plasma excitations. The dispersion equation (\ref{eq:2u}) as well as (\ref{eq:2e}) is obtained from the pole of the corresponding ``scalar'' part $\,^{\ast}\!\triangle_{-}(q)$ of the effective quark propagator (\ref{ap:B11}), which in turn includes all effects of the medium to leading order in $g$. The equation (\ref{eq:2u}) is only valid when the external frequency $q_{0}$ and the momentum ${\bf q}$ are of order $g\hspace{0.02cm}T$. When the strong coupling constant $g$ is small we will have $q_{0},\,|{\bf q}|\ll T$.\\
\indent As it was said already above, the amplitudes $a^{\phantom{\ast}\!\!a}_{{\bf k}}$ and $a^{\ast\ \!\!a}_{{\bf k}}$ in the expansion for the longitudinal mode of oscillations (\ref{eq:2q}) are usual (commuting) normal variables of the gauge field satisfying the canonical (super)Poisson bracket relations
\begin{equation}
\bigl\{a^{\phantom{\ast}\!\!a}_{{\bf k}},\,a^{\phantom{\ast}\!\!b}_{{\bf k}^{\prime}}\bigr\}_{\rm SPB} = 0,
\quad\!
\bigl\{a^{\ast\ \!\!a}_{{\bf k}},\,a^{\ast\ \!\!b}_{{\bf k}^{\prime}}\bigr\}_{\rm SPB} = 0, 
\quad\!
\bigl\{a^{\phantom{\ast}\!\!a}_{{\bf k}},\,a^{\ast\ \!\!b}_{{\bf k}^{\prime}}\bigr\}_{\rm SPB}
=
\delta^{\hspace{0.02cm} ab}\ \!\! (2\pi)^{3}\hspace{0.03cm}\delta({\bf k} - {\bf k}^{\prime}).
\label{eq:2i}
\end{equation}
From the other hand, we consider the amplitudes $b^{\;i}_{{\bf q}}(\lambda)\, \bigl(\,\equiv b^{\;i\hspace{0.03cm}(-)}_{{\bf q}}(\lambda)\bigr)$ and $b^{\hspace{0.03cm}\ast\,i}_{{\bf q}}(\lambda) \,\bigl(\,\equiv b^{\hspace{0.03cm}\ast\,i\hspace{0.03cm}(-)}_{{\bf q}}(\lambda)\bigr)$ in the expansion (\ref{eq:2r}) for plasmino oscillation mode as  Grassman-valued (anticommuting) variables, the Poisson superbrackets $({\rm SPB})$ of which has the following standard form:
\begin{equation}
\Bigl\{b^{\,i}_{{\bf q}^{\phantom\prime}}(\lambda),\hspace{0.03cm} b^{\,j}_{{\bf q}^{\prime}}(\lambda^{\prime})\!\hspace{0.03cm} \Bigr\}_{\rm SPB}\! = 0,
\;
\Bigl\{b^{\,\ast\, i}_{{\bf q}^{\phantom\prime}}(\lambda),\hspace{0.03cm} b^{\hspace{0.03cm} \ast\hspace{0.03cm} j}_{{\bf q}^{\prime}}(\lambda^{\prime})\!\hspace{0.03cm} \Bigr\}_{\rm SPB}\! = 0, 
\;
\Bigl\{b^{\,i}_{{\bf q}^{\phantom\prime}}(\lambda),\hspace{0.03cm} b^{\hspace{0.03cm} \ast\hspace{0.03cm} j}_{{\bf q}^{\prime}}(\lambda^{\prime})\!\hspace{0.03cm} \Bigr\}_{\rm SPB}\!
=
\delta^{\,ij}\hspace{0.01cm}\delta_{\lambda\lambda^{\prime}}\hspace{0.02cm}(2\pi)^{3}\hspace{0.02cm}
\delta({\bf q}\, -\, {\bf q}^{\prime}).
\label{eq:2o}
\end{equation}
For the case of a continuous media we take the following expression as the definition of the Poisson superbracket
\begin{equation}
\bigl\{F,\,G\bigr\}_{\rm SPB} 
\label{eq:2p}
\end{equation}
\[
=\!
\int\! d\hspace{0.02cm}{\bf k\hspace{0.01cm}}'\!\hspace{0.02cm}
\left\{\frac{\delta\hspace{0.01cm} F}{\delta\hspace{0.02cm} a^{\phantom{\ast}\!\!c}_{{\bf k}'}}
\,\frac{\delta\hspace{0.01cm}  G}{\delta\hspace{0.02cm} a^{\ast\ \!\!c}_{{\bf k}'}}
\,-\,
\frac{\delta\hspace{0.01cm}  F}{\delta\hspace{0.02cm} a^{\ast\ \!\!c}_{{\bf k}'}}\,
\frac{\delta\hspace{0.01cm}  G}{\delta\hspace{0.02cm} a^{\phantom{\ast}\!\!c}_{{\bf k}'}}\right\}
+
\int\! d{\bf q\hspace{0.01cm}}'\!\hspace{0.02cm}
\left\{\frac{\!\!\overleftarrow{\delta}\! F}{\,\delta\hspace{0.02cm} b^{\phantom{\ast}\!\!i}_{{\bf q}^{\prime}}}\,
\frac{\!\!\overrightarrow{\delta}\! G}{\,\delta\hspace{0.02cm} b^{\hspace{0.03cm} \ast\ \!\!i}_{{\bf q}^{\prime}}}
\,+\,(-1)^{P_{F} + P_{G}}\,
\frac{\!\!\overrightarrow{\delta}\! F}{\,\delta\hspace{0.02cm} b^{\hspace{0.03cm} \ast\ \!\!i}_{{\bf q}^{\prime}}}\,
\frac{\!\!\overleftarrow{\delta}\! G}{\,\delta\hspace{0.02cm} b^{\phantom{\ast}\!\!i}_{{\bf q}^{\prime}}}\right\}.
\]
Here, $\overrightarrow{\delta}\!/\delta\hspace{0.02cm} b^{\,*\hspace{0.03cm}i}_{{\bf q}}$ and $\overleftarrow{\delta}\!/\delta\hspace{0.02cm} b^{\,i}_{{\bf q}}$ are the right and left functional derivatives\footnote{\,In our notations of the right and left variational derivatives we follow the notations accepted for the right and left derivatives adopted in \cite{casalbuoni_1976_1, casalbuoni_1976_2, berezin_1987}, therefore, 
\[
\delta F = 
\int\! d\hspace{0.02cm}{\bf k\hspace{0.01cm}}'\!\hspace{0.02cm}
\left\{\frac{\delta F}{\delta\hspace{0.02cm} a^{\phantom{\ast}\!\!c}_{{\bf k}'}}\, 
\delta\hspace{0.02cm} a^{\phantom{\ast}\!\!c}_{{\bf k}'}
\,+\,
\frac{\delta  F}{\delta\hspace{0.02cm} a^{\ast\ \!\!c}_{{\bf k}'}}\,
\delta\hspace{0.02cm} a^{\ast\ \!\!c}_{{\bf k}'}\right\}
\,+\,
\int\! d{\bf q\hspace{0.01cm}}'\!\hspace{0.02cm}
\left\{\frac{\!\!\overleftarrow{\delta}\! F}{\,\delta\hspace{0.02cm} b^{\phantom{\ast}\!\!i}_{{\bf q}^{\prime}}}\,
\delta\hspace{0.02cm} b^{\phantom{\ast}\!\!i}_{{\bf q}^{\prime}}
\,+\,
\delta\hspace{0.02cm} b^{\,\ast\ \!\!i}_{{\bf q}^{\prime}}\,
\frac{\!\!\overrightarrow{\delta}\! F}{\,\delta\hspace{0.02cm} b^{\,\ast\ \!\!i}_{{\bf q}^{\prime}}}\right\}.
\]
Moreover, we adopt the following rule for the variations of amplitudes $b^{\phantom{\ast}\!\!i}_{{\bf q}}(\lambda)$ and $c^{\phantom{\ast}\!\!a}_{{\bf k}}$:
\[
\frac{\!\!\overleftarrow{\delta}\! b^{\phantom{\ast}\!\!i}_{{\bf q}}(\lambda)}
{\,\delta\hspace{0.02cm} b^{\phantom{\ast}\!\!i^{\prime}}_{{\bf q}^{\prime}}(\lambda^{\prime})}
=
\delta^{\hspace{0.02cm}i\hspace{0.02cm}i^{\prime}\!}\delta_{\lambda\lambda^{\prime}}\hspace{0.03cm}(2\pi)^{3}\hspace{0.03cm}\delta({\bf q} - {\bf q}^{\prime}),
\qquad
\frac{\delta\hspace{0.01cm} c^{\phantom{\ast}\!\!a}_{{\bf k}}}{\delta\hspace{0.02cm} c^{\phantom{\ast}\!\!a^{\prime}}_{{\bf k}^{\prime}}}
=
\delta^{\hspace{0.02cm}a\hspace{0.02cm}a^{\prime}}(2\pi)^{3}\hspace{0.03cm}\delta({\bf k} - {\bf k}^{\prime})
\]
and so on, i.e. on the right-hand side we introduce the factor $(2\pi)^{3}$.}, $P_{F}$ and $P_{G}$ designate Grassmann evenness of the functions $F$ and $G$, correspondingly. For the sake of brevity, hereafter we do not write out the explicit dependence of the $b$-functions on helicity, always keeping in mind that the summation over color indices in the fundamental representation implies also the summation over helicity states. Besides, for the sake of simplicity of notation the abbreviation ${\rm SPB}$ will be omitted, thereby suggesting that by the braces $\{\cdot\,,\cdot\}$ we always mean the Poisson superbracket.\\
\indent Let us write the Hamilton equations for the functions  $a^{\phantom{\ast}\!\!a}_{{\bf k}}$, $b^{\,i}_{{\bf q}}$ and their complex conjugation 
\begin{equation}
\frac{\partial\hspace{0.02cm}a^{\phantom{\ast}\!\!a}_{{\bf k}}}{\partial\hspace{0.02cm} t}
=
-i\hspace{0.02cm}\bigl\{a^{\phantom{\ast}\!\!a}_{{\bf k}}, H\bigr\} \equiv  -i\,\frac{\delta H}{\delta\hspace{0.01cm} a^{\ast\ \!\!a}_{{\bf k}}},
\qquad
\frac{\partial\hspace{0.02cm}a^{\ast\ \!\!a}_{{\bf k}}}{\partial\hspace{0.02cm} t}
=
-i\bigl\{a^{\ast\ \!\!a}_{{\bf k}}, H\bigr\} \equiv  i\,\frac{\delta H}{\delta\hspace{0.01cm} a^{\phantom{\ast}\!\!a}_{{\bf k}}},
\label{eq:2a}
\end{equation}
\begin{equation}
\frac{\partial\hspace{0.02cm}b^{\,i}_{{\bf q}}}{\partial\hspace{0.02cm} t}
=
-i\left\{b^{\,i}_{\bf q}, H\right\} \equiv  -i\,\frac{\!\!\!\!\overrightarrow{\delta}\! H}{\,\delta\hspace{0.015cm} b^{\,*\hspace{0.03cm}i}_{{\bf q}}},
\qquad
\frac{\partial\hspace{0.02cm}b^{\,\ast\,i}_{{\bf q}}}{\!\!\partial\hspace{0.02cm} t}
=
-i\left\{b^{\,\ast\,i}_{\bf q}, H\right\} \equiv  i\,\frac{\!\!\overleftarrow{\delta}\! H}
{\,\delta\hspace{0.015cm} b^{\,i}_{{\bf q}}}.
\label{eq:2s}
\end{equation}
Here, the function $H$ represents a Hamiltonian for the system of plasmons and plasminos, which is equal to a sum $H =  H^{(0)} + H_{int}$, where
\begin{equation}
H^{(0)} =  \!\int\!\frac{d\hspace{0.02cm}{\bf k}}{(2\pi)^{3}}\ \omega^{l}_{{\bf k}}\ \!
a^{\ast\hspace{0.03cm}a}_{{\bf k}} a^{ a}_{{\bf k}}
+
\int\!\frac{d{\bf q}}{(2\pi)^{3}}\ \omega_{\bf q}^{-}\ \!
b^{\,\ast\hspace{0.04cm}i}_{{\bf q}}\ \!b^{\,i}_{{\bf q}}\
\label{eq:2d}
\end{equation}
is the Hamiltonian of noninteracting plasmons and plasminos, and ${H}_{int}$ is the interaction Hamiltonian.\\
\indent In the approximation of small amplitudes, the interaction Hamiltonian can be presented in the form of a formal   integro-power series in the bosonic functions ${a}^{a}_{{\bf k}}$ and ${a}^{\ast\hspace{0.03cm} a}_{{\bf k}}$, and in the fermionic functions $b^{\, i}_{{\bf q}}$ and $b^{\,\ast\, i}_{{\bf q}}$:
\[
H_{int} = H^{(3)} + H^{(4)} +\, \ldots\,\,,
\]
where the third-order interaction Hamiltonian has the following structure:
\begin{align}
H^{(3)} 
=  \int\frac{d\hspace{0.02cm}{\bf k}\, d\hspace{0.02cm}{\bf k}_{1}\hspace{0.03cm} d\hspace{0.02cm}{\bf k}_{2}}{(2\pi)^{9}}
&\Bigl\{\hspace{0.02cm}{\mathcal V}^{\; a\, a_{1}\, a_{2}}_{{\bf k},\, {\bf k}_{1},\, {\bf k}_{2}}\, a^{\ast\hspace{0.03cm}  a}_{{\bf k}}\,
a^{\,a_{1}}_{{\bf k}_{1}}\, a^{\,a_{2}}_{{\bf k}_{2}}
\,+\,
{\mathcal V}^{\,*\,a\, a_{1}\, a_{2}}_{{\bf k},\, {\bf k}_{1},\, {\bf k}_{2}}\, 
a^{\!\phantom{\ast}a}_{{\bf k}}\, a^{\ast\,a_{1}}_{{\bf k}_{1}}\hspace{0.03cm}  a^{\ast\, a_{2}}_{{\bf k}_{2}}
\Bigr\} \notag\\[0.7ex]
&\times
(2\pi)^{3}\hspace{0.03cm}\delta({\bf k} - {\bf k}_{1} - {\bf k}_{2}) \notag\\[0.7ex]
+\, \frac{1}{3}\int\frac{d\hspace{0.02cm}{\bf k}\, d\hspace{0.02cm}{\bf k}_{1}\hspace{0.03cm}  d\hspace{0.02cm}{\bf k}_{2}}{(2\pi)^{9}}
&\Bigl\{\hspace{0.02cm}{\mathcal U}^{\; a\, a_{1}\, a_{2}}_{\,{\bf k},\, {\bf k}_{1},\, {\bf k}_{2}}\, a^{a}_{{\bf k}}\, a^{a_{1}}_{{\bf k}_{1}}\,
a^{a_{2}}_{{\bf k}_{2}}
\,+\,
{\mathcal U}^{\,*\,a\, a_{1}\, a_{2}}_{\; {\bf k},\,{\bf k}_{1},\, {\bf k}_{2}}\, a^{\ast\ \!a}_{{\bf k}}\,
a^{\ast\,a_{1}}_{{\bf k}_{1}}\, a^{\ast\,a_{2}}_{{\bf k}_{2}}
\Bigr\} \notag\\[0.7ex]
&\times
(2\pi)^{3}\hspace{0.03cm}\delta({\bf k} + {\bf k}_{1} + {\bf k}_{2}) \notag\\[0.7ex]
+\int\frac{d\hspace{0.02cm}{\bf k}_{1}\hspace{0.03cm}  d{\bf q}\, d{\bf q}_{1}}{(2\pi)^{9}}
&\Bigl\{\hspace{0.02cm}{\mathcal G}^{\; a_{1}\, i\,  i_{1}}_{\,{\bf k}_1,\, {\bf q},\, {\bf q}_{1}}\, {a}^{\, a_1}_{{\bf k}_1}\,
b^{\,\ast\, i}_{{\bf q}}\, b^{\,{\ast}\,i_{1}}_{{\bf q}_{1}}
\,-\,
{\mathcal G}^{\,\ast\, a_{1}\, i\;  i_{1}}_{\,{\bf k}_1,\, {\bf q},\, {\bf q}_{1}}\, 
{a}^{\,\ast\, a_{1}}_{{\bf k}_{1}}\hspace{0.03cm}  b^{\; i}_{{\bf q}}\, b^{\; i_{1}}_{{\bf q}_{1}}
\Bigr\}
\label{eq:2f}\\[0.7ex]
&\times
(2\pi)^{3}\hspace{0.03cm}\delta({\bf k}_{1} - {\bf q} - {\bf q}_{1})
\notag \\[0.7ex]
+ \int\frac{d\hspace{0.02cm}{\bf k}_{1}\hspace{0.03cm}  d{\bf q}\, d{\bf q}_{1}}{(2\pi)^{9}}
&\Bigl\{\hspace{0.02cm}{\mathcal P}^{\; a_{1}\, i\;  i_{1}}_{{\bf k}_{1},\, {\bf q},\, {\bf q}_{1}}\; {a}^{\,a_{1}}_{{\bf k}_{1}}\,
b^{\,\ast\,i}_{{\bf q}}\, b^{\; i_{1}}_{{\bf q}_{1}}\,
(2\pi)^{3}\hspace{0.03cm}\delta({\bf k}_1 - {\bf q} + {\bf q}_{1})
\notag\\[0.7ex]
+\!\!
&\;\;{\mathcal P}^{\,\ast\, a_{1}\, i_{1}\,  i}_{{\bf k}_1,\, {\bf q}_{1},\, {\bf q}}\, {a}^{\,\ast\, a_{1}}_{{\bf k}_1}\,
b^{\,\ast\, i}_{{\bf q}}\, b^{\; i_{1}}_{{\bf q}_{1}}
(2\pi)^{3}\hspace{0.03cm}\delta({\bf k}_1 + {\bf q} - {\bf q}_{1})
\Bigr\}
\notag\\[1ex]
+ \int\frac{d\hspace{0.02cm}{\bf k}_{1}\hspace{0.03cm} d{\bf q}\, d{\bf q}_{1}}{(2\pi)^{9}}
&\Bigl\{\hspace{0.02cm}{\mathcal K}^{\; a_{1}\, i\,  i_{1}}_{{\bf k}_1,\, {\bf q},\, {\bf q}_{1}}\, 
{a}^{\, a_1}_{{\bf k}_1}\, b^{\,i}_{{\bf q}}\ b^{\,\,\,i_{1}}_{{\bf q}_{1}}
\,-\,
{\mathcal K}^{\,\ast\, a_{1}\, i\,  i_{1}}_{{\bf k}_1,\, {\bf q},\, {\bf q}_{1}}\, 
{a}^{{\ast}\, a_1}_{{\bf k}_1}\, b^{\,\ast\, i}_{{\bf q}}\, b^{\,\ast\,i_{1}}_{{\bf q}_{1}}
\Bigr\} \notag\\[0.7ex]
&\times
(2\pi)^{3}\hspace{0.03cm}\delta({\bf k}_1 + {\bf q} + {\bf q}_{1})
\notag
\end{align}
and, correspondingly, the fourth-order interaction Hamiltonian is
\begin{equation}
\begin{split}
H^{(4)}\! 
=
&\int\frac{d{\bf q}\, d{\bf q}_{1}\hspace{0.03cm} d\hspace{0.02cm}{\bf k}_{1}\hspace{0.03cm} d\hspace{0.02cm}{\bf k}_{2}}{(2\pi)^{12}}\ 
T^{\, (2)\, i\, i_{1}\, a_{1}\, a_{2}}_{\ {\bf q},\, {\bf q}_{1},\, {\bf k}_{1},\, {\bf k}_{2}}\, 
b^{\,\ast\, i}_{{\bf q}}\, b^{\;i_{1}}_{{\bf q}_{1}}\hspace{0.03cm} {a}^{\ast\ \!\!a_{1}}_{{\bf k}_{1}}\hspace{0.03cm}  
{a}^{\!\phantom{\ast}a_{2}}_{{\bf k}_{2}}\ (2\pi)^{3}\hspace{0.03cm}\delta({\bf q} + {\bf k}_1 - {\bf q}_{1} - {\bf k}_{2})
\\[1ex]
+\;
\frac{1}{2}&\int\frac{d{\bf q}\, d{\bf q}_{1}\hspace{0.03cm} d{\bf q}_{2}\hspace{0.04cm} d{\bf q}_{3}}{(2\pi)^{12}}\, 
T^{\,(2)\, i\, i_{1}\, i_{2}\, i_{3}}_{{\bf q},\, {\bf q}_{1},\, {\bf q}_{2},\, {\bf q}_{3}}\, 
b^{\,\ast\, i}_{{\bf q}}\hspace{0.04cm} b^{\,\ast\, i_{1}}_{{\bf q}_{1}}\hspace{0.03cm} b^{\;i_{2}}_{{\bf q}_{2}}\, b^{\;i_{3}}_{{\bf q}_{3}}\,
(2\pi)^{3}\hspace{0.03cm}\delta({\bf q} + {\bf q}_{1} - {\bf q}_{2} - {\bf q}_{3}).
\end{split}
\label{eq:2g}
\end{equation}
Here we have not given a purely bosonic contribution proportional to  $a^{\ast\ \!a}_{{\bf k}}\, a^{\ast\
\!a_{1}}_{{\bf k}_{1}} a^{\phantom{\ast}\!\!a_{2}}_{{\bf k}_{2}}\, a^{\phantom{\ast}\!\!a_{3}}_{{\bf k}_{3}}$, since we considered it earlier in  \cite{markov_2020}.
%
%
For a reason which will become obvious  in sections \ref{section_13} and \ref{section_14} we have adopted the designation $``(2)''$ in the definition of the four-point vertex functions $T^{\, (2)\, i\, i_{1}\, a_{1}\, a_{2}}_{\ {\bf q},\, {\bf q}_{1},\, {\bf k}_{1},\, {\bf k}_{2}}$ and $T^{\,(2)\, i\, i_{1}\, i_{2}\, i_{3}}_{{\bf q},\, {\bf q}_{1},\, {\bf q}_{2},\, {\bf q}_{3}}$. Recall that for simplicity we do not write out in an explicit form dependence of the vertex functions in (\ref{eq:2f}) and (\ref{eq:2g}), and of the variables $b^{\,i}_{{\bf q}},\hspace{0.02cm} b^{\,\ast\,i}_{{\bf q}}$ on the helicity $\lambda$ implying that the summation over color indices in the fundamental representation $i,\,i_{1},\,\ldots\,$ also corresponds the summation over the helicity states $\lambda,\,\lambda_{1},\,\ldots\,.$ In the expression (\ref{eq:2g}) we have kept only the ``essential'' terms by virtue of the fact that the resonance conditions for four-wave processes \cite{markov_2006}
\[
\left\{
\begin{array}{l}
{\bf q} + {\bf q}_{1} + {\bf q}_{2} + {\bf q}_{3} = 0\\[5pt]
\omega^{-}_{{\bf q}} + \omega^{-}_{{\bf q}_{1}} + \omega^{-}_{{\bf q}_{2}} + \omega^{-}_{{\bf q}_{3}}=0,
\end{array}\right.\ \ \
\left\{\begin{array}{l}{\bf q} = {\bf q}_{1} + {\bf q}_{2} + {\bf q}_{3} \\[5pt]
\omega^{-}_{{\bf q}} = \omega^{-}_{{\bf q}_{1}} + \omega^{-}_{{\bf q}_{2}} + \omega^{-}_{{\bf q}_{3}},
\end{array}
\right.
\]
\vspace{0.2cm}
\[
\left
\{\begin{array}{l}
{\bf q} + {\bf q}_{1} + {\bf k}_{1} + {\bf k}_{2} = 0\\[5pt]
\omega^{-}_{{\bf q}} + \omega^{-}_{{\bf q}_{1}} + \omega^{l}_{{\bf k}_{1}} + \omega^{l}_{{\bf k}_{2}} = 0,
\end{array}
\right.\ \ \
\left\{
\begin{array}{l}{\bf q} = {\bf q}_{1} + {\bf k}_{1} + {\bf k}_{2} \\[5pt]
\omega^{-}_{{\bf q}} = \omega^{-}_{{\bf q}_{1}} + \omega^{l}_{{\bf k}_{1}} + \omega^{l}_{{\bf k}_{2}},
\end{array}
\right.
\] 
\vspace{0.2cm}
\[
\left\{
\begin{array}{l}
{\bf k}_{1} = {\bf q} + {\bf q}_{1} + {\bf k}_{2} \\[5pt]
\omega^{l}_{{\bf k}_{1}} = \omega^{-}_{{\bf q}} + \omega^{-}_{{\bf q}_{1}} + \omega^{l}_{{\bf k}_{2}},
\end{array}
\right.\ \ \ \qquad
\left\{
\begin{array}{l}{\bf q} + {\bf q}_{1} = {\bf k}_{1} + {\bf k}_{2} \\[5pt]
\omega^{-}_{{\bf q}} + \omega^{-}_{{\bf q}_{1}} = \omega^{l}_{{\bf k}_{1}} + \omega^{l}_{{\bf k}_{2}}
\end{array}
\right.
\]
have no solutions for plasmon and plasmino spectra defined by the dispersion equations (\ref{eq:2e}) and  (\ref{eq:2u}).\\
\indent The vertex functions ${\mathcal V}^{\; a\, a_{1}\, a_{2}}_{{\bf k},\, {\bf k}_{1},\, {\bf k}_{2}},\, {\mathcal U}^{\; a\, a_{1}\, a_{2}}_{{\bf k},\, {\bf k}_{1},\, {\bf k}_{2}},\, {\mathcal G}^{\; a_{1}\, i\;  i_{1}}_{{\bf k}_1,\, {\bf q},\, {\bf q}_{1}},\, {\mathcal K}^{\; a_{1}\, i\;  i_{1}}_{{\bf k}_1,\, {\bf q},\, {\bf q}_{1}}$, and
$T^{\, (2)\, i\; i_{1}\; i_{2}\; i_{3}}_{\ {\bf q},\, {\bf q}_{1},\, {\bf q}_{2},\, {\bf q}_{3}}$ satisfy the ``conditions of natural symmetry'', which specify that the integrals in Eqs.\,(\ref{eq:2f}) and (\ref{eq:2g}) are unaffected by relabeling of the dummy color indices and integration variables. These conditions have the following form:
\[
{\mathcal V}^{\; a\, a_{1}\, a_{2}}_{{\bf k},\, {\bf k}_{1},\, {\bf k}_{2}} = {\mathcal V}^{\; a\, a_{2}\, a_{1}}_{{\bf k},\, {\bf k}_{2},\, {\bf k}_{1}},
\quad
{\mathcal U}^{\; a\, a_{1}\, a_{2}}_{{\bf k},\, {\bf k}_{1},\, {\bf k}_{2}} = {\mathcal U}^{\; a\, a_{2}\, a_{1}}_{{\bf k},\, {\bf k}_{2},\, {\bf k}_{1}}
= {\mathcal U}^{\, a_{1}\, a_{2}\, a}_{{\bf k}_{1},\, {\bf k}_{2},\, {\bf k}}\ ,
\]
\begin{equation}
{\mathcal G}^{\; a_{1}\, i\;  i_{1}}_{{\bf k}_1,\; {\bf q},\; {\bf q}_{1}} = -\hspace{0.04cm}{\mathcal G}^{\; a_{1}\, i_{1}\,  i}_{{\bf k}_1,\, {\bf q}_{1},\, {\bf q}},
\quad
{\mathcal K}^{\; a_{1}\, i\;  i_{1}}_{{\bf k}_{1},\, {\bf q},\, {\bf q}_{1}} = -\hspace{0.04cm}{\mathcal K}^{\; a_{1}\, i_{1}\,  i}_{{\bf k}_1,\, {\bf q}_{1},\, {\bf q}},
\label{eq:2h}
\vspace{0.2cm}
\end{equation}
\begin{equation}
T^{(2)\, i\; i_{1}\; i_{2}\; i_{3}}_{{\bf q},\, {\bf q}_{1},\, {\bf q}_{2},\, {\bf q}_{3}} 
= 
 -\hspace{0.04cm}T^{(2)\, i_{1}\;  i\ i_{2}\; i_{3}}_{{\bf q}_{1},\, {\bf q},\, {\bf q}_{2},\, {\bf q}_{3}}
=
 -\hspace{0.04cm}T^{(2)\, i\; i_{1}\; i_{3}\; i_{2}}_{{\bf q},\, {\bf q}_{1},\, {\bf q}_{3},\, {\bf q}_{2}}.
\label{eq:2j}
\vspace{0.1cm}
\end{equation}
The real nature of the Hamiltonian (\ref{eq:2f}) is obvious. A  reality of the Hamiltonian (\ref{eq:2g}) entails a validity of
additional relations for the vertex functions $T^{\, (2)\, i\, i_{1}\, a_{1}\, a_{2}}_{\,{\bf q},\, {\bf q}_{1},\, {\bf k}_{1},\, {\bf k}_{2}}$ and $T^{\,(2)\, i\; i_{1}\; i_{2}\; i_{3}}_{\ {\bf q},\, {\bf q}_{1},\, {\bf q}_{2},\, {\bf q}_{3}}$:
\begin{equation}
T^{\,(2)\, i\; i_{1}\, a_{1}\, a_{2}}_{\ {\bf q},\, {\bf q}_{1},\, {\bf k}_{1},\, {\bf k}_{2}}
=
T^{\hspace{0.03cm}\ast\hspace{0.03cm}(2)\, i_{1}\, i\; a_{2}\; a_{1}}_{\ {\bf q}_{1},\, {\bf q},\, {\bf k}_{2},\, {\bf k}_{1}},
\qquad
T^{\,(2)\, i\; i_{1}\; i_{2}\; i_{3}}_{\ {\bf q},\, {\bf q}_{1},\, {\bf q}_{2},\, {\bf q}_{3}}
=
T^{\hspace{0.03cm}\ast\hspace{0.03cm}(2)\, i_{2}\, i_{3}\; i\; i_{1}}_{\; {\bf q}_{2},\, {\bf q}_{3},\, {\bf q},\, {\bf q}_{1}}. 
\label{eq:2k}
\end{equation}
\indent The vertex functions in the Hamiltonians $H^{(3)}$ and $H^{(4)}$ are defined by specific properties of the system under study, in our case by a high-temperature quark-gluon plasma. An explicit form of the three-point amplitudes ${\mathcal V}^{\; a\, a_{1}\, a_{2}}_{{\bf k},\, {\bf k}_{1},\, {\bf k}_{2}}$ and ${\mathcal U}^{\; a\, a_{1}\, a_{2}}_{{\bf k},\, {\bf k}_{1},\, {\bf k}_{2}}$ within the hard thermal loop approximation was obtain in \cite{markov_2020}. They have the following color and momentum structures: 
\begin{equation}
{\mathcal V}^{\ \! a\, a_{1}\hspace{0.03cm} a_{2}}_{{\bf k},\, {\bf k}_{1},\, {\bf k}_{2}}
=
f^{\hspace{0.03cm} a\, a_{1}\hspace{0.03cm} a_{2}\,}\hspace{0.02cm}{\mathcal V}_{\, {\bf k},\, {\bf k}_{1},\, {\bf k}_{2}},
\qquad
{\mathcal U}^{\ \! a\, a_{1}\hspace{0.03cm} a_{2}}_{{\bf k},\, {\bf k}_{1},\, {\bf k}_{2}}
=
f^{\hspace{0.03cm} a\, a_{1}\hspace{0.03cm} a_{2}\,}\hspace{0.02cm}{\mathcal U}_{\, {\bf k},\, {\bf k}_{1},\, {\bf k}_{2}},
\label{eq:2l}
\end{equation}
where
\begin{equation}
{\mathcal V}_{\, {\bf k},\, {\bf k}_{1},\, {\bf k}_{2}} = 
\frac{1}{2^{3/4}}\,g\hspace{0.03cm}
\Biggl(\frac{ \epsilon^{l}_{\mu}({\bf k})}{\sqrt{2\omega^l_{{\bf k}_{\phantom{1}}}}}\Biggr)\!
\Biggl(\frac{\epsilon^{l}_{\mu_{1}}({\bf k}_{1})}{\sqrt{2\omega^l_{{\bf k}_{1}}}}\Biggr)\!
\Biggl(\frac{\epsilon^{l}_{\mu_{2}}({\bf k}_{2})}{\sqrt{2\omega^l_{{\bf k}_{2}}}}\Biggr)\!
\,^{\ast}\Gamma^{\mu\mu_1\mu_2}(k,- k_{1},- k_{2})\Bigr|_{\rm \,on-shell}
\hspace{0.4cm} 
\label{eq:2z}
\end{equation}
and
\begin{equation}
{\mathcal U}_{\, {\bf k},\, {\bf k}_{1},\, {\bf k}_{2}} =
\frac{1}{2^{3/4}}\,g\hspace{0.03cm}
\Biggl(\frac{ \epsilon^{l}_{\mu}({\bf k})}{\sqrt{2\omega^l_{{\bf k}_{\phantom{1}}}}}\Biggr)\!
\Biggl(\frac{\epsilon^{l}_{\mu_{1}}({\bf k}_{1})}{\sqrt{2\omega^l_{{\bf k}_{1}}}}\Biggr)\!
\Biggl(\frac{\epsilon^{l}_{\mu_{2}}({\bf k}_{2})}{\sqrt{2\omega^l_{{\bf k}_{2}}}}\Biggr)\!
\,^{\ast}\Gamma^{\mu\mu_1\mu_2}(- k,- k_{1},- k_{2})\Bigr|_{\rm \,on-shell}.
\label{eq:2x}
\end{equation}
The explicit form of the effective three-gluon vertex $\,^{\ast}\Gamma^{\mu\mu_1\mu_2}(k, k_{1}, k_{2})$ on the right-hand side of these expressions is defined by formulas (\ref{ap:A1})\,--\,(\ref{ap:A3}). In the expressions (\ref{eq:2z}) and (\ref{eq:2x}) we corrected the noticed inaccuracy in the numeric factor.


\section{Canonical transformations}
\setcounter{equation}{0}
\label{section_3}

Let us consider the transformation from the initial bosonic and fermionic functions $a^{a}_{\bf k}$ and $b^{\,i}_{\bf q}$ to the new bosonic and fermionic ones $c^{a}_{\hspace{0.02cm}\bf k}$ and $f^{\,i}_{\bf q}$:
\begin{align}
&a^{a}_{\bf k} = a^{a}_{{\bf k}}(c^{a}_{\hspace{0.02cm}\bf k},\, c^{\ast\ \!\!a}_{{\hspace{0.02cm}\bf k}}\!,\, f^{\,i}_{\bf q},\, f^{\,\ast\ \!\!i}_{{\bf q}}),
\label{eq:3q}\\[0.8ex]
&b^{\,i}_{\bf q} = b^{\,i}_{{\bf q}}\hspace{0.02cm}(\hspace{0.02cm}c^{a}_{\hspace{0.02cm}\bf k},\, c^{\ast\ \!\!a}_{{\hspace{0.02cm}\bf k}}\!,\, f^{\,i}_{\bf q},\,f^{\,\ast\ \!\!i}_{{\bf q}}\hspace{0.02cm}).
\label{eq:3w}
\end{align}
We shall demand that the Hamilton equations in terms of new variables have the form (\ref{eq:2a}) and (\ref{eq:2s}) with the same Hamiltonian $H$. Straightforward but rather cumbersome calculations result in two systems of integral relations. The first of them has the following form:
\begin{subequations} 
\label{eq:3e}
\begin{align}
&\int\! d\hspace{0.02cm}{\bf k\hspace{0.01cm}}'\!\hspace{0.01cm}
\left\{\frac{\delta\hspace{0.01cm}  a^{\phantom{\ast}\!\!a}_{{\bf k}}}{\delta c^{\phantom{\ast}\!\!c}_{\hspace{0.02cm}{\bf k}'}}
\,\frac{\delta\hspace{0.01cm}  a^{\ast\ \!\!b}_{{\bf k}''}}{\delta c^{\ast\ \!\!c}_{\hspace{0.02cm}{\bf k}'}}
\,-\,
\frac{\delta\hspace{0.01cm}  a^{\phantom{\ast}\!\!a}_{{\bf k}}}
{\delta c^{\ast\ \!\!c}_{\hspace{0.02cm}{\bf k}'}}\,
\frac{\delta\hspace{0.01cm}  a^{\ast\ \!\!b}_{{\bf k}''}}
{\delta c^{\phantom{\ast}\!\!c}_{\hspace{0.02cm}{\bf k}'}}\right\}
+
\int\! d{\bf q\hspace{0.01cm}}'\!\hspace{0.01cm}
\left\{\frac{\overleftarrow{\delta}\!\!\hspace{0.04cm} a^{\phantom{\ast}\!\!a}_{{\bf k}}}{\delta\!\hspace{0.02cm} f^{\phantom{\ast}\!\!i}_{{\bf q}^{\prime}}}\,
\frac{\overrightarrow{\delta}\!\!\hspace{0.04cm} a^{\ast\ \!\! b}_{{\bf k}''}}{\delta\!\hspace{0.02cm} f^{\, \ast\ \!\!i}_{{\bf q}^{\prime}}}
\,+\,
\frac{\overrightarrow{\delta}\!\!\hspace{0.04cm} a^{\phantom{\ast}\!\!a}_{{\bf k}}}{\delta\!\hspace{0.02cm} f^{\, \ast\ \!\!i}_{{\bf q}^{\prime}}}\,
\frac{\overleftarrow{\delta}\!\!\hspace{0.04cm} a^{\ast\ \!\! b}_{{\bf k}''}}{\delta\!\hspace{0.02cm} f^{\phantom{\ast}\!\!i}_{{\bf q}^{\prime}}}\right\}
\!=
\delta^{ab}\delta ({\bf k}-{\bf k}\!\ ''),
\label{eq:3ea}
\\[0.8ex]
&\int\! d\hspace{0.02cm}{\bf k\hspace{0.01cm}}'\!\hspace{0.01cm}\left\{\frac{\delta\hspace{0.02cm}  a^{\phantom{\ast}\!\!a}_{{\bf k}}}
{\delta c^{\phantom{\ast}\!\!c}_{\hspace{0.02cm}{\bf k}^{\prime}}}
\,\frac{\delta\hspace{0.02cm}  a^{\phantom{\ast}\!\!b}_{{\bf k}''}}
{\delta c^{\ast\ \!\!c}_{\hspace{0.02cm}{\bf k}^{\prime}}}
\,-\,
\frac{\delta\hspace{0.02cm}  a^{\phantom{\ast}\!\!a}_{{\bf k}}}
{\delta c^{\ast\ \!\!c}_{\hspace{0.02cm}{\bf k}^{\prime}}}\,
\frac{\delta\hspace{0.02cm}  a^{\phantom{\ast}\!\!b}_{{\bf k}''}}
{\delta c^{\phantom{\ast}\!\!c}_{\hspace{0.02cm}{\bf k}^{\prime}}}\right\}
+
\int\! d{\bf q\hspace{0.01cm}}'\!\hspace{0.01cm}
\left\{\frac{\overleftarrow{\delta}\!\!\hspace{0.04cm} a^{\phantom{\ast}\!\!a}_{{\bf k}}}{\delta\!\hspace{0.02cm} f^{\phantom{\ast}\!\!i}_{{\bf q}^{\prime}}}\,
\frac{\overrightarrow{\delta}\!\!\hspace{0.04cm} a^{b}_{{\bf k}''}}{\delta\!\hspace{0.02cm} f^{\, \ast\ \!\!i}_{{\bf q}^{\prime}}}
\,+\,
\frac{\overrightarrow{\delta}\!\!\hspace{0.04cm} a^{\phantom{\ast}\!\!a}_{{\bf k}}}{\delta\!\hspace{0.02cm} f^{\, \ast\ \!\!i}_{{\bf q}^{\prime}}}\,
\frac{\overleftarrow{\delta}\!\!\hspace{0.04cm} a^{b}_{{\bf k}''}}{\delta\!\hspace{0.02cm} f^{\phantom{\ast}\!\!i}_{{\bf q}^{\prime}}}\right\} = 0,
\label{eq:3eb}
\\[0.8ex]
&\int\! d\hspace{0.02cm}{\bf k\hspace{0.01cm}}'\!\hspace{0.01cm}\left\{\frac{\delta\hspace{0.02cm}  a^{\phantom{\ast}\!\!a}_{{\bf k}}}
{\delta c^{\phantom{\ast}\!\!c}_{\hspace{0.02cm}{\bf k}^{\prime}}}
\,\frac{\delta\hspace{0.02cm}  b^{\,i}_{{\bf q}''}}
{\delta c^{\ast\ \!\!c}_{\hspace{0.02cm}{\bf k}^{\prime}}}
\,-\,
\frac{\delta\hspace{0.02cm}  a^{\phantom{\ast}\!\!a}_{{\bf k}}}
{\delta c^{\ast\ \!\!c}_{\hspace{0.02cm}{\bf k}^{\prime}}}\,
\frac{\delta\hspace{0.02cm}  b^{\,i}_{{\bf q}''}}
{\delta c^{\phantom{\ast}\!\!c}_{\hspace{0.02cm}{\bf k}^{\prime}}}\right\}
+
\int\! d{\bf q\hspace{0.01cm}}'\!\hspace{0.01cm}
\left\{\frac{\overleftarrow{\delta}\!\!\hspace{0.04cm} a^{\phantom{\ast}\!\!a}_{{\bf k}}}{\delta\!\hspace{0.02cm} f^{\phantom{\ast}\!\!k}_{{\bf q}^{\prime}}}\,
\frac{\overrightarrow{\delta}\! b^{\,i}_{{\bf q}''}}{\delta\!\hspace{0.02cm} f^{\, \ast\ \!\!k}_{{\bf q}^{\prime}}}
\,-\,
\frac{\overrightarrow{\delta}\!\!\hspace{0.04cm} a^{\phantom{\ast}\!\!a}_{{\bf k}}}{\delta\!\hspace{0.02cm} f^{\, \ast\ \!\!k}_{{\bf q}^{\prime}}}\,
\frac{\overleftarrow{\delta}\! b^{\,i}_{{\bf q}''}}{\delta\!\hspace{0.02cm} f^{\phantom{\ast}\!\!k}_{{\bf q}^{\prime}}}\right\} = 0,
\label{eq:3ec}\\[0.8ex]
&\int\! d\hspace{0.02cm}{\bf k\hspace{0.01cm}}'\!\hspace{0.01cm}\left\{\frac{\delta\hspace{0.02cm}  a^{\phantom{\ast}\!\!a}_{{\bf k}}}
{\delta c^{\phantom{\ast}\!\!c}_{\hspace{0.02cm}{\bf k}^{\prime}}}
\,\frac{\delta\hspace{0.02cm}  b^{\,\ast\hspace{0.02cm} i}_{{\bf q}''}}
{\delta c^{\ast\ \!\!c}_{\hspace{0.02cm}{\bf k}^{\prime}}}
\,-\,
\frac{\delta\hspace{0.02cm}  a^{\phantom{\ast}\!\!a}_{{\bf k}}}
{\delta c^{\ast\ \!\!c}_{\hspace{0.02cm}{\bf k}^{\prime}}}\,
\frac{\delta\hspace{0.02cm}  b^{\,\ast\hspace{0.02cm} i}_{{\bf q}''}}
{\delta c^{\phantom{\ast}\!\!c}_{\hspace{0.02cm}{\bf k}^{\prime}}}\right\}
+
\int\! d{\bf q\hspace{0.01cm}}'\!\hspace{0.01cm}
\left\{\frac{\overleftarrow{\delta}\!\!\hspace{0.04cm} a^{\phantom{\ast}\!\!a}_{{\bf k}}}{\delta\!\hspace{0.02cm} f^{\phantom{\ast}\!\!k}_{{\bf q}^{\prime}}}\,
\frac{\overrightarrow{\delta}\! b^{\,\ast\hspace{0.02cm}i}_{{\bf q}''}}{\delta\!\hspace{0.02cm} f^{\, \ast\ \!\!k}_{{\bf q}^{\prime}}}
\,-\,
\frac{\overrightarrow{\delta}\!\!\hspace{0.04cm} a^{\phantom{\ast}\!\!a}_{{\bf k}}}{\delta\!\hspace{0.02cm} f^{\, \ast\ \!\!k}_{{\bf q}^{\prime}}}\,
\frac{\overleftarrow{\delta}\! b^{\,\ast\hspace{0.02cm} i}_{{\bf q}''}}{\delta\!\hspace{0.02cm} f^{\phantom{\ast}\!\!k}_{{\bf q}^{\prime}}}\right\} = 0
\label{eq:3ed}
\end{align}
\end{subequations}
and, correspondingly, the second system is
\begin{subequations} 
\label{eq:3r}
\begin{align}
&\int\! d\hspace{0.02cm}{\bf k\hspace{0.01cm}}'\!\hspace{0.01cm}
\left\{\frac{\delta\hspace{0.02cm}  b^{\,i}_{{\bf q}}}{\delta c^{\phantom{\ast}\!\!c}_{\hspace{0.02cm}{\bf k}^{\prime}}}
\,\frac{\delta\hspace{0.02cm}  b^{\,\ast j}_{{\bf q}''}}{\delta c^{\ast\ \!\!c}_{\hspace{0.02cm}{\bf k}^{\prime}}}
\,-\,
\frac{\delta\hspace{0.02cm}  b^{\,i}_{{\bf q}}}{\delta c^{\ast\ \!\!c}_{\hspace{0.02cm}{\bf k}^{\prime}}}\,
\frac{\delta\hspace{0.02cm}  b^{\,\ast j}_{{\bf q}''}}{\delta c^{\phantom{\ast}\!\!c}_{\hspace{0.02cm}{\bf k}^{\prime}}}\right\}
+
\int\! d{\bf q\hspace{0.01cm}}'\!\hspace{0.01cm}
\left\{\frac{\overleftarrow{\delta}\! b^{\,i}_{{\bf q}}}{\delta\!\hspace{0.02cm} f^{\phantom{\ast}\!\!k}_{{\bf q}^{\prime}}}\,
\frac{\overrightarrow{\delta}\! b^{\,\ast\hspace{0.02cm}j}_{{\bf q}''}}{\delta\!\hspace{0.02cm} f^{\, \ast\ \!\!k}_{{\bf q}^{\prime}}}
\,+\,
\frac{\overrightarrow{\delta}\! b^{\,i}_{{\bf q}}}{\delta\!\hspace{0.02cm} f^{\, \ast\ \!\!k}_{{\bf q}^{\prime}}}\,
\frac{\overleftarrow{\delta}\! b^{\,\ast\hspace{0.02cm} j}_{{\bf q}''}}{\delta\!\hspace{0.02cm} f^{\phantom{\ast}\!\!k}_{{\bf q}^{\prime}}}\right\}
\!= \delta^{ij}\delta ({\bf q} - {\bf q}\!\ ''),
\label{eq:3ra}
\\[0.8ex]
&\int\! d\hspace{0.02cm}{\bf k\hspace{0.01cm}}'\!\hspace{0.01cm}
\left\{\frac{\delta\hspace{0.02cm}  b^{\,i}_{{\bf q}}}{\delta c^{\phantom{\ast}\!\!c}_{\hspace{0.02cm}{\bf k}^{\prime}}}
\,\frac{\delta\hspace{0.02cm}  b^{\,j}_{{\bf q}''}}{\delta c^{\ast\ \!\!c}_{\hspace{0.02cm}{\bf k}^{\prime}}}
\,-\,
\frac{\delta\hspace{0.02cm}  b^{\,i}_{{\bf q}}}{\delta c^{\ast\ \!\!c}_{\hspace{0.02cm}{\bf k}^{\prime}}}\,
\frac{\delta\hspace{0.02cm}  b^{\,j}_{{\bf q}''}}{\delta c^{\phantom{\ast}\!\!c}_{\hspace{0.02cm}{\bf k}^{\prime}}}\right\}
+
\int\! d{\bf q\hspace{0.01cm}}'\!\hspace{0.01cm}
\left\{\frac{\overleftarrow{\delta}\! b^{\,i}_{{\bf q}}}{\delta\!\hspace{0.02cm} f^{\phantom{\ast}\!\!k}_{{\bf q}^{\prime}}}\,
\frac{\overrightarrow{\delta}\! b^{\,j}_{{\bf q}''}}{\delta\!\hspace{0.02cm} f^{\, \ast\ \!\!k}_{{\bf q}^{\prime}}}
\,+\,
\frac{\overrightarrow{\delta}\! b^{\,i}_{{\bf q}}}{\delta\!\hspace{0.02cm} f^{\, \ast\ \!\!k}_{{\bf q}^{\prime}}}\,
\frac{\overleftarrow{\delta}\! b^{\,j}_{{\bf q}''}}{\delta\!\hspace{0.02cm} f^{\phantom{\ast}\!\!k}_{{\bf q}^{\prime}}}\right\} = 0,
\label{eq:3rb}
\\[0.8ex]
&\int\! d\hspace{0.02cm}{\bf k\hspace{0.01cm}}'\!\hspace{0.01cm}
\left\{\frac{\delta\hspace{0.02cm}  b^{\,i}_{{\bf q}}}{\delta c^{\phantom{\ast}\!\!c}_{\hspace{0.02cm}{\bf k}^{\prime}}}
\,\frac{\delta\hspace{0.02cm}  a^{\phantom{\ast}\!\!a}_{{\bf k}''}}{\delta c^{\ast\ \!\!c}_{\hspace{0.02cm}{\bf k}^{\prime}}}
\,-\,
\frac{\delta\hspace{0.02cm}  b^{\,i}_{{\bf q}}}{\delta c^{\ast\ \!\!c}_{\hspace{0.02cm}{\bf k}^{\prime}}}\,
\frac{\delta\hspace{0.02cm}  a^{\phantom{\ast}\!\!a}_{{\bf k}''}}{\delta c^{\phantom{\ast}\!\!c}_{\hspace{0.02cm}{\bf k}^{\prime}}}\right\}
+
\int\! d{\bf q\hspace{0.01cm}}'\!\hspace{0.01cm}
\left\{\frac{\overleftarrow{\delta}\! b^{\,i}_{{\bf q}}}{\delta\!\hspace{0.02cm} f^{\phantom{\ast}\!\!k}_{{\bf q}^{\prime}}}
\,\frac{\overrightarrow{\delta}\!  a^{\phantom{\ast}\!\!a}_{{\bf k}''}}{\delta\!\hspace{0.02cm} f^{\,\ast\ \!\!k}_{{\bf q}^{\prime}}}
\,-\,
\frac{\overrightarrow{\delta}\!  b^{\,i}_{{\bf q}}}{\delta\!\hspace{0.02cm} f^{\,\ast\ \!\!k}_{{\bf q}^{\prime}}}\,
\frac{\overleftarrow{\delta}\!\!\hspace{0.04cm} a^{\phantom{\ast}\!\!a}_{{\bf k}''}}{\delta\!\hspace{0.02cm} f^{\phantom{\ast}\!\!k}_{{\bf q}^{\prime}}}\right\}
= 0,
\label{eq:3rc}
\\[0.8ex]
&\int\! d\hspace{0.02cm}{\bf k\hspace{0.01cm}}'\!\hspace{0.01cm}
\left\{\frac{\delta\hspace{0.02cm}  b^{\,i}_{{\bf q}}}{\delta c^{\phantom{\ast}\!\!c}_{\hspace{0.02cm}{\bf k}^{\prime}}}
\,\frac{\delta\hspace{0.02cm}  a^{\ast a}_{{\bf k}''}}{\delta c^{\ast\ \!\!c}_{\hspace{0.02cm}{\bf k}^{\prime}}}
\,-\,
\frac{\delta\hspace{0.02cm}  b^{\,i}_{{\bf q}}}{\delta c^{\ast\ \!\!c}_{\hspace{0.02cm}{\bf k}^{\prime}}}\,
\frac{\delta\hspace{0.02cm}  a^{\ast a}_{{\bf k}''}}{\delta c^{\phantom{\ast}\!\!c}_{\hspace{0.02cm}{\bf k}^{\prime}}}\right\}
+
\int\! d{\bf q\hspace{0.01cm}}'\!\hspace{0.01cm}
\left\{\frac{\overleftarrow{\delta}\!  b^{\,i}_{{\bf q}}}{\delta\!\hspace{0.02cm} f^{\phantom{\ast}\!\!k}_{{\bf q}^{\prime}}}
\,\frac{\overrightarrow{\delta}\!  a^{\ast\hspace{0.02cm} a}_{{\bf k}''}}{\delta\!\hspace{0.02cm} f^{\,\ast\ \!\!k}_{{\bf q}^{\prime}}}
\,-\,
\frac{\overrightarrow{\delta}\!  b^{\,i}_{{\bf q}}}{\delta\!\hspace{0.02cm} f^{\,\ast\ \!\!k}_{{\bf q}^{\prime}}}\,
\frac{\overleftarrow{\delta}\!  a^{\ast\hspace{0.02cm}  a}_{{\bf k}''}}{\delta\!\hspace{0.02cm} f^{\phantom{\ast}\!\!k}_{{\bf q}^{\prime}}}\right\}
= 0.
\label{eq:3rd}
\end{align}
\end{subequations}
These canonicity conditions can be written in a very compact form if we make use of the definition of the Poisson superbracket (\ref{eq:2p}) and replace the variation variables by the new ones: $a^{a}_{\bf k}\rightarrow c^{a}_{\hspace{0.02cm}\bf k}$ and $b^{\,i}_{\bf q} \rightarrow f^{\,i}_{\bf q}$. In this case the superbrackets for the original variables $a^{a}_{\bf k}$ and $b^{\,i}_{\bf q}$, Eqs.\,(\ref{eq:2i}) and (\ref{eq:2o}), turn to the canonicity conditions (\ref{eq:3e}) and (\ref{eq:3r}), which impose certain restrictions on the functional dependencies (\ref{eq:3q}) and (\ref{eq:3w}). Let us present the right-hand sides of  (\ref{eq:3q}) and (\ref{eq:3w}) in the form of integer-degree series in the normal variables $c^{a}_{\hspace{0.02cm}\bf k}$ and $f^{\,i}_{\bf q}$. The most common dependence of the transformation (\ref{eq:3q}) up to cubic terms in $c^{a}_{\hspace{0.02cm}\bf k}$ and $f^{\,i}_{\bf q}$ has the following form:

\begin{equation}
a^{a}_{{\bf k}} = c^{a}_{\hspace{0.02cm}{\bf k}}\,+
\label{eq:3t}
\end{equation}
\[
+ \int\frac{d\hspace{0.02cm}{\bf k}_{1}\hspace{0.02cm}  d\hspace{0.02cm}{\bf k}_{2}}{(2\pi)^{6}} 
\left[V^{(1)\, a\, a_{1}\, a_{2}}_{\ {\bf k},\, {\bf k}_{1},\, {\bf k}_{2}}\, c^{a_{1}}_{{\bf k}_{1}}\, c^{a_{2}}_{{\bf k}_{2}}
\,+\,
V^{(2)\, a\, a_{1}\, a_{2}}_{\ {\bf k},\, {\bf k}_{1},\, {\bf k}_{2}}\, c^{\ast\, a_{1}}_{{\bf k}_{1}}\, c^{\phantom{\ast}\!\!a_{2}}_{{\bf k}_{2}}
\,+\,
V^{(3)\ a\, a_{1}\, a_{2}}_{\ {\bf k},\, {\bf k}_{1},\, {\bf k}_{2}}\; 
c^{\ast\, a_{1}}_{{\bf k}_{1}} c^{\ast\, a_{2}}_{{\bf k}_{2}}\right] 
\]
\[
+\int\frac{d{\bf q}_{1}\hspace{0.02cm}  d{\bf q}_{2}}{(2\pi)^{6}}
\left[
F^{(1)\, a\, i_{1}\, i_{2}}_{\ {\bf k},\, {\bf q}_{1},\, {\bf q}_{2}}\; f^{\,i_{1}}_{{\bf q}_{1}} f^{\,i_{2}}_{{\bf q}_{2}}
\,+\,
F^{(2)\, a\, i_{1}\, i_{2}}_{\ {\bf k},\, {\bf q}_{1},\, {\bf q}_{2}}\ f^{\,\ast\, i_{1}}_{{\bf q}_{1}} f^{\phantom{\ast}\!i_{2}}_{{\bf q}_{2}}
+
F^{(3)\, a\, i_{1}\, i_{2}}_{\ {\bf k},\, {\bf q}_{1},\, {\bf q}_{2}} f^{\,\ast\hspace{0.04cm} i_{1}}_{{\bf q}_{1}} f^{\,\ast\hspace{0.04cm} i_{2}}_{{\bf q}_{2}}\right] 
\hspace{0.6cm}
\]
\begin{align}
+ \int\frac{d\hspace{0.02cm}{\bf k}_{1}\hspace{0.02cm} d\hspace{0.02cm}{\bf k}_{2}\hspace{0.03cm} d\hspace{0.02cm}{\bf k}_{3}}{(2\pi)^{9}}
\Bigl[\hspace{0.04cm}&W^{(1)\, a\, a_{1}\, a_{2}\, a_{3}}_{\ {\bf k},\, {\bf k}_{1},\, {\bf k}_{2},\, {\bf k}_{3}}\, c^{a_{1}}_{{\bf k}_{1}}\, c^{a_{2}}_{{\bf k}_{2}}\, c^{a_{3}}_{{\bf k}_{3}}
\,+\,
W^{(2)\, a\, a_{1}\, a_{2}\, a_{3}}_{\ {\bf k},\, {\bf k}_{1},\, {\bf k}_{2},\, {\bf k}_{3}}\, c^{\ast\, a_{1}}_{{\bf k}_{1}}\hspace{0.03cm} c^{a_{2}}_{{\bf k}_{2}}\, c^{a_{3}}_{{\bf k}_{3}} 
\notag\\[0.5ex]
+\;
&W^{(3)\, a\, a_{1}\, a_{2}\, a_{3}}_{\ {\bf k},\, {\bf k}_{1},\, {\bf k}_{2},\, {\bf k}_{3}}\, c^{\ast\, a_{1}}_{{\bf k}_{1}}\hspace{0.03cm} c^{\ast\, a_{2}}_{{\bf k}_{2}}\hspace{0.03cm} c^{a_{3}}_{{\bf k}_{3}}
+\,
W^{(4)\, a\, a_{1}\, a_{2}\, a_{3}}_{\ {\bf k},\, {\bf k}_{1},\, {\bf k}_{2},\, {\bf k}_{3}}\hspace{0.03cm} c^{\ast\, a_{1}}_{{\bf k}_{1}}\hspace{0.03cm} c^{\ast\, a_{2}}_{{\bf k}_{2}}\hspace{0.03cm} c^{\ast\, a_{3}}_{{\bf k}_{3}}\Bigr]
\hspace{0.7cm}
\notag
\end{align}
\vspace{-0.5cm}
\begin{align}
\hspace{0.5cm}
+
\!\int\frac{d\hspace{0.02cm}{\bf k}_{1}\hspace{0.02cm} d{\bf q}_{2}\hspace{0.03cm} d{\bf q}_{3}}{(2\pi)^{9}}
\Bigl[\hspace{0.03cm} 
&J^{\hspace{0.03cm}(1)\, a\, a_{1}\, i_{1}\, i_{2}}_{\ {\bf k},\, {\bf k}_{1},\, {\bf q}_{1},\, {\bf q}_{2}}\, c^{a_{1}}_{{\bf k}_{1}}\, f^{\, i_{1}}_{{\bf q}_{1}} f^{\, i_{2}}_{{\bf q}_{2}}
\,+\,
J^{\hspace{0.03cm}(2)\, a\, a_{1}\, i_{1}\, i_{2}}_{\ {\bf k},\, {\bf k}_{1},\, {\bf q}_{1},\, {\bf q}_{2}}\, c^{a_{1}}_{{\bf k}_{1}}\,
f^{\,\ast\, i_{1}}_{{\bf q}_{1}} f^{\phantom{\ast}\!\!i_{2}}_{{\bf q}_{2}}
\notag\\[0.5ex]
+\;
&J^{\hspace{0.03cm}(3)\, a\, a_{1}\, i_{1}\, i_{2}}_{\ {\bf k},\, {\bf k}_{1},\, {\bf q}_{1},\, {\bf q}_{2}}\ c^{a_{1}}_{{\bf k}_{1}}\,
f^{\,\ast\, i_{1}}_{{\bf q}_{1}} f^{\,\ast\,i_{2}}_{{\bf q}_{2}}
\,+\,
J^{\hspace{0.03cm}(4)\, a\, a_{1}\, i_{1}\, i_{2}}_{\ {\bf k},\, {\bf k}_{1},\, {\bf q}_{1},\, {\bf q}_{2}}\, c^{\ast\, a_{1}}_{{\bf k}_{1}}\,
f^{\, i_{1}}_{{\bf q}_{1}}\, f^{\, i_{2}}_{{\bf q}_{2}}
\notag\\[1ex]
+\;
&J^{\hspace{0.03cm}(5)\, a\, a_{1}\, i_{1}\, i_{2}}_{\ {\bf k},\, {\bf k}_{1},\, {\bf q}_{1},\, {\bf q}_{2}}\, c^{\ast\, a_{1}}_{{\bf k}_{1}}
f^{\,\ast\, i_{1}}_{{\bf q}_{1}}\hspace{0.03cm} f^{\, i_{2}}_{{\bf q}_{2}}
\,+\,
J^{\hspace{0.03cm}(6)\, a\, a_{1}\, i_{1}\, i_{2}}_{\ {\bf k},\, {\bf k}_{1},\, {\bf q}_{1},\, {\bf q}_{2}}\, c^{\ast\, a_{1}}_{{\bf k}_{1}}
f^{\,\ast\, i_{1}}_{{\bf q}_{1}} f^{\,\ast\, i_{2}}_{{\bf q}_{2}}\Bigr] +\,\ldots\;\ _{.}
\notag
\end{align}
Similarly, the most common dependence for the transformation (\ref{eq:3w}) up to cubic terms is
\begin{equation}
b^{\,i}_{{\bf q}} = f^{\,i}_{{\bf q}}\,+
\label{eq:3y}
\end{equation}
\[
+\int\frac{d\hspace{0.02cm}{\bf k}_{1}\hspace{0.02cm}d{\bf q}_{1}}{(2\pi)^{6}}\hspace{0.02cm}
\Bigl[\,
Q^{(1)\, i\, a_{1}\, i_{1}}_{\ {\bf q},\, {\bf k}_{1},\, {\bf q}_{1}}\, c^{a_{1}}_{{\bf k}_{1}}\, f^{\,i_{1}}_{{\bf q}_{1}}
+\,
Q^{(2)\, i\, a_{1}\, i_{1}}_{\ {\bf q},\, {\bf k}_{1},\, {\bf q}_{1}}\, c^{a_{1}}_{{\bf k}_{1}}\, f^{\, \ast\,i_{1}}_{{\bf q}_{1}}
+\,
Q^{(3)\, i\, a_{1}\, i_{1}}_{\ {\bf q},\, {\bf k}_{1},\, {\bf q}_{1}}\, c^{\ast\, a_{1}}_{{\bf k}_{1}}\hspace{0.03cm} f^{\,i_{1}}_{{\bf q}_{1}}
+\,
Q^{(4)\, i\, a_{1}\, i_{1}}_{\ {\bf q},\, {\bf k}_{1},\, {\bf q}_{1}}\, c^{\ast\, a_{1}}_{{\bf k}_{1}}\hspace{0.03cm} f^{\, \ast\,i_{1}}_{{\bf q}_{1}}\,\Bigr]
\]
\vspace{-0.3cm}
\begin{align}
+\!\int\frac{d\hspace{0.02cm}{\bf k}_{1}\hspace{0.02cm}d\hspace{0.02cm}{\bf k}_{2}\hspace{0.02cm}d{\bf q}_{1}}{(2\pi)^{9}}\hspace{0.02cm}
\Bigl[&R^{\,(1)\, i\, a_{1}\, a_{2}\, i_{1}}_{\ {\bf q},\, {\bf k}_{1},\, {\bf k}_{2},\, {\bf q}_{1}}\, c^{a_{1}}_{{\bf k}_{1}}\, c^{a_{2}}_{{\bf k}_{2}}\, f^{\,i_{1}}_{{\bf q}_{1}}
\,+\,
R^{\,(2)\, i\, a_{1}\, a_{2}\, i_{1}}_{\ {\bf q},\, {\bf k}_{1},\, {\bf k}_{2},\, {\bf q}_{1}}\, c^{\ast\, a_{1}}_{{\bf k}_{1}}\, c^{a_{2}}_{{\bf k}_{2}}\, f^{\,i_{1}}_{{\bf q}_{1}} 
\notag\\[0.5ex]
+\, &R^{\,(3)\, i\, a_{1}\, a_{2}\, i_{1}}_{\ {\bf q},\, {\bf k}_{1},\, {\bf k}_{2},\, {\bf q}_{1}}\, c^{\ast\, a_{1}}_{{\bf k}_{1}}\hspace{0.03cm} c^{\ast\, a_{2}}_{{\bf k}_{2}}\hspace{0.03cm} f^{\,i_{1}}_{{\bf q}_{1}} 
\,+\,
R^{\,(4)\, i\, a_{1}\, a_{2}\, i_{1}}_{\ {\bf q},\, {\bf k}_{1},\, {\bf k}_{2},\, {\bf q}_{1}}\, c^{a_{1}}_{{\bf k}_{1}}\, c^{a_{2}}_{{\bf k}_{2}}\, f^{\, \ast\,i_{1}}_{{\bf q}_{1}}
\notag\\[1.7ex]
+\, &R^{\,(5)\, i\, a_{1}\, a_{2}\, i_{1}}_{\ {\bf q},\, {\bf k}_{1},\, {\bf k}_{2},\, {\bf q}_{1}}\, c^{\ast\, a_{1}}_{{\bf k}_{1}}\hspace{0.03cm} c^{a_{2}}_{{\bf k}_{2}}\, f^{\, \ast\,i_{1}}_{{\bf q}_{1}}
\,+\,
R^{\,(6)\, i\, a_{1}\, a_{2}\, i_{1}}_{\ {\bf q},\, {\bf k}_{1},\, {\bf k}_{2},\, {\bf q}_{1}}\, c^{\ast\, a_{1}}_{{\bf k}_{1}}\, c^{\ast\, a_{2}}_{{\bf k}_{2}}\, f^{\, \ast\,i_{1}}_{{\bf q}_{1}}\Bigr] 
\notag
\end{align}
\vspace{-0.7cm}
\begin{align}
\hspace{1.4cm}
+\!\int\frac{d{\bf q}_{1}\hspace{0.02cm}d{\bf q}_{2}\hspace{0.02cm}d{\bf q}_{3}}{(2\pi)^{9}}\hspace{0.02cm}
\Bigl[&S^{\,(1)\, i\; i_{1}\, i_{2}\, i_{3}}_{\ {\bf q},\, {\bf q}_{1},\, {\bf q}_{2},\, {\bf q}_{3}}\, f^{\,i_{1}}_{{\bf q}_{1}}\, f^{\,i_{2}}_{{\bf q}_{2}}\, f^{\,i_{3}}_{{\bf q}_{3}}
\,+\,
S^{\,(2)\, i\; i_{1}\, i_{2}\, i_{3}}_{\ {\bf q},\, {\bf q}_{1},\, {\bf q}_{2},\, {\bf q}_{3}}\, f^{\,\ast\, i_{1}}_{{\bf q}_{1}}\hspace{0.03cm} f^{\,i_{2}}_{{\bf q}_{2}}\, f^{\,i_{3}}_{{\bf q}_{3}}
\notag\\[0.6ex]
+\,
&S^{\,(3)\, i\; i_{1}\, i_{2}\, i_{3}}_{\ {\bf q},\, {\bf q}_{1},\, {\bf q}_{2},\, {\bf q}_{3}}\, f^{\,\ast\, i_{1}}_{{\bf q}_{1}}\hspace{0.03cm} f^{\,\ast\, i_{2}}_{{\bf q}_{2}}\hspace{0.03cm} f^{\,i_{3}}_{{\bf q}_{3}}
+
S^{\,(4)\, i\; i_{1}\, i_{2}\, i_{3}}_{\ {\bf q},\, {\bf q}_{1},\, {\bf q}_{2},\, {\bf q}_{3}}\, f^{\,\ast\, i_{1}}_{{\bf q}_{1}}\hspace{0.03cm} f^{\,\ast\, i_{2}}_{{\bf q}_{2}}\hspace{0.03cm} f^{\,\ast\, i_{3}}_{{\bf q}_{3}}\Bigr] +\,\ldots\;\ _{.}
\notag
\end{align}
Note first of all that the coefficient functions $V^{(1)\, a\, a_{1}\, a_{2}}_{\ {\bf k},\, {\bf k}_{1},\, {\bf k}_{2}}$, $V^{(3)\, a\, a_{1}\, a_{2}}_{\ {\bf k},\, {\bf k}_{1},\, {\bf k}_{2}}$, $F^{(1,3)\, a\, i_{1}\, i_{2}}_{\ {\bf k},\, {\bf q}_{1},\, {\bf q}_{2}}$, $J^{\hspace{0.03cm}(1,3,4,6)\, a\, a_{1}\, i_{1}\, i_{2}}_{\ {\bf k},\, {\bf k}_{1},\, {\bf q}_{1},\, {\bf q}_{2}}$, $R^{\hspace{0.03cm}(1,3,4,6)\, i\, a_{1}\, a_{2}\, i_{1}}_{\ {\bf q},\, {\bf k}_{1},\, {\bf k}_{2},\, {\bf q}_{1}}$ and $S^{\hspace{0.03cm}(1,2,3,4)\, i\; i_{1}\, i_{2}\, i_{3}}_{\ {\bf q},\, {\bf q}_{1},\, {\bf q}_{2},\, {\bf q}_{3}}$ should satisfy the following symmetry conditions:
\begin{equation}
\begin{array}{llll}
&V^{(1)\, a\, a_{1}\, a_{2}}_{\ {\bf k},\, {\bf k}_{1},\, {\bf k}_{2}} = V^{(1)\, a\, a_{2}\, a_{1}}_{\ {\bf k},\, {\bf k}_{2},\, {\bf k}_{1}},
\qquad
&V^{(3)\, a\, a_{1}\, a_{2}}_{\ {\bf k},\, {\bf k}_{1},\, {\bf k}_{2}} = V^{(3)\, a\, a_{2}\, a_{1}}_{\ {\bf k},\, {\bf k}_{2},\, {\bf k}_{1}}, 
\\[2.5ex]
&F^{(1)\, a\, i_{1}\, i_{2}}_{\ {\bf k},\, {\bf q}_{1},\, {\bf q}_{2}} = - F^{(1)\, a\, i_{2}\, i_{1}}_{\ {\bf k},\, {\bf q}_{2},\, {\bf q}_{1}},
\qquad
&F^{(3)\, a\, i_{1}\, i_{2}}_{\ {\bf k},\, {\bf q}_{1},\, {\bf q}_{2}} = - F^{(3)\, a\, i_{2}\, i_{1}}_{\ {\bf k},\, {\bf q}_{2},\, {\bf q}_{1}},
\end{array}
\vspace{0.25cm}
\label{eq:3u} 
\end{equation}
\[
\begin{array}{llll}
&J^{\hspace{0.03cm}(1)\, a\, a_{1}\, i_{1}\, i_{2}}_{\ {\bf k},\, {\bf k}_{1},\, {\bf q}_{1},\, {\bf q}_{2}} = - J^{\hspace{0.03cm}(1)\, a\, a_{1}\, i_{2}\, i_{1}}_{\ {\bf k},\, {\bf k}_{1},\, {\bf q}_{2},\, {\bf q}_{1}},
\quad
&J^{\hspace{0.03cm}(3)\, a\, a_{1}\, i_{1}\, i_{2}}_{\ {\bf k},\, {\bf k}_{1},\, {\bf q}_{1},\, {\bf q}_{2}} = - J^{\hspace{0.03cm}(3)\, a\, a_{1}\, i_{2}\, i_{1}}_{\ {\bf k},\, {\bf k}_{1},\, {\bf q}_{2},\, {\bf q}_{1}}, 
\notag\\[2.5ex]
&J^{\hspace{0.03cm}(4)\, a\, a_{1}\, i_{1}\, i_{2}}_{\ {\bf k},\, {\bf k}_{1},\, {\bf q}_{1},\, {\bf q}_{2}} = - J^{\hspace{0.03cm}(4)\, a\, a_{1}\, i_{2}\, i_{1}}_{\ {\bf k},\, {\bf k}_{1},\, {\bf q}_{2},\, {\bf q}_{1}},
\quad
&J^{\hspace{0.03cm}(6)\, a\, a_{1}\, i_{1}\, i_{2}}_{\ {\bf k},\, {\bf k}_{1},\, {\bf q}_{1},\, {\bf q}_{2}} = - J^{\hspace{0.03cm}(6)\, a\, a_{1}\, i_{2}\, i_{1}}_{\ {\bf k},\, {\bf k}_{1},\, {\bf q}_{2},\, {\bf q}_{1}},
\notag\\[2.5ex]
&R^{\hspace{0.03cm}(1)\, i\, a_{1}\, a_{2}\, i_{1}}_{\ {\bf q},\, {\bf k}_{1},\, {\bf k}_{2},\, {\bf q}_{1}} = R^{\hspace{0.03cm}(1)\, \, i\, a_{2}\, a_{1}\, i_{1}}_{\ {\bf q},\, {\bf k}_{2},\, {\bf k}_{1},\, {\bf q}_{1}},
\quad
&R^{\hspace{0.03cm}(3)\, i\, a_{1}\, a_{2}\, i_{1}}_{\ {\bf q},\, {\bf k}_{1},\, {\bf k}_{2},\, {\bf q}_{1}} = R^{\hspace{0.03cm}(3)\, \, i\, a_{2}\, a_{1}\, i_{1}}_{\ {\bf q},\, {\bf k}_{2},\, {\bf k}_{1},\, {\bf q}_{1}},
\notag\\[2.5ex]
&R^{\hspace{0.03cm}(4)\, i\, a_{1}\, a_{2}\, i_{1}}_{\ {\bf q},\, {\bf k}_{1},\, {\bf k}_{2},\, {\bf q}_{1}} = R^{\hspace{0.03cm}(4)\, \, i\, a_{2}\, a_{1}\, i_{1}}_{\ {\bf q},\, {\bf k}_{2},\, {\bf k}_{1},\, {\bf q}_{1}},
\quad
&R^{\hspace{0.03cm}(6)\, i\, a_{1}\, a_{2}\, i_{1}}_{\ {\bf q},\, {\bf k}_{1},\, {\bf k}_{2},\, {\bf q}_{1}} = R^{\hspace{0.03cm}(6)\, \, i\, a_{2}\, a_{1}\, i_{1}}_{\ {\bf q},\, {\bf k}_{2},\, {\bf k}_{1},\, {\bf q}_{1}},
\notag\\[2.5ex]
&S^{\hspace{0.03cm}(2)\, i\; i_{1}\, i_{2}\, i_{3}}_{\ {\bf q},\, {\bf q}_{1},\, {\bf q}_{2},\, {\bf q}_{3}} = - S^{\hspace{0.03cm}(2)\, i\; i_{1}\, i_{3}\, i_{2}}_{\ {\bf q},\, {\bf q}_{1},\, {\bf q}_{3},\, {\bf q}_{2}},
\quad
&S^{\hspace{0.03cm}(3)\, i\; i_{1}\, i_{2}\, i_{3}}_{\ {\bf q},\, {\bf q}_{1},\, {\bf q}_{2},\, {\bf q}_{3}} = - S^{\hspace{0.03cm}(3)\, i\; i_{2}\, i_{1}\, i_{3}}_{\ {\bf q},\, {\bf q}_{2},\, {\bf q}_{1},\, {\bf q}_{3}},
\notag
\end{array}
\vspace{0.1cm}
\]
\begin{align}
\hspace{1.6cm}&S^{\hspace{0.03cm}(1)\, i\; i_{1}\, i_{2}\, i_{3}}_{\ {\bf q},\, {\bf q}_{1},\, {\bf q}_{2},\, {\bf q}_{3}} = - S^{\hspace{0.03cm}(1)\, i\; i_{2}\, i_{1}\, i_{3}}_{\ {\bf q},\, {\bf q}_{2},\, {\bf q}_{1},\, {\bf q}_{3}}
= - S^{\hspace{0.03cm}(1)\, i\; i_{1}\, i_{3}\, i_{2}}_{\ {\bf q},\, {\bf q}_{1},\, {\bf q}_{3},\, {\bf q}_{2}} = S^{\hspace{0.03cm}(1)\, i\; i_{2}\, i_{3}\, i_{1}}_{\ {\bf q},\, {\bf q}_{2},\, {\bf q}_{3},\, {\bf q}_{1}} =\, \ldots\,,
\label{eq:3i}\\[1.5ex]
&S^{\hspace{0.03cm}(4)\, i\; i_{1}\, i_{2}\, i_{3}}_{\ {\bf q},\, {\bf q}_{1},\, {\bf q}_{2},\, {\bf q}_{3}} = - S^{\hspace{0.03cm}(4)\, i\; i_{2}\, i_{1}\, i_{3}}_{\ {\bf q},\, {\bf q}_{2},\, {\bf q}_{1},\, {\bf q}_{3}}
= - S^{\hspace{0.03cm}(4)\, i\; i_{1}\, i_{3}\, i_{2}}_{\ {\bf q},\, {\bf q}_{1},\, {\bf q}_{3},\, {\bf q}_{2}} = S^{\hspace{0.03cm}(4)\, i\; i_{2}\, i_{3}\, i_{1}}_{\ {\bf q},\, {\bf q}_{2},\, {\bf q}_{3},\, {\bf q}_{1}} =\, \ldots\,.
\label{eq:3o}
\end{align}
\indent Further, substituting the expansions (\ref{eq:3t}) and  (\ref{eq:3y}) into the system of the canonicity conditions (\ref{eq:3e}) and (\ref{eq:3r}), we obtain rather nontrivial integral relations connecting various coefficient functions among themselves. A complete list of the integral relations connecting the coefficient functions of the second and third orders is given in Appendices \ref{appendix_C} and \ref{appendix_D}. Here, we have provided only algebraic relations for the second-order coefficient functions:
\begin{equation}
\begin{split}
&V^{(2)\, a\, a_{1}\, a_{2}}_{\ {\bf k},\, {\bf k}_{1},\, {\bf k}_{2}} = -\hspace{0.01cm}2\hspace{0.03cm}V^{\,\ast\hspace{0.03cm}(1)\, a_{2}\, a_{1}\, a}_{\ {\bf k}_{2},\, {\bf k}_{1},\, {\bf k}},
\quad
V^{(3)\, a\, a_{1}\, a_{2}}_{\ {\bf k},\, {\bf k}_{1},\, {\bf k}_{2}} = V^{(3)\, a_{1}\, a\, a_{2}}_{\ {\bf k}_{1},\, {\bf k},\, 
{\bf k}_{2}},\\[1ex]
&Q^{(1)\, i_{1}\, a\, i_{2}}_{\ {\bf q}_{1},\, {\bf k},\, {\bf q}_{2}} = -\hspace{0.01cm}F^{\,\ast\hspace{0.03cm} (2)\, a\, i_{2}\, i_{1}}_{\ {\bf k},\, {\bf q}_{2},\, {\bf q}_{1}},
\qquad
Q^{(2)\, i_{1}\, a\, i_{2}}_{\ {\bf q}_{1},\, {\bf k},\, {\bf q}_{2}} = 2\hspace{0.01cm} F^{\,\ast\hspace{0.03cm} (1)\, a\, i_{1}\, i_{2}}_{\ {\bf k},\, {\bf q}_{1},\, {\bf q}_{2}},
\\[1ex]
&Q^{(3)\, i_{1}\, a\, i_{2}}_{\ {\bf q}_{1},\, {\bf k},\, {\bf q}_{2}} =  F^{(2)\, a\, i_{1}\, i_{2}}_{\ {\bf k},\, {\bf q}_{1},\, {\bf q}_{2}},
\qquad\quad\,
Q^{(4)\, i_{1}\, a\, i_{2}}_{\ {\bf q}_{1},\, {\bf k},\, {\bf q}_{2}} =  2\hspace{0.01cm}F^{(3)\, a\, i_{1}\, i_{2}}_{\ {\bf k},\, {\bf q}_{1},\, {\bf q}_{2}}.
\end{split}
\label{eq:3p}
\end{equation}
Let us note in passing that a useful consequence of the first relations in (\ref{eq:3u}) and (\ref{eq:3p}) is
\begin{equation}
V^{(2)\, a\, a_{1}\, a_{2}}_{\ {\bf k},\, {\bf k}_{1},\, {\bf k}_{2}} = V^{(2)\, a_{1}\, a\, a_{2}}_{\ {\bf k}_{1},\, {\bf k},\, {\bf k}_{2}}.
\label{eq:3a}
\end{equation}


\section{Eliminating `non-essential' Hamiltonian $H^{(3)}$}
\setcounter{equation}{0}
\label{section_4}

In this section we consider the issue of eliminating the third-order interaction Hamiltonian $H^{(3)}$, Eq.\,(\ref{eq:2f}) upon switching from the original bosonic and fermionic functions $a^{\hspace{0.03cm}a}_{\bf k}$ and $b^{\,i}_{\bf q}$ to the new functions $c^{\hspace{0.03cm}a}_{\bf k}$ and $f^{\,i}_{\bf q}$ as a result of the canonical transformations (\ref{eq:3t}) and (\ref{eq:3y}). To achieve this, we substitute the expansions (\ref{eq:3t}) and (\ref{eq:3y}) into the 
free-field Hamiltonian $H^{(0)}$, given by the expression (\ref{eq:2d}) and keep only the terms cubic in $c^{\!\phantom{\ast}a}_{{\bf k}}$ and $f^{\,i}_{\bf q}$. As a consequence of performing appropriate symmetrization or antisymmetrization we get the following expression:
\begin{align}
 \int\frac{d\hspace{0.02cm} {\bf k}\hspace{0.03cm} d\hspace{0.02cm} {\bf k}_{1}\hspace{0.02cm}  d\hspace{0.02cm} {\bf k}_{2}}{(2\pi)^{9}}
\Bigl\{\omega^{l}_{{\bf k}}
&\Bigl(V^{(1)\, a\, a_{1}\, a_{2}}_{\ {\bf k},\, {\bf k}_{1},\, {\bf k}_{2}}\, 
c^{\ast\hspace{0.03cm} a}_{{\bf k}}\, c^{\,a_{1}}_{{\bf k}_{1}}\, c^{\,a_{2}}_{{\bf k}_{2}}
\,+\,
V^{\,\ast\,(1)\, a\, a_{1}\, a_{2}}_{\ {\bf k},\, {\bf k}_{1},\, {\bf k}_{2}}\, 
c^{\!\phantom{\ast}a}_{{\bf k}}\, c^{\ast\,a_{1}}_{{\bf k}_{1}}\hspace{0.03cm} c^{\ast\,a_{2}}_{{\bf k}_{2}}
\Bigr)
\notag\\[1ex]
+\;  
\frac{1}{2}\,(\omega^{l}_{{\bf k}_{1}} + \omega^{l}_{{\bf k}_{2}})
&\Bigl(V^{(2)\, a_{1}\, a_{2}\, a}_{\ {\bf k}_{1},\, {\bf k}_{2},\, {\bf k}}\, 
c^{\!\phantom{\ast}\;\! a}_{{\bf k}}\, c^{\ast\,a_{1}}_{{\bf k}_{1}}\hspace{0.03cm} c^{\ast\,a_{2}}_{{\bf k}_{2}}
\,+\,
V^{\,\ast\,(2)\, a_{1}\, a_{2}\, a}_{\ {\bf k}_{1},\, {\bf k}_{2},\, {\bf k}}\, 
 c^{\ast\,a}_{{\bf k}} \hspace{0.03cm} c^{\!\phantom{\ast} a_{1}}_{{\bf k}_{1}}\hspace{0.03cm} c^{a_{2}}_{{\bf k}_{2}}
\Bigr)
\notag\\[1ex]
+\; 
\frac{1}{3}\,(\omega^{l}_{{\bf k}} + \omega^{l}_{{\bf k}_{1}}\! + \omega^{l}_{{\bf k}_{2}})
&\Bigl(V^{(3)\, a\, a_{1}\, a_{2}}_{\ {\bf k},\, {\bf k}_{1},\, {\bf k}_{2}}\,
c^{\ast\ \!a}_{{\bf k}}\,c^{\ast\,a_{1}}_{{\bf k}_{1}}\hspace{0.03cm} c^{\ast\,a_{2}}_{{\bf k}_{2}}
\,+\,
V^{\,\ast\,(3)\, a\, a_{1}\, a_{2}}_{\ {\bf k},\, {\bf k}_{1},\, {\bf k}_{2}}\, 
c^{\!\phantom{\ast}a}_{{\bf k}}\, c^{a_{1}}_{{\bf k}_{1}}\, c^{a_{2}}_{{\bf k}_{2}}
\Bigr)\!\Bigr\} 
\label{eq:4q}\\[2ex]
+\int\frac{d\hspace{0.02cm}{\bf k}_{1}\hspace{0.02cm} d{\bf q}\hspace{0.03cm}  d{\bf q}_{1}}{(2\pi)^{9}}
\Bigl\{
-\Bigl(\omega^{l}_{{\bf k}_{1}}&F^{\,\ast\,(1)\, a_{1}\, i\, i_{1}}_{\ {\bf k}_{1},\, {\bf q},\, {\bf q}_{1}}
-\,
\frac{1}{2}\,\Bigl[\,\omega^{-}_{{\bf q}}\hspace{0.03cm}Q^{(2)\, i\, a_{1}\, i_{1}}_{\ {\bf q},\, {\bf k}_{1},\, {\bf q}_{1}}
- \omega^{-}_{{\bf q}_{1}}Q^{(2)\, i_{1}\, a_{1}\, i}_{\ {\bf q}_{1},\, {\bf k}_{1},\, {\bf q}}
\Bigr]\Bigr) 
c^{\, a_1}_{{\bf k}_1}\hspace{0.03cm} f^{\,\ast\, i}_{{\bf q}}\hspace{0.03cm} f^{\,{\ast}\,i_{1}}_{{\bf q}_{1}}
\notag\\[0.7ex]
+\;
\Bigl(\omega^{l}_{{\bf k}_{1}}&F^{(1)\, a_{1}\, i\, i_{1}}_{\ {\bf k}_{1},\, {\bf q},\, {\bf q}_{1}}
+\,
\frac{1}{2}\,\Bigl[\,\omega^{-}_{{\bf q}}\hspace{0.03cm}Q^{\hspace{0.03cm}\ast\,(2)\, i_{1}\, a_{1}\, i}_{\ {\bf q}_{1},\, {\bf k}_{1},\, {\bf q}}
\,-\, 
\omega^{-}_{{\bf q}_{1}}Q^{\hspace{0.03cm}\ast\,(2)\, i\, a_{1}\, i_{1}}_{\ {\bf q},\, {\bf k}_{1},\, {\bf q}_{1}}
\Bigr]\Bigr) 
c^{\,\ast\, a_1}_{{\bf k}_1} f^{\; i}_{{\bf q}}\hspace{0.03cm} f^{\; i_{1}}_{{\bf q}_{1}}
\notag\\[2ex]
+\;
\Bigl(\omega^{l}_{{\bf k}_{1}}&F^{\,\ast\,(2)\, a_{1}\, i_{1}\hspace{0.02cm} i}_{\ {\bf k}_{1},\, {\bf q}_{1},\, {\bf q}}
\,+\,
\omega^{-}_{{\bf q}}\hspace{0.03cm}Q^{(1)\, i\, a_{1}\, i_{1}}_{\ {\bf q},\, {\bf k}_{1},\, {\bf q}_{1}}
\,+\, 
\omega^{-}_{{\bf q}_{1}}Q^{\hspace{0.03cm}\ast\,(3)\, i_{1}\, a_{1}\, i}_{\ {\bf q}_{1},\, {\bf k}_{1},\, {\bf q}}
\Bigr) 
c^{\,a_1}_{{\bf k}_1}\hspace{0.03cm} f^{\,\ast\, i}_{{\bf q}}\hspace{0.03cm} f^{\; i_{1}}_{{\bf q}_{1}}
\notag\\[0.7ex]
+\;
\Bigl(\omega^{l}_{{\bf k}_{1}}&F^{(2)\, a_{1}\, i\, i_{1}}_{\ {\bf k}_{1},\, {\bf q},\, {\bf q}_{1}}
\,+\,
\omega^{-}_{{\bf q}}\hspace{0.03cm}Q^{(3)\, i\, a_{1}\, i_{1}}_{\ {\bf q},\, {\bf k}_{1},\, {\bf q}_{1}}
\,+\, 
\omega^{-}_{{\bf q}_{1}}Q^{\hspace{0.03cm}\ast\,(1)\, i_{1}\, a_{1}\, i}_{\ {\bf q}_{1},\, {\bf k}_{1},\, {\bf q}}
\Bigr) 
c^{\,\ast\, a_{1}}_{{\bf k}_1}\hspace{0.02cm} f^{\,\ast\, i}_{{\bf q}}\hspace{0.03cm} f^{\; i_{1}}_{{\bf q}_{1}}
\notag\\[2ex]
-\;
\Bigl(\hspace{0.03cm}\omega^{l}_{{\bf k}_{1}}\hspace{0.03cm}
&F^{\,\ast\,(3)\, a_{1}\, i\, i_{1}}_{\ {\bf k}_{1},\, {\bf q},\, {\bf q}_{1}}
-\,
\frac{1}{2}\,\Bigl[\,\omega^{-}_{{\bf q}}\hspace{0.03cm}Q^{\hspace{0.03cm}\ast\,(4)\, i_{1}\, a_{1}\, i}_{\ {\bf q}_{1},\, {\bf k}_{1},\, {\bf q}}
\,-\, 
\omega^{-}_{{\bf q}_{1}}Q^{\hspace{0.03cm}\ast\,(4)\, i\, a_{1}\, i_{1}}_{\ {\bf q},\, {\bf k}_{1},\, {\bf q}_{1}}
\Bigr]\Bigr) 
c^{\, a_1}_{{\bf k}_1}\hspace{0.03cm} f^{\,i}_{{\bf q}}\, f^{\,\,\,i_{1}}_{{\bf q}_{1}}
\notag\\[0.7ex]
+\,
\Bigl(\omega^{l}_{{\bf k}_{1}\!}&F^{(3)\, a_{1}\, i\, i_{1}}_{\ {\bf k}_{1},\, {\bf q},\, {\bf q}_{1}}
+\,
\frac{1}{2}\,\Bigl[\,\omega^{-}_{{\bf q}}\hspace{0.03cm}Q^{(4)\, i\, a_{1}\, i_{1}}_{\ {\bf q},\, {\bf k}_{1},\, {\bf q}_{1}}
\,-\, 
\omega^{-}_{{\bf q}_{1}}Q^{(4)\, i_{1}\, a_{1}\, i}_{\ {\bf q}_{1},\, {\bf k}_{1},\, {\bf q}}
\Bigr]\Bigr) 
c^{{\ast}\, a_1}_{{\bf k}_1}\hspace{0.03cm} f^{\,\ast\, i}_{{\bf q}}\hspace{0.03cm} f^{\,\ast\,i_{1}}_{{\bf q}_{1}}
\Bigr\}.
\notag
\end{align}
Here, we have taken into account the symmetry conditions (\ref{eq:3u}) and (\ref{eq:3a}). In the Hamiltonian $H^{(3)}$, Eq.\,(\ref{eq:2f}), we perform the replacements:  $a^{a}_{\bf k}\rightarrow c^{a}_{\bf k}$ and $b^{\,i}_{\bf q} \rightarrow f^{\,i}_{\bf q}$. Adding the expression thus obtained to (\ref{eq:4q}), collecting similar terms and using the relations (\ref{eq:3p}), finally we obtain an explicit form of the coefficient functions in the quadratic part of the canonical transformations (\ref{eq:3t}) and (\ref{eq:3y}), which exclude the cubic terms of the interaction Hamiltonian:
\begin{equation}
\left\{\!\!\!
\begin{array}{ll}
&V^{(1)\, a\, a_{1}\, a_{2}}_{\ {\bf k},\, {\bf k}_{1},\, {\bf k}_{2}} 
=
-\hspace{0.02cm}\displaystyle\frac{{\mathcal V}^{\, a\, a_{1}\, a_{2}}_{\ {\bf k},\, {\bf k}_{1},\, {\bf k}_{2}}}
{\omega^{l}_{{\bf k}} - \omega^{l}_{{\bf k}_{1}} - \omega^{l}_{{\bf k}_{2}}}\,
(2\pi)^{3}\delta({\bf k}-{\bf k}_{1}-{\bf k}_{2}), 
\\[3ex]
&V^{(3)\, a\, a_{1}\, a_{2}}_{\ {\bf k},\, {\bf k}_{1},\, {\bf k}_{2}}
= 
-\hspace{0.02cm}\displaystyle\frac{{\mathcal U}^{\hspace{0.03cm}*\, a\, a_{1}\, a_{2}}_{\ {\bf k},\, {\bf k}_{1},\, {\bf k}_{2}}}{\omega^{l}_{{\bf k}} + \omega^{l}_{{\bf k}_{1}} + \omega^{l}_{{\bf k}_{2}}}\,
(2\pi)^{3}\delta({\bf k}+{\bf k}_{1}+{\bf k}_{2}),
\end{array}\
\right.
\label{eq:4w}
\end{equation}
\vspace{0.15cm}
\begin{equation}
\left\{\!\!\!
\begin{array}{ll}
&F^{(1)\, a_{1}\, i\;  i_{1}}_{\;{\bf k}_1,\, {\bf q},\, {\bf q}_{1}}
\,=\,
\displaystyle\frac{{\mathcal G}^{\hspace{0.03cm}\ast\; a_{1}\, i\;  i_{1}}_{\;{\bf k}_1,\, {\bf q},\, {\bf q}_{1}}}
{\omega^{l}_{{\bf k}_{1}} - \omega^{-}_{{\bf q}} -  \omega^{-}_{{\bf q}_{1}}}\
(2\pi)^{3}\hspace{0.03cm}\delta({\bf k}_1 - {\bf q} - {\bf q}_{1}),
\\[3ex]
&F^{(2)\, a_{1}\, i\;  i_{1}}_{\;{\bf k}_1,\, {\bf q},\, {\bf q}_{1}}
=
-\hspace{0.02cm}\displaystyle\frac{{\mathcal P}^{\hspace{0.03cm}\ast\, a_{1}\, i_{1}\,  i}_{\;{\bf k}_1,\, {\bf q}_{1},\, {\bf q}}}
{ \omega^{l}_{{\bf k}_{1}} - \omega^{-}_{{\bf q}_{1}} + \omega^{-}_{{\bf q}}}\
(2\pi)^{3}\hspace{0.03cm}\delta({\bf k}_{1} - {\bf q}_{1} + {\bf q}),
\\[3ex]
&F^{(3)\, a_{1}\, i\;  i_{1}}_{\;{\bf k}_1,\, {\bf q},\, {\bf q}_{1}}
\,=\,
\displaystyle\frac{{\mathcal K}^{\hspace{0.03cm}\ast\; a_{1}\, i\;  i_{1}}_{\;{\bf k}_1,\, {\bf q},\, {\bf q}_{1}}}
{\omega^{l}_{{\bf k}_{1}} + \omega^{-}_{{\bf q}} + \omega^{-}_{{\bf q}_{1}}}\
(2\pi)^{3}\hspace{0.03cm}\delta({\bf k}_1 + {\bf q} + {\bf q}_{1}).
\end{array}\
\right.
\label{eq:4e}
\end{equation}
The coefficients $V^{(2)}$ and $Q^{(n)},\,n = 1,\,2,\,3,\,4$ are found from Eq.\,(\ref{eq:3p}). We have previously obtained the relations (\ref{eq:4w}) in \cite{markov_2020}. These expressions imply that due to specific character of the dispersion equations for bosonic (\ref{eq:2e}) and for fermionic (\ref{eq:2u}) excitations in the hot quark-gluon plasma, the resonance conditions for three-wave processes with plasmons
\[
\left\{
\begin{array}{ll}
{\bf k} = {\bf k}_{1} + {\bf k}_{2}, \\[1.5ex]
\omega^{l}_{{\bf k}} = \omega^{l}_{{\bf k}_{1}} + \omega^{l}_{{\bf k}_{2}},
\end{array}
\right.
\quad
\left\{
\begin{array}{ll}
{\bf k} + {\bf k}_{1} + {\bf k}_{2} = 0, \\[1.5ex]
\omega^{l}_{{\bf k}} + \omega^{l}_{{\bf k}_{1}} + \omega^{l}_{{\bf k}_{2}} = 0,
\end{array}\
\right.
\]
and with plasmons and plasminos
\[
\left\{
\begin{array}{ll}
{\bf q}+{\bf q}_{1}+{\bf k}_{1} = 0,\\[1.5ex]
\omega^{-}_{{\bf q}} + \omega^{-}_{{\bf q}_{1}} + \omega^{l}_{{\bf k}_{1}} = 0,
\end{array}
\right.
\quad
\left\{
\begin{array}{ll}
{\bf q}={\bf q}_{1} + {\bf k}_{1}, \\[1.5ex]
\omega^{-}_{{\bf q}} = \omega^{-}_{{\bf q}_{1}} + \omega^{l}_{{\bf k}_{1}},
\end{array}\
\right.
\quad
\left\{
\begin{array}{ll}
{\bf k}_{1} = {\bf q} + {\bf q}_{1}, \\[1.5ex]
\omega^{l}_{{\bf k}_{1}} = \omega^{-}_{{\bf q}} + \omega^{-}_{{\bf q}_{1}}.
\end{array}\
\right.
\]
have no solutions. In other words, the decay and merge processes involving three collective excitations that lie on the resonance surface $\omega^{l} = \omega^{l}_{\bf k}$ and $\omega^{-} = \omega^{-}_{\bf q}$ are forbidden.


\section{Effective fourth-order Hamiltonian. Elastic scatte\-ring of plasmino off plasmon}
\setcounter{equation}{0}
\label{section_5}

\begin{figure}[hbtp]
\begin{center}
\includegraphics[width=0.95\textwidth]{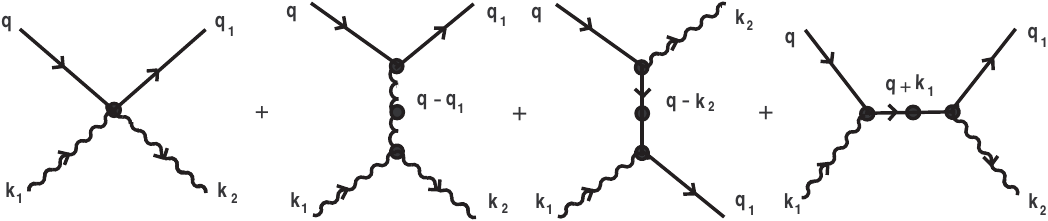}
\end{center}
\vspace{-0.5cm}
\caption{The matrix element for the elastic scattering process of plasmino off plasmons. The straight and wave lines denote soft-quark and soft-gluon excitations, correspondingly}
\label{fig1}
\end{figure}

Now we can move to the construction of an explicit form of two effective fourth-order Hamiltoni\-ans, which describe the elastic scattering of plasmino off plasmino and plasmino off plasmon. In this section we consider the derivation of the 
effective Hamiltonian for the second scattering process. In terms of the original variables $a^{a}_{\bf k}$ and $b^{\,i}_{\bf q}$, the Hamiltonian for the scattering process is defined by the first term on the right-hand side of (\ref{eq:2g}). In this term we make the substitution $a^{a}_{\bf k}\rightarrow c^{a}_{\bf k}$ and $b^{\,i}_{\bf q} \rightarrow f^{\,i}_{\bf q}$. Further we define all similar terms of fourth-order in $f^{\,\ast\, i}_{{\bf q}} f^{\;i_{1}}_{{\bf q}_{1}} c^{\ast\ \!\!a_{1}}_{{\bf k}_{1}} c^{\!\phantom{\ast} a_{2}}_{{\bf k}_{2}}$ from the free-field Hamiltonian $H^{(0)}$, Eq.\,(\ref{eq:2d}), and from the Hamiltonian $H^{(3)}$, Eq.\,(\ref{eq:2f}) to be developed under the canonical transformations (\ref{eq:3t}) and (\ref{eq:3y}). Taking into account the relations (\ref{eq:3p}) we have the following contributions from $H^{(0)}$:
\begin{equation}
\int\!\frac{d{\bf q}\, d{\bf q}_{1}\hspace{0.03cm} d\hspace{0.02cm}{\bf k}_{1}\hspace{0.03cm} d\hspace{0.02cm}{\bf k}_{2}}{(2\pi)^{12}}\, 
\biggl\{\!\Bigl(\omega^{l}_{{\bf k}_{1}}J^{\hspace{0.03cm}(2)\, a_{1}\hspace{0.03cm} a_{2}\, i\; i_{1}}_{\ {\bf k}_{1},\, {\bf k}_{2},\, {\bf q},\, {\bf q}_{1}}
+\,
\omega^{l}_{{\bf k}_{2}}J^{\hspace{0.03cm}\ast\hspace{0.03cm}(2)\, a_{2}\, a_{1}\, i_{1}\, i}_{\ {\bf k}_{2},\, {\bf k}_{1},\, {\bf q}_{1},\, 
{\bf q}}
+\,
\omega^{-}_{{\bf q}}R^{\hspace{0.03cm}(2)\, i\, a_{1}\, a_{2}\, i_{1}}_{\ {\bf q},\, {\bf k}_{1},\, {\bf k}_{2},\, {\bf q}_{1}}
+\,
\omega^{-}_{{\bf q}_{1}}R^{\hspace{0.03cm}\ast\hspace{0.03cm}(2)\, i_{1}\, a_{2}\, a_{1}\, i}_{\ {\bf q}_{1},\, {\bf k}_{2},\, {\bf k}_{1},\,  {\bf q}}
\Bigr)
\label{eq:5q}
\end{equation}
\begin{align}
-\;
2\!\int\!\frac{d\hspace{0.02cm}{\bf k}^{\prime}}{(2\pi)^{3}}\,\omega^{l}_{{\bf k}^{\prime}}\Bigl[\hspace{0.03cm}
&V^{(1)\hspace{0.03cm} a_{1}\, a_{2}\, a^{\prime}}_{\ {\bf k}_{1},\, {\bf k}_{2},\, {\bf k}^{\prime}}\hspace{0.01cm} 
F^{(2)\, a^{\prime}\, i\, i_{1}}_{\ {\bf k}^{\prime},\, {\bf q},\, {\bf q}_{1}}
+
V^{\hspace{0.03cm}\ast\hspace{0.03cm} (1)\hspace{0.03cm} a_{2}\, a_{1}\, a^{\prime}}_{\ {\bf k}_{2},\, {\bf k}_{1},\, {\bf k}^{\prime}}\hspace{0.01cm} F^{\hspace{0.03cm}\ast\hspace{0.03cm}(2)\, a^{\prime}\, i_{1}\, i}_{\ {\bf k}^{\prime},\, {\bf q}_{1},\, {\bf q}}\hspace{0.03cm}\Bigr]
\notag\\[1ex]
+\,
\!\int\!\frac{d{\bf q}^{\prime}}{(2\pi)^{3}}\,\omega^{-}_{{\bf q}^{\prime}}\Bigl[\hspace{0.02cm}
4\hspace{0.02cm}&F^{(1)\, a_{1}\, i_{1}\, i^{\prime}}_{\ {\bf k}_{1},\, {\bf q}_{1},\, {\bf q}^{\prime}}\, 
F^{\hspace{0.03cm}\ast\,(1)\hspace{0.03cm} a_{2}\, i^{\prime}\, i}_{\ {\bf k}_{2},\,{\bf q}^{\prime},\, {\bf q}}
+
F^{(2)\, a_{1}\, i\, i^{\prime}}_{\ {\bf k}_{1},\, {\bf q},\, {\bf q}^{\prime}}\, 
F^{\hspace{0.03cm}\ast\hspace{0.03cm}(2)\, a_{2}\, i_{1}\, i^{\prime}}_{\ {\bf k}_{2},\, {\bf q}_{1},\, {\bf q}^{\prime}}
\notag\\[1ex]
&\hspace{0.1cm}+
F^{(2)\, a_{1}\, i^{\prime}\, i_{1}}_{\ {\bf k}_{1},\, {\bf q}^{\prime},\, {\bf q}_{1}}\, 
F^{\hspace{0.03cm}\ast\hspace{0.03cm}(2)\hspace{0.03cm} a_{2}\, i^{\prime}\, i}_{\ {\bf k}_{2}, \, {\bf q}^{\prime},\, {\bf q}}
+
4\hspace{0.02cm}F^{(3)\, a_{1}\, i^{\prime}\, i}_{\ {\bf k}_{1},\, {\bf q}^{\prime},\, {\bf q}}\, 
F^{\hspace{0.03cm}\ast\,(3)\, a_{2}\, i_{1}\, i^{\prime}}_{\ {\bf k}_{2},\, {\bf q}_{1},\, {\bf q}^{\prime}}\hspace{0.02cm}\Bigr]\biggr\}
f^{\hspace{0.03cm} \ast\hspace{0.03cm} i}_{{\bf q}} f^{\;i_{1}}_{{\bf q}_{1}} c^{\ast\ \!\!a_{1}}_{{\bf k}_{1}} c^{\!\phantom{\ast} a_{2}}_{{\bf k}_{2}}. 
\notag 
\end{align}
Let us analyze the contribution in  parentheses in (\ref{eq:5q}). For this purpose we use the integral relations given in Appendix \ref{appendix_C}. To exclude the functions $J^{\hspace{0.03cm}\ast\hspace{0.03cm}(2)\, a_{2}\, a_{1}\, i_{1}\, i}_{\ {\bf k}_{2},\, {\bf k}_{1},\, {\bf q}_{1},\, {\bf q}}$, $R^{\hspace{0.03cm}(2)\, i\, a_{1}\, a_{2}\, i_{1}}_{\ {\bf q},\, {\bf k}_{1},\, {\bf k}_{2},\, {\bf q}_{1}}$ and $R^{\hspace{0.03cm}\ast\hspace{0.03cm}(2)\, i_{1}\, a_{2}\, a_{1}\, i}_{\ {\bf q}_{1},\, {\bf k}_{2},\, {\bf k}_{1},\,  {\bf q}}$ we use the relations (\ref{ap:C1b}),  (\ref{ap:C3a}) and (\ref{ap:C4b}), correspondingly. As a result we obtain
\begin{equation}
\omega^{l}_{{\bf k}_{1}}J^{\hspace{0.03cm}(2)\, a_{1}\, a_{2}\, i\, i_{1}}_{\ {\bf k}_{1},\, {\bf k}_{2},\, {\bf q},\, {\bf q}_{1}}
+\,
\omega^{l}_{{\bf k}_{2}}J^{\hspace{0.03cm}\ast\hspace{0.03cm}(2)\, a_{2}\, a_{1}\, i_{1}\, i}_{\ {\bf k}_{2},\, {\bf k}_{1},\, {\bf q}_{1},\, 
{\bf q}}
+\,
\omega^{-}_{{\bf q}}R^{\hspace{0.03cm}(2)\, i\, a_{1}\, a_{2}\, i_{1}}_{\ {\bf q},\, {\bf k}_{1},\, {\bf k}_{2},\, {\bf q}_{1}}
+\hspace{0.03cm}
\omega^{-}_{{\bf q}_{1}}R^{\hspace{0.03cm}\ast\hspace{0.03cm}(2)\, i_{1}\, a_{2}\, a_{1}\, i}_{\ {\bf q}_{1},\, {\bf k}_{2},\, {\bf k}_{1},\,  {\bf q}}
\label{eq:5w}
\end{equation}
\[
=
\bigl(\omega^{-}_{{\bf q}} + \omega^{l}_{{\bf k}_{1}} - \omega^{-}_{{\bf q}_{1}} - \omega^{l}_{{\bf k}_{2}}\bigr)
J^{\hspace{0.03cm}(2)\, a_{1}\, a_{2}\, i\, i_{1}}_{\ {\bf k}_{1},\, {\bf k}_{2},\, {\bf q},\, {\bf q}_{1}}
\]
\vspace{-0.3cm}
\begin{align}
-\,
2\,\bigl(\omega^{-}_{\bf q} - \omega^{-}_{{\bf q}_{1}}\bigr)
\!&\int\!\frac{d\hspace{0.02cm}{\bf k}^{\prime}}{(2\pi)^{3}}\Bigl[\hspace{0.03cm}
V^{(1)\,a_{1}\, a_{2}\, a^{\prime}}_{\ {\bf k}_{1},\, {\bf k}_{2},\, {\bf k}^{\prime}}\, F^{(2)\, a^{\prime}\, i\, i_{1}}_{\ {\bf k}^{\prime},\, {\bf q},\, {\bf q}_{1}}
-
V^{\hspace{0.03cm}\ast\,(1)\,a_{2}\, a_{1}\, a^{\prime}}_{\ {\bf k}_{2},\, {\bf k}_{1},\, {\bf k}^{\prime}}\, F^{\hspace{0.03cm}\ast\hspace{0.03cm}(2)\, a^{\prime}\, i_{1}\, i}_{\ {\bf k}^{\prime},\, {\bf q}_{1},\, {\bf q}}\hspace{0.03cm}\Bigr]
\notag\\[1ex]
-\;
\omega^{-}_{{\bf q}}\!&\int\!\frac{d{\bf q}^{\prime}}{(2\pi)^{3}}\,
\Bigl[\hspace{0.02cm}
4\hspace{0.01cm}F^{(1)\, a_{1}\, i_{1}\, i^{\prime}}_{\ {\bf k}_{1},\, {\bf q}_{1},\, {\bf q}^{\prime}}\, 
F^{\hspace{0.03cm}\ast\,(1)\, a_{2}\, i\, i^{\prime}}_{\ {\bf k}_{2},\, {\bf q},\, {\bf q}^{\prime}}
+
F^{(2)\, a_{1}\, i^{\prime}\, i_{1}}_{\ {\bf k}_{1},\, {\bf q}^{\prime},\, {\bf q}_{1}}\, 
F^{\hspace{0.03cm}\ast\hspace{0.03cm}(2)\, a_{2}\, i^{\prime}\, i}_{{\bf k}_{2},\, {\bf q}^{\prime},\, {\bf q}}\hspace{0.02cm}\Bigr]
\notag\\[1ex]
-\;
\omega^{-}_{{\bf q}_{1}}\!&\int\!\frac{d{\bf q}^{\prime}}{(2\pi)^{3}}\,
\Bigl[\hspace{0.02cm}
F^{(2)\, a_{1}\, i\, i^{\prime}}_{\ {\bf k}_{1},\, {\bf q},\, {\bf q}^{\prime}}\, 
F^{\hspace{0.03cm}\ast\hspace{0.03cm}(2)\, a_{2}\, i_{1}\, i^{\prime}}_{\ {\bf k}_{2},\, {\bf q}_{1},\, {\bf q}^{\prime}}
-
4\hspace{0.01cm}F^{(3)\, a_{1}\, i^{\prime}\, i}_{\ {\bf k}_{1},\, {\bf q}^{\prime},\, {\bf q}}\, 
F^{\hspace{0.03cm}\ast\,(3)\, a_{2}\, i_{1}\, i^{\prime}}_{{\bf k}_{2},\, {\bf q}_{1},\, {\bf q}^{\prime}}\hspace{0.02cm}\Bigr]
\notag\\[1ex]
-\;
\omega^{l}_{{\bf k}_{2}}\!&\int\!\frac{d{\bf q}^{\prime}}{(2\pi)^{3}}\,
\Bigl[\hspace{0.02cm}
4\hspace{0.01cm}F^{(1)\, a_{1}\, i_{1}\, i^{\prime}}_{\ {\bf k}_{1},\, {\bf q}_{1},\, {\bf q}^{\prime}}\, 
F^{\hspace{0.03cm}\ast\,(1)\, a_{2}\, i^{\prime}\, i}_{\ {\bf k}_{2},\,{\bf q}^{\prime},\, {\bf q}}
+
F^{(2)\, a_{1}\, i\, i^{\prime}}_{\ {\bf k}_{1},\, {\bf q},\, {\bf q}^{\prime}}\, 
F^{\hspace{0.03cm}\ast\hspace{0.03cm}(2)\, a_{2}\, i_{1}\, i^{\prime}}_{\ {\bf k}_{2},\, {\bf q}_{1},\, {\bf q}^{\prime}}
\notag\\[1ex]
&\hspace{2cm}-
F^{(2)\, a_{1}\, i^{\prime}\, i_{1}}_{\ {\bf k}_{1},\, {\bf q}^{\prime},\, {\bf q}_{1}}\, 
F^{\hspace{0.03cm}\ast\hspace{0.03cm}(2)\, a_{2}\, i^{\prime}\, i}_{\ {\bf k}_{2}, \, {\bf q}^{\prime},\, {\bf q}}
-
4\hspace{0.01cm}F^{(3)\, a_{1}\, i^{\prime}\, i}_{\ {\bf k}_{1},\, {\bf q}^{\prime},\, {\bf q}}\, 
F^{\hspace{0.03cm}\ast\,(3)\, a_{2}\, i_{1}\, i^{\prime}}_{{\bf k}_{2},\, {\bf q}_{1},\, {\bf q}^{\prime}}\hspace{0.02cm}\Bigr].
\notag 
\end{align}
\indent Further, we analyze the fourth-order contributions from the Hamiltonian $H^{(3)}$, Eq.\,(\ref{eq:2f}). Here, for the sake of convenience of the subsequent considerations we pass from the vertex functions ${\mathcal V},\, {\mathcal G},\, {\mathcal P}$ and ${\mathcal K}$ to the coefficient functions $V^{(1)}$ and $F^{(i)},\,i = 1,2,3$ by the rules (\ref{eq:4w}) and (\ref{eq:4e}). As a consequence we have
\begin{equation}
\int\!\frac{d{\bf q}\, d{\bf q}_{1}\hspace{0.03cm} d\hspace{0.02cm}{\bf k}_{1}\hspace{0.03cm} d\hspace{0.02cm}{\bf k}_{2}}{(2\pi)^{12}}\, 
f^{\,\ast\, i}_{{\bf q}} f^{\;i_{1}}_{{\bf q}_{1}} c^{\ast\ \!\!a_{1}}_{{\bf k}_{1}} c^{\!\phantom{\ast} a_{2}}_{{\bf k}_{2}}
\label{eq:5e}
\end{equation}
\begin{align}
\times\,\biggl\{
-\,
2\!\int\!\frac{d\hspace{0.02cm}{\bf k}^{\prime}}{(2\pi)^{3}}\,
\Bigl[
&\bigl(\omega^{l}_{{\bf k}_{1}} - \omega^{l}_{{\bf k}^{\prime}} - \omega^{l}_{{\bf k}_{2}}\bigr)
V^{(1)\,a_{1}\, a_{2}\, a^{\prime}}_{\ {\bf k}_{1},\, {\bf k}_{2},\, {\bf k}^{\prime}}\hspace{0.01cm} 
F^{(2)\, a^{\prime}\, i\, i_{1}}_{\ {\bf k}^{\prime},\, {\bf q},\, {\bf q}_{1}}
+
\bigl(\omega^{l}_{{\bf k}_{2}} - \omega^{l}_{{\bf k}^{\prime}} - \omega^{l}_{{\bf k}_{1}}\bigr)
V^{\hspace{0.03cm}\ast\,(1)\,a_{2}\, a_{1}\, a^{\prime}}_{\ {\bf k}_{2},\, {\bf k}_{1},\, {\bf k}^{\prime}}\hspace{0.01cm}
F^{\hspace{0.03cm}\ast\hspace{0.03cm}(2)\, a^{\prime}\, i_{1}\, i}_{\ {\bf k}^{\prime},\, {\bf q}_{1},\, {\bf q}}
\notag\\[1ex]
-\;
&\bigl(\omega^{l}_{{\bf k}^{\prime}} + \omega^{-}_{{\bf q}} - \omega^{-}_{{\bf q}_{1}}\bigr)
V^{(1)\,a_{1}\, a_{2}\, a^{\prime}}_{\ {\bf k}_{1},\, {\bf k}_{2},\, {\bf k}^{\prime}}\hspace{0.01cm}
F^{(2)\, a^{\prime}\, i\, i_{1}}_{\ {\bf k}^{\prime},\, {\bf q},\, {\bf q}_{1}}
-
\bigl(\omega^{l}_{{\bf k}^{\prime}} - \omega^{-}_{{\bf q}} + \omega^{-}_{{\bf q}_{1}}\bigr)
V^{\hspace{0.03cm}\ast\,(1)\,a_{2}\, a_{1}\, a^{\prime}}_{\ {\bf k}_{2},\, {\bf k}_{1},\, {\bf k}^{\prime}}\hspace{0.01cm} F^{\hspace{0.03cm}\ast\hspace{0.03cm}(2)\, a^{\prime}\, i_{1}\, i}_{\ {\bf k}^{\prime},\, {\bf q}_{1},\, {\bf q}}
\,\Bigr]
\notag\\[1ex]
+\,
\!\int\!\frac{d{\bf q}^{\prime}}{(2\pi)^{3}}\,
\Bigl[&\bigl(\omega^{l}_{{\bf k}_{2}} - \omega^{-}_{{\bf q}^{\prime}} + \omega^{-}_{{\bf q}_{1}}\bigr)
F^{(2)\, a_{1}\, i\, i^{\prime}}_{\ {\bf k}_{1},\, {\bf q},\, {\bf q}^{\prime}}\, 
F^{\hspace{0.03cm}\ast\hspace{0.03cm}(2)\, a_{2}\, i_{1}\, i^{\prime}}_{\ {\bf k}_{2},\, {\bf q}_{1},\, {\bf q}^{\prime}}
+
\bigl(\omega^{l}_{{\bf k}_{1}} - \omega^{-}_{{\bf q}^{\prime}} + \omega^{-}_{{\bf q}}\bigr)
F^{(2)\, a_{1}\, i\, i^{\prime}}_{\ {\bf k}_{1},\, {\bf q},\, {\bf q}^{\prime}}\, 
F^{\hspace{0.03cm}\ast\hspace{0.03cm}(2)\, a_{2}\, i_{1}\, i^{\prime}}_{\ {\bf k}_{2},\, {\bf q}_{1},\, {\bf q}^{\prime}}
\notag\\[1ex]
-\;
&\bigl(\omega^{l}_{{\bf k}_{1}} + \omega^{-}_{{\bf q}^{\prime}} - \omega^{-}_{{\bf q}_{1}}\bigr)
F^{(2)\, a_{1}\, i^{\prime}\, i_{1}}_{\ {\bf k}_{1},\, {\bf q}^{\prime},\, {\bf q}_{1}}\, 
F^{\hspace{0.03cm}\ast\hspace{0.03cm}(2)\, a_{2}\, i^{\prime}\, i}_{\ {\bf k}_{2}, \, {\bf q}^{\prime},\, {\bf q}}
-
\bigl(\omega^{l}_{{\bf k}_{2}} - \omega^{-}_{{\bf q}} + \omega^{-}_{{\bf q}^{\prime}}\bigr)
F^{(2)\, a_{1}\, i^{\prime}\, i_{1}}_{\ {\bf k}_{1},\, {\bf q}^{\prime},\, {\bf q}_{1}}\, 
F^{\hspace{0.03cm}\ast\hspace{0.03cm}(2)\, a_{2}\, i^{\prime}\, i}_{\ {\bf k}_{2}, \, {\bf q}^{\prime},\, {\bf q}}
\,\Bigr]
\notag\\[1ex]
-\,4
\!\int\!\frac{d{\bf q}^{\prime}}{(2\pi)^{3}}\,
\Bigl[&\bigl( \omega^{-}_{{\bf q}} - \omega^{l}_{{\bf k}_{2}} + \omega^{-}_{{\bf q}^{\prime}}\bigr)
F^{(1)\, a_{1}\, i_{1}\, i^{\prime}}_{\ {\bf k}_{1},\, {\bf q}_{1},\, {\bf q}^{\prime}}\, 
F^{\hspace{0.03cm}\ast\,(1)\, a_{2}\, i^{\prime}\, i}_{\ {\bf k}_{2},\,{\bf q}^{\prime},\, {\bf q}}
-
\bigl(\omega^{-}_{{\bf q}^{\prime}} - \omega^{l}_{{\bf k}_{1}} + \omega^{-}_{{\bf q}_{1}}\bigr)
F^{(1)\, a_{1}\, i^{\prime}\, i_{1}}_{\ {\bf k}_{1},\, {\bf q}^{\prime},\, {\bf q}_{1}}\, 
F^{\hspace{0.03cm}\ast\,(1)\, a_{2}\, i^{\prime}\, i}_{\ {\bf k}_{2}, \, {\bf q}^{\prime},\, {\bf q}}
\notag\\[1ex]
+\;
&\bigl(\omega^{l}_{{\bf k}_{2}} + \omega^{-}_{{\bf q}^{\prime}} + \omega^{-}_{{\bf q}_{1}}\bigr)
F^{(3)\, a_{1}\, i^{\prime}\, i}_{\ {\bf k}_{1},\, {\bf q}^{\prime},\, {\bf q}}\, 
F^{\hspace{0.03cm}\ast\,(3)\, a_{2}\, i_{1}\, i^{\prime}}_{\ {\bf k}_{2},\, {\bf q}_{1},\, {\bf q}^{\prime}}
+
\bigl(\omega^{l}_{{\bf k}_{1}} + \omega^{-}_{{\bf q}^{\prime}} + \omega^{-}_{{\bf q}}\bigr)
F^{(3)\, a_{1}\, i^{\prime}\, i}_{\ {\bf k}_{1},\, {\bf q}^{\prime},\, {\bf q}}\, 
F^{\hspace{0.03cm}\ast\,(3)\, a_{2}\, i_{1}\, i^{\prime}}_{\ {\bf k}_{2},\, {\bf q}_{1},\, {\bf q}^{\prime}}
\,\Bigr]\biggr\}.
\notag
\end{align}
Here we did not collect similar terms since the given expression is more suitable on the last step of our calculations. Let us put together (\ref{eq:5q}), (\ref{eq:5e}) and the first contribution in (\ref{eq:2g}) taking into account (\ref{eq:5w}) and collect similar terms. Most terms are mutually reduced. The final step is to pass from the functions $V^{(1)}$ and $F^{(i)},\,i = 1,2,3$ to the ``physical'' functions ${\mathcal V},\, {\mathcal G},\, {\mathcal P}$ and ${\mathcal K}$ by the rules (\ref{eq:4w}) and (\ref{eq:4e}). Thereby we result in the effective fourth-order Hamiltonian describing the elastic scattering process of plasmino off plasmon:  
\begin{equation}
{\mathcal H}^{(4)}_{qg\rightarrow qg} 
=
\int\frac{d{\bf q}\, d{\bf q}_{1}\hspace{0.03cm} d\hspace{0.02cm}{\bf k}_{1}\hspace{0.03cm} d\hspace{0.02cm}{\bf k}_{2}}{(2\pi)^{12}}
\label{eq:5r}
\end{equation}
\[
\times\,\Bigl\{\!\hspace{0.02cm}
\bigl(\omega^{-}_{{\bf q}} + \omega^{l}_{{\bf k}_{1}} - \omega^{-}_{{\bf q}_{1}} - \omega^{l}_{{\bf k}_{2}}\bigr)
J^{\hspace{0.03cm}(2)\, a_{1}\, a_{2}\, i\, i_{1}}_{\, {\bf k}_{1},\, {\bf k}_{2},\, {\bf q},\, {\bf q}_{1}}
\hspace{0.03cm}+\hspace{0.03cm}
\widetilde{T}^{(2)\, i\, i_{1}\, a_{1}\, a_{2}}_{\, {\bf q},\, {\bf q}_{1},\, {\bf k}_{1},\, {\bf k}_{2}}\, 
(2\pi)^{3}\hspace{0.03cm}\delta({\bf q} + {\bf k}_1 - {\bf q}_{1} - {\bf k}_{2})
\Bigr\}
f^{\,\ast\, i}_{{\bf q}} f^{\;i_{1}}_{{\bf q}_{1}}\, c^{\ast\ \!\!a_{1}}_{{\bf k}_{1}} c^{\hspace{0.03cm}a_{2}}_{{\bf k}_{2}},
\]
where the effective amplitude $\widetilde{T}^{(2)}$ has the following structure:
\begin{equation}
\widetilde{T}^{(2)\, i\, i_{1}\, a_{1}\, a_{2}}_{\, {\bf q},\, {\bf q}_{1},\, {\bf k}_{1},\, {\bf k}_{2}}
=
T^{(2)\, i\, i_{1}\, a_{1}\, a_{2}}_{\, {\bf q},\, {\bf q}_{1},\, {\bf k}_{1},\, {\bf k}_{2}}
\label{eq:5t}
\end{equation}
\vspace{-0.5cm}
\begin{align}
-\;
4\, &\frac{{\mathcal G}^{\; a_{2}\;  i\; i^{\hspace{0.02cm}\prime}}_{{\bf k}_{2},\, {\bf q},\, {\bf k}_{2} - {\bf q}}\ 
{\mathcal G}^{\hspace{0.03cm}\ast\, a_{1}\, i_{1}\, i^{\hspace{0.02cm}\prime}}_{{\bf k}_{1},\, {\bf q}_{1},\, {\bf k}_{1} - {\bf q}_{1}}}{\omega^{l}_{{\bf k}_{2}} - \omega^{-}_{{\bf q}}- \omega^{-}_{{\bf k}_{2} - {\bf q}}} 
\;+\;
\frac{{\mathcal P}^{\; a_{2}\, i^{\hspace{0.02cm}\prime}\,  i_{1}}_{{\bf k}_{2},\, {\bf k}_{2} + {\bf q}_{1},\, {\bf q}_{1}}\, 
{\mathcal P}^{\hspace{0.03cm}\ast\, a_{1}\, i^{\hspace{0.02cm}\prime}\,  i}_{{\bf k}_{1},\, {\bf k}_{1} + {\bf q},\,  {\bf q}}}
{\omega^{l}_{{\bf k}_{2}} - \omega^{-}_{{\bf k}_{2} + {\bf q}_{1}} + \omega^{-}_{{\bf q}_{1}}}
\;-\; 
\frac{{\mathcal P}^{\; a_{2}\, i\;  i^{\hspace{0.02cm}\prime}}_{{\bf k}_{2},\, {\bf q},\, {\bf q} - {\bf k}_{2}}\, 
{\mathcal P}^{\hspace{0.03cm}\ast\, a_{1}\, i_{1}\, i^{\hspace{0.02cm}\prime}}_{{\bf k}_{1},\, {\bf q}_{1},\,  {\bf q}_{1} - {\bf k}_{1}}}
{\omega^{l}_{{\bf k}_{2}} - \omega^{-}_{{\bf q}} + \omega^{-}_{{\bf q} - {\bf k}_{2}}}
\notag\\[1.5ex]
+\;
4\, &\frac{{\mathcal K}^{\; a_{2}\, i_{1}\, i^{\hspace{0.02cm}\prime}}_{{\bf k}_{2},\, {\bf q}_{1},\, -{\bf k}_{2} - {\bf q}_{1}}\, 
{\mathcal K}^{\hspace{0.03cm}\ast\, a_{1}\,  i\; i^{\hspace{0.02cm}\prime}}_{{\bf k}_{1},\, {\bf q},\, -{\bf k}_{1} - {\bf q}}}
{\omega^{l}_{{\bf k}_{2}} + \omega^{-}_{{\bf q}_{1}} + \omega^{-}_{-{\bf k}_{2} - {\bf q}_{1}}}
-
2\,\frac{{\mathcal V}^{\; a_{1}\, a_{2}\,  a^{\prime}}_{{\bf k}_{1}, {\bf k}_{2},\, {\bf k}_{1} - {{\bf k}_{2}}}\, 
{\mathcal P}^{\hspace{0.03cm}\ast\, a^{\prime}\, i_{1}\, i}_{\;{\bf q}_{1} - {\bf q},\, {\bf q}_{1},\, {\bf q}}}
{\omega^{l}_{{\bf q}_{1} - {\bf q}} - \omega^{-}_{{\bf q}_{1}} + \omega^{-}_{{\bf q}}}
-
2\, \frac{{\mathcal P}^{\; a^{\prime}\, i\;  i_{1}}_{\;{\bf q} - {\bf q}_{1},\, {\bf q},\, {\bf q}_{1}}\, 
{\mathcal V}^{\hspace{0.03cm}\ast\, a_{2}\, a_{1}\;  a^{\prime}}_{{\bf k}_{2}, {\bf k}_{1},\, {\bf k}_{2} - {\bf k}_{1}}}
{\omega^{l}_{{\bf q} - {\bf q}_{1}} - \omega^{-}_{{\bf q}} + \omega^{-}_{{\bf q}_{1}}}.
\notag
\end{align}
Recall that we do not explicitly write out the dependence on spin variable. Summing over the color index $i^{\hspace{0.02cm}\prime}$ in four terms on the right-hand side of (\ref{eq:5t}) implies also summing over the helicity variable $\lambda^{\prime}$. For example, in the numerator of the third term we have to write 
\begin{equation}
\sum\limits_{\lambda^{\prime}\hspace{0.02cm} =\hspace{0.02cm}\pm \hspace{0.02cm} 1}
{\mathcal P}^{\; a_{2}\, i^{\hspace{0.02cm}\prime}\,  i_{1}}_{{\bf k}_{2},\, {\bf k}_{2} + {\bf q}_{1},\, {\bf q}_{1}}(\lambda^{\prime},\lambda_{1})\,
{\mathcal P}^{\hspace{0.03cm}\ast\, a_{1}\, i^{\hspace{0.02cm}\prime}\,  i}_{{\bf k}_{1},\, {\bf k}_{1} + {\bf q},\,  {\bf q}}(\lambda^{\prime},\lambda)
\label{eq:5y}
\end{equation} 
etc. The first term in  braces in (\ref{eq:5r}) has the factor $(\omega^{-}_{{\bf q}} + \omega^{l}_{{\bf k}_{1}} - \omega^{-}_{{\bf q}_{1}} - \omega^{l}_{{\bf k}_{2}})$,  which in fact represents the energy conservation law in the scattering process under investigation (see (\ref{eq:12i})). If this law of conservation is approximately satisfied, then the contribution of this term to the effective Hamiltonian can be completely neglected. In section \ref{section_14} we will discuss in more detail the case when the ``resonance frequency difference'' 
\[
\Delta\hspace{0.02cm}\omega \equiv
\omega^{-}_{{\bf q}} + \omega^{l}_{{\bf k}_{1}} - \omega^{-}_{{\bf q}_{1}} - \omega^{l}_{{\bf k}_{2}}
\]
can be arbitrary and not necessarily small. Hereinafter, the effective Hamiltonians will be de\-signated by the calligraphic letter ${\mathcal H}$, including also the Hamiltonian ${\mathcal H}^{(0)}$ for non-interacting plasmons and plasmi\-nos in the new variables:
\begin{equation}
{\mathcal H}^{(0)} =  \!\int\!\frac{d\hspace{0.02cm}{\bf k}}{(2\pi)^{3}}\ \omega^{l}_{{\bf k}}\ \!
c^{\ast\hspace{0.03cm}a}_{{\bf k}}\hspace{0.03cm} c^{\phantom{\ast}\!\! a}_{{\bf k}}
+
\int\!\frac{d{\bf q}}{(2\pi)^{3}}\ \omega_{\bf q}^{-}\hspace{0.02cm}
f^{\,\ast\hspace{0.03cm}i}_{{\bf q}}\hspace{0.01cm} f^{\,i}_{{\bf q}}.
\label{eq:5u}
\end{equation}
\indent Figure\,\ref{fig1} gives the diagrammatic interpretation of different terms in the effective amplitude  
$\widetilde{T}^{(2)\, i\, i_{1}\, a_{1}\, a_{2}}_{\, {\bf q},\, {\bf q}_{1},\, {\bf k}_{1},\, {\bf k}_{2}}$.
The first graph represents a direct interaction of plasmino with plasmon induced by the amplitude $T^{(2)\, i\, i_{1}\, a_{1}\, a_{2}}_{\, {\bf q},\, {\bf q}_{1},\, {\bf k}_{1},\, {\bf k}_{2}}$ in the general expression (\ref{eq:5t}). The remaining graphs are connected with the interaction of two plasminos with plasmons and of three plasmons among themselves generated by the amplitudes ${\mathcal G}^{\; a\, i\;  i_{1}}_{\;{\bf k},\, {\bf q},\, {\bf q}_{1}}$, ${\mathcal P}^{\; a\, i\;  i_{1}}_{\;{\bf k},\, {\bf q},\, {\bf q}_{1}}$, ${\mathcal K}^{\; a\, i\;  i_{1}}_{\;{\bf k},\, {\bf q},\, {\bf q}_{1}}$ and ${\mathcal V}^{\; a\, a_{1}\;  a_{2}}_{\;{\bf k},\, {\bf k}_{1},\, {\bf k}_{2}}$ with intermediate ``virtual'' oscillations and represent the contributions $s$\,- and $t$\,-\,channels. The effective amplitude $\widetilde{T}^{(2)\, i\, i_{1}\, a_{1}\, a_{2}}_{\, {\bf q},\, {\bf q}_{1},\, {\bf k}_{1},\, {\bf k}_{2}}$ also includes contribution from  $u$\,-\,channel, which is not presented in fig.\,\ref{fig1}.


\section{Effective fourth-order Hamiltonian. Elastic scattering of plasmino off plasmino}
\setcounter{equation}{0}
\label{section_6}

\begin{figure}[hbtp]
\begin{center}
\includegraphics[width=0.5\textwidth]{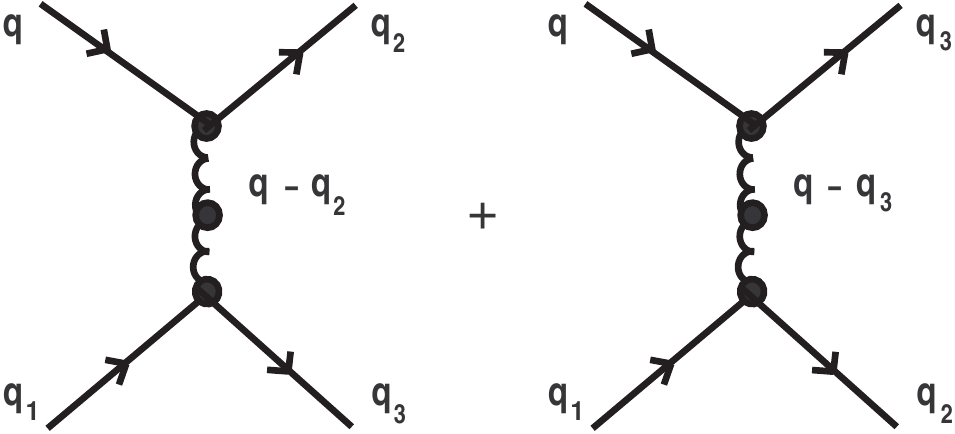}
\end{center}
\vspace{-0.5cm}
\caption{M$\o$ller-like  elastic scattering of plasmino off plasmino} 
\label{fig2}
\end{figure}

We now proceed to defining the effective Hamiltonian for the elastic scattering process of  plasmino off plasmino. Similar to the previous case, as a first step we need to obtain all contributions proportional to 
$f^{\,\ast\, i}_{{\bf q}} f^{\,\ast\, i_{1}}_{{\bf q}_{1}} f^{\;i_{2}}_{{\bf q}_{2}} f^{\;i_{3}}_{{\bf q}_{3}}$ from the free-field  Hamiltonian $H^{(0)}$ given by the equation (\ref{eq:2d}) as a result of the canonical transformations of the boson $a^{\phantom{\ast}\!\! a}_{{\bf k}}$ and fermion $b^{\,i}_{{\bf q}}$ variables, Eqs.\,(\ref{eq:3t}) and (\ref{eq:3y}). Taking into account the relations (\ref{eq:3p}), we have the following contributions from $H^{(0)}$:
\[
\int\!\frac{d{\bf q}\, d{\bf q}_{1}\, d{\bf q}_{2}\, d{\bf q}_{3}}{(2\pi)^{12}}\, 
\biggl\{\hspace{0.03cm} \frac{1}{2}\hspace{0.04cm} 
\Bigl(\omega^{-}_{{\bf q}}S^{\,(2)\, i\; i_{1}\, i_{2}\, i_{3}}_{\ {\bf q},\, {\bf q}_{1},\, {\bf q}_{2},\, {\bf q}_{3}}
-
\omega^{-}_{{\bf q}_{1}}S^{\,(2)\, i_{1}\, i\; i_{2}\, i_{3}}_{\ {\bf q}_{1},\, {\bf q},\, {\bf q}_{2},\, {\bf q}_{3}}
+\,
\omega^{-}_{{\bf q}_{3}}S^{\hspace{0.03cm}\ast\,(2)\, i_{3}\, i_{2}\, i_{1}\, i}_{\ {\bf q}_{3},\, {\bf q}_{2},\, {\bf q}_{1},\, {\bf q}}
-
\omega^{-}_{{\bf q}_{2}}S^{\hspace{0.03cm}\ast\,(2)\, i_{2}\, i_{3}\, i_{1}\, i}_{\ {\bf q}_{2},\, {\bf q}_{3},\, {\bf q}_{1},\, {\bf q}}
\Bigr)
\]
\vspace{-0.5cm}
\begin{align}
-\,
\!\int\!\frac{d\hspace{0.03cm}{\bf k}^{\prime}}{(2\pi)^{3}}\;\omega^{l}_{{\bf k}^{\prime}}
\Bigl[\hspace{0.03cm}
&F^{(1)\, a^{\prime}\, i_{2}\, i_{3}}_{\ {\bf k}^{\prime},\, {\bf q}_{2},\, {\bf q}_{3}}\, 
F^{\hspace{0.03cm}\ast\,(1)\, a^{\prime}\, i\, i_{1}}_{\ {\bf k}^{\prime},\,{\bf q},\, {\bf q}_{1}}
\,+\,
F^{(3)\, a^{\prime}\, i\, i_{1}}_{\ {\bf k}^{\prime},\, {\bf q},\, {\bf q}_{1}}\, 
F^{\hspace{0.03cm}\ast\,(3)\, a^{\prime}\, i_{2}\, i_{3}}_{\ {\bf k}^{\prime},\, {\bf q}_{2},\, {\bf q}_{3}}\hspace{0.03cm}\Bigr]
\label{eq:6q}\\[0.7ex]
-\,\frac{1}{4}
\int\!\frac{d\hspace{0.03cm}{\bf k}^{\prime}}{(2\pi)^{3}}\;\omega^{l}_{{\bf k}^{\prime}}
\Bigl[\hspace{0.02cm}
&F^{(2)\, a^{\prime}\, i_{1}\, i_{3}}_{\ {\bf k}^{\prime},\, {\bf q}_{1},\, {\bf q}_{3}}\, 
F^{\hspace{0.03cm}\ast\,(2)\, a^{\prime}\, i_{2}\, i}_{\ {\bf k}^{\prime},\,{\bf q}_{2},\, {\bf q}}
+
F^{(2)\, a^{\prime}\, i\, i_{2}}_{\ {\bf k}^{\prime},\, {\bf q},\, {\bf q}_{2}}\, 
F^{\hspace{0.03cm}\ast\,(2)\, a^{\prime}\, i_{3}\, i_{1}}_{\ {\bf k}^{\prime},\,{\bf q}_{3},\, {\bf q}_{1}}
\notag\\[0.7ex]
&\hspace{0.1cm}-
F^{(2)\, a^{\prime}\, i\, i_{3}}_{\ {\bf k}^{\prime},\, {\bf q},\, {\bf q}_{3}}\, 
F^{\hspace{0.03cm}\ast\,(2)\, a^{\prime}\, i_{2}\, i_{1}}_{\ {\bf k}^{\prime},\,{\bf q}_{2},\, {\bf q}_{1}}
-
F^{(2)\, a^{\prime}\, i_{1}\, i_{2}}_{\ {\bf k}^{\prime},\, {\bf q}_{1},\, {\bf q}_{2}}\, 
F^{\hspace{0.03cm}\ast\,(2)\, a^{\prime}\, i_{3}\, i}_{\ {\bf k}^{\prime},\,{\bf q}_{3},\, {\bf q}}\hspace{0.02cm}\Bigr]\biggr\}
f^{\,\ast\, i}_{{\bf q}} f^{\,\ast\, i_{1}}_{{\bf q}_{1}} f^{\;i_{2}}_{{\bf q}_{2}} f^{\;i_{3}}_{{\bf q}_{3}}.
\notag 
\end{align}
The antisymmetrization has been performed, where necessary, as indicated by the corresponding factors $1/2$ and $1/4$. Let us analyze the expression in  parentheses in (\ref{eq:6q}). For this purpose we make use of the integral relations from the Appendix \ref{appendix_D}. For eliminating the functions $S^{(2)\, i_{1}\, i\; i_{2}\, i_{3}}_{\ {\bf q}_{1},\, {\bf q},\, {\bf q}_{2},\, {\bf q}_{3}}$,  $S^{\hspace{0.03cm}\ast\,(2)\, i_{3}\, i_{2}\, i_{1}\, i}_{\ {\bf q}_{3},\, {\bf q}_{2},\, {\bf q}_{1},\, {\bf q}}$ and $S^{\hspace{0.03cm}\ast\,(2)\, i_{2}\, i_{3}\, i_{1}\, i}_{\ {\bf q}_{2},\, {\bf q}_{3},\, {\bf q}_{1},\, {\bf q}}$ we employ the relations (\ref{ap:D4a}) and (\ref{ap:D2b}), correspondingly. In the end, we obtain
\begin{equation}
\omega^{-}_{{\bf q}}S^{\,(2)\, i\; i_{1}\, i_{2}\, i_{3}}_{\ {\bf q},\, {\bf q}_{1},\, {\bf q}_{2},\, {\bf q}_{3}}
-
\omega^{-}_{{\bf q}_{1}}S^{\,(2)\, i_{1}\, i\; i_{2}\, i_{3}}_{\ {\bf q}_{1},\, {\bf q},\, {\bf q}_{2},\, {\bf q}_{3}}
+\,
\omega^{-}_{{\bf q}_{3}}S^{\hspace{0.03cm}\ast\,(2)\, i_{3}\, i_{2}\, i_{1}\, i}_{\ {\bf q}_{3},\, {\bf q}_{2},\, {\bf q}_{1},\, {\bf q}}
-
\omega^{-}_{{\bf q}_{2}}S^{\hspace{0.03cm}\ast\,(2)\, i_{2}\, i_{3}\, i_{1}\, i}_{\ {\bf q}_{2},\, {\bf q}_{3},\, {\bf q}_{1},\, {\bf q}}
\label{eq:6w}
\end{equation}
\[
=
\bigl(\omega^{-}_{{\bf q}} + \omega^{-}_{{\bf q}_{1}} - \omega^{-}_{{\bf q}_{2}} - \omega^{-}_{{\bf q}_{3}}\bigr)
S^{\,(2)\, i\; i_{1}\, i_{2}\, i_{3}}_{\ {\bf q},\, {\bf q}_{1},\, {\bf q}_{2},\, {\bf q}_{3}}
\]
\begin{align}
-\,\frac{1}{2}\,
\omega^{-}_{{\bf q}_{1}}\!&\int\!\frac{d\hspace{0.03cm}{\bf k}^{\prime}}{(2\pi)^{3}}\,
\Bigl[\hspace{0.02cm}
F^{(2)\, a^{\prime}\, i_{1}\, i_{3}}_{\ {\bf k}^{\prime},\, {\bf q}_{1},\, {\bf q}_{3}}\, 
F^{\hspace{0.03cm}\ast\,(2)\, a^{\prime}\, i_{2}\, i}_{\ {\bf k}^{\prime},\,{\bf q}_{2},\, {\bf q}}
+
F^{(2)\, a^{\prime}\, i\, i_{3}}_{\ {\bf k}^{\prime},\, {\bf q},\, {\bf q}_{3}}\, 
F^{\hspace{0.03cm}\ast\,(2)\, a^{\prime}\, i_{2}\, i_{1}}_{\ {\bf k}^{\prime},\,{\bf q}_{2},\, {\bf q}_{1}}
\notag\\[0.7ex]
&\hspace{2cm}-
F^{(2)\, a^{\prime}\, i\, i_{2}}_{\ {\bf k}^{\prime},\, {\bf q},\, {\bf q}_{2}}\, 
F^{\hspace{0.03cm}\ast\,(2)\, a^{\prime}\, i_{3}\, i_{1}}_{\ {\bf k}^{\prime},\,{\bf q}_{3},\, {\bf q}_{1}}
-
F^{(2)\, a^{\prime}\, i_{1}\, i_{2}}_{\ {\bf k}^{\prime},\, {\bf q}_{1},\, {\bf q}_{2}}\, 
F^{\hspace{0.03cm}\ast\,(2)\, a^{\prime}\, i_{3}\, i}_{\ {\bf k}^{\prime},\,{\bf q}_{3},\, {\bf q}}\hspace{0.02cm}\Bigr]
\notag\\[1ex]
+\,\frac{1}{2}\,
\omega^{-}_{{\bf q}_{2}}\!&\int\!\frac{d\hspace{0.03cm}{\bf k}^{\prime}}{(2\pi)^{3}}\,
\Bigl[\hspace{0.02cm}
4\hspace{0.02cm}F^{(1)\, a^{\prime}\, i_{2}\, i_{3}}_{\ {\bf k}^{\prime},\, {\bf q}_{2},\, {\bf q}_{3}}\, 
F^{\hspace{0.03cm}\ast\,(1)\, a^{\prime}\, i\, i_{1}}_{\ {\bf k}^{\prime},\,{\bf q},\, {\bf q}_{1}}
+
F^{(2)\, a^{\prime}\, i\, i_{3}}_{\ {\bf k}^{\prime},\, {\bf q},\, {\bf q}_{3}}\, 
F^{\hspace{0.03cm}\ast\,(2)\, a^{\prime}\, i_{2}\, i_{1}}_{\ {\bf k}^{\prime},\,{\bf q}_{2},\, {\bf q}_{1}}
\notag\\[0.7ex]
&\hspace{2cm}-
4\hspace{0.02cm}F^{(3)\, a^{\prime}\, i\, i_{1}}_{\ {\bf k}^{\prime},\, {\bf q},\, {\bf q}_{1}}\, 
F^{\hspace{0.03cm}\ast\,(3)\, a^{\prime}\, i_{2}\, i_{3}}_{\ {\bf k}^{\prime},\,{\bf q}_{2},\, {\bf q}_{3}}
-
F^{(2)\, a^{\prime}\, i_{1}\, i_{2}}_{\ {\bf k}^{\prime},\, {\bf q}_{1},\, {\bf q}_{2}}\, 
F^{\hspace{0.03cm}\ast\,(2)\, a^{\prime}\, i_{3}\, i}_{\ {\bf k}^{\prime},\,{\bf q}_{3},\, {\bf q}}\hspace{0.02cm}\Bigr]
\notag\\[1ex]
-\,\frac{1}{2}\,
\omega^{-}_{{\bf q}_{3}}\!&\int\!\frac{d\hspace{0.03cm}{\bf k}^{\prime}}{(2\pi)^{3}}\,
\Bigl[\hspace{0.02cm}
4\hspace{0.02cm}F^{(1)\, a^{\prime}\, i_{3}\, i_{2}}_{\ {\bf k}^{\prime},\, {\bf q}_{3},\, {\bf q}_{2}}\, 
F^{\hspace{0.03cm}\ast\,(1)\, a^{\prime}\, i\, i_{1}}_{\ {\bf k}^{\prime},\,{\bf q},\, {\bf q}_{1}}
+
F^{(2)\, a^{\prime}\, i\, i_{2}}_{\ {\bf k}^{\prime},\, {\bf q},\, {\bf q}_{2}}\, 
F^{\hspace{0.03cm}\ast\,(2)\, a^{\prime}\, i_{3}\, i_{1}}_{\ {\bf k}^{\prime},\,{\bf q}_{3},\, {\bf q}_{1}}
\notag\\[0.7ex]
&\hspace{2cm}-
4\hspace{0.02cm}F^{(3)\, a^{\prime}\, i\, i_{1}}_{\ {\bf k}^{\prime},\, {\bf q},\, {\bf q}_{1}}\, 
F^{\hspace{0.03cm}\ast\,(3)\, a^{\prime}\, i_{3}\, i_{2}}_{\ {\bf k}^{\prime},\,{\bf q}_{3},\, {\bf q}_{2}}
-
F^{(2)\, a^{\prime}\, i_{1}\, i_{3}}_{\ {\bf k}^{\prime},\, {\bf q}_{1},\, {\bf q}_{3}}\, 
F^{\hspace{0.03cm}\ast\,(2)\, a^{\prime}\, i_{2}\, i}_{\ {\bf k}^{\prime},\,{\bf q}_{2},\, {\bf q}}\hspace{0.02cm}\Bigr].
\notag
\end{align} 
\indent To proceed further we analyze the relevant fourth-order contributions from the Hamiltonian $H^{(3)}$, Eq.\, (\ref{eq:2f}). Here, we again pass from the physical vertex functions ${\mathcal G},\, {\mathcal P}$, and 
${\mathcal K}$ to the coefficient functions $F^{(i)},\,i = 1,2,3$, by the rules (\ref{eq:4e}). Performing the corresponding antisymmetriza\-tion where it is required, as a result we obtain 
\begin{equation}
\int\!\frac{d{\bf q}\, d{\bf q}_{1}\hspace{0.03cm} d{\bf q}_{2}\hspace{0.03cm} d{\bf q}_{3}}{(2\pi)^{12}}\, 
f^{\,\ast\, i}_{{\bf q}} f^{\,\ast\, i_{1}}_{{\bf q}_{1}} f^{\;i_{2}}_{{\bf q}_{2}} f^{\;i_{3}}_{{\bf q}_{3}}
\label{eq:6e}
\end{equation}
\vspace{-0.6cm}
\begin{align}
\times\,\biggl\{
\frac{1}{4}\int\!\frac{d\hspace{0.03cm}{\bf k}^{\prime}}{(2\pi)^{3}}\,
\Bigl[\hspace{0.03cm}
&\bigl(\omega^{l}_{{\bf k}^{\prime}} - \omega^{-}_{{\bf q}_{1}} + \omega^{-}_{{\bf q}_{3}}\bigr)
F^{(2)\, a^{\prime}\, i\, i_{2}}_{\ {\bf k}^{\prime},\, {\bf q},\, {\bf q}_{2}}\, 
F^{\hspace{0.03cm}\ast\,(2)\, a^{\prime}\, i_{3}\, i_{1}}_{\ {\bf k}^{\prime},\,{\bf q}_{3},\, {\bf q}_{1}}
-
\bigl(\omega^{l}_{{\bf k}^{\prime}} - \omega^{-}_{{\bf q}} + \omega^{-}_{{\bf q}_{3}}\bigr)
F^{(2)\, a^{\prime}\, i_{1}\, i_{2}}_{\ {\bf k}^{\prime},\, {\bf q}_{1},\, {\bf q}_{2}}\, 
F^{\hspace{0.03cm}\ast\,(2)\, a^{\prime}\, i_{3}\, i}_{\ {\bf k}^{\prime},\,{\bf q}_{3},\, {\bf q}}
\notag\\[1ex]
+\;
&\bigl(\omega^{l}_{{\bf k}^{\prime}} - \omega^{-}_{{\bf q}} + \omega^{-}_{{\bf q}_{2}}\bigr)
F^{(2)\, a^{\prime}\, i_{1}\, i_{3}}_{\ {\bf k}^{\prime},\, {\bf q}_{1},\, {\bf q}_{3}}\, 
F^{\hspace{0.03cm}\ast\,(2)\, a^{\prime}\, i_{2}\, i}_{\ {\bf k}^{\prime},\,{\bf q}_{2},\, {\bf q}}
\;-
\bigl(\omega^{l}_{{\bf k}^{\prime}} - \omega^{-}_{{\bf q}_{1}} + \omega^{-}_{{\bf q}_{2}}\bigr)
F^{(2)\, a^{\prime}\, i\, i_{3}}_{\ {\bf k}^{\prime},\, {\bf q},\, {\bf q}_{3}}\, 
F^{\hspace{0.03cm}\ast\,(2)\, a^{\prime}\, i_{2}\, i_{1}}_{\ {\bf k}^{\prime},\,{\bf q}_{2},\, {\bf q}_{1}}
\notag\\[1.5ex]
+\;
&\bigl(\omega^{l}_{{\bf k}^{\prime}} + \omega^{-}_{{\bf q}_{1}} - \omega^{-}_{{\bf q}_{3}}\bigr)
F^{(2)\, a^{\prime}\, i_{1}\, i_{3}}_{\ {\bf k}^{\prime},\, {\bf q}_{1},\, {\bf q}_{3}}\, 
F^{\hspace{0.03cm}\ast\,(2)\, a^{\prime}\, i_{2}\, i}_{\ {\bf k}^{\prime},\,{\bf q}_{2},\, {\bf q}}
-
\bigl(\omega^{l}_{{\bf k}^{\prime}} + \omega^{-}_{{\bf q}} - \omega^{-}_{{\bf q}_{3}}\bigr)
F^{(2)\, a^{\prime}\, i\, i_{3}}_{\ {\bf k}^{\prime},\, {\bf q},\, {\bf q}_{3}}\, 
F^{\hspace{0.03cm}\ast\,(2)\, a^{\prime}\, i_{2}\, i_{1}}_{\ {\bf k}^{\prime},\,{\bf q}_{2},\, {\bf q}_{1}}
\notag\\[1ex]
+\;
&\bigl(\omega^{l}_{{\bf k}^{\prime}} + \omega^{-}_{{\bf q}} - \omega^{-}_{{\bf q}_{2}}\bigr)
F^{(2)\, a^{\prime}\, i\, i_{2}}_{\ {\bf k}^{\prime},\, {\bf q},\, {\bf q}_{2}}\, 
F^{\hspace{0.03cm}\ast\,(2)\, a^{\prime}\, i_{3}\, i_{1}}_{\ {\bf k}^{\prime},\,{\bf q}_{3},\, {\bf q}_{1}}
\,-
\bigl(\omega^{l}_{{\bf k}^{\prime}} + \omega^{-}_{{\bf q}_{1}} - \omega^{-}_{{\bf q}_{2}}\bigr)
F^{(2)\, a^{\prime}\, i_{1}\, i_{2}}_{\ {\bf k}^{\prime},\, {\bf q}_{1},\, {\bf q}_{2}}\, 
F^{\hspace{0.03cm}\ast\,(2)\, a^{\prime}\, i_{3}\, i}_{\ {\bf k}^{\prime},\,{\bf q}_{3},\, {\bf q}}
\,\Bigr]
\notag\\[1.5ex]
+
\int\!\frac{d\hspace{0.03cm}{\bf k}^{\prime}}{(2\pi)^{3}}\,
\Bigl[
&\bigl(\omega^{l}_{{\bf k}^{\prime}} - \omega^{-}_{{\bf q}} - \omega^{-}_{{\bf q}_{1}}\bigr)
F^{(1)\, a^{\prime}\, i_{2}\, i_{3}}_{\ {\bf k}^{\prime},\, {\bf q}_{2},\, {\bf q}_{3}}\, 
F^{\hspace{0.03cm}\ast\,(1)\, a^{\prime}\, i\, i_{1}}_{\ {\bf k}^{\prime},\,{\bf q},\, {\bf q}_{1}}
+
\bigl(\omega^{l}_{{\bf k}^{\prime}} - \omega^{-}_{{\bf q}_{2}} - \omega^{-}_{{\bf q}_{3}}\bigr)
F^{(1)\, a^{\prime}\, i_{2}\, i_{3}}_{\ {\bf k}^{\prime},\, {\bf q}_{2},\, {\bf q}_{3}}\, 
F^{\hspace{0.03cm}\ast\,(1)\, a^{\prime}\, i\, i_{1}}_{\ {\bf k}^{\prime},\,{\bf q},\, {\bf q}_{1}}
\notag\\
+\;
&\bigl(\omega^{l}_{{\bf k}^{\prime}} + \omega^{-}_{{\bf q}} + \omega^{-}_{{\bf q}_{1}}\bigr)
F^{(3)\, a^{\prime}\, i\, i_{1}}_{\ {\bf k}^{\prime},\, {\bf q},\, {\bf q}_{1}}\, 
F^{\hspace{0.03cm}\ast\,(3)\, a^{\prime}\, i_{2}\, i_{3}}_{\ {\bf k}^{\prime},\,{\bf q}_{2},\, {\bf q}_{3}}
+
\bigl(\omega^{l}_{{\bf k}^{\prime}} + \omega^{-}_{{\bf q}_{2}} + \omega^{-}_{{\bf q}_{3}}\bigr)
F^{(3)\, a^{\prime}\, i\, i_{1}}_{\ {\bf k}^{\prime},\, {\bf q},\, {\bf q}_{1}}\, 
F^{\hspace{0.03cm}\ast\,(3)\, a^{\prime}\, i_{2}\, i_{3}}_{\ {\bf k}^{\prime},\,{\bf q}_{2},\, {\bf q}_{3}}
\,\Bigr]\!\hspace{0.01cm}\biggr\}.
\notag
\end{align}
As in the case of (\ref{eq:5e}), here we did not collect similar terms since the given expression is more convenient for the subsequent calculations. In the second term on the right-hand side of (\ref{eq:2g}) we make the substitutions $b^{\,\ast\,i}_{\bf q} \rightarrow f^{\,\ast\,i}_{\bf q}$ and $b^{\,i}_{\bf q} \rightarrow f^{\,i}_{\bf q}$. We add (\ref{eq:6q}) and (\ref{eq:6e}) to the expression thus obtained with allowance for (\ref{eq:6w}) and collect similar terms. Here, most terms are mutually canceled. By the last step we move from the coefficient functions $F^{(i)},\,i = 1,2,3$ to the vertex functions ${\mathcal G},\, {\mathcal P}$ and ${\mathcal K}$ by the rules (\ref{eq:4e}). In doing so we result in the fourth-order effective Hamiltonian describing the elastic scattering process of plasmino off plasmino  
\begin{equation}
{\mathcal H}^{(4)}_{qq\rightarrow qq} 
\,=
\frac{1}{2}\int\frac{d{\bf q}\, d{\bf q}_{1}\hspace{0.03cm} d{\bf q}_{2}\hspace{0.03cm} d{\bf q}_{3}}{(2\pi)^{12}}
\label{eq:6r}
\end{equation}
\[
\times\,\biggl\{
\bigl(\hspace{0.02cm}\omega^{-}_{{\bf q}} + \omega^{-}_{{\bf q}_{1}} - \omega^{-}_{{\bf q}_{2}} - \omega^{-}_{{\bf q}_{3}}\bigr)
S^{\,(2)\, i\; i_{1}\, i_{2}\, i_{3}}_{\, {\bf q},\, {\bf q}_{1},\, {\bf q}_{2},\, {\bf q}_{3}}
+
\widetilde{T}^{\, (2)\, i\; i_{1}\, i_{2}\, i_{3}}_{\, {\bf q},\, {\bf q}_{1},\, {\bf q}_{2},\, {\bf q}_{3}}\, 
(2\pi)^{3}\hspace{0.03cm}\delta({\bf q} + {\bf q}_1 - {\bf q}_{2} - {\bf q}_{3})
\biggr\}
f^{\,\ast\, i}_{{\bf q}} f^{\,\ast\, i_{1}}_{{\bf q}_{1}} f^{\;i_{2}}_{{\bf q}_{2}} f^{\;i_{3}}_{{\bf q}_{3}},
\]
where the effective amplitude $\widetilde{T}^{\,(2)}$ has the following structure:
\begin{align}
\widetilde{T}^{\,(2)\, i\; i_{1}\, i_{2}\, i_{3}}_{\, {\bf q},\, {\bf q}_{1},\, {\bf q}_{2},\, {\bf q}_{3}}
=
T^{\,(2)\, i\; i_{1}\, i_{2}\, i_{3}}_{\, {\bf q},\, {\bf q}_{1},\, {\bf q}_{2},\, {\bf q}_{3}}
\;+\,
2\,&\frac{{\mathcal G}^{\; a\, i\;  i_{1}}_{{\bf q} + {\bf q}_{1},\, {\bf q},\, {\bf q}_{1}}\, 
{\mathcal G}^{\hspace{0.03cm}\ast\, a\, i_{2}\, i_{3}}_{{\bf q}_{2} + {\bf q}_{3},\, {\bf q}_{2},\, {\bf q}_{3}}}
{\omega^{l}_{{\bf q}_{2} + {\bf q}_{3}} - \omega^{-}_{{\bf q}_{2}} - \omega^{-}_{{\bf q}_{3}}} 
\;+\;
2\,\frac{{\mathcal K}^{\; a\, i_{2}\, i_{3}}_{-{\bf q}_{2} - {\bf q}_{3},\, {\bf q}_{2},\, {\bf q}_{3}}\,
{\mathcal K}^{\hspace{0.03cm}\ast\, a\, i\;  i_{1}}_{-{\bf q} - {\bf q}_{1},\, {\bf q},\, {\bf q}_{1}}}
{\omega^{l}_{-{\bf q}_{2} - {\bf q}_{3}} + \omega^{-}_{{\bf q}_{2}} + \omega^{-}_{{\bf q}_{3}}} 
\label{eq:6t}\\[1.5ex]
\;+\;
\frac{1}{2}\,\Biggl[\Biggl(
&\frac{{\mathcal P}^{\; a\, i\; i_{2}}_{{\bf q} - {\bf q}_{2},\, {\bf q},\, {\bf q}_{2}}\, 
{\mathcal P}^{\hspace{0.03cm}\ast\, a\, i_{3}\, i_{1}}_{{\bf q}_{3} - {\bf q}_{1},\, {\bf q}_{3},\,  {\bf q}_{1}}}
{\omega^{l}_{{\bf q}_{3} - {\bf q}_{1}} - \omega^{-}_{{\bf q}_{3}} + \omega^{-}_{{\bf q}_{1}}}
\;+\; 
\frac{{\mathcal P}^{\; a\, i_{1}\, i_{3}}_{{\bf q}_{1} - {\bf q}_{3},\, {\bf q}_{1},\, {\bf q}_{3}}\, 
{\mathcal P}^{\hspace{0.03cm}\ast\, a\, i_{2}\, i}_{{\bf q}_{2} - {\bf q},\, {\bf q}_{2},\, {\bf q}}}
{\omega^{l}_{{\bf q}_{1} - {\bf q}_{3}} - \omega^{-}_{{\bf q}_{1}} + \omega^{-}_{{\bf q}_{3}}}
\Biggr)
\notag\\[1.5ex]
-\;
\Biggl(
&\frac{{\mathcal P}^{\; a\, i_{1}\, i_{2}}_{{\bf q}_{1} - {\bf q}_{2},\, {\bf q}_{1},\, {\bf q}_{2}}\, 
{\mathcal P}^{\hspace{0.03cm}\ast\, a\, i_{3}\, i}_{{\bf q}_{3} - {\bf q},\, {\bf q}_{3},\, {\bf q}}}
{\omega^{l}_{{\bf q}_{1} - {\bf q}_{2}} - \omega^{-}_{{\bf q}_{1}} + \omega^{-}_{{\bf q}_{2}}}
\;+\; 
\frac{{\mathcal P}^{\; a\, i\, i_{3}}_{{\bf q} - {\bf q}_{3},\, {\bf q},\, {\bf q}_{3}}\, 
{\mathcal P}^{\hspace{0.03cm}\ast\, a\, i_{2}\, i_{1}}_{{\bf q}_{2} - {\bf q}_{1},\, {\bf q}_{2},\, {\bf q}_{1}}}
{\omega^{l}_{{\bf q}_{2} - {\bf q}_{1}} - \omega^{-}_{{\bf q}_{2}} + \omega^{-}_{{\bf q}_{1}}}
\Biggr)\Biggr].
\notag
\end{align}
The first term in braces in (\ref{eq:6r}) has the factor which represents the conservation energy law in this scattering process. The same reasoning as in the previous section applies here for the effective Hamiltonian (\ref{eq:6r}), namely, if the conservation law approximately satisfied, then the contribution of this term into the effective Hamiltonian can be completely neglected. In section \ref{section_13} we will discuss the general case when the ``resonance frequency difference'' 
\begin{equation}
\Delta\hspace{0.02cm}\omega \equiv
\omega^{-}_{{\bf q}} + \omega^{-}_{{\bf q}_{1}} - \omega^{-}_{{\bf q}_{2}} - \omega^{-}_{{\bf q}_{3}}
\label{eq:6y}
\end{equation}
can be arbitrary. Recall that the effective amplitude (\ref{eq:6t}) depends also on four helicity variables $\lambda,\, \lambda_{1},\, \lambda_{2}$ and $\lambda_{3}$.\\
\indent The effective amplitude obtained has a simple diagrammatic interpretation. On fig.\,\ref{fig2} the graphs of the elastic scattering processes of plasmino off plasmino are presented.
\vspace{0.4cm}
As it will be shown further within the framework of the (semi)classical description of the quark-gluon plasma the four-plasmino amplitude $T^{\,(2)\, i\, i_{1}\, i_{2}\, i_{3}}_{{\bf q},\, {\bf q}_{1},\, {\bf q}_{2},\, {\bf q}_{3}}$ in the expression (\ref{eq:6t}) is absent. For this reason among the graphs there is no graph defining a direct interaction of four plasminos. The condition of smallness of amplitudes in this pure ``fermion'' case means fulfillment of the condition
\begin{equation}
|\ \!\widetilde{T}^{(2)}| |\ \!f\ \!|^{\,2} \ll\! \Bigl({\bf q}\cdot\partial\omega^{-}_{\bf q}/\partial{\bf q}\Bigr).
\label{eq:6u}
\end{equation}


\section{\bf Fourth-order correlation function for soft quark-anti\-quark exci\-tations}
\label{section_7}
\setcounter{equation}{0}

\indent As stated above, the Hamiltonians (\ref{eq:5r}) and (\ref{eq:6r}) describe the elastic scattering processes of plasminos off plasmons and plasminos among themselves correspondingly, i.e. the processes $2\rightarrow 2$. The equations of motion for the fermionic $f^{\,i}_{{\bf q}}, \,f^{\,\ast\, i}_{{\bf q}}$ and bosonic $c^{\phantom{\hspace{0.03cm}\ast} \!a}_{{\bf k}}, \,c^{\hspace{0.03cm}\ast\,b}_{{\bf k}}$ normal variables are defined  by the corresponding Hamilton equations. Let us consider first the Hamiltonian ${\mathcal H}^{(4)}_{qq\rightarrow qq}$, Eq.\,(\ref{eq:6r}). In this case we have
\begin{equation}
	\hspace{0.08cm}
	\frac{\partial\!\hspace{0.035cm} f^{\hspace{0.03cm}\ast\,i}_{{\bf q}}}{\partial\hspace{0.03cm} t}
	=
	-\hspace{0.03cm}i\hspace{0.04cm}\Bigl\{f^{\hspace{0.03cm}\ast\,i}_{{\bf q}}\hspace{0.03cm},\hspace{0.03cm} {\mathcal H}^{(0)\!} + {\mathcal H}^{(4)}_{qq\rightarrow qq}\Bigr\}
	=
	i\hspace{0.04cm}\omega^{-}_{{\bf q}}\, f^{\hspace{0.03cm}\ast\,i}_{{\bf q}}
	\label{eq:7q}
\end{equation}
\[
\hspace{0.25cm}
-\, i\!\int\!\frac{d{\bf q}_{1}\hspace{0.03cm} d{\bf q}_{2}\hspace{0.03cm} d{\bf q}_{3}}{(2\pi)^{9}}\, 
\widetilde{T}^{\hspace{0.03cm}\ast\hspace{0.03cm}(2)\, i\, i_{1}\, i_{2}\, i_{3}}_{\ {\bf q},\, {\bf q}_{1},\; {\bf q}_{2},\ {\bf q}_{3}}\, 
f^{\phantom{\hspace{0.03cm}\ast}\!\!i_{1}}_{{\bf q}_{1}}\, f^{\hspace{0.03cm}\ast\,i_{2}}_{{\bf q}_{2}} 
f^{\hspace{0.03cm}\ast\,i_{3}}_{{\bf q}_{3}}\,
(2\pi)^{3}\hspace{0.03cm}\delta({\bf q} + {\bf q}_{1} - {\bf q}_{2} - {\bf q}_{3}).
\]
\begin{equation}
\frac{\partial\!\hspace{0.035cm} f^{\,j}_{{\bf q}^{\prime}}}{\partial\hspace{0.03cm} t}
=
-\hspace{0.03cm}i\hspace{0.04cm}\Bigl\{f^{\,j}_{{\bf q}^{\prime}}\hspace{0.03cm},\hspace{0.03cm} {\mathcal H}^{(0)\!} + {\mathcal H}^{(4)}_{qq\rightarrow qq}\Bigr\}
=
- i\hspace{0.04cm}\omega^{-}_{{\bf q}^{\prime}}\,f^{\,i}_{{\bf q}^{\prime}}
\label{eq:7w}
\end{equation}
\[
-\, i\!\int\frac{d{\bf q}_{1}\hspace{0.03cm} d{\bf q}_{2}\hspace{0.03cm} d{\bf q}_{3}}{(2\pi)^{9}}\ 
\widetilde{T}^{\,(2)\, j\; i_{1}\, i_{2}\, i_{3}}_{{\bf q}^{\prime},\, {\bf q}_{1},\, {\bf q}_{2},\ {\bf q}_{3}}\ 
f^{\hspace{0.03cm}\ast\,i_{1}}_{{\bf q}_{1}} f^{\phantom{\hspace{0.03cm}\ast}\!\!i_{2}}_{{\bf q}_{2}}\,
f^{\phantom{\hspace{0.03cm}\ast}\!\!i_{3}}_{{\bf q}_{3}}\,
(2\pi)^{3}\hspace{0.03cm}\delta({\bf q}^{\prime} + {\bf q}_{1} - {\bf q}_{2} - {\bf q}_{3}).
\]
In the first equation we have taken into account the Grassmann nature of the functions $f^{\,i}_{{\bf q}},\, f^{\hspace{0.03cm}\ast\,i}_{{\bf q}}$ and the symmetry property for the effective scattering amplitude $\widetilde{T}^{\,(2)}$ with respect to permutation of the first and the second pairs of indices (and momenta) among themselves or in other words under the interchange of incoming and outgoing external plasmino lines
\begin{equation}
\widetilde{T}^{\hspace{0.03cm}\ast\hspace{0.03cm}(2)\, i\, i_{1}\, i_{2}\, i_{3}}_{\ {\bf q},\, {\bf q}_{1},\, {\bf q}_{2},\, {\bf q}_{3}}
=
\widetilde{T}^{(2)\, i_{2}\, i_{3}\, i\; i_{1}}_{{\bf q}_{2},\, {\bf q}_{3},\, {\bf q},\, {\bf q}_{1}}.
\label{eq:7e}
\end{equation}
\indent The exact equations (\ref{eq:7q}) and  (\ref{eq:7w}), in the absence of an external field in QGP enable us to define a kinetic equation for the number density of colorless plasminos  $n^{ij\,-}_{{\bf q}\!\!}\! \equiv \delta^{\hspace{0.02cm}ij} n^{-}_{{\bf q}}$. If the ensemble of interacting Fermi-excitations for a low nonlinearity level (\ref{eq:6u}) has random phases, then this ensemble can be statistically described  by introducing the fermion correlation function:
\begin{equation}
\bigl\langle\,\!f^{\hspace{0.03cm}\ast\,i}_{{\bf q}}(\lambda) f^{\,j}_{{\bf q}^{\prime}}(\lambda^{\prime})\bigr\rangle
=
\delta^{\hspace{0.02cm}ij}\delta_{\lambda\lambda^{\prime}}\hspace{0.02cm}(2\pi)^{3}\hspace{0.03cm}
\delta({\bf q} - {\bf q}^{\prime})\hspace{0.03cm}n^{-}_{\bf q}.
\label{eq:7r}
\end{equation}
Here, we have explicitly highlighted the dependence on helicity. Its evolution due to the influence of nonlinear resonant four-wave interactions is described by a proper kinetic equation which will be derived below. We can find the relationship between two representations of the plasmino number density, namely, between the representations determined by the correlation functions (\ref{eq:2y}) and (\ref{eq:7r}). First of all we note that the plasmino number density $n^{-}_{\bf q}$ in fact plays a role of a {\it spectrum} of the random fermionic wave field of the system under consideration. A connection between the spectrum ${\mathfrak n}^{-}_{\bf q}$ of the physically real random wave field in (\ref{eq:2y}) and the ``spectrum''  $n^{-}_{\bf q}$ in (\ref{eq:7r}) is defined by the canonical transformation (\ref{eq:3y}) and a statistical hypothesis similar to that will be employed below in the derivation of the kinetic equation. In particular,
the concepts of the Kolmogorov power-law spectra of weak turbulence \cite{zakharov_book_1992} apply to $n^{-}_{\bf q}$. In practical applications we need specifically the physical spectrum ${\mathfrak n}^{-}_{\bf q}$, so that we have to consider its relationship to $n^{-}_{\bf q}$. This rather nontrivial circumstance was first pointed out by Krasitskii \cite{krasitskii_1990}.\\
\indent Using Eq.\,(\ref{eq:3y}), we have to calculate the correlation function $\!\bigl\langle\hspace{0.02cm} b^{\hspace{0.02cm}\ast\,i\hspace{0.03cm}(-)}_{{\bf q}}(\lambda) \hspace{0.04cm} b^{\,j\hspace{0.03cm}(-)}_{{\bf q}^{\prime}}(\lambda^{\prime})\bigr\rangle$, and to apply the Gaussian hypothesis (equivalent to the random phase approximation) to the correla\-tion functions of higher orders in the fermionic variable $f^{\,i}_{{\bf q}}$ (and, generally speaking, in the bosonic variable $c^{a}_{\bf k}$), which appear on the right-hand side of the equation. According to this hypothesis all of the odd moments vanish, whereas the even moments can be expressed in terms of the second moments (among the even moments the only nonzero moments are those which contain the same numbers of the factors $f^{\,i}_{{\bf q}}$ and $f^{\hspace{0.03cm}\ast\,i}_{{\bf q}}$, and also of the factors $c^{a}_{\bf k}$ and $c^{\hspace{0.03cm}\ast\,a}_{{\bf k}}$). This calculation procedure yields the nonlinear functional relationship between two representations:
\begin{equation}
{\mathfrak n}^{-}_{\bf q} = {\mathfrak n}^{-}_{\bf q}\hspace{0.03cm} [\hspace{0.03cm} n^{-}_{\bf q}, N^{l}_{{\bf k}}\hspace{0.03cm} ],
\label{eq:7t}
\end{equation}
which also implicates the plasmons number density $N^{l}_{{\bf k}}$ (see below Eq.\,(\ref{eq:10y})). Here, we do not give an explicit form of this relationship. We note only that the difference between ${\mathfrak n}^{-}_{\bf q}$ and $n^{-}_{\bf q}$ has to be taken into account in analyses of the Kolmogorov power-law spectra of weak turbulence in a quark-gluon plasma.\\ 
\indent It is necessary to note that the introduction of the distribution function of fermionic quasipar\-tic\-les (plasminos) $n^{-}_{\bf q}\equiv n^{-}({\bf q}, {\bf x}, t)$, depending both on plasmino momentum $\hbar\hspace{0.04cm}{\bf q}$  and on coordi\-na\-te ${\bf x}$ and time $t$, has a sense only in the case when plasminos number is changing slowly in space and in time. This means that change of the function at distances of the order of a wavelength $\lambda = 2\pi/q$ and over time intervals of the order of an oscillation period $T = 2\pi/\omega^{-}_{\bf q}$ must be much smaller than the function $n^{-}_{{\bf q}}$.\\
\indent Based on the Hamilton equations (\ref{eq:7q}) and (\ref{eq:7w}), we can define the kinetic equation for the plasminos number density $n^{-}_{\bf q}$. For this purpose we multiply the equations (\ref{eq:7q}) and (\ref{eq:7w}) by $f^{\hspace{0.03cm}\ast\,j}_{{\bf q}^{\prime}}$ and $f^{\phantom{\ast}\!\!i}_{{\bf q}}$, from the right and from the left, respectively,
\[
\frac{\partial \hspace{0.01cm}f^{\,i}_{{\bf q}}}{\partial\hspace{0.03cm} t}\,f^{\hspace{0.03cm}\ast\,j}_{{\bf q}^{\prime}}
=
-\hspace{0.03cm} i\hspace{0.04cm}\omega^{-}_{{\bf q}}
f^{\phantom{\hspace{0.03cm}\ast}\!\!i}_{{\bf q}}\hspace{0.03cm}f^{\hspace{0.03cm}\ast\,j}_{{\bf q}^{\prime}}
\]
\[
+\; i\!\int\frac{d{\bf q}_{1}\hspace{0.03cm} d{\bf q}_{2}\hspace{0.04cm} d{\bf q}_{3}}{(2\pi)^{9}}\ 
\widetilde{T}^{(2)\, i\; i_{1}\, i_{2}\, i_{3}}_{{\bf q},\, {\bf q}_{1},\, {\bf q}_{2},\, {\bf q}_{3}}\ 
f^{\hspace{0.03cm}\ast\,j}_{{\bf q}^{\prime}}\hspace{0.03cm}
f^{\hspace{0.03cm}\ast\,i_{1}}_{{\bf q}_{1}}\hspace{0.02cm} 
f^{\phantom{\hspace{0.03cm}\ast}\!\!i_{2}}_{{\bf q}_{2}}\, 
f^{\phantom{\hspace{0.03cm}\ast}\!\!i_{3}}_{{\bf q}_{3}}\,
(2\pi)^{3}\hspace{0.03cm}\delta({\bf q} + {\bf q}_{1} - {\bf q}_{2} - {\bf q}_{3}),
\]
\[
f^{\phantom{\hspace{0.03cm}\ast}\!\!i}_{{\bf q}}\,\frac{\partial \hspace{0.01cm}f^{\hspace{0.03cm}\ast\,j}_{{\bf q}^{\prime}}}{\partial\hspace{0.03cm} t}
=
i\hspace{0.04cm}\omega^{-}_{{\bf q}^{\prime}} f^{\phantom{\hspace{0.03cm}\ast}\!\!i}_{{\bf q}^{\phantom{`}}}\!
\hspace{0.03cm}f^{\hspace{0.03cm}\ast\,j}_{{\bf q}^{\prime}}
\]
\[
-\; i\!\int\frac{d{\bf q}_{1}\hspace{0.03cm} d{\bf q}_{2}\hspace{0.04cm} d{\bf q}_{3}}{(2\pi)^{9}}\ 
\widetilde{T}^{\hspace{0.03cm}\ast\,(2)\, j\, i_{1}\, i_{2}\, i_{3}}_{{\bf q}^{\prime},\, {\bf q}_{1},\, {\bf q}_{2},\, {\bf q}_{3}}
\, 
f^{\phantom{\hspace{0.03cm}\ast}\!i}_{{\bf q}}\,
f^{\phantom{\hspace{0.03cm}\ast}\!\!i_{1}}_{{\bf q}_{1}}\hspace{0.03cm}
f^{\hspace{0.03cm}\ast\,i_{2}}_{{\bf q}_{2}}\hspace{0.03cm} 
f^{\hspace{0.03cm}\ast\,i_{3}}_{{\bf q}_{3}}\,
(2\pi)^{3}\hspace{0.03cm}\delta({\bf q}^{\prime} + {\bf q}_{1} - {\bf q}_{2} - {\bf q}_{3}).
\]
Summing and averaging two last equations, we obtain further
\begin{equation}
\delta^{\hspace{0.02cm}ij}(2\pi)^{3}\hspace{0.03cm}\delta({\bf q} - {\bf q}\!\ ')\,
\frac{\partial\hspace{0.03cm} n^{-}_{{\bf q}}}{\partial\hspace{0.03cm} t}
\label{eq:7y}
\end{equation}
\[
=
-\hspace{0.02cm}i\!\int\frac{d{\bf q}_{1}\hspace{0.03cm} d{\bf q}_{2}\hspace{0.03cm} d{\bf q}_{3}}{(2\pi)^{9}}\
\biggl\{\widetilde{T}^{(2)\, i\; i_{1}\, i_{2}\, i_{3}}_{{\bf q},\, {\bf q}_{1},\, {\bf q}_{2},\, {\bf q}_{3}}\,
I^{\,j\; i_{1}\, i_{2}\, i_{3}}_{{\bf q}^{\prime},\, {\bf q}_{1},\, {\bf q}_{2},\, {\bf q}_{3}}\,
(2\pi)^{3}\hspace{0.03cm}\delta({\bf q} + {\bf q}_{1} - {\bf q}_{2} - {\bf q}_{3})
\]
\[
-\ 
\widetilde{T}^{\hspace{0.03cm}\ast\,(2)\, j\, i_{1}\, i_{2}\, i_{3}}_{\ {\bf q}^{\prime},\, {\bf q}_{1},\, {\bf q}_{2},\, {\bf q}_{3}}\,
I^{\, i_{2}\, i_{3}\, i\; i_{1}}_{{\bf q}_{2},\, {\bf q}_{3},\, {\bf q},\, {\bf q}_{1}}\,
(2\pi)^{3}\hspace{0.03cm}\delta({\bf q}^{\prime} + {\bf q}_{1} - {\bf q}_{2} - {\bf q}_{3})\biggr\},
\]
where
\begin{equation}
I^{\, i\; i_{1}\, i_{2}\, i_{3}}_{{\bf q},\, {\bf q}_{1},\, {\bf q}_{2},\, {\bf q}_{3}}
=
\bigl\langle f^{\hspace{0.03cm}\ast\,i}_{{\bf q}}\, f^{\hspace{0.03cm}\ast\,i_{1}}_{{\bf q}_{1}}\hspace{0.03cm}
f^{\phantom{\hspace{0.03cm}\ast}\!\!i_{2}}_{{\bf q}_{2}}\, f^{\phantom{\hspace{0.03cm}\ast}\!\!i_{3}}_{{\bf q}_{3}}\,\bigr\rangle
\label{eq:7yy}
\end{equation}
is the four-point correlation function. Differentiation of the correlation function $I^{\, i\; i_{1}\; i_{2}\; i_{3}}_{{\bf q},\, {\bf q}_{1},\, {\bf q}_{2},\, {\bf q}_{3}}$ with respect to $t$,  with allowance made for (\ref{eq:7q}) and (\ref{eq:7w}), yields the equation on the right-hand side of which contains the sixth-order correlation functions in the variables $f^{\hspace{0.03cm}\ast\, i}_{{\bf q}}$ and $f^{\phantom{\hspace{0.03cm}\ast}\!\!i}_{{\bf q}}$:
\begin{equation}
\frac{\partial\hspace{0.01cm} I^{\, i\; i_{1}\, i_{2}\, i_{3}}_{{\bf q},\, {\bf q}_{1},\, {\bf q}_{2},\, {\bf q}_{3}}}{\partial\hspace{0.03cm} t}
=
i\hspace{0.03cm}\bigl[\,\omega^{-}_{{\bf q}} + \omega^{-}_{{\bf q}_{1}} - \omega^{-}_{{\bf q}_{2}} -
\omega^{-}_{{\bf q}_{3}}\bigr]
\, I^{\, i\; i_{1}\, i_{2}\, i_{3}}_{{\bf q},\, {\bf q}_{1},\, {\bf q}_{2},\, {\bf q}_{3}}
\label{eq:7u}
\end{equation}
\begin{align}
-\; i\!&\int\frac{d{\bf q}^{\prime}_{1}\, d{\bf q}^{\prime}_{2}\, d{\bf q}^{\prime}_{3}}{(2\pi)^{9}}\
\widetilde{T}^{\hspace{0.03cm}\ast\hspace{0.03cm}(2)\, i\; i^{\prime}_{1}\, i^{\prime}_{2}\, i^{\prime}_{3}}_{{\bf q},\, 
{\bf q}^{\prime}_{1},\, {\bf q}^{\prime}_{2},\, {\bf q}^{\prime}_{3}}\,
\bigl\langle
f^{\, i^{\prime}_{1}}_{{\bf q}^{\prime}_{1}}
\,f^{\hspace{0.04cm}\ast\,i^{\prime}_{2}}_{{\bf q}^{\prime}_{2}}
\hspace{0.03cm} f^{\hspace{0.04cm}\ast\,i^{\prime}_{3}}_{{\bf q}^{\prime}_{3}}
\hspace{0.03cm} f^{\hspace{0.04cm}\ast\,\!i_{1}}_{{\bf q}^{\phantom{\prime}}_{1}}
\hspace{0.03cm} f^{\phantom{\hspace{0.03cm}\ast}\!\!\!i^{\phantom{\prime}}_{2}}_{{\bf q}^{\phantom{\prime}}_{2}}
\,f^{\phantom{\hspace{0.03cm}\ast}\!i^{\phantom{\prime}}_{3}}_{{\bf q}^{\phantom{\prime}}_{3}}
\,\bigr\rangle\, 
(2\pi)^{3}\hspace{0.03cm}\delta({\bf q} + {\bf q}^{\prime}_{1} - {\bf q}^{\prime}_{2} - {\bf q}^{\prime}_{3})
\notag\\[1ex]
-\; i\!&\int\frac{d{\bf q}^{\prime}_{1}\, d{\bf q}^{\prime}_{2}\, d{\bf q}^{\prime}_{3}}{(2\pi)^{9}}\
\widetilde{T}^{\hspace{0.03cm}\ast\hspace{0.03cm}(2)\, i_{1}\, i^{\prime}_{1}\, i^{\prime}_{2}\, i^{\prime}_{3}}_{{\bf q}_{1},\, {\bf q}^{\prime}_{1},\, {\bf q}^{\prime}_{2},\, {\bf q}^{\prime}_{3}}\,
\bigl\langle
f^{\hspace{0.04cm}\ast\,i^{\phantom{\prime}}\!}_{{\bf q}}
\hspace{0.03cm} f^{\,i^{\prime}_{1}}_{{\bf q}^{\prime}_{1}}
\,f^{\hspace{0.04cm}\ast\,i^{\prime}_{2}}_{{\bf q}^{\prime}_{2}}
\hspace{0.03cm} f^{\hspace{0.04cm}\ast\,i^{\prime}_{3}}_{{\bf q}^{\prime}_{3}}
\,f^{\phantom{\hspace{0.03cm}\ast}\!\!i_{2}}_{{\bf q}^{\phantom{\prime}}_{2}}
\,f^{\phantom{\hspace{0.03cm}\ast}\!\!i_{3}}_{{\bf q}^{\phantom{\prime}}_{3}}
\,\bigr\rangle\, 
(2\pi)^{3}\hspace{0.03cm}\delta({\bf q}_{1} + {\bf q}^{\prime}_{1} - {\bf q}^{\prime}_{2} - {\bf q}^{\prime}_{3})
\notag\\[1ex]
-\; i\!&\int\frac{d{\bf q}^{\prime}_{1}\, d{\bf q}^{\prime}_{2}\, d{\bf q}^{\prime}_{3}}{(2\pi)^{9}}\
\widetilde{T}^{\hspace{0.03cm}(2)\, i_{2}\, i^{\prime}_{1}\, i^{\prime}_{2}\, i^{\prime}_{3}}_{{\bf q}_{2},\, 
{\bf q}^{\prime}_{1},\, {\bf q}^{\prime}_{2},\, {\bf q}^{\prime}_{3}}\,
\bigl\langle
f^{\hspace{0.04cm}\ast\, i}_{{\bf q}^{\phantom{\prime}}}
\hspace{0.03cm} f^{\hspace{0.04cm}\ast\, i_{1}}_{{\bf q}_{1}}
\hspace{0.03cm} f^{\hspace{0.04cm}\ast\, i^{\prime}_{1}}_{{\bf q}^{\prime}_{1}}
\,f^{\, i^{\prime}_{2}}_{{\bf q}^{\prime}_{2}}
\,f^{\, i^{\prime}_{3}}_{{\bf q}^{\prime}_{3}}
\,f^{\phantom{\hspace{0.03cm}\ast}\!\!i_{3}}_{{\bf q}^{\phantom{\prime}}_{3}}
\,\bigr\rangle\, 
(2\pi)^{3}\hspace{0.03cm}\delta({\bf q}_{2} + {\bf q}^{\prime}_{1} - {\bf q}^{\prime}_{2} - {\bf q}^{\prime}_{3})
\notag\\[1ex]
-\; i\!&\int\frac{d{\bf q}^{\prime}_{1}\, d{\bf q}^{\prime}_{2}\, d{\bf q}^{\prime}_{3}}{(2\pi)^{9}}\
\widetilde{T}^{\hspace{0.03cm}(2)\, i_{3}\, i^{\prime}_{1}\, i^{\prime}_{2}\, i^{\prime}_{3}}_{{\bf q}_{3},\, 
{\bf q}^{\prime}_{1},\, {\bf q}^{\prime}_{2},\, {\bf q}^{\prime}_{3}}\,
\bigl\langle
f^{\hspace{0.04cm}\ast\, i}_{{\bf q}^{\phantom{\prime}}}
\hspace{0.03cm} f^{\hspace{0.04cm}\ast\, i_{1}}_{{\bf q}^{\phantom{\prime}}_{1}}
\hspace{0.03cm} f^{\,i_{2}}_{{\bf q}^{\phantom{\prime}}_{2}}
\,f^{\hspace{0.04cm}\ast\, i^{\prime}_{1}}_{{\bf q}^{\prime}_{1}}
\hspace{0.03cm} f^{\, i^{\prime}_{2}}_{{\bf q}^{\prime}_{2}}
\,f^{\, i^{\prime}_{3}}_{{\bf q}^{\prime}_{3}}
\,\bigr\rangle\, 
(2\pi)^{3}\hspace{0.03cm}\delta({\bf q}_{3} + {\bf q}^{\prime}_{1} - {\bf q}^{\prime}_{2} - {\bf q}^{\prime}_{3}).
\notag
\end{align}
Let us close the chain of equations for the correlation functions by expressing the sixth-order correlation functions in terms of pairwise correlation functions. Thus, for example, the first correlation function on the right-hand side of 
(\ref{eq:7u}) has the following structure:
\begin{equation}
\bigl\langle
f^{\, i^{\prime}_{1}}_{{\bf q}^{\prime}_{1}}
\,f^{\hspace{0.04cm}\ast\,i^{\prime}_{2}}_{{\bf q}^{\prime}_{2}}
\hspace{0.03cm} f^{\hspace{0.04cm}\ast\,i^{\prime}_{3}}_{{\bf q}^{\prime}_{3}}
\hspace{0.03cm} f^{\hspace{0.04cm}\ast\,i_{1}}_{{\bf q}^{\phantom{\prime}}_{1}}
\hspace{0.03cm} f^{\phantom{\hspace{0.03cm}\ast}\!\!i^{\phantom{\prime}}_{2}}_{{\bf q}^{\phantom{\prime}}_{2}}
\,f^{\phantom{\hspace{0.03cm}\ast}\!\!i^{\phantom{\prime}}_{3}}_{{\bf q}^{\phantom{\prime}}_{3}}
\bigr\rangle 
=
\label{eq:7i}
\end{equation}
\begin{align}
=
-\hspace{0.03cm}(2\pi)^{9}\hspace{0.03cm}\Bigl\{
&\delta^{\;\!i^{\prime}_{3}i^{\phantom{\prime}}_{2}}
\delta^{\;\!i^{\prime}_{1}i^{\phantom{\prime}}_{1}}
\delta^{\;\!i^{\prime}_{2}i^{\phantom{\prime}}_{3}}\,
\delta({\bf q}^{\prime}_{3} - {\bf q}^{\phantom{\prime}}_{2})
\delta({\bf q}^{\prime}_{1} - {\bf q}^{\phantom{\prime}}_{1})
\delta({\bf q}^{\prime}_{2} - {\bf q}^{\phantom{\prime}}_{3})\,
n^{-}_{{\bf q}_{2}} n^{-}_{{\bf q}_{1}} n^{-}_{{\bf q}_{3}}
\notag\\[1ex]
-\
&\delta^{\;\!i^{\prime}_{3}i^{\phantom{\prime}}_{3}}
\delta^{\;\!i^{\prime}_{2}i^{\phantom{\prime}}_{2}}
\delta^{\;\!i^{\phantom{\prime}}_{1}i^{\prime}_{1}}\,
\delta({\bf q}^{\prime}_{3} - {\bf q}^{\phantom{\prime}}_{3})
\delta({\bf q}^{\prime}_{2} - {\bf q}^{\phantom{\prime}}_{2})
\delta({\bf q}^{\prime}_{1} - {\bf q}^{\phantom{\prime}}_{1})\,
n^{-}_{{\bf q}_{3}} n^{-}_{{\bf q}_{2}} n^{-}_{{\bf q}_{1}}
\notag\\[1ex]
+\
&\delta^{\;\!i^{\prime}_{3}i^{\prime}_{1}}
\delta^{\;\!i_{1}i_{3}}
\delta^{\;\!i^{\prime}_{2}i^{\phantom{\prime}}_{2}}\,
\delta({\bf q}^{\prime}_{3} - {\bf q}^{\prime}_{1})
\delta({\bf q}_{1} - {\bf q}_{3})
\delta({\bf q}^{\prime}_{2} - {\bf q}^{\phantom{\prime}}_{2})\,
n^{-}_{{\bf q}^{\prime}_{1}} n^{-}_{{\bf q}_{3}} n^{-}_{{\bf q}_{2}}
\notag\\[1ex]
-\
&\delta^{\;\!i^{\prime}_{3}i^{\prime}_{1}}
\delta^{\;\!i_{1}i_{2}}
\delta^{\;\!i^{\prime}_{2}i^{\phantom{\prime}}_{3}}\,
\delta({\bf q}^{\prime}_{3} - {\bf q}^{\prime}_{1})
\delta({\bf q}_{1} - {\bf q}_{2})
\delta({\bf q}^{\prime}_{2} - {\bf q}^{\phantom{\prime}}_{3})\,
n^{-}_{{\bf q}^{\prime}_{1}} n^{-}_{{\bf q}_{2}} n^{-}_{{\bf q}_{3}}
\notag\\[1ex]
+\
&\delta^{\;\!i^{\prime}_{3}i^{\phantom{\prime}}_{3}}
\delta^{\;\!i_{1}i_{2}}
\delta^{\;\!i^{\prime}_{2}i^{\prime}_{1}}\,
\delta({\bf q}^{\prime}_{3} - {\bf q}^{\phantom{\prime}}_{3})
\delta({\bf q}_{1} - {\bf q}_{2})
\delta({\bf q}^{\prime}_{2} - {\bf q}^{\prime}_{1})\,
n^{-}_{{\bf q}_{3}} n^{-}_{{\bf q}_{2}} n^{-}_{{\bf q}^{\prime}_{1}}
\notag\\[1ex]
-\
&\delta^{\;\!i^{\prime}_{3}i^{\phantom{\prime}}_{2}}
\delta^{\;\!i_{1}i_{3}}
\delta^{\;\!i^{\prime}_{2}i^{\prime}_{1}}\,
\delta({\bf q}^{\prime}_{3} - {\bf q}^{\phantom{\prime}}_{2})
\delta({\bf q}_{1} - {\bf q}_{3})
\delta({\bf q}^{\prime}_{2} - {\bf q}^{\prime}_{1})\,
n^{-}_{{\bf q}_{2}} n^{-}_{{\bf q}_{3}} n^{-}_{{\bf q}^{\prime}_{1}}\Bigr\}.
\notag
\end{align}
In this expression only first two terms give the required contribution to the desired kinetic equation. Therefore, by substituting these terms into the first integral on the right-hand side of (\ref{eq:7u}), summing over the color indices $i^{\prime}_{1},\, i^{\prime}_{2}$, $i^{\prime}_{3}$, and integrating over the momenta ${\bf q}^{\prime}_{1},\, {\bf q}^{\prime}_{2}$, ${\bf q}^{\prime}_{3}$, as a result we get
\begin{equation}
-\hspace{0.03cm}i\hspace{0.04cm}\Bigl\{
\widetilde{T}^{\hspace{0.03cm}\ast\hspace{0.03cm}(2)\, i\; i_{1}\, i_{2}\, i_{3}}_{\ {\bf q},\, {\bf q}_{1},\, {\bf q}_{2},\, 
{\bf q}_{3}}\; n^{-}_{{\bf q}_{1}} n^{-}_{{\bf q}_{2}} n^{-}_{{\bf q}_{3}}
-
\widetilde{T}^{\hspace{0.03cm}\ast\hspace{0.03cm}(2)\, i\; i_{1}\, i_{3}\, i_{2}}_{\ {\bf q},\, {\bf q}_{1},\, {\bf q}_{3},\, 
{\bf q}_{2}}\; n^{-}_{{\bf q}_{1}} n^{-}_{{\bf q}_{2}} n^{-}_{{\bf q}_{3}}
\Bigr\}
\hspace{0.03cm}(2\pi)^{3}\hspace{0.03cm}
\delta({\bf q} + {\bf q}_{1} - {\bf q}_{2} - {\bf q}_{3})
\label{eq:7o}
\end{equation}
\[
= -\hspace{0.03cm}2\hspace{0.03cm}i\,
\widetilde{T}^{\hspace{0.03cm}\ast\hspace{0.03cm}(2)\, i\; i_{1}\, i_{2}\, i_{3}}_{\ {\bf q},\, {\bf q}_{1},\, {\bf q}_{2},\, 
{\bf q}_{3}}\; n^{-}_{{\bf q}_{1}} n^{-}_{{\bf q}_{2}} n^{-}_{{\bf q}_{3}}
\hspace{0.03cm}(2\pi)^{3}\hspace{0.03cm}
\delta({\bf q} + {\bf q}_{1} - {\bf q}_{2} - {\bf q}_{3}).
\]
Here, in the last step we have taken into account the property of antisymmetry of the effective amplitude $\widetilde{T}^{\hspace{0.03cm}(2)}$ under interchange of two last color indices and corresponding momenta
\begin{equation}
\widetilde{T}^{\hspace{0.03cm}\ast\hspace{0.03cm}(2)\, i\, i_{1}\, i_{2}\, i_{3}}_{\ {\bf q},\, {\bf q}_{1},\, {\bf q}_{2},\, {\bf q}_{3}}
=
-\hspace{0.03cm}
\widetilde{T}^{\hspace{0.03cm}\ast\hspace{0.03cm}(2)\, i\, i_{1}\, i_{3}\, i_{2}}_{\ {\bf q},\, {\bf q}_{1},\, {\bf q}_{3},\, {\bf q}_{2}}.
\label{eq:7p}
\end{equation}
\indent Further we consider the second six-point correlation function in (\ref{eq:7u}). In this correlator, we write out explicitly only the proper terms:
\[
\bigl\langle f^{\,\ast\,i}_{{\bf q}^{\phantom{\prime}}}
\hspace{0.03cm} f^{\,i^{\prime}_{1}}_{{\bf q}^{\prime}_{1}}
\, f^{\,\ast\,i^{\prime}_{2}}_{{\bf q}^{\prime}_{2}}
\hspace{0.03cm} f^{\,\ast\,i^{\prime}_{3}}_{{\bf q}^{\prime}_{3}}
\hspace{0.03cm} f^{\phantom{\hspace{0.03cm}\ast}\!\!i_{2}}_{{\bf q}^{\phantom{\prime}}_{2}}
\,f^{\phantom{\hspace{0.03cm}\ast}\!\!i_{3}}_{{\bf q}^{\phantom{\prime}}_{3}}
\bigr\rangle
=
\]
\[
=
-\hspace{0.03cm}(2\pi)^{9}\Bigl\{\hspace{0.03cm}
\delta^{\;\!i^{\phantom{\prime}}\!i^{\prime}_{1}}\hspace{0.03cm}
\delta^{\;\!i^{\phantom{\prime}}_{2}i^{\prime}_{2}}\hspace{0.03cm}
\delta^{\;\!i^{\phantom{\prime}}_{3}i^{\prime}_{3}}\,
\delta({\bf q}^{\prime}_{1} - {\bf q}^{\phantom{\prime}})\hspace{0.03cm}
\delta({\bf q}^{\prime}_{2} - {\bf q}^{\phantom{\prime}}_{2})\hspace{0.03cm}
\delta({\bf q}^{\prime}_{3} - {\bf q}^{\phantom{\prime}}_{3})\,
n^{-}_{{\bf q}} n^{-}_{{\bf q}_{2}} n^{-}_{{\bf q}_{3}}
-\, (2^{\prime}\rightleftarrows 3^{\prime})\, +\, \ldots\,\Bigr\}.
\]
Substituting this expression into the second integral in (\ref{eq:7u}), we obtain the expression which is similar to the expression (\ref{eq:7o}):
\[
2\hspace{0.03cm}i\,
\widetilde{T}^{\hspace{0.03cm}\ast\hspace{0.03cm}(2)\, i_{1}\, i\; i_{2}\, i_{3}}_{\ {\bf q}_{1},\, {\bf q},\, {\bf q}_{2},\, 
{\bf q}_{3}}\; n^{-}_{\bf q} n^{-}_{{\bf q}_{2}} n^{-}_{{\bf q}_{3}}
\,(2\pi)^{3}\hspace{0.03cm}
\delta({\bf q} + {\bf q}_{1} - {\bf q}_{2} - {\bf q}_{3}).
\]
The same reasoning for the third and fourth correlators on right-hand side of (\ref{eq:7u}) gives us two remaining contributions, correspondingly,
\[
\hspace{0.35cm}
2\hspace{0.03cm}i\,
\widetilde{T}^{\,(2)\, i_{2}\, i_{3}\, i\; i_{1}}_{\, {\bf q}_{3},\, {\bf q}_{2},\, {\bf q},\, {\bf q}_{1}}\;
n^{-}_{\bf q} n^{-}_{{\bf q}_{1}} n^{-}_{{\bf q}_{3}}
\,(2\pi)^{3}\hspace{0.03cm}
\delta({\bf q} + {\bf q}_{1} - {\bf q}_{2} - {\bf q}_{3}).
\]
and
\[
-\hspace{0.03cm}2\hspace{0.03cm}i\,
\widetilde{T}^{\,(2)\, i_{3} \, i_{2}\, i\; i_{1}}_{\, {\bf q}_{3},\, {\bf q}_{2},\, {\bf q},\, {\bf q}_{1}}\;
n^{-}_{\bf q} n^{-}_{{\bf q}_{1}} n^{-}_{{\bf q}_{2}}
\,(2\pi)^{3}\hspace{0.03cm}
\delta({\bf q} + {\bf q}_{1} - {\bf q}_{2} - {\bf q}_{3}).
\]
Considering the symmetry relations (\ref{eq:7e}) and (\ref{eq:7p}) for the scattering amplitude $\widetilde{T}^{\,(2)}$, we obtain the equation for the fourth-order correlation function, instead of (\ref{eq:7u}),
\begin{equation}
\frac{\partial\hspace{0.01cm} I^{\;i\; i_{1}\, i_{2}\, i_{3}}_{{\bf q},\, {\bf q}_{1},\, {\bf q}_{2},\, {\bf q}_{3}}}{\partial\hspace{0.03cm} t}
=
i\hspace{0.03cm}\bigl[\,\omega^{-}_{{\bf q}} + \omega^{-}_{{\bf q}_{1}} - \omega^{-}_{{\bf q}_{2}} -
\omega^{-}_{{\bf q}_{3}}\bigr]
\hspace{0.03cm} I^{\; i\; i_{1}\, i_{2}\, i_{3}}_{{\bf q},\, {\bf q}_{1},\, {\bf q}_{2},\, {\bf q}_{3}}
\label{eq:7a}
\end{equation}
\[
-\;2\hspace{0.03cm}i\,
\widetilde{T}^{\,\ast\hspace{0.03cm}(1)\, i\, i_{1}\, i_{2}\, i_{3}}_{\ {\bf q},\, {\bf q}_{1},\, {\bf q}_{2},\, {\bf q}_{3}}\, 
\Bigl(
n^{-}_{\bf q} n^{-}_{{\bf q}_{2}} n^{-}_{{\bf q}_{3}} +
n^{-}_{{\bf q}_{1}} n^{-}_{{\bf q}_{2}} n^{-}_{{\bf q}_{3}} -
n^{-}_{{\bf q}} n^{-}_{{\bf q}_{1}} n^{-}_{{\bf q}_{3}} -
n^{-}_{{\bf q}} n^{-}_{{\bf q}_{1}} n^{-}_{{\bf q}_{2}}
\Bigr)
\]
\[
\times
\,(2\pi)^{3}\hspace{0.03cm}
\delta({\bf q} + {\bf q}_{1} - {\bf q}_{2} - {\bf q}_{3}).
\]


\section{\bf Kinetic equation for plasmino-plasmino scattering}
\label{section_8}
\setcounter{equation}{0}

Let us consider the derivation of the kinetic equation describing the elastic scattering process of plasminos among themselves. We will follow the line of reasoning, which was used for the pure bosonic case in \cite{markov_2020}. In  Eq.\,(\ref{eq:7a}) we neglect the term with the time derivative in comparison with the term containing the difference of 
the eigenfrequencies of wave packets. Instead of (\ref{eq:7a}), we have
\begin{equation}
I^{\;i\; i_{1}\, i_{2}\, i_{3}}_{{\bf q},\, {\bf q}_{1},\, {\bf q}_{2},\, {\bf q}_{3}} 
\simeq\,
n^{-}_{{\bf q}} n^{-}_{{\bf q}_{1}}\ \!(2\pi)^{6}\hspace{0.03cm}
\Bigl[\ \! 
\delta^{\hspace{0.03cm}i\ \!\!i_{3}} \delta^{\hspace{0.03cm}i_{1}\hspace{0.01cm} i_{2}}\ \! \delta({\bf q} - {\bf q}_{3}) \delta({\bf q}_{1} - {\bf q}_{2})
-
\delta^{\hspace{0.03cm}i\ \!\!i_{2}} \delta^{\hspace{0.03cm}i_{1}\hspace{0.01cm} i_{3}}\ \! \delta({\bf q} - {\bf q}_{2}) \delta({\bf q}_{1} - {\bf q}_{3})
\Bigr]
\label{eq:8_1q}
\end{equation}
\[
+\; \frac{2}{\Delta\hspace{0.03cm}\omega - i\hspace{0.03cm}0}\
\widetilde{T}^{\hspace{0.03cm}\ast\hspace{0.03cm}(2)\hspace{0.03cm} i\; i_{1}\, i_{2}\, i_{3}}_{\ {\bf q},\, {\bf q}_{1},\, {\bf q}_{2},\, {\bf q}_{3}}\ \!
\Bigl(
n^{-}_{\bf q} n^{-}_{{\bf q}_{2}} n^{-}_{{\bf q}_{3}} +
n^{-}_{{\bf q}_{1}} n^{-}_{{\bf q}_{2}} n^{-}_{{\bf q}_{3}} -
n^{-}_{{\bf q}} n^{-}_{{\bf q}_{1}} n^{-}_{{\bf q}_{3}} -
n^{-}_{{\bf q}} n^{-}_{{\bf q}_{1}} n^{-}_{{\bf q}_{2}}\Bigr)
\]
\[
\times
\ \!(2\pi)^{3}\hspace{0.03cm}
\delta({\bf q} + {\bf q}_{1} - {\bf q}_{2} - {\bf q}_{3}),
\]
where for brevity we have designated the resonance frequency difference
\begin{equation}
\Delta\hspace{0.03cm}\omega =
\omega^{-}_{{\bf q}} + \omega^{-}_{{\bf q}_{1}} - \omega^{-}_{{\bf q}_{2}} - \omega^{-}_{{\bf q}_{3}}.
\label{eq:8q}
\end{equation}
The first term on the right-hand side of (\ref{eq:8_1q}) corresponding to the completely uncorrelated waves (purely Gaussian fluctuations) is the solution of the homogeneous equation for the fourth-order correlation function $I^{\,i\; i_{1}\, i_{2}\, i_{3}}_{{\bf q},\, {\bf q}_{1},\, {\bf q}_{2},\, {\bf q}_{3}}$. The structure of this contribution follows from the definitions of the four- and two-point correlation functions (\ref{eq:7yy}) и (\ref{eq:7r}):
\[
I^{\, i\; i_{1}\, i_{2}\, i_{3}}_{{\bf q},\, {\bf q}_{1},\, {\bf q}_{2},\, {\bf q}_{3}}
=
\bigl\langle f^{\hspace{0.03cm}\ast\,i}_{{\bf q}}\, f^{\hspace{0.03cm}\ast\,i_{1}}_{{\bf q}_{1}}\hspace{0.03cm}
f^{\phantom{\hspace{0.03cm}\ast}\!\!i_{2}}_{{\bf q}_{2}}\, f^{\phantom{\hspace{0.03cm}\ast}\!\!i_{3}}_{{\bf q}_{3}}\,\bigr\rangle
\simeq\,
\bigl\langle f^{\hspace{0.03cm}\ast\,i}_{{\bf q}}\hspace{0.03cm}
f^{\phantom{\hspace{0.03cm}\ast}\!\!i_{3}}_{{\bf q}_{3}}\,\bigr\rangle
\bigl\langle f^{\hspace{0.03cm}\ast\,i_{1}}_{{\bf q}_{1}}\,
f^{\phantom{\hspace{0.03cm}\ast}\!\!i_{2}}_{{\bf q}_{2}}\,\bigr\rangle
-
\bigl\langle f^{\hspace{0.03cm}\ast\,i}_{{\bf q}}\,
f^{\phantom{\hspace{0.03cm}\ast}\!\!i_{2}}_{{\bf q}_{2}}\,\bigr\rangle
\bigl\langle f^{\hspace{0.03cm}\ast\,i_{1}}_{{\bf q}_{1}}\hspace{0.03cm}
f^{\phantom{\hspace{0.03cm}\ast}\!\!i_{3}}_{{\bf q}_{3}}\,\bigr\rangle
\]
\[
=
\delta^{\hspace{0.03cm}i\ \!\!i_{3}}(2\pi)^{3}\hspace{0.03cm}\delta({\bf q} - {\bf q}_{3})\hspace{0.03cm}n^{-}_{{\bf q}}\hspace{0.03cm}
\delta^{\hspace{0.03cm}i_{1}\ \!\!i_{2}}(2\pi)^{3}\hspace{0.03cm}\delta({\bf q}_{1} - {\bf q}_{2})\hspace{0.03cm}n^{-}_{{\bf q}_{1}}
-
\delta^{\hspace{0.03cm}i\ \!\!i_{2}}(2\pi)^{3}\hspace{0.03cm}\delta({\bf q} - {\bf q}_{2})\hspace{0.03cm}n^{-}_{{\bf q}}\hspace{0.03cm}
\delta^{\hspace{0.03cm}i_{1}\ \!\!i_{3}}(2\pi)^{3}\hspace{0.03cm}\delta({\bf q}_{1} - {\bf q}_{3})\hspace{0.03cm}n^{-}_{{\bf q}_{1}}.
\]
The second term in (\ref{eq:8_1q}) defines the deviation of four-point correlator from the Gaussian approximation for weak nonlinearity level of interacting waves.\\
\indent Let us substitute the first term into the right-hand side of the equation for $n^{-}_{\bf q}$ (\ref{eq:7y}):
\begin{align}
-\hspace{0.03cm}i\ \!(2\pi)^{3}\hspace{0.03cm}n^{-}_{\bf q}\!\int\!\frac{d{\bf q}_{1}}{(2\pi)^{3}}\
n^{-}_{{\bf q}_{1}}
\Bigl\{&\widetilde{T}^{\,(2)\, i\, i_{1}\, j\, i_{1}}_{{\bf q},\, {\bf q}_{1},\, {\bf q}^{\prime},\, {\bf q}_{1}}\ \!
\delta({\bf q} - {\bf q}^{\prime})
\,-\,
\widetilde{T}^{\,(2)\,i\,i_{1}\, i_{1}\, j}_{{\bf q},\, {\bf q}_{1},\, {\bf q}_{1},\, {\bf q}^{\prime}}\ \!
\delta({\bf q} - {\bf q}^{\prime})
\notag\\[1ex]
&-\;
\widetilde{T}^{\,\ast\hspace{0.03cm}(2)\, j\, i_{1}\, i\, i_{1}}_{\ {\bf q}^{\prime},\, {\bf q}_{1},\, {\bf q},\, {\bf q}_{1}}\ \!
\delta({\bf q}^{\prime} - {\bf q})
\,+\,
\widetilde{T}^{\,\ast\hspace{0.03cm}(2)\, j\, i_{1}\, i_{1}\, i}_{\ {\bf q}^{\prime},\, {\bf q}_{1},\, {\bf q}_{1},\, {\bf q}}\ \!
\delta({\bf q}^{\prime} - {\bf q})\Bigr\} 
\notag
\end{align}
\begin{equation}
= -\hspace{0.03cm}2\hspace{0.03cm} i\ \!(2\pi)^{3}\hspace{0.03cm}\delta({\bf q} - {\bf q}^{\prime})\hspace{0.03cm}
n^{-}_{\bf q}\!\int\!\frac{d{\bf q}_{1}}{(2\pi)^{3}}\ n^{-}_{{\bf q}_{1}}
\Bigl\{\widetilde{T}^{\,(2)\, i\, i_{1}\, j\, i_{1}}_{{\bf q},\, {\bf q}_{1},\, {\bf q},\, {\bf q}_{1}}
\,-\,
\widetilde{T}^{\,\ast\hspace{0.03cm}(2)\, j\, i_{1}\, i\, i_{1}}_{\ {\bf q},\; {\bf q}_{1},\; {\bf q},\; {\bf q}_{1}}\Bigr\}.
\label{eq:8w}
\end{equation}
Substituting further the second term into the right-hand side of Eq.\,(\ref{eq:7y}), we obtain
\[
-\hspace{0.03cm}2\hspace{0.035cm}i\!\int\frac{d{\bf q}_{1}\hspace{0.02cm} d{\bf q}_{2}\hspace{0.03cm} d{\bf q}_{3}}{(2\pi)^{9}}\;
\biggl\{
\widetilde{T}^{\,(2)\, i\, i_{1}\, i_{2}\, i_{3}}_{{\bf q},\, {\bf q}_{1},\, {\bf q}_{2},\, {\bf q}_{3}}
\biggl(\frac{1}{\omega^{-}_{{\bf q}^{\prime}} + \omega^{-}_{{\bf q}_{1}} - \omega^{-}_{{\bf q}_{2}} - \omega^{-}_{{\bf q}_{3}} - i\hspace{0.03cm}0}\biggr)\ \!
\widetilde{T}^{\,\ast\hspace{0.03cm}(2)\, j\; i_{1}\, i_{2}\, i_{3}}_{\ {\bf q}^{\prime},\, {\bf q}_{1},\, {\bf q}_{2},\, {\bf q}_{3}}
\]
\[
\times\ \!
\Bigl[\ \!n^{-}_{{\bf q}^{\prime}}n^{-}_{{\bf q}_{2}} n^{-}_{{\bf q}_{3}} +
n^{-}_{{\bf q}_{1}} n^{-}_{{\bf q}_{2}} n^{-}_{{\bf q}_{3}} -
n^{-}_{{\bf q}^{\prime}} n^{-}_{{\bf q}_{1}} n^{-}_{{\bf q}_{3}} -
n^{-}_{{\bf q}} n^{-}_{{\bf q}_{1}} n^{-}_{{\bf q}_{2}} \Bigr] 
\vspace{0.2cm}
\]
\[
\times\ \!
(2\pi)^{3}\hspace{0.03cm} \delta({\bf q}^{\prime} + {\bf q}_{1} - {\bf q}_{2} - {\bf q}_{3})\ \!
(2\pi)^{3}\hspace{0.03cm} \delta({\bf q} + {\bf q}_{1} - {\bf q}_{2} - {\bf q}_{3}) 
\vspace{0.35cm}
\]
\[
\hspace{3.6cm}
-\,
\widetilde{T}^{\,\ast\hspace{0.03cm}(2)\, j\, i_{1}\, i_{2}\, i_{3}}_{\ {\bf q}^{\prime},\, {\bf q}_{1},\, {\bf q}_{2},\, {\bf q}_{3}}
\biggl(\frac{1}{\omega^{-}_{{\bf q}_{2}} + \omega^{-}_{{\bf q}_{3}} - \omega^{-}_{{\bf q}} - \omega^{-}_{{\bf q}_{1}} - i\hspace{0.03cm}0}\biggr)\ \!
\widetilde{T}^{\hspace{0.03cm}\ast\hspace{0.03cm}(2)\hspace{0.03cm} i_{2}\, i_{3}\, i\, i_{1}}_{\ {\bf q}_{2},\, {\bf q}_{3},\, {\bf q},\, {\bf q}_{1}}
\vspace{0.2cm}
\]
\[
\times\ \!
\Bigl[\ \!n^{-}_{{\bf q}}n^{-}_{{\bf q}_{1}} n^{-}_{{\bf q}_{2}} +
n^{-}_{{\bf q}} n^{-}_{{\bf q}_{1}} n^{-}_{{\bf q}_{3}} -
n^{-}_{{\bf q}_{1}} n^{-}_{{\bf q}_{2}} n^{-}_{{\bf q}_{3}} -
n^{-}_{{\bf q}} n^{-}_{{\bf q}_{2}} n^{-}_{{\bf q}_{3}} \Bigr]
\]
\[
\times\ \!
(2\pi)^{3}\hspace{0.03cm} \delta({\bf q} + {\bf q}_{1} - {\bf q}_{2} - {\bf q}_{3})\ \!
(2\pi)^{3}\hspace{0.03cm} \delta({\bf q}^{\prime} + {\bf q}_{1} - {\bf q}_{2} - {\bf q}_{3})\biggr\}.
\]
With allowance made for the property (\ref{eq:7e}) and for the equality
\[
\delta({\bf q}^{\prime} + {\bf q}_{1} - {\bf q}_{2} - {\bf q}_{3})\ \!
\delta({\bf q} + {\bf q}_{1} - {\bf q}_{2} - {\bf q}_{3}) =
\delta({\bf q} - {\bf q}^{\prime})\ \!
\delta({\bf q} + {\bf q}_{1} - {\bf q}_{2} - {\bf q}_{3}),
\]
the previous expression can be written in a more compact form:
\[
-\hspace{0.03cm}2\hspace{0.03cm}i\hspace{0.03cm} (2\pi)^{3}\hspace{0.03cm}\delta({\bf q} - {\bf q}^{\prime})
\!\int\frac{d{\bf q}_{1}\hspace{0.02cm} d{\bf q}_{2}\hspace{0.03cm} d{\bf q}_{3}}{(2\pi)^{9}}\,
(2\pi)^{3}\hspace{0.03cm} \delta({\bf q} + {\bf q}_{1} - {\bf q}_{2} - {\bf q}_{3})
\left\{
\frac{1}{\Delta\hspace{0.03cm}\omega - i\hspace{0.03cm}0}
\, -\,
\frac{1}{\Delta\hspace{0.03cm}\omega + i\hspace{0.03cm}0} \hspace{0.03cm}
\right\}
\]
\begin{equation}
\times\,
\widetilde{T}^{\,(2)\, i\, i_{1}\, i_{2}\, i_{3}}_{{\bf q},\, {\bf q}_{1},\, {\bf q}_{2},\, {\bf q}_{3}}\;
\widetilde{T}^{\,\ast\hspace{0.03cm}(2)\, j\, i_{1}\, i_{2}\, i_{3}}_{\ {\bf q}^{\prime},\, {\bf q}_{1},\, {\bf q}_{2},\, {\bf q}_{3}}\,
\bigl[\hspace{0.03cm} n^{-}_{\bf q} n^{-}_{{\bf q}_{2}} n^{-}_{{\bf q}_{3}} +
n^{-}_{{\bf q}_{1}} n^{-}_{{\bf q}_{2}} n^{-}_{{\bf q}_{3}} -
n^{-}_{{\bf q}} n^{-}_{{\bf q}_{1}} n^{-}_{{\bf q}_{3}} -
n^{-}_{{\bf q}} n^{-}_{{\bf q}_{1}} n^{-}_{{\bf q}_{2}} \hspace{0.03cm}\bigr].
\label{eq:8e}
\end{equation}
We recall that $\Delta\hspace{0.03cm}\omega$ is defined by the expression (\ref{eq:8q}). Contracting further the obtained expressions (\ref{eq:7y}), (\ref{eq:8w}) and (\ref{eq:8e}) with $\delta^{\hspace{0.03cm}i j}$, taking into account Sokhotsky's formula 
\begin{equation}
\frac{1}{\Delta\hspace{0.03cm}\omega - i\hspace{0.03cm}0}
 -
\frac{1}{\Delta\hspace{0.03cm}\omega + i\hspace{0.03cm}0} = 2\hspace{0.02cm}i\hspace{0.02cm}\pi\hspace{0.03cm}\delta(\Delta\hspace{0.03cm}\omega),
\label{eq:8r}
\end{equation}
and reducing the factor $(2\pi)^{3}\hspace{0.03cm}\delta({\bf q} - {\bf q}^{\prime})$, we find the desired kinetic equation for the colorless abnormal quark excitations
\begin{equation}
\frac{d\hspace{0.03cm} n^{-}_{{\bf q}}}{d\hspace{0.03cm} t}\
=
-\hspace{0.03cm}\frac{4}{C_{A}}\ \!
n^{-}_{\bf q}\!\int\!\frac{d\hspace{0.03cm}{\bf q}_{1}}{(2\pi)^{3}}\ n^{-}_{{\bf q}_{1}}\ \!
{\rm Im}
\Bigl[\hspace{0.03cm}\widetilde{T}^{\hspace{0.03cm}(2)\, i\, i_{1}\, i\, i_{1}}_{{\bf q},\, {\bf q}_{1},\, {\bf q},\, {\bf q}_{1}}\Bigr]
\label{eq:8t}
\end{equation}
\[
-\, \frac{2}{C_{A}}
\!\int\frac{d\hspace{0.03cm}{\bf q}_{1}\hspace{0.02cm} d\hspace{0.03cm}{\bf q}_{2}\hspace{0.03cm} d\hspace{0.03cm}{\bf q}_{3}}
{(2\pi)^{9}}\ (2\pi)^{4}\hspace{0.03cm}
\delta(\omega^{-}_{{\bf q}} + \omega^{-}_{{\bf q}_{1}} - \omega^{-}_{{\bf q}_{2}} -
\omega^{-}_{{\bf q}_{3}})\ \! \delta({\bf q} + {\bf q}_{1} - {\bf q}_{2} - {\bf q}_{3})\ \!
\]
\[
\times\, 
\widetilde{T}^{\,(2)\, i\; i_{1}\, i_{2}\, i_{3}}_{{\bf q},\, {\bf q}_{1},\, {\bf q}_{2},\, {\bf q}_{3}}\;
\widetilde{T}^{\hspace{0.03cm} \ast\hspace{0.03cm}(2)\hspace{0.03cm} i\, i_{1}\, i_{2}\, i_{3}}_{\ {\bf q},\, {\bf q}_{1},\, {\bf q}_{2},\, {\bf q}_{3}}\hspace{0.02cm}
\Bigl(
n^{-}_{\bf q} n^{-}_{{\bf q}_{2}} n^{-}_{{\bf q}_{3}} +
n^{-}_{{\bf q}_{1}} n^{-}_{{\bf q}_{2}} n^{-}_{{\bf q}_{3}} -
n^{-}_{{\bf q}} n^{-}_{{\bf q}_{1}} n^{-}_{{\bf q}_{3}} -
n^{-}_{{\bf q}} n^{-}_{{\bf q}_{1}} n^{-}_{{\bf q}_{2}}\Bigr).
\]
Here, $C_{A} = N_{c}$ is the invariant for the group $SU(N_{c})$. The first term on the right-hand side of (\ref{eq:8t}) describes the process of nonlinear Landau damping of collective Fermi-excitations, a decrement of which represents a linear functional of the plasmino number density $n^{-}_{\bf q}$:
\begin{equation}
\hat{\gamma}_{\cal F}[\hspace{0.03cm} n_{\bf q}^{-}]  \equiv \gamma^{-}({\bf q})
=
\frac{4}{C_{A}}\!\hspace{0.03cm}\int\!\frac{d\hspace{0.03cm}{\bf q}_{1}}{(2\pi)^{3}}\ n^{-}_{{\bf q}_{1}}\ \!
{\rm Im} \Bigl[\hspace{0.03cm}\widetilde{T}^{\,(2)\, i\, i_{1}\, i\, i_{1}}_{{\bf q},\, {\bf q}_{1},\, {\bf q},\, {\bf q}_{1}}\Bigr].
\label{eq:8y}
\end{equation}
The second term in (\ref{eq:8t}) is connected with the process of elastic plasmino-plasmino scattering. The equation (\ref{eq:8t}) can be also presented in a more visual form:
\begin{equation}
\frac{d\hspace{0.03cm} n_{\bf q}^{-}}{d\hspace{0.03cm} t} \equiv \frac{\partial\hspace{0.03cm} n_{\bf q}^{-}}{\partial\hspace{0.03cm} t} +
{\bf v}_{\bf q}^{-}\cdot\frac{\partial\hspace{0.03cm} n_{\bf q}^{-}}
{\partial\hspace{0.04cm} {\bf x}}
=
-\,\hat{\gamma}_{\cal F}[\hspace{0.03cm} n_{\bf q}^{-}] \, n_{\bf q}^{-} - n_{\bf q}^{-}\hspace{0.02cm} \Gamma_{\rm d} [ n_{\bf q}^{-} ] + ( 1 - n_{\bf q}^{-})\hspace{0.03cm}\Gamma_{\rm i}[ n_{\bf q}^{-} ],
\label{eq:8u}
\end{equation}
where
\[
{\bf v}_{\bf q}^{-} = \frac{\partial\hspace{0.03cm} \omega_{\bf q}^{-}}{\partial\hspace{0.03cm} {\bf q}} = 
- \Biggl[\left(\frac{\partial\hspace{0.02cm} {\rm Re}\,^{\ast}\!\Delta^{\!-1}_{-}(q)}{\partial\hspace{0.02cm} {\bf q}} \right)\!
\left( \frac{\partial\hspace{0.02cm} {\rm Re}\,^{\ast}\!\Delta^{\!-1}_{-}(q)}
{\partial\hspace{0.02cm} \omega} \right)^{\!\!-1}\,\Biggr]
\Bigg\vert_{\omega\hspace{0.03cm} = \hspace{0.03cm} \omega_{\bf q}^{-}}
\]
is the group velocity of fermionic abnormal oscillation modes, and generalized decay rate $\Gamma_{\rm d}$ and inverse regeneration rate $\Gamma_{\rm i}$ represent nonlinear functionals of the plasmino number density:
\[
\Gamma_{\rm d}[n_{\bf q}^{-}] =\!
\int\!d\hspace{0.02cm}{\cal T}_{qq\hspace{0.02cm}\rightarrow\hspace{0.02cm} qq}
\hspace{0.04cm}{\it w}_{4}({\bf q}, {\bf q}_{1}; {\bf q}_{2}, {\bf q}_{3})\,
n_{{\bf q}_1}^{-} (1 - n_{{\bf q}_{2}}^{-}) (1 - n_{{\bf q}_{3}}^{-})
\]
and correspondingly,
\[
\Gamma_{\rm i} [n_{\bf q}^{-}] =\!
\int\!d\hspace{0.02cm}{\cal T}_{qq\hspace{0.02cm}\rightarrow\hspace{0.02cm} qq}
\hspace{0.04cm}{\it w}_{4}({\bf q}, {\bf q}_{1}; {\bf q}_{2}, {\bf q}_{3})\,
(1 -  n_{{\bf q}_1}^{-}) n_{{\bf q}_{2}}^{-} n_{{\bf q}_{3}}^{-}\,.
\hspace{0.85cm}
\]
Here,
\begin{equation}
{\it w}_{4}({\bf q}, {\bf q}_{1}; {\bf q}_{2},{\bf q}_{3}) = 
\frac{\!2}{C_{A}}\,\! \sum\limits_{\lambda,\, \lambda_{1},\,\ldots\, =\,\pm}\!\!
\widetilde{T}^{\,(2)\, i\; i_{1}\, i_{2}\, i_{3}}_{{\bf q},\, {\bf q}_{1},\, {\bf q}_{2},\, {\bf q}_{3}}\;
\widetilde{T}^{\hspace{0.03cm}\ast\hspace{0.03cm}(2)\, i\, i_{1}\, i_{2}\, i_{3}}_{\ {\bf q},\, {\bf q}_{1},\, {\bf q}_{2},\, {\bf q}_{3}}
\label{eq:8i}
\end{equation}
is the scattering probability for the process of elastic scattering of two colorless plasminos, and the integration measure is defined as
\[
d\hspace{0.02cm}{\cal T}_{qq\hspace{0.02cm}\rightarrow\hspace{0.02cm} qq} \equiv
(2\pi)^{4}\,
\delta(\omega^{-}_{{\bf q}} + \omega^{-}_{{\bf q}_{1}} - \omega^{-}_{{\bf q}_{2}} - \omega^{-}_{{\bf q}_{3}})\ \!
\delta({\bf q} + {\bf q}_{1} - {\bf q}_{2} - {\bf q}_{3})\ \!
\frac{d{\bf q}_{1}\hspace{0.02cm} d{\bf q}_{2}\hspace{0.03cm} d{\bf q}_{3}}{(2\pi)^{9}}.
\]
In conclusion of this section, we note that on the left-hand side of the kinetic equation (\ref{eq:8u}), in addition to taking into account transfer with a group velocity, an additional term associated for classical nonlinear waves with a nonlinear correction to the frequency should be taken into account (see, for example, the review \cite{zakharov_1985}).
%


\section{Explicit form of the vertex functions $T^{\hspace{0.03cm} (2)\, i\; i_{1}\, i_{2}\, i_{3}}_{{\bf q},\, {\bf q}_{1},\, {\bf q}_{2},\, {\bf q}_{3}}$,  ${\mathcal G}^{\; a\, i\;  i_{1}}_{\,{\bf k},\, {\bf q},\, {\bf q}_{1}}$, ${\mathcal K}^{\; a_{1}\, i\;  i_{1}}_{\,{\bf k}_1,\, {\bf q},\, {\bf q}_{1}}$, and ${\mathcal P}^{\; a_{1}\, i\;  i_{1}}_{\,{\bf k}_1,\, {\bf q},\, {\bf q}_{1}}$}
\label{section_9}
\setcounter{equation}{0}

We pass to the determination of an explicit form of the vertex functions $T^{\,(2)\, i\, i_{1}\, i_{2}\, i_{3}}_{{\bf q},\, {\bf q}_{1},\, {\bf q}_{2},\, {\bf q}_{3}}$,  ${\mathcal G}^{\; a\, i\,  i_{1}}_{\,{\bf k},\, {\bf q},\, {\bf q}_{1}}$, ${\mathcal K}^{\; a_{1}\, i\,  i_{1}}_{\,{\bf k}_1,\, {\bf q},\, {\bf q}_{1}}$, and ${\mathcal P}^{\; a_{1}\, i\,  i_{1}}_{\,{\bf k}_1,\, {\bf q},\, {\bf q}_{1}}$, which enter into the effective amplitude (\ref{eq:6t}). We define these functions in the approximation of the so-called {\it hard thermal loops} (HTL) \cite{blaizot_2002, kraemmer_2004, ghiglieri_2020}. In the paper \cite{markov_2006}, within the HTL-approximation the probability of plasmino-plasmino scattering
\begin{equation}
{\it w}_{qq\rightarrow qq}^{\!(--;--)}
({\bf q}, {\bf q}_1; {\bf q}_2, {\bf q}_3) = \frac{1}{C_{A}}
\sum\limits_{\lambda,\,\lambda_{1},\,\ldots\,=\,\pm}\!\!
{\rm T}^{\; i\; i_{1}\, i_{2}\, i_{3}}_{\lambda\hspace{0.02cm}\lambda_{1}\hspace{0.02cm}\lambda_{2}\hspace{0.02cm}\lambda_{3}}
({\bf q},{\bf q}_1;-{\bf q}_2,-{\bf q}_3)
\bigl({\rm T}^{\; i\; i_{1}\, i_{2}\, i_{3}}_{\lambda\hspace{0.02cm}\lambda_{1}\hspace{0.02cm}\lambda_{2}\hspace{0.02cm}\lambda_{3}}
({\bf q},{\bf q}_1;-{\bf q}_2,-{\bf q}_3)\bigr)^{\!\ast}
\label{eq:9q}
\end{equation}
was obtained. Here, the matrix element of elastic plasmino-plasmino scattering has the following structure:
\begin{equation}
{\rm T}^{\; i\; i_{1}\, i_{2}\, i_{3}}_{\lambda\hspace{0.02cm}\lambda_{1}\hspace{0.02cm}\lambda_{2}\hspace{0.02cm}\lambda_{3}}({\bf q},{\bf q}_1;-{\bf q}_2,-{\bf q}_3)
\equiv g^2
\left(\frac{Z_{-}({\bf q})}{2}\right)^{\!\!1/2}\, \prod_{i=1}^{3} \left(\frac{Z_{-}({\bf q}_i)}{2}\right)^{\!\!1/2}
\label{eq:9w}
\end{equation}
\[
\times\,
v^{(-)}_{\alpha}(\hat{\bf q},\lambda)\hspace{0.02cm}
v^{(-)}_{\alpha_1}(\hat{\bf q}_1,\lambda_{1})\hspace{0.02cm}
\bar{v}^{(-)}_{\alpha_2}(\hat{\bf q}_2,\lambda_{2})\hspace{0.02cm}
\bar{v}^{(-)}_{\alpha_3}(\hat{\bf q}_3,\lambda_{3})\,
\!\,^{\ast}\tilde{\bar{\Gamma}}^{\; i\, i_{1}\, i_{2}\, i_{3}}_{\alpha\hspace{0.02cm}\alpha_1\alpha_2\alpha_3}
(q,q_1,-q_2,-q_3)
\]
and, in turn, the effective amplitude $\!\,^{\ast}\tilde{\bar{\Gamma}}^{\; i\; i_{1}\, i_{2}\, i_{3}}_{\alpha\alpha_1\alpha_2\alpha_3}(q,q_1,-q_2,-q_3)$ is defined as
\begin{equation}
\!\,^{\ast}\tilde{\bar{\Gamma}}^{\; i\; i_{1}\, i_{2}\, i_{3}}_{\alpha\hspace{0.02cm}\alpha_1\alpha_2\alpha_3}
(q,q_1,-q_2,-q_3) =
\label{eq:9e}
\vspace{-0.2cm}
\end{equation}
\begin{align}
= &- (t^a)^{i_{2}\hspace{0.015cm}i}(t^a)^{i_{3}\hspace{0.015cm}i_{1}}
\,^{\ast}\Gamma^{(Q)\mu}_{\alpha_{2}\hspace{0.02cm}\alpha}(-q + q_{2}; -q_{2}, q)
\,^{\ast}{\cal D}_{\mu\mu^{\prime}}(- q + q_{2})
\,^{\ast}\Gamma^{(G)\mu^{\prime}}_{\alpha_{3}\hspace{0.02cm}\alpha_{1}}(q_{1} - q_{3}; q_{3}, -q_{1})
\notag\\[1.5ex]
&+(t^a)^{i_{2}\hspace{0.015cm}i_{1}}(t^a)^{i_{3}\hspace{0.015cm}i}
\,^{\ast}\Gamma^{(Q)\mu}_{\alpha_{3}\hspace{0.02cm}\alpha}(-q + q_{3}; -q_{3}, q)
\,^{\ast}{\cal D}_{\mu\mu^{\prime}}(-q + q_{3})
\,^{\ast}\Gamma^{(G)\mu^{\prime}}_{\alpha_{2}\hspace{0.02cm}\alpha_{1}}(q_{1} - q_{2}; q_{2}, -q_{1}).
\notag
\end{align}
The form of the gluon propagator $\,^{\ast}\tilde{\cal D}_{\mu\mu^{\prime}}(k)$ and the vertex functions $\,^{\ast}\Gamma^{(Q)\mu}_{\alpha\hspace{0.02cm}\beta}(q-q_1;q_1,-q)$ and $\,^{\ast}\Gamma^{(G)\mu}_{\alpha\hspace{0.02cm}\beta}(q-q_1;q_1,-q)$,  is given in Appendices \ref{appendix_A} and \ref{appendix_B} in the HTL-approximation, Eqs.\,(\ref{ap:A8})\,--\,(\ref{ap:A10}) and (\ref{ap:B4})\,--\,(\ref{ap:B6}), correspondingly. For the effective amplitude (\ref{eq:9e}), the following proper\-ty of antisymmetry with respect to the permutation of two last soft-quark legs holds:
\[
\!\,^{\ast}\tilde{\bar{\Gamma}}^{\, i\; i_{1}\, i_{2}\, i_{3}}_{\alpha\hspace{0.02cm} \alpha_{1} \alpha_{2} \alpha_{3}}
(q, q_{1}, -q_{2}, -q_{3})=
-\,
\!\,^{\ast}\tilde{\bar{\Gamma}}^{\, i\; i_{1}\, i_{3}\, i_{2}}_{\alpha\hspace{0.02cm} \alpha_{1} \alpha_{3} \alpha_{2}}
(q, q_{1}, -q_{3}, -q_{2}).
\]
There is no a similar property in general case for the permutation of two first soft-quark legs by virtue of the fact that the effective two-quarks--one-gluon vertex functions with different time ordering of external legs are not equal to each other
\[
\,^{\ast}\Gamma^{(Q)\mu}_{\alpha\hspace{0.02cm}\beta}(q - q_{1}; q_{1}, -q) 
\ne
\,^{\ast}\Gamma^{(G)\mu}_{\alpha\hspace{0.02cm}\beta}(q - q_{1}; q_{1}, -q).
\]
\indent Comparing two expressions (\ref{eq:8i}) and (\ref{eq:9q}) for the plasmino–plasmino scattering probability, we see that the effective amplitude $\widetilde{T}^{\hspace{0.03cm}(2)\, i\; i_{1}\, i_{2}\, i_{3}}_{{\bf q},\, {\bf q}_{1},\, {\bf q}_{2},\, {\bf q}_{3}}$ defined by the expression (\ref{eq:6t}) should be identified up to a numerical factor with the matrix element ${\rm T}^{\, i\; i_{1}\, i_{2}\, i_{3}}_{\lambda\hspace{0.02cm}\lambda_{1}\hspace{0.02cm}\lambda_{2}\hspace{0.02cm}\lambda_{3}} ({\bf q},{\bf q}_1;-{\bf q}_2, -{\bf q}_3)$ (more exactly with its complex conjugation) calculated using the high-temperature quantum field theory, i.e.,
\begin{equation}
\widetilde{T}^{\,(2)\, i\; i_{1}\, i_{2}\, i_{3}}_{{\bf q},\, {\bf q}_{1},\, {\bf q}_{2},\, {\bf q}_{3}}
(\lambda, \lambda_{1}, \lambda_{2}, \lambda_{3})
=
-\hspace{0.02cm}\frac{1}{\sqrt{2}}\,\bigl({\rm T}^{\, i\; i_{1}\, i_{2}\, i_{3}}_{\lambda\hspace{0.02cm}\lambda_{1}\hspace{0.02cm}\lambda_{2}\hspace{0.02cm}\lambda_{3}}({\bf q},{\bf q}_1;-{\bf q}_2,-{\bf q}_3)\bigr)^{\!\ast}.
\label{eq:9r}
\end{equation}
From expressions for the effective amplitudes (\ref{eq:6t}) and (\ref{eq:9w}), (\ref{eq:9e}), we can immediately obtain
\begin{equation}
T^{\,(2)\, i\; i_{1}\, i_{2}\, i_{3}}_{{\bf q},\, {\bf q}_{1},\, {\bf q}_{2},\, {\bf q}_{3}} = 0,
\label{eq:9t}
\end{equation}
i.e., the vertex function defining direct interaction of four plasminos in the HTL-approximation equals zero.\\
\indent Let us now determine an explicit form of the three-point vertex functions ${\mathcal G}^{\; a\, i\,  i_{1}}_{\,{\bf k},\, {\bf q},\, {\bf q}_{1}}$, ${\mathcal K}^{\; a_{1}\, i\,  i_{1}}_{\,{\bf k}_1,\, {\bf q},\, {\bf q}_{1}}$ and ${\mathcal P}^{\; a_{1}\, i\,  i_{1}}_{\,{\bf k}_1,\, {\bf q},\, {\bf q}_{1}}$ in the integrands of the third-order Hamiltonian $ H^{(3)}$ (\ref{eq:2f}). In contrast to the previous case, however, here we have a more nontrivial situation. Considering the definitions (\ref{eq:6t}),  (\ref{eq:9w}), (\ref{eq:9e}) and the relation (\ref{eq:9t}), from formula (\ref{eq:9r}) we arrive at the following initial expression for an analysis:
\begin{align}
2\,&\frac{{\mathcal G}^{\; a\, i\;  i_{1}}_{{\bf q} + {\bf q}_{1},\, {\bf q},\, {\bf q}_{1}}\ 
{\mathcal G}^{\hspace{0.03cm}\ast\, a\, i_{2}\, i_{3}}_{{\bf q}_{2} + {\bf q}_{3},\, {\bf q}_{2},\, {\bf q}_{3}}}
{\omega^{\hspace{0.02cm}l}_{{\bf q}_{2} + {\bf q}_{3}} - \omega^{-}_{{\bf q}_{2}} - \omega^{-}_{{\bf q}_{3}}} 
\;+\;
2\,\frac{{\mathcal K}^{\; a\, i_{2}\,  i_{3}}_{-{\bf q}_{2} - {\bf q}_{3},\, {\bf q}_{2},\, {\bf q}_{3}}\
{\mathcal K}^{\hspace{0.03cm}\ast\, a\, i\;  i_{1}}_{-{\bf q} - {\bf q}_{1},\, {\bf q},\, {\bf q}_{1}}}
{\omega^{\hspace{0.02cm}l}_{-{\bf q}_{2} - {\bf q}_{3}} + \omega^{-}_{{\bf q}_{2}} + \omega^{-}_{{\bf q}_{3}}} 
\notag\\[1.5ex]
\;+\;
\frac{1}{2}\,\biggl[\biggl(
&\frac{{\mathcal P}^{\; a\, i\,  i_{2}}_{{\bf q} - {\bf q}_{2},\, {\bf q},\, {\bf q}_{2}}\ 
{\mathcal P}^{\hspace{0.03cm}\ast\, a\, i_{3}\, i_{1}}_{{\bf q}_{3} - {\bf q}_{1},\, {\bf q}_{3},\,  {\bf q}_{1}}}
{\omega^{\hspace{0.02cm}l}_{{\bf q}_{3} - {\bf q}_{1}} - \omega^{-}_{{\bf q}_{3}} + \omega^{-}_{{\bf q}_{1}}}
\;+\; 
\frac{{\mathcal P}^{\; a\, i_{1}\,  i_{3}}_{{\bf q}_{1} - {\bf q}_{3},\, {\bf q}_{1},\, {\bf q}_{3}}\ 
{\mathcal P}^{\hspace{0.03cm}\ast\, a\, i_{2}\,  i}_{{\bf q}_{2} - {\bf q},\, {\bf q}_{2},\, {\bf q}}}
{\omega^{\hspace{0.02cm}l}_{{\bf q}_{1} - {\bf q}_{3}} - \omega^{-}_{{\bf q}_{1}} + \omega^{-}_{{\bf q}_{3}}}
\biggr)
\notag\\[1.5ex]
-\;
\biggl(
&\frac{{\mathcal P}^{\; a\, i_{1}\,  i_{2}}_{{\bf q}_{1} - {\bf q}_{2},\, {\bf q}_{1},\, {\bf q}_{2}}\ 
{\mathcal P}^{\hspace{0.03cm}\ast\, a\, i_{3}\, i}_{{\bf q}_{3} - {\bf q},\, {\bf q}_{3},\, {\bf q}}}
{\omega^{\hspace{0.02cm}l}_{{\bf q}_{1} - {\bf q}_{2}} - \omega^{-}_{{\bf q}_{1}} + \omega^{-}_{{\bf q}_{2}}}
\;+\; 
\frac{{\mathcal P}^{\; a\, i\,  i_{3}}_{{\bf q} - {\bf q}_{3},\, {\bf q},\, {\bf q}_{3}}\ 
{\mathcal P}^{\hspace{0.03cm}\ast\, a\, i_{2}\,  i_{1}}_{{\bf q}_{2} - {\bf q}_{1},\, {\bf q}_{2},\, {\bf q}_{1}}}
{\omega^{\hspace{0.02cm}l}_{{\bf q}_{2} - {\bf q}_{1}} - \omega^{-}_{{\bf q}_{2}} + \omega^{-}_{{\bf q}_{1}}}
\biggr)\biggr]
\label{eq:9y}
\end{align}
\[
= \frac{1}{\sqrt{2}}\,g^2
\left(\frac{Z_{-}({\bf q})}{2}\right)^{\!1/2}\, \prod_{i=1}^{3} \left(\frac{Z_{-}({\bf q}_i)}{2}\right)^{\!1/2}
\!\bar{v}^{(-)}_{\alpha}(\hat{\bf q},\lambda)\hspace{0.02cm}
\bar{v}^{(-)}_{\alpha_{1}}(\hat{\bf q}_{1},\lambda_{1})\hspace{0.02cm}
v^{(-)}_{\alpha_{2}}(\hat{\bf q}_{2},\lambda_{2})\hspace{0.02cm}
v^{(-)}_{\alpha_{3}}(\hat{\bf q}_{3},\lambda_{3})
\vspace{-0.4cm}
\]
\begin{align}
\hspace{1cm}
\times\,\Bigl\{&(t^a)^{i\hspace{0.03cm}i_{2}}(t^a)^{i_{1}i_{3}}
\,^{\ast}\Gamma^{(Q)\mu}_{\alpha\hspace{0.02cm}\alpha_{2}}(-q + q_{2}; q, -q_{2})
\,^{\ast}{\cal D}_{\mu\mu^{\prime}}(q - q_{2})
\,^{\ast}\Gamma^{(G)\mu^{\prime}}_{\alpha_{1}\alpha_{3}}(-q_{1} + q_{3}; q_{1}, -q_{3})
\notag\\[1.5ex]
-\, &(t^a)^{i\hspace{0.03cm}i_{3}}(t^a)^{i_{1}i_{2}}
\,^{\ast}\Gamma^{(Q)\mu}_{\alpha\hspace{0.02cm}\alpha_{3}}(-q + q_{3}; q, -q_{3})
\,^{\ast}{\cal D}_{\mu\mu^{\prime}}(q - q_{3})
\,^{\ast}\Gamma^{(G)\mu^{\prime}}_{\alpha_{1}\alpha_{2}}(-q_{1} + q_{2}; q_{1}, -q_{2})\Bigr\}\Bigr|_{\rm \,on-shell},
\notag
\end{align}
where we have used the conjugation rules for the HTL-resumed vertex functions $\,^{\ast}\Gamma^{(Q)\mu}$ and $\,^{\ast}\Gamma^{(G)\mu}$, Eqs.\,(\ref{ap:B8}) and (\ref{ap:B9}), and for the effective gluon propagator $\!\,^{\ast}\widetilde{\cal D}_{\mu\mu^{\prime}}$, Eq.\,(\ref{ap:A12}).\\
\indent At the first step, in the effective gluon propagators $\!\,^{\ast}\widetilde{\cal D}_{\mu\mu^{\prime}}$ on the right-hand side of the expression (\ref{eq:9y}) we retain only the terms with the longitudinal projector $\widetilde{Q}_{\mu\mu^{\prime}}$. For example, for the first propagator $\,^{\ast}\widetilde{\cal D}_{\mu\mu^{\prime}} (q - q_{2})$ we make the substitution
\begin{equation}
\,^{\ast}\widetilde{\cal D}_{\mu\mu^{\prime}}(q - q_{2})
\Rightarrow
-\, \widetilde{Q}_{\mu\mu^{\prime}}(q - q_{2}) \,^{\ast}\!\Delta^{l}(q - q_{2}).
\label{eq:9u}
\end{equation}
Here, an explicit form of the right-hand side due to the definitions (\ref{ap:A9}) and (\ref{ap:A10}) is 
\[
-\,\frac{\tilde{u}_{\mu}(q - q_{2})\, \tilde{u}_{\mu^{\prime}}(q - q_{2})}
{\bar{u}^2(q - q_{2})}\ \frac{1}{(q - q_{2})^{2} - \Pi^{l}(q - q_{2})}.
\]
Analogous operations are performed for the second propagator $\,^{\ast}\widetilde{\cal D}_{\mu\mu^{\prime}}(q - q_{3})$. In the vicinity of the pole $\omega\sim\omega^{\hspace{0.02cm}l}_{\bf k}$, the longitudinal scalar propagator $\,^{\ast}\!\Delta^l(k) = \,^{\ast}\!\Delta^l(\omega,\,{\bf k})$ behaves as (see, for example, \cite{weldon_1998} and \cite{blaizot_1994(1)})
\[
\,^{\ast}\!\Delta^l(\omega,\,{\bf k}) = \frac{1}{\omega^{2} - {\bf k}^{2} - \Pi^{l}(\omega,\,{\bf k})}
\simeq
\frac{Z_l({\bf k})}{\omega^{2} - (\omega^{\hspace{0.02cm}l}_{\bf k})^{2}}
=
\biggl(\frac{Z_l({\bf k})}{2\hspace{0.03cm}\omega^{\hspace{0.02cm}l}_{\bf k}}\biggr)
\biggl[\,\frac{1}{\omega - \omega^{\hspace{0.02cm}l}_{\bf k}} - \frac{1}{\omega + \omega^{\hspace{0.02cm}l}_{\bf k}}\,\biggr] .
\]
Making use of this approximation, we obtain the following expressions for the first scalar propagator:
\begin{equation}
^{\ast}\!\Delta^{l}(q - q_{2})
\label{eq:9i}
\end{equation}
\[
\simeq
-\,\biggl(\frac{Z_l({\bf q} - {\bf q}_{2})}{2\ \!\omega^{\hspace{0.02cm}l}_{{\bf q} - {\bf q}_{2}}}\biggr)^{\!\!1/2}
\biggl(\frac{Z_l(-{\bf q}_{1} + {\bf q}_{3})}{2\ \!\omega^{\hspace{0.02cm}l}_{-{\bf q}_{1} + {\bf q}_{3}}}\biggr)^{\!\!1/2}\,
\biggl[\,\frac{1}{\omega^{\hspace{0.02cm}l}_{{\bf q}_{1} - {\bf q}_{3}}\! - \omega^{-}_{{\bf q}_{1}} + \omega^{-}_{{\bf q}_{3}}}
\,+\,
\frac{1}{\omega^{\hspace{0.02cm}l}_{{\bf q}_{3} - {\bf q}_{1}}\! - \omega^{-}_{{\bf q}_{3}} + \omega^{-}_{{\bf q}_{1}}}\,\biggr]
\]
and for the second scalar one
\begin{equation}
^{\ast}\!\Delta^{l}(-q + q_{3})
\label{eq:9o}
\end{equation}
\[
\simeq
-\,\biggl(\frac{Z_l({\bf q} - {\bf q}_{3})}{2\ \!\omega^{\hspace{0.02cm}l}_{{\bf q} - {\bf q}_{3}}}\biggr)^{\!\!1/2}
\biggl(\frac{Z_l(-{\bf q}_{1} + {\bf q}_{2})}{2\ \!\omega^{\hspace{0.02cm}l}_{-{\bf q}_{1} - {\bf q}_{2}}}\biggr)^{\!\!1/2}\,
\biggl[\,\frac{1}{\omega^{\hspace{0.02cm}l}_{{\bf q}_{1} - {\bf q}_{2}}\! - \omega^{-}_{{\bf q}_{1}} + \omega^{-}_{{\bf q}_{2}}}
\,+\,
\frac{1}{\omega^{\hspace{0.02cm}l}_{{\bf q}_{2} - {\bf q}_{1}}\! - \omega^{-}_{{\bf q}_{2}} + \omega^{-}_{{\bf q}_{1}}}\,\biggr].
\]
In deriving (\ref{eq:9i}) and (\ref{eq:9o}) we have taken into account the evenness of the dispersion relation (i.e., $\omega^l_{-{\bf k}} =  \omega^l_{{\bf k}}$) and the conservation of energy and momentum, which take place for the elastic scattering of plasmino off plasmino
\begin{equation}
\left\{
\begin{array}{ll}
{\bf q} + {\bf q}_{1} = {\bf q}_{2} + {\bf q}_{3}, \\[1.5ex]
\omega^{-}_{{\bf q}} + \omega^{-}_{{\bf q}_{1}} = \omega^{-}_{{\bf q}_{2}} + \omega^{-}_{{\bf q}_{3}}.
\end{array}
\right.
\label{eq:9p}
\end{equation}
\indent We see from the approximation of scalar propagators (\ref{eq:9i}) and (\ref{eq:9o}) the contributions propor\-tion\-al to the factors $1/(\omega^{\hspace{0.02cm}l}_{{\bf q}_{2} + {\bf q}_{3}} - \omega^{-}_{{\bf q}_{2}} - \omega^{-}_{{\bf q}_{3}})$ and $1/(\omega^{\hspace{0.02cm}l}_{{\bf q}_{2} + {\bf q}_{3}} + \omega^{-}_{{\bf q}_{2}} + \omega^{-}_{{\bf q}_{3}})$ to be absent. However, these factors occur on the left-hand side of (\ref{eq:9y}). For this reason within the framework of the hard thermal loop approximation we need, in addition to the condition (\ref{eq:9t}), to require the fulfillment of the following equalities:
\begin{equation}
{\mathcal G}^{\; a\, i\;  i_{1}}_{{\bf k},\, {\bf q},\, {\bf q}_{1}} = 
{\mathcal K}^{\; a\, i\;  i_{1}}_{{\bf k} ,\, {\bf q},\, {\bf q}_{1}} = 0.
\label{eq:9a}
\end{equation}
\indent Further, as the second step, we have to ``untangle'' the color structure of the left-hand side of (\ref{eq:9y}). For this, we set for the three-point vertex function ${\mathcal P}^{\; a\, i\,  i_{1}}_{{\bf k},\, {\bf q},\, {\bf q}_{1}}$
\begin{equation}
{\mathcal P}^{\; a\, i\, i_{1}}_{{\bf k},\, {\bf q},\, {\bf q}_{1}}
=
(t^{a})^{i\hspace{0.03cm}i_{1}}\hspace{0.04cm}{\mathcal P}_{{\bf k},\, {\bf q},\, {\bf q}_{1}},
\quad
{\mathcal P}^{\hspace{0.03cm}\ast\, a\, i\, i_{1}}_{{\bf k},\, {\bf q},\,  {\bf q}_{1}}
=
(t^{a})^{i_{1}i}\hspace{0.04cm}{\mathcal P}^{\hspace{0.03cm}\ast}_{{\bf k},\, {\bf q},\, {\bf q}_{1}}.
\label{eq:9s}
\end{equation}
Taking into account all above-mentioned we can put the equality (\ref{eq:9y}) in the following form:
 \begin{align}
\frac{1}{2}\,\biggl[\hspace{0.03cm} (t^a)^{i\hspace{0.03cm}i_{2}}(t^a)^{i_{1}i_{3}} \biggl(
&\frac{{\mathcal P}_{{\bf q} - {\bf q}_{2},\, {\bf q},\, {\bf q}_{2}}\, 
{\mathcal P}^{\hspace{0.03cm}\ast}_{{\bf q}_{3} - {\bf q}_{1},\, {\bf q}_{3},\,  {\bf q}_{1}}}
{\omega^{\hspace{0.02cm}l}_{{\bf q}_{3} - {\bf q}_{1}} - \omega^{-}_{{\bf q}_{3}} + \omega^{-}_{{\bf q}_{1}}}
\;+\; 
\frac{{\mathcal P}_{{\bf q}_{1} - {\bf q}_{3},\, {\bf q}_{1},\, {\bf q}_{3}}\, 
{\mathcal P}^{\hspace{0.03cm}\ast}_{{\bf q}_{2} - {\bf q},\, {\bf q}_{2},\, {\bf q}}}
{\omega^{\hspace{0.02cm}l}_{{\bf q}_{1} - {\bf q}_{3}} - \omega^{-}_{{\bf q}_{1}} + \omega^{-}_{{\bf q}_{3}}}
\biggr)
\notag\\[1.5ex]
-\;
(t^a)^{i\hspace{0.03cm}i_{3}}(t^a)^{i_{1}i_{2}}\biggl(
&\frac{{\mathcal P}_{{\bf q} - {\bf q}_{3},\, {\bf q},\, {\bf q}_{3}}\ 
{\mathcal P}^{\hspace{0.03cm}\ast}_{{\bf q}_{2} - {\bf q}_{1},\, {\bf q}_{2},\, {\bf q}_{1}}}
{\omega^{\hspace{0.02cm}l}_{{\bf q}_{2} - {\bf q}_{1}} - \omega^{-}_{{\bf q}_{2}} + \omega^{-}_{{\bf q}_{1}}}
\;+\; 
\frac{{\mathcal P}_{{\bf q}_{1} - {\bf q}_{2},\, {\bf q}_{1},\, {\bf q}_{2}}\ 
{\mathcal P}^{\hspace{0.03cm}\ast}_{{\bf q}_{3} - {\bf q},\, {\bf q}_{3},\, {\bf q}}}
{\omega^{\hspace{0.02cm}l}_{{\bf q}_{1} - {\bf q}_{2}} - \omega^{-}_{{\bf q}_{1}} + \omega^{-}_{{\bf q}_{2}}}
\biggr)\biggr] 
\notag
\end{align}
\begin{equation}
= \frac{1}{\sqrt{2}}\,g^2
\left(\frac{Z_{-}({\bf q})}{2}\right)^{\!1/2}\! \prod_{i=1}^{3} \left(\frac{Z_{-}({\bf q}_i)}{2}\right)^{\!1/2}
\!\!\!\bar{v}^{(-)}_{\alpha}(\hat{\bf q},\lambda)\hspace{0.02cm}
\bar{v}^{(-)}_{\alpha_{1}}(\hat{\bf q}_{1},\lambda_{1})\hspace{0.02cm}
v^{(-)}_{\alpha_{2}}(\hat{\bf q}_{2},\lambda_{2})\hspace{0.02cm}
v^{(-)}_{\alpha_{3}}(\hat{\bf q}_{3},\lambda_{3})
\label{eq:9d}
\vspace{-0.3cm}
\end{equation}
\begin{align}
&\times\!\hspace{0.03cm}\biggl\{(t^a)^{i\hspace{0.03cm}i_{2}}(t^a)^{i_{1}i_{3}}
\biggl(\frac{Z_l({\bf q} - {\bf q}_{2})}{2\ \!\omega^{\hspace{0.02cm}l}_{{\bf q} - {\bf q}_{2}}}\biggr)^{\!\!1/2}
\biggl(\frac{\tilde{u}_{\mu}(q - q_{2})}{\sqrt{\bar{u}^2(q - q_{2})}}\biggr)
\biggl(\frac{Z_l(-{\bf q}_{1} + {\bf q}_{3})}{2\ \!\omega^{\hspace{0.02cm}l}_{-{\bf q}_{1} + {\bf q}_{3}}}\biggr)^{\!\!1/2}
\biggl(\frac{\tilde{u}_{\mu^{\prime}}(-q_{1} + q_{3})}{\sqrt{\bar{u}^2(-q_{1} + q_{3})}}\biggr)
\notag\\[1.5ex]
&\!\,^{\ast}\Gamma^{(Q)\mu}_{\alpha\hspace{0.02cm}\alpha_{2}}(-q + q_{2}; q, -q_{2})
\biggl(
\frac{1}{\omega^{\hspace{0.02cm}l}_{{\bf q}_{3} - {\bf q}_{1}}\! - \omega^{-}_{{\bf q}_{3}} + \omega^{-}_{{\bf q}_{1}}}
\,+\,
\frac{1}{\omega^{\hspace{0.02cm}l}_{{\bf q}_{1} - {\bf q}_{3}}\! - \omega^{-}_{{\bf q}_{1}} + \omega^{-}_{{\bf q}_{3}}}
\biggr)
\!\,^{\ast}\Gamma^{(G)\mu^{\prime}}_{\alpha_{1}\alpha_{3}}(-q_{1} + q_{3}; q_{1}, -q_{3})
\notag\\[2ex]
&\hspace{0.3cm}
-(t^a)^{i\hspace{0.03cm}i_{3}}(t^a)^{i_{1}i_{2}}
\biggl(\frac{Z_l({\bf q} - {\bf q}_{3})}{2\ \!\omega^{\hspace{0.02cm}l}_{{\bf q} - {\bf q}_{3}}}\biggr)^{\!\!1/2}
\biggl(\frac{\tilde{u}_{\mu}(q - q_{3})}{\sqrt{\bar{u}^2(q - q_{3})}}\biggr)
\biggl(\frac{Z_l(-{\bf q}_{1} + {\bf q}_{2})}{2\ \!\omega^{\hspace{0.02cm}l}_{-{\bf q}_{1} + {\bf q}_{2}}}\biggr)^{\!\!1/2}
\biggl(\frac{\tilde{u}_{\mu^{\prime}}(-q_{1} + q_{2})}{\sqrt{\bar{u}^2(-q_{1} + q_{2})}}\biggr)
\times
\notag\\[1.5ex]
&\!\,^{\ast}\Gamma^{(Q)\mu}_{\alpha\hspace{0.02cm}\alpha_{3}}(-q + q_{3}; q, -q_{3})
\biggl(
\frac{1}{\omega^{\hspace{0.02cm}l}_{{\bf q}_{2} - {\bf q}_{1}}\! - \omega^{-}_{{\bf q}_{2}} + \omega^{-}_{{\bf q}_{1}}}
+
\frac{1}{\omega^{\hspace{0.02cm}l}_{{\bf q}_{1} - {\bf q}_{2}}\! - \omega^{-}_{{\bf q}_{1}} + \omega^{-}_{{\bf q}_{2}}}
\biggr)
\!\,^{\ast}\Gamma^{(G)\mu^{\prime}}_{\alpha_{1}\alpha_{2}}(-q_{1} + q_{2}; q_{1}, -q_{2})\biggr\}\biggr|_{\rm \,on-shell},
\notag
\end{align}
where we have used the property of rearrangement of the external quark momenta (\ref{ap:B8}) for the vertex function $\!\,^{\ast}\Gamma^{(Q)}_{\mu}$. In order to turn the relation (\ref{eq:9d}) into an identity one needs to set
\begin{equation}
{\mathcal P}^{\hspace{0.03cm}\ast}_{{\bf q}_{2} - {\bf q},\, {\bf q}_{2},\, {\bf q}}(\lambda_{2},\lambda)
=
-\hspace{0.02cm}{\mathcal P}_{{\bf q} - {\bf q}_{2},\, {\bf q},\, {\bf q}_{2}}(\lambda,\lambda_{2}),
\quad
{\mathcal P}^{\hspace{0.03cm}\ast}_{{\bf q}_{3} - {\bf q}_{1},\, {\bf q}_{3},\,  {\bf q}_{1}}(\lambda_{3},\lambda_{1})
=
-\hspace{0.02cm}{\mathcal P}_{{\bf q}_{1} - {\bf q}_{3},\, {\bf q}_{1},\, {\bf q}_{3}}(\lambda_{1},\lambda_{3}),
\;\ldots
\label{eq:9f}
\end{equation}
and
\begin{equation}
{\mathcal P}_{{\bf q} - {\bf q}_{2},\, {\bf q},\, {\bf q}_{2}}(\lambda,\lambda_{2})
=
\hspace{0.5cm}
\label{eq:9g}
\end{equation}
\[
{2}^{1/4}\hspace{0.02cm}i\hspace{0.01cm}g\!
\left(\frac{Z_{-}({\bf q})}{2}\right)^{\!\!1/2}\!\! \left(\frac{Z_{-}({\bf q}_{2})}{2}\right)^{\!\!1/2}
\!\!\!\bar{v}^{(-)}_{\alpha}(\hat{\bf q},\lambda)\hspace{0.03cm}  v^{(-)}_{\alpha_{2}}(\hat{\bf q}_{2},\lambda_{2})
\Biggl(\frac{\epsilon^{l}_{\mu}({\bf q} - {\bf q}_{2})}
{\sqrt{2\hspace{0.03cm}\omega^{\hspace{0.02cm}l}_{{\bf q}\hspace{0.02cm} - \hspace{0.02cm} {\bf q}_{2}}}}\Biggr)
\!\,^{\ast}\Gamma^{(Q)\mu}_{\alpha\hspace{0.02cm}\alpha_{2}}(-q \,+\, q_{2}; q, -q_{2})
\biggr|_{\rm \,on-shell},
\vspace{0.5cm}
\]
\begin{equation}
{\mathcal P}_{{\bf q}_{1} - {\bf q}_{3},\, {\bf q}_{1},\,  {\bf q}_{3}}(\lambda_{1},\lambda_{3})
=
\label{eq:9h}
\end{equation}
\[
{2}^{1/4}\hspace{0.02cm}i\hspace{0.01 cm}g\!
\left(\frac{Z_{-}({\bf q}_{1})}{2}\right)^{\!\!1/2}\!\! \left(\frac{Z_{-}({\bf q}_{3})}{2}\right)^{\!\!1/2}\!
\!\!\!\bar{v}^{(-)}_{\alpha_{1}}(\hat{\bf q}_{1},\lambda_{1})\hspace{0.03cm} v^{(-)}_{\alpha_{3}}(\hat{\bf q}_{3},\lambda_{3})
\Biggl(\frac{\epsilon^{l}_{\mu^{\prime}}(-{\bf q}_{1} \!+\! {\bf q}_{3})}{\sqrt{2\omega^l_{-{\bf q}_{1}\hspace{0.02cm} +\hspace{0.02cm} {\bf q}_{3}}}}\Biggr)
\!\,^{\ast}\Gamma^{(G)\mu^{\prime}}_{\alpha_{1}\alpha_{3}}(-q_{1} +\, q_{3}; q_{1}, -q_{3})
\biggr|_{\rm \,on-shell},
\]
\[
\ldots\,.
\]
Here, four-vectors of the form
\begin{equation}
\left(\frac{Z_l({\bf k})}{2\hspace{0.03cm}\omega_{\bf k}^l}\right)^{\!\!1/2}\!\!\!
\left.\frac{\tilde{u}_{\mu}(k)}{\sqrt{\bar{u}^{2}(k)}}\ \!\right|_{\rm on-shell}\!
\equiv\,
\frac{1}{\sqrt{2\hspace{0.03cm}\omega^l_{\bf k}}}\ \epsilon^{l}_{\mu}({\bf k})
\label{eq:9j}
\end{equation}
on the right-hand side of Eqs.\,(\ref{eq:9g}), (\ref{eq:9h}) and so on are the ordinary wavefunctions of a longitudinal physical gluon in the $A_{0}$\hspace{0.03cm}-\hspace{0.03cm}gauge, where factor $\sqrt{Z_l({\bf k})}$ ensures renormalization of the gluon wavefunction due to thermal effects. For the definitions of vertex function ${\mathcal P}_{{\bf k},\, {\bf q},\, {\bf q}_{1}}$, Eqs.\,(\ref{eq:9g}) and (\ref{eq:9h}), to be compatible it is necessary that the equality of the kind of
\begin{equation}
\!\,^{\ast}\Gamma^{(Q)\mu}_{\alpha\hspace{0.02cm}\alpha_{2}}(-q + q_{2}; q, -q_{2})\bigr|_{\rm \,on-shell}
\,=\,
\!\,^{\ast}\Gamma^{(G)\mu}_{\alpha\hspace{0.02cm}\alpha_{2}}(-q + q_{2}; q, -q_{2})\bigr|_{\rm \,on-shell}.
\label{eq:9k}
\end{equation}
be true. As we have shown in \cite{markov_2001} because of the absence of the linear Landau damping for the
plasmino mode from the definitions of two-quark--one-gluon vertex functions (\ref{ap:B4})\,--\,(\ref{ap:B6}) the equality (\ref{eq:9k}) really holds. By using the conjugation rule of the HTL-resumed vertex function $\!\,^{\ast}\Gamma^{(Q)}_{\mu}$, Eq.\,(\ref{ap:B8}), it is not difficult to make sure that the definition (\ref{eq:9g}) satisfies the first requirement in (\ref{eq:9f}). In conclusion we write out completely the conjugation rule of the vertex function ${\mathcal P}^{\; a\, i\;  i_{1}}_{{\bf k},\, {\bf q},\, {\bf q}_{1}}$. Taking into account (\ref{eq:9s}) and (\ref{eq:9f}), we have
\[
{\mathcal P}^{\,\ast\; a\, i\;  i_{1}}_{{\bf k},\, {\bf q},\, {\bf q}_{1}} = 
-\hspace{0.02cm} {\mathcal P}^{\; a\; i_{1}\,  i}_{\!\!-{\bf k},\, {\bf q}_{1},\, {\bf q}}.
\]
This property should be added to the symmetry relations (\ref{eq:2h}) with the proviso that it is true only in the hard thermal loop approximation.


\section{\bf Fourth-order correlation function for soft quark and gluon excitations}
\label{section_10}
\setcounter{equation}{0}

Let us consider the construction of a system of kinetic equations describing the elastic scattering process of plasmino off plasmon and visa versa. As the interaction Hamiltonian here, we take the effective Hamiltonian ${\mathcal H}^{(4)}_{qg\rightarrow qg}$, Eq.\,(\ref{eq:5r}). The equations of motion for the fermionic $f^{\,i}_{{\bf q}}, \,f^{\,\ast\hspace{0.02cm} j}_{{\bf q}^{\prime}}$ and bosonic $c^{\phantom{\hspace{0.03cm}\ast} \!\!a}_{{\bf k}}, \,c^{\hspace{0.03cm}\ast\,b}_{{\bf k}^{\prime}}$ normal variables are defined by the corresponding Hamilton equations. For soft fermionic excitations we have
\begin{equation}
\frac{\partial\hspace{0.04cm} \! f^{\hspace{0.03cm}\ast\,i}_{{\bf q}}}{\partial\hspace{0.03cm} t}
=
-\hspace{0.03cm}i\hspace{0.04cm}\Bigl\{f^{\hspace{0.03cm}\ast\,i}_{{\bf q}}\hspace{0.03cm},\hspace{0.03cm} {\mathcal H}^{(0)\!} + {\mathcal H}^{(4)}_{qg\rightarrow qg}\Bigr\}
=
i\hspace{0.04cm}\omega^{-}_{{\bf q}}\,f^{\,\ast\hspace{0.03cm}i}_{{\bf q}}
\label{eq:10q}
\end{equation}
\[
+\; i\!\int\frac{d{\bf q}_{1}\hspace{0.02cm} d{\bf k}_{1}\hspace{0.03cm} d{\bf k}_{2}}{(2\pi)^{9}}\ 
\widetilde{T}^{\hspace{0.03cm}\ast\,(2)\, i\, i_{1}\, a_{1}\, a_{2}}_{\ {\bf q},\, {\bf q}_{1},\, {\bf k}_{1},\, {\bf k}_{2}}\, 
f^{\hspace{0.03cm}\ast\;i_{1}}_{{\bf q}_{1}}\, 
c^{\,a_{1}}_{{\bf k}_{1}}\hspace{0.03cm} 
c^{\hspace{0.03cm}\ast\hspace{0.03cm}a_{2}}_{{\bf k}_{2}}\,
(2\pi)^{3}\hspace{0.03cm}\delta({\bf q} + {\bf k}_1 - {\bf q}_{1} - {\bf k}_{2}),
\]
\begin{equation}
\hspace{0.3cm}
\frac{\partial\hspace{0.04cm} \! f^{\hspace{0.03cm}j}_{{\bf q}^{\prime}}}{\partial\hspace{0.03cm} t}
=
-\hspace{0.03cm}i\hspace{0.04cm}\Bigl\{f^{\hspace{0.03cm}j}_{{\bf q}^{\prime}}\hspace{0.03cm},\hspace{0.03cm} {\mathcal H}^{(0)\!} + {\mathcal H}^{(4)}_{qg\rightarrow qg}\Bigr\}
=
-\hspace{0.03cm}i\hspace{0.04cm}\omega^{-}_{{\bf q}^{\prime}}\, f^{\hspace{0.03cm}j}_{{\bf q}^{\prime}}
\label{eq:10w}
\end{equation}
\[
\hspace{0.6cm}
-\; i\!\int\frac{d{\bf q}_{1}\hspace{0.02cm} d{\bf k}_{1}\hspace{0.03cm} d{\bf k}_{2}}{(2\pi)^{9}}\ 
\widetilde{T}^{\hspace{0.03cm}(2)\, j\, i_{1}\, a_{1}\, a_{2}}_{\ {\bf q}^{\prime},\, {\bf q}_{1},\; {\bf k}_{1},\, {\bf k}_{2}}\, 
f^{\hspace{0.03cm}i_{1}}_{{\bf q}_{1}}\hspace{0.03cm} 
c^{\hspace{0.03cm}\ast\hspace{0.03cm} a_{1}}_{{\bf k}_{1}}\hspace{0.03cm} 
c^{\,a_{2}}_{{\bf k}_{2}}\, 
(2\pi)^{3}\hspace{0.03cm}\delta({\bf q}^{\prime} + {\bf k}_{1} - {\bf q}_{1} - {\bf k}_{2}).
\]
In the latter equation we have taken into account the symmetry condition for the effective scattering amplitude
\begin{equation}
\widetilde{T}^{\hspace{0.03cm} (2)\hspace{0.03cm} i\, i_{1}\, a_{1}\, a_{2}}_{\, {\bf q},\, {\bf q}_{1},\, {\bf k}_{1},\, {\bf k}_{2}}
=
\widetilde{T}^{\,\ast\hspace{0.03cm}(2)\, i_{1}\, i\, a_{2}\, a_{1}}_{\ {\bf q}_{1},\, {\bf q},\; {\bf k}_{2},\ 
{\bf k}_{1}}.
\label{eq:10e}
\end{equation}
This relation is a consequence of the requirement of the reality of the effective Hamiltonian ${\mathcal H}^{(4)}_{qg\rightarrow qg}$. Further, for soft Bose-excitations we define the second pair of the canonical equations of motions with the same Hamiltonian
\begin{equation}
\frac{\partial \hspace{0.02cm}c^{\hspace{0.03cm}\ast\hspace{0.03cm}a}_{{\bf k}}}{\partial\hspace{0.03cm} t}
=
i\hspace{0.04cm}\Bigl\{c^{\hspace{0.03cm}\ast\hspace{0.03cm}a}_{{\bf k}}\hspace{0.03cm},\hspace{0.03cm} {\mathcal H}^{(0)\!} + {\mathcal H}^{(4)}_{qg\rightarrow qg}\Bigr\}
=
i\hspace{0.04cm}\omega^{\hspace{0.02cm}l}_{{\bf k}}\,c^{\hspace{0.03cm}\ast\hspace{0.03cm}a}_{{\bf k}}
\label{eq:10r}
\end{equation}
\[
-\; i\!\int\frac{d{\bf q}_{1}\hspace{0.02cm} d{\bf q}_{2}\hspace{0.03cm} d{\bf k}_{1}}{(2\pi)^{9}}\ 
\widetilde{T}^{\hspace{0.03cm}\ast\,(2)\, i_{1}\, i_{2}\, a\; a_{1}}_{\, {\bf q}_{1},\, {\bf q}_{2},\, {\bf k},\, {\bf k}_{1}}\, 
f^{\;i_{1}}_{{\bf q}_{1}}\hspace{0.03cm} 
f^{\hspace{0.03cm}\ast\,i_{2}}_{{\bf q}_{2}}\hspace{0.03cm} 
c^{\hspace{0.03cm}\ast \hspace{0.03cm}a_{1}}_{{\bf k}_{1}}\,
(2\pi)^{3}\hspace{0.03cm}\delta({\bf k} + {\bf q}_{1} - {\bf k}_{1} - {\bf q}_{2}),
\]
\begin{equation}
\hspace{0.3cm}
\frac{\partial \hspace{0.02cm}c^{\,b}_{{\bf k}^{\prime}}}{\partial\hspace{0.03cm} t}
=
-\hspace{0.03cm}i\hspace{0.04cm}\Bigl\{c^{\,b}_{{\bf k}^{\prime}}\hspace{0.03cm},\hspace{0.03cm} {\mathcal H}^{(0)\!} + {\mathcal H}^{(4)}_{qg\rightarrow qg}\Bigr\}
=
-\hspace{0.03cm}i\hspace{0.04cm}\omega^{\hspace{0.02cm}l}_{{\bf k}^{\prime}}\, c^{\,b}_{{\bf k}^{\prime}}
\label{eq:10t}
\end{equation}
\[
\hspace{0.5cm}
-\; i\!\int\frac{d{\bf q}_{1}\hspace{0.02cm} d{\bf q}_{2}\hspace{0.03cm} d{\bf k}_{1}}{(2\pi)^{9}}\ 
\widetilde{T}^{\hspace{0.03cm}(2)\, i_{1}\, i_{2}\, b\, a_{1}}_{\ {\bf q}_{1},\, {\bf q}_{2},\; {\bf k}^{\prime},\, {\bf k}_{1}}\, 
f^{\hspace{0.03cm}\ast\,i_{1}}_{{\bf q}_{1}}\hspace{0.03cm} 
f^{\;i_{2}}_{{\bf q}_{2}}\hspace{0.03cm} 
c^{\,a_{1}}_{{\bf k}_{1}}\, 
(2\pi)^{3}\hspace{0.03cm}\delta({\bf k}^{\prime} + {\bf q}_{1} - {\bf k}_{1} - {\bf q}_{2}).
\]
In the case when an external gauge field is absent in the system, the exact equations (\ref{eq:10q}), (\ref{eq:10w}), (\ref{eq:10r}), and (\ref{eq:10t}) enable us to define the kinetic equations for the colorless plasmino number density $n^{ij\,-}_{{\bf q}\!\!}\! \equiv \delta^{\hspace{0.03cm}ij} n^{-}_{{\bf q}}$ and for the colorless plasmon number density  $N^{ab\,l}_{{\bf k}\!\!}\! \equiv \delta^{ab} N^{l}_{{\bf k}}$. If the ensemble of interacting Bose-excitations at low nonlinearity level has  random phases, then it can be statistically described by introducing (in addition to the fermionic correlation function (\ref{eq:7r})) the bosonic correlation function \cite{markov_2020}:
\begin{equation}
\bigl\langle\hspace{0.03cm}c^{\ast\ \!\!a}_{{\bf k}}\hspace{0.03cm} c^{\phantom{\ast}\!\!b}_{{\bf k}^{\prime}}\bigr\rangle
=
\delta^{a\hspace{0.015cm} b}(2\pi)^{3}\hspace{0.03cm}\delta({\bf k} - {\bf k}^{\prime})N^{l}_{{\bf k}}.
\label{eq:10y}
\end{equation}
As well as in the fermionic case we can find the relationship between the spectrum ${\mathcal N}^{\hspace{0.03cm} l}_{\bf k}$ of a physically real random bosonic wave field in (\ref{eq:2w}) and the ``spectrum'' $N^{\hspace{0.03cm}l}_{\bf k}$ in (\ref{eq:10y}). We need to use the canonical transformation (\ref{eq:3t}) at this time. Using Eq.\,(\ref{eq:3t}), we have to calculate the correlation function $\bigl\langle\hspace{0.03cm}a^{\ast\ \!a}_{{\bf k}}\hspace{0.03cm} a^{\phantom{\ast}\!\!b}_{{\bf k}^{\prime}}\bigr\rangle$, and to apply the Gaussian hypothesis to the correlation functions of higher orders in the bosonic variable $a^{a}_{{\bf k}}$ (and also in the fermionic variable $f^{\,i}_{{\bf q}}$). In addition to (\ref{eq:7t}), this calculation procedure yields the second nonlinear functional relationship between two representations of the spectrum
\[
{\mathcal N}^{\hspace{0.03cm} l}_{\bf k} = {\mathcal N}^{\hspace{0.03cm} l}_{\bf k}\hspace{0.03cm} [\hspace{0.03cm} n^{-}_{\bf q}, N^{l}_{{\bf k}}\hspace{0.03cm}],
\]
which also includes the plasmino number density $n^{-}_{{\bf q}}$ from the representation (\ref{eq:7r}). The difference between ${\mathcal N}^{\hspace{0.03cm} l}_{\bf k}$ and $N^{l}_{{\bf k}}$ is to be taken into consideration in analyses of the Kolmogorov power-law spectra of weak turbulence in QGP.\\ 
\indent Further, as in the fermionic case it is necessary to note that introducing the distribution function of quasiparticles (plasmons) $N^{\hspace{0.03cm}l}_{\bf k}\equiv N^{l}({\bf k}, {\bf x}, t)$ depending both on plasmon momentum $\hbar\hspace{0.04cm}{\bf k}$ and on coordinate ${\bf x}$ and time $t$ has a sense only in the case when the number density varies slowly in space and time. This means a change of the function $N^{l}_{{\bf k}}$ at distances of the order of a wavelength $\lambda = 2\pi/k$ and for time intervals of the order of the oscillation period $T = 2\pi/\omega^{\hspace{0.02cm}l}_{\bf k}$ should be much smaller than the function $N^{l}_{{\bf k}}$ itself.\\
\indent The first step is to define the kinetic equations for the plasmino $n^{-}_{\bf q}$ and plasmon $N^{\hspace{0.03cm}l}_{\bf k}$ number densities employing the Hamilton equations (\ref{eq:10q}), (\ref{eq:10w}), (\ref{eq:10r}) and (\ref{eq:10t}). Using precisely the same reasoning as in section \ref{section_7}, we get
\begin{equation}
\delta^{\hspace{0.03cm}ij}(2\pi)^{3}\hspace{0.03cm}\delta({\bf q} - {\bf q}\!\ ')\,
\frac{\partial\hspace{0.03cm} n^{-}_{{\bf q}}}{\partial\hspace{0.03cm} t}\
=
\label{eq:10u}
\end{equation}
\[
=
-\hspace{0.03cm}i\!\int\frac{d{\bf q}_{1}\hspace{0.02cm} d{\bf k}_{1}\hspace{0.03cm} d{\bf k}_{2}}{(2\pi)^{9}}\
\biggl\{\widetilde{T}^{\hspace{0.03cm} (2)\hspace{0.03cm} j\, i_{1}\, a_{1}\, a_{2}}_{\, {\bf q}^{\prime},\, {\bf q}_{1},\, {\bf k}_{1},\, {\bf k}_{2}}\, 
I^{\, i\, i_{1}\, a_{1}\, a_{2}}_{{\bf q},\, {\bf q}_{1},\, {\bf k}_{1},\, {\bf k}_{2}}\,
(2\pi)^{3}\hspace{0.03cm}\delta({\bf q}^{\prime} + {\bf k}_{1} - {\bf q}_{1} - {\bf k}_{2})
\]
\[
-\ 
\widetilde{T}^{\hspace{0.03cm}\ast\hspace{0.03cm}(2)\hspace{0.03cm} i\, i_{1}\, a_{1}\, a_{2}}_{\ {\bf q},\, {\bf q}_{1},\, {\bf k}_{1},\, {\bf k}_{2}}\, 
I^{\, i_{1}\, j\; a_{2}\, a_{1}}_{{\bf q}_{1},\, {\bf q}^{\prime},\, {\bf k}_{2},\, {\bf k}_{1}}\,
(2\pi)^{3}\hspace{0.02cm}\delta({\bf q} + {\bf k}_{1} - {\bf q}_{1} - {\bf k}_{2})\biggr\}
\]
and
\begin{equation}
\delta^{ab}(2\pi)^{3}\hspace{0.02cm}\delta({\bf k} - {\bf k}\!\ ')\,
\frac{\partial\hspace{0.01cm} N^{l}_{{\bf k}}}{\partial\hspace{0.03cm} t}\
=
\label{eq:10i}
\end{equation}
\[
=
-\hspace{0.03cm}i\!\int\frac{d{\bf q}_{1}\hspace{0.02cm} d{\bf q}_{2}\hspace{0.03cm} d{\bf k}_{1}}{(2\pi)^{9}}\
\biggl\{\widetilde{T}^{\,(2)\, i_{1}\, i_{2}\, b\, a_{1}}_{\, {\bf q}_{1},\, {\bf q}_{2},\, {\bf k}^{\prime},\, {\bf k}_{1}}\, 
I^{\;i_{1}\, i_{2}\, a\, a_{1}}_{{\bf q}_{1},\, {\bf q}_{2},\, {\bf k},\, {\bf k}_{1}}\,
(2\pi)^{3}\hspace{0.03cm}\delta({\bf k}^{\prime} + {\bf q}_{1} - {\bf k}_{1} - {\bf q}_{2})
\]
\[
-\ 
\widetilde{T}^{\,\ast\hspace{0.03cm}(2)\, i_{1}\, i_{2}\, a\, a_{1}}_{\ {\bf q}_{1},\, {\bf q}_{2},\, {\bf k},\, {\bf k}_{1}}\, 
I^{\, i_{2}\, i_{1}\, a_{1}\, b}_{{\bf q}_{2},\, {\bf q}_{1},\, {\bf k}_{1},\, {\bf k}^{\prime}}\,
(2\pi)^{3}\hspace{0.03cm}\delta({\bf k} + {\bf q}_{1} - {\bf k}_{1} - {\bf q}_{2})\biggr\},
\]
where
\[
I^{\, i\; i_{1}\, a_{1}\, a_{2}}_{{\bf q},\, {\bf q}_{1},\, {\bf k}_{1},\, {\bf k}_{2}}
=
\bigl\langle f^{\hspace{0.03cm}\ast\,i}_{{\bf q}}\, 
f^{\,i_{1}}_{{\bf q}_{1}}\, 
c^{\ast\ \!\!a_{1}}_{{\bf k}_{1}}\hspace{0.03cm} 
c^{\hspace{0.03cm}a_{2}}_{{\bf k}_{2}}\,\bigr\rangle
\]
is the four-point correlation function. By differentiating the correlation function $I^{\, i\ i_{1}\ a_{1}\ a_{2}}_{{\bf q},\, {\bf q}_{1},\, {\bf k}_{1},\, {\bf k}_{2}}$ with respect to $t$ with allowance made for (\ref{eq:10q}), (\ref{eq:10w}), (\ref{eq:10r}) and (\ref{eq:10t}), we derive the equation the right-hand side of which will contain the six-order correlation function in the variables $f^{\hspace{0.03cm}\ast\, i}_{{\bf q}}\!,\, f^{\phantom{\hspace{0.03cm}\ast}\!\!i}_{{\bf q}}$ and $c^{\phantom{\hspace{0.03cm}\ast} \!\!a}_{{\bf k}}, \,c^{\hspace{0.03cm}\ast\,a}_{{\bf k}}$:
\begin{equation}
\frac{\partial\hspace{0.01cm} I^{\, i\; i_{1}\, a_{1}\, a_{2}}_{{\bf q},\, {\bf q}_{1},\, {\bf k}_{1},\, {\bf k}_{2}}}{\partial\hspace{0.03cm} t}
=
i\hspace{0.03cm}\bigl[\,\omega^{-}_{{\bf q}} + \omega^{\hspace{0.02cm}l}_{{\bf k}_{1}} - \omega^{-}_{{\bf q}_{1}} -
\omega^{\hspace{0.02cm}l}_{{\bf k}_{2}}\bigr]
\, I^{\, i\; i_{1}\, a_{1}\, a_{2}}_{{\bf q},\, {\bf q}_{1},\, {\bf k}_{1},\, {\bf k}_{2}}\ +
\label{eq:10o}
\end{equation}
\begin{align}
+\; i\!&\int\frac{d{\bf q}^{\prime}_{1}\hspace{0.03cm} d{\bf k}^{\prime}_{1}\hspace{0.03cm} d{\bf k}^{\prime}_{2}}{(2\pi)^{9}}\
\widetilde{T}^{\,\ast\hspace{0.03cm}(2)\, i\; i^{\prime}_{1}\, a^{\prime}_{1}\, a^{\prime}_{2}}_{\ {\bf q},\, 
{\bf q}^{\prime}_{1},\, {\bf k}^{\prime}_{1},\, {\bf k}^{\prime}_{2}}\,
\bigl\langle
f^{\hspace{0.04cm}\ast\, i^{\prime}_{1}}_{{\bf q}^{\prime}_{1}} f^{\,i_{1}}_{{\bf q}^{\phantom{\prime}}_{1}}\,
c^{\ast\ \!\!a^{\prime}_{2}}_{{\bf k}^{\prime}_{2}}\, 
c^{\hspace{0.03cm}a^{\prime}_{1}}_{{\bf k}^{\prime}_{1}}\,
c^{\ast\ \!\!a_{1}}_{{\bf k}^{\phantom{\prime}}_{1}}\hspace{0.03cm} 
c^{\hspace{0.03cm}a_{2}}_{{\bf k}^{\phantom{\prime}}_{2}}\,\bigr\rangle\, 
(2\pi)^{3}\hspace{0.03cm}\delta({\bf q}^{\prime}_{1} + {\bf k}^{\prime}_{2} - {\bf q} - {\bf k}^{\prime}_{1})
\notag\\[1ex]
-\; i\!&\int\frac{d{\bf q}^{\prime}_{1}\hspace{0.03cm} d{\bf k}^{\prime}_{1}\hspace{0.03cm} d{\bf k}^{\prime}_{2}}{(2\pi)^{9}}\
\widetilde{T}^{\,(2)\, i_{1}\, i^{\prime}_{1}\, a^{\prime}_{1}\, a^{\prime}_{2}}_{\ {\bf q}_{1},\, 
{\bf q}^{\prime}_{1},\, {\bf k}^{\prime}_{1},\, {\bf k}^{\prime}_{2}}\,
\bigl\langle
f^{\hspace{0.04cm}\ast\, i}_{{\bf q}^{\phantom{\prime}}} 
f^{\;i^{\prime}_{1}}_{{\bf q}^{\prime}_{1}}\,
c^{\ast\ \!\!a^{\prime}_{1}}_{{\bf k}^{\prime}_{1}}\hspace{0.03cm} 
c^{\hspace{0.03cm}a^{\prime}_{2}}_{{\bf k}^{\prime}_{2}}\,
c^{\ast\ \!\!a_{1}}_{{\bf k}^{\phantom{\prime}}_{1}}\hspace{0.03cm} 
c^{\hspace{0.03cm}a_{2}}_{{\bf k}^{\phantom{\prime}}_{2}}\,\bigr\rangle\, 
(2\pi)^{3}\hspace{0.03cm}\delta({\bf q}^{\phantom{\prime}}_{1} + {\bf k}^{\prime}_{1} - {\bf q}^{\prime}_{1} - {\bf k}^{\prime}_{2})
\notag\\[1ex]
+\; i\!&\int\frac{d{\bf q}^{\prime}_{1}\hspace{0.03cm} d{\bf q}^{\prime}_{2}\hspace{0.04cm} d{\bf k}^{\prime}_{1}}{(2\pi)^{9}}\
\widetilde{T}^{\,\ast\hspace{0.03cm}(2)\, i^{\prime}_{2}\, i^{\prime}_{1}\, a_{1}\, a^{\prime}_{1}}_{\ {\bf q}^{\prime}_{2},\, 
{\bf q}^{\prime}_{1},\, {\bf k}_{1},\, {\bf k}^{\prime}_{1}}\,
\bigl\langle
f^{\hspace{0.04cm}\ast\,i^{\phantom{\prime}}\!\!}_{{\bf q}}\hspace{0.03cm}
f^{\,i_{1}}_{{\bf q}_{1}}\hspace{0.03cm}
f^{\hspace{0.04cm}\ast\,i^{\prime}_{1}}_{{\bf q}^{\prime}_{1}}\hspace{0.03cm}
f^{\phantom{\hspace{0.02cm}\ast}\!i^{\prime}_{2}}_{{\bf q}^{\prime}_{2}}\,
c^{\ast\ \!\!a^{\prime}_{1}}_{{\bf k}^{\prime}_{1}}\hspace{0.03cm} 
c^{\hspace{0.03cm}a_{2}}_{{\bf k}^{\phantom{\prime}}_{2}}
\,\bigr\rangle\, 
(2\pi)^{3}\hspace{0.03cm}\delta({\bf q}^{\prime}_{1} + {\bf k}^{\prime}_{1} - {\bf q}^{\prime}_{2} - 
{\bf k}^{\phantom{\prime}}_{1})
\notag\\[1ex]
-\; i\!&\int\frac{d{\bf q}^{\prime}_{1}\hspace{0.03cm} d{\bf q}^{\prime}_{2}\hspace{0.04cm} d{\bf k}^{\prime}_{1}}{(2\pi)^{9}}\
\widetilde{T}^{\,(2)\, i^{\prime}_{1}\, i^{\prime}_{2}\, a_{2}\; a^{\prime}_{1}}_{\ {\bf q}^{\prime}_{1},\, {\bf q}^{\prime}_{2},\, {\bf k}_{2},\, {\bf k}^{\prime}_{1}}\,
\bigl\langle
f^{\hspace{0.04cm}\ast\,i^{\phantom{\prime}}\!\!}_{{\bf q}}\hspace{0.03cm}
f^{\,i_{1}}_{{\bf q}_{1}}\,
f^{\hspace{0.04cm}\ast\,i^{\prime}_{1}}_{{\bf q}^{\prime}_{1}}\hspace{0.03cm}
f^{\phantom{\hspace{0.03cm}\ast}\!i^{\prime}_{2}}_{{\bf q}^{\prime}_{2}}
\,c^{\ast\ \!\!a_{1}}_{{\bf k}^{\phantom{\prime}}_{1}}\hspace{0.03cm} 
c^{\hspace{0.03cm}a^{\prime}_{1}}_{{\bf k}^{\prime}_{1}}
\,\bigr\rangle\, 
(2\pi)^{3}\hspace{0.03cm}\delta({\bf q}^{\prime}_{1} + {\bf k}^{\phantom{\prime}}_{2} - {\bf q}^{\prime}_{2} - 
{\bf k}^{\prime}_{1}).
\notag
\end{align}
As in the pure fermionic case, we close the chain of equations by expressing the six-order correlation functions in terms of the pair correlation functions. We keep only the terms which will give the proper contributions to the required kinetic equations:
\begin{align}
&\bigl\langle
f^{\hspace{0.04cm}\ast\, i^{\prime}_{1}}_{{\bf q}^{\prime}_{1}}\! 
f^{\,i_{1}}_{{\bf q}^{\phantom{\prime}}_{1}}\,
c^{\ast\ \!\!a^{\prime}_{2}}_{{\bf k}^{\prime}_{2}}\hspace{0.03cm} 
c^{\hspace{0.03cm}a^{\prime}_{1}}_{{\bf k}^{\prime}_{1}}\,
c^{\ast\ \!\!a_{1}}_{{\bf k}^{\phantom{\prime}}_{1}}\hspace{0.03cm} 
c^{\hspace{0.03cm}a_{2}}_{{\bf k}^{\phantom{\prime}}_{2}}\bigr\rangle
\,\simeq\,
\delta^{\hspace{0.03cm}i^{\prime}_{1}i^{\phantom{\prime}}_{1}}
\delta^{\hspace{0.03cm}a^{\prime}_{1}a^{\phantom{\prime}}_{1}}
\delta^{\hspace{0.03cm}a^{\prime}_{2}a^{\phantom{\prime}}_{2}}\,
(2\pi)^{9}\hspace{0.03cm}
\delta({\bf q}^{\prime}_{1} - {\bf q}^{\phantom{\prime}}_{1})
\delta({\bf k}^{\prime}_{1} - {\bf k}^{\phantom{\prime}}_{1})
\delta({\bf k}^{\prime}_{2} - {\bf k}^{\phantom{\prime}}_{2})\,
n^{-}_{{\bf q}_{1}} N^{l}_{{\bf k}_{1}} N^{l}_{{\bf k}_{2}},
\notag\\[1.5ex]
&\bigl\langle
f^{\hspace{0.04cm}\ast\, i}_{{\bf q}^{\phantom{\prime}}} 
f^{\,i^{\prime}_{1}}_{{\bf q}^{\prime}_{1}}\,
c^{\ast\ \!\!a^{\prime}_{1}}_{{\bf k}^{\prime}_{1}}\hspace{0.03cm} 
c^{\hspace{0.03cm}a^{\prime}_{2}}_{{\bf k}^{\prime}_{2}}\,
c^{\ast\ \!\!a_{1}}_{{\bf k}^{\phantom{\prime}}_{1}}\hspace{0.02cm} 
c^{\hspace{0.03cm}a_{2}}_{{\bf k}^{\phantom{\prime}}_{2}}\,\bigr\rangle
\,\simeq\,
\delta^{\hspace{0.03cm}i^{\phantom{\prime}}\!i^{\prime}_{1}}
\delta^{\hspace{0.03cm}a^{\prime}_{1}a^{\phantom{\prime}}_{2}}
\delta^{\hspace{0.03cm}a^{\prime}_{2}a^{\phantom{\prime}}_{1}}\,
(2\pi)^{9}\hspace{0.03cm}
\delta({\bf q}^{\prime}_{1} - {\bf q})
\hspace{0.03cm}
\delta({\bf k}^{\prime}_{1} - {\bf k}^{\phantom{\prime}}_{2})
\hspace{0.03cm}
\delta({\bf k}^{\prime}_{2} - {\bf k}^{\phantom{\prime}}_{1})\,
n^{-}_{{\bf q}} N^{l}_{{\bf k}_{1}} N^{l}_{{\bf k}_{2}},
\notag\\[1.5ex]
&\bigl\langle
f^{\hspace{0.04cm}\ast\,i^{\phantom{\prime}}\!\!}_{{\bf q}}\hspace{0.03cm}
f^{\,i_{1}}_{{\bf q}^{\phantom{\prime}}_{1}}\,
f^{\hspace{0.04cm}\ast\,i^{\prime}_{1}}_{{\bf q}^{\prime}_{1}}\hspace{0.03cm} 
f^{\phantom{\hspace{0.03cm}\ast}\!i^{\prime}_{2}}_{{\bf q}^{\prime}_{2}}\,
c^{\ast\ \!\!a^{\prime}_{1}}_{{\bf k}^{\prime}_{1}}\hspace{0.03cm} 
c^{\hspace{0.03cm}a_{2}}_{{\bf k}^{\phantom{\prime}}_{2}}
\,\bigr\rangle
\simeq
-\hspace{0.03cm}
\delta^{\hspace{0.03cm}i^{\prime}_{2}i}
\delta^{\hspace{0.03cm}i_{1}i^{\prime}_{1}}
\delta^{\hspace{0.03cm}a^{\prime}_{1}a^{\phantom{\prime}}_{2}}\,
(2\pi)^{9}\hspace{0.03cm}
\delta({\bf q}^{\prime}_{2} - {\bf q})
\hspace{0.03cm}
\delta({\bf q}^{\prime}_{1} - {\bf q}_{1})
\hspace{0.03cm}
\delta({\bf k}^{\prime}_{1} - {\bf k}^{\phantom{\prime}}_{2})\,
n^{-}_{{\bf q}} n^{-}_{{\bf q}_{1}} N^{l}_{{\bf k}_{2}},
\notag\\[1.5ex]
&\bigl\langle
f^{\hspace{0.04cm}\ast\,i^{\phantom{\prime}}\!\!}_{{\bf q}}\hspace{0.03cm}
f^{\,i_{1}}_{{\bf q}^{\phantom{\prime}}_{1}}\,
f^{\hspace{0.04cm}\ast\,i^{\prime}_{1}}_{{\bf q}^{\prime}_{1}}\hspace{0.03cm}
f^{\phantom{\hspace{0.03cm}\ast}\!i^{\prime}_{2}}_{{\bf q}^{\prime}_{2}}\,
c^{\ast\ \!\!a_{1}}_{{\bf k}^{\phantom{\prime}}_{1}}\hspace{0.03cm} 
c^{\hspace{0.03cm}a^{\prime}_{1}}_{{\bf k}^{\prime}_{1}}
\,\bigr\rangle
\simeq
-\hspace{0.03cm}
\delta^{\hspace{0.03cm}i^{\prime}_{2}i}
\delta^{\hspace{0.03cm}i_{1}i^{\prime}_{1}}
\delta^{\hspace{0.03cm}a^{\prime}_{1}a^{\phantom{\prime}}_{1}}\,
(2\pi)^{9}\hspace{0.03cm}
\delta({\bf q}^{\prime}_{2} - {\bf q})
\hspace{0.03cm}
\delta({\bf q}^{\prime}_{1} - {\bf q}_{1})
\hspace{0.03cm}
\delta({\bf k}^{\prime}_{1} - {\bf k}^{\phantom{\prime}}_{1})\,
n^{-}_{{\bf q}} n^{-}_{{\bf q}_{1}} N^{l}_{{\bf k}_{1}}.
\notag
\end{align}
Substituting these expressions into the right-hand side of (\ref{eq:10o}) and considering the symmetry condition (\ref{eq:10e}) for the scattering amplitude, instead of (\ref{eq:10o}) we derive the equation for the fourth-order correlation function 
\begin{equation}
\frac{\partial\hspace{0.01cm} I^{\, i\; i_{1}\, a_{1}\, a_{2}}_{{\bf q},\, {\bf q}_{1},\, {\bf k}_{1},\, {\bf k}_{2}}}{\partial\hspace{0.03cm} t}
=
i\hspace{0.03cm}\bigl[\,\omega^{-}_{{\bf q}} + \omega^{\hspace{0.02cm}l}_{{\bf k}_{1}} - \omega^{-}_{{\bf q}_{1}} -
\omega^{\hspace{0.02cm}l}_{{\bf k}_{2}}\bigr]
\, I^{\, i\; i_{1}\, a_{1}\, a_{2}}_{{\bf q},\, {\bf q}_{1},\, {\bf k}_{1},\, {\bf k}_{2}}
\label{eq:10p}
\end{equation}
\[
-\,i\,
\widetilde{T}^{\hspace{0.03cm} \ast\hspace{0.03cm} (2)\hspace{0.03cm} i\, i_{1}\, a_{1}\, a_{2}}_{\ {\bf q},\, {\bf q}_{1},\; {\bf k}_{1},\, {\bf k}_{2}}\, 
\Bigl(
n^{-}_{{\bf q}_{1}} N^{l}_{{\bf k}_{1}} N^{l}_{{\bf k}_{2}} -
n^{-}_{{\bf q}} N^{l}_{{\bf k}_{1}} N^{l}_{{\bf k}_{2}} -
n^{-}_{{\bf q}} n^{-}_{{\bf q}_{1}} N^{l}_{{\bf k}_{2}} +
n^{-}_{{\bf q}} n^{-}_{{\bf q}_{1}} N^{l}_{{\bf k}_{1}}
\Bigr)
\]
\[
\times
\,(2\pi)^{3}\hspace{0.03cm}
\delta({\bf q} + {\bf k}_{1} - {\bf q}_{1} - {\bf k}_{2}).
\]


\section{\bf Kinetic equations for soft quark and gluon excitations}
\label{section_11}
\setcounter{equation}{0}

The self-consistent set of equations (\ref{eq:10u}), (\ref{eq:10i}) and (\ref{eq:10p}) determines, in principle, the evolution of plasmino and plasmon number densities $n^{-}_{{\bf q}}$ and $N^{\hspace{0.03cm}l}_{\bf k}$. However, we introduce one more simplification: in Eq.\,(\ref{eq:10p}), we disregard the term with the time derivative as compared to the term containing the difference in the eigenfrequencies of wave packets. Instead of equation (\ref{eq:10p}), we have
\begin{equation}
I^{\, i\; i_{1}\, a_{1}\, a_{2}}_{{\bf q},\, {\bf q}_{1},\, {\bf k}_{1},\, {\bf k}_{2}} \simeq
(2\pi)^{6}\hspace{0.03cm} \delta^{\hspace{0.03cm}i\ \!\!i_{1}} \delta^{\hspace{0.03cm}a_{1}\hspace{0.02cm} a_{2}}\ \! 
\delta({\bf q} - {\bf q}_{1})\hspace{0.03cm} \delta({\bf k}_{1} - {\bf k}_{2})\,
n^{-}_{{\bf q}} N^{l}_{{\bf k}_{1}}
\label{eq:11q}
\end{equation}
\[
+\; \frac{1}{\Delta\hspace{0.02cm}\omega - i\hspace{0.03cm}0}\
\widetilde{T}^{\,\ast\hspace{0.03cm}(2)\, i\, i_{1}\, a_{1}\, a_{2}}_{\ {\bf q},\, {\bf q}_{1},\; {\bf k}_{1},\, {\bf k}_{2}}\, 
\Bigl(
n^{-}_{{\bf q}_{1}} N^{l}_{{\bf k}_{1}} N^{l}_{{\bf k}_{2}} -
n^{-}_{{\bf q}} N^{l}_{{\bf k}_{1}} N^{l}_{{\bf k}_{2}} -
n^{-}_{{\bf q}} n^{-}_{{\bf q}_{1}} N^{l}_{{\bf k}_{2}} +
n^{-}_{{\bf q}} n^{-}_{{\bf q}_{1}} N^{l}_{{\bf k}_{1}}
\Bigr)
\]
\[
\times
\,(2\pi)^{3}\hspace{0.03cm}
\delta({\bf q} + {\bf k}_{1} - {\bf q}_{1} - {\bf k}_{2}),
\]
where now the resonance frequency difference is
\[
\Delta\hspace{0.02cm}\omega \equiv
\omega^{-}_{{\bf q}} + \omega^{l}_{{\bf k}_{1}} - \omega^{-}_{{\bf q}_{1}} - \omega^{l}_{{\bf k}_{2}}.
\]
Here, the first term on the right-hand side, which corresponds to completely uncorrelated waves (Gaussian fluctuations) is the solution to the homogeneous equation for the fourth-order correlation function
$ I^{\, i\; i_{1}\, a_{1}\, a_{2}}_{{\bf q},\, {\bf q}_{1},\, {\bf k}_{1},\, {\bf k}_{2}}$. The structure of this contribution is defined in the same way as that of the first term in (\ref{eq:8_1q}). The second term determines the deviation of the four-point correlator from the Gaussian approximation for a low nonlinearity level of interacting waves.\\
\indent We substitute the first term from (\ref{eq:11q}) into the right-hand side of Eq.\,(\ref{eq:10u}) for $n^{-}_{\bf q}$. As a result we obtain
\begin{equation}
i\hspace{0.04cm}(2\pi)^{3}\hspace{0.03cm}\delta({\bf q} - {\bf q}^{\prime})\hspace{0.03cm}
n^{-}_{\bf q}\!\int\!\frac{d{\bf k}_{1}}{(2\pi)^{3}}\, N^{l}_{{\bf k}_{1}}
\Bigl\{\widetilde{T}^{\,(2)\, i\, j\, a_{1}\, a_{1}}_{\, {\bf q},\, {\bf q},\, {\bf k}_{1},\, {\bf k}_{1}}
-
\widetilde{T}^{\,\ast\hspace{0.03cm}(2)\, i\, j\, a_{1}\, a_{1}}_{\ {\bf q},\, {\bf q},\; {\bf k}_{1},\, {\bf k}_{1}}
\Bigr\}.
\label{eq:11w}
\end{equation}
Further, we substitute the second term into the right-hand side of Eq.\,(\ref{eq:10u}). Simple algebraic transformations, in view of the symmetry condition (\ref{eq:10e}), lead us to
\[
i\hspace{0.04cm} (2\pi)^{3}\hspace{0.03cm}\delta({\bf q} - {\bf q}^{\prime})
\!\int\frac{d{\bf q}_{1}\hspace{0.02cm} d{\bf k}_{1}\hspace{0.03cm} d{\bf k}_{2}}{(2\pi)^{9}}\,
(2\pi)^{3}\hspace{0.03cm} \delta({\bf q} + {\bf k}_{1} - {\bf q}_{1} - {\bf k}_{2})
\left\{\ \!
\frac{1}{\Delta\hspace{0.02cm}\omega - i\hspace{0.03cm}0}
\ -\
\frac{1}{\Delta\hspace{0.02cm}\omega + i\hspace{0.03cm}0}\hspace{0.03cm} \right\}
\]
\begin{equation}
\times\ \!
\widetilde{T}^{\,(2)\, i\, i_{1}\, a_{1}\, a_{2}}_{\, {\bf q},\, {\bf q}_{1},\, {\bf k}_{1},\, {\bf k}_{2}}\ \!
\widetilde{T}^{\,\ast\hspace{0.03cm}(2)\, j\, i_{1}\, a_{1}\, a_{2}}_{\ {\bf q},\, {\bf q}_{1},\; {\bf k}_{1},\, 
{\bf k}_{2}}
\Bigl[\,n^{-}_{{\bf q}_{1}} N^{l}_{{\bf k}_{1}} N^{l}_{{\bf k}_{2}} -\  \ldots\ \Bigr].
\label{eq:11e}
\end{equation}
Next, performing the contraction of the obtained expressions (\ref{eq:10u}), (\ref{eq:11w}), and (\ref{eq:11e}) with 
$\delta^{\hspace{0.02cm}ij}$, considering Sokhotsky's formula (\ref{eq:8r}) and canceling out the factor $(2\pi)^{3}\hspace{0.03cm}\delta({\bf q} - {\bf q}^{\prime})$, we get the desired kinetic equation for abnormal quark excitations 
\begin{equation}
\frac{d\hspace{0.03cm} n^{-}_{{\bf q}}}{d\hspace{0.03cm} t}\
=
\frac{2}{C_{A}}\ \!
n^{-}_{\bf q}\!\int\!\frac{d{\bf k}}{(2\pi)^{3}}\, N^{l}_{{\bf k}}\,
{\rm Im}
\Bigl[\hspace{0.03cm}\widetilde{T}^{\,(2)\, i\, i\, a\, a}_{\, {\bf q},\, {\bf q},\, {\bf k},\, {\bf k}}\Bigr]
\label{eq:11r}
\end{equation}
\[
+\; \frac{1}{C_{A}}
\!\int\frac{d\hspace{0.03cm}{\bf q}_{1}\hspace{0.02cm} d\hspace{0.03cm}{\bf k}_{1}\hspace{0.03cm} d\hspace{0.03cm}{\bf k}_{2}}
{(2\pi)^{9}}\ (2\pi)^{4}\hspace{0.03cm}
\delta(\omega^{-}_{{\bf q}} + \omega^{l}_{{\bf k}_{1}} - \omega^{-}_{{\bf q}_{1}} - \omega^{l}_{{\bf k}_{2}})\ \! 
\delta({\bf q} + {\bf k}_{1} - {\bf q}_{1} - {\bf k}_{2})
\]
\[
\times\, 
\widetilde{T}^{\hspace{0.03cm} (2)\, i\, i_{1}\, a_{1}\, a_{2}}_{\, {\bf q},\, {\bf q}_{1},\, {\bf k}_{1},\, 
{\bf k}_{2}}\ \!
\widetilde{T}^{\hspace{0.03cm}\ast\hspace{0.03cm}(2)\hspace{0.03cm} i\, i_{1}\, a_{1}\, a_{2}}_{\ {\bf q},\, 
{\bf q}_{1},\; {\bf k}_{1},\, {\bf k}_{2}}\hspace{0.02cm}
\Bigl(
n^{-}_{{\bf q}_{1}} N^{l}_{{\bf k}_{1}} N^{l}_{{\bf k}_{2}} -
n^{-}_{{\bf q}} N^{l}_{{\bf k}_{1}} N^{l}_{{\bf k}_{2}} -
n^{-}_{{\bf q}} n^{-}_{{\bf q}_{1}} N^{l}_{{\bf k}_{2}} +
n^{-}_{{\bf q}} n^{-}_{{\bf q}_{1}} N^{l}_{{\bf k}_{1}}\Bigr).
\]
The first term on the right-hand side of Eq.\,(\ref{eq:11r}) describes the nonlinear Landau damping of soft quark excitations where now, unlike (\ref{eq:8y}), the decrement is a linear functional of plasmon number density $N^{\hspace{0.03cm}l}_{\bf k}$:
\[
\hat{\gamma}_{\cal F}[\hspace{0.03cm} N_{\bf k}^{\hspace{0.03cm}l}]  \equiv \gamma^{-}({\bf q})
=
-\hspace{0.03cm}\frac{2}{C_{A}}\hspace{0.03cm}
\!\int\!\frac{d{\bf k}}{(2\pi)^{3}}\, N^{l}_{{\bf k}}\,
{\rm Im}
\Bigl[\hspace{0.03cm}\widetilde{T}^{\,(2)\, i\, i\, a\, a}_{\, {\bf q},\, {\bf q},\, {\bf k},\, {\bf k}}\Bigr].
\]
The second term in Eq.\,(\ref{eq:11r}) is associated with an elastic plasmino-plasmon scattering. We can also write Eq.\,(\ref{eq:11r}) in a more compact form:
\begin{equation}
\frac{d\hspace{0.03cm} n_{\bf q}^{-}}{d\hspace{0.03cm} t} \equiv \frac{\partial\hspace{0.03cm} n_{\bf q}^{-}}{\partial\hspace{0.03cm} t} +
{\bf v}_{\bf q}^{-}\cdot\frac{\partial\hspace{0.03cm} n_{\bf q}^{-}}
{\partial\hspace{0.04cm} {\bf x}}
=
-\,\hat{\gamma}_{\cal F}[\hspace{0.03cm} N_{\bf k}^{\hspace{0.03cm}l}] \, n_{\bf q}^{-} - n_{\bf q}^{-}\hspace{0.02cm} 
\Gamma^{(f)}_{\rm d} 
[n_{\bf q}^{-},N^{\hspace{0.03cm}l}_{\bf k}] + ( 1 - n_{\bf q}^{-})\hspace{0.03cm}\Gamma^{(f)}_{\rm i}[ n_{\bf q}^{-}, N^{\hspace{0.03cm}l}_{\bf k}],
\label{eq:11t}
\end{equation}
where ${\bf v}_{\bf q}^{-}$ is the group velocity of abnormal soft-quark mode, and the generalized decay rate 
$\Gamma^{(f)}_{\rm d}$ and the inverse regeneration rate $\Gamma^{(f)}_{\rm i}$ are nonlinear functionals of the plasmino and plasmon number densities:
\[
\Gamma^{(f)}_{\rm d}[n_{\bf q}^{-},N^{\hspace{0.03cm}l}_{\bf k}]=\!
\int\!d\hspace{0.02cm}{\cal T}_{\hspace{0.03cm}qg\hspace{0.02cm}\rightarrow\hspace{0.02cm} qg}
\hspace{0.04cm}{\it w}_{\hspace{0.03cm}qg\hspace{0.02cm}\rightarrow\hspace{0.02cm} qg}({\bf q}, {\bf k}_{1}; {\bf q}_{1}, {\bf k}_{2})\hspace{0.03cm}
(1 - n_{{\bf q}_{1}}^{-})\hspace{0.03cm}N^{l}_{{\bf k}_{1}}(1 + N^{l}_{{\bf k}_{2}})
\]
and, accordingly,
\[
\Gamma^{(f)}_{\rm i} [n_{\bf q}^{-},N^{\hspace{0.03cm}l}_{\bf k}] =\!
\int\!d\hspace{0.02cm}{\cal T}_{qg\hspace{0.02cm}\rightarrow\hspace{0.02cm} qg}
\hspace{0.04cm}{\it w}_{qg\hspace{0.02cm}\rightarrow\hspace{0.02cm} qg}({\bf q}, {\bf k}_{1}; {\bf q}_{1}, {\bf k}_{2})\hspace{0.03cm}
n_{{\bf q}_1}^{-} (1 + N^{l}_{{\bf k}_{1}})\hspace{0.03cm}N^{l}_{{\bf k}_{2}}\,.
\hspace{0.8cm}
\]
Here,
\begin{equation}
{\it w}_{\hspace{0.03cm}qg\hspace{0.02cm}\rightarrow\hspace{0.02cm} qg}({\bf q}, {\bf k}_{1}; {\bf q}_{1}, {\bf k}_{2}) 
=\frac{1}{C_{A}}\,\! \sum\limits_{\lambda,\, \lambda_{1}\, =\,\pm}\!\!
\widetilde{T}^{\hspace{0.03cm} (2)\hspace{0.03cm} i\, i_{1}\, a_{1}\, a_{2}}_{\, {\bf q},\, {\bf q}_{1},\, {\bf k}_{1},\, {\bf k}_{2}}\ \!
\widetilde{T}^{\hspace{0.03cm}\ast\hspace{0.03cm}(2)\hspace{0.03cm} i\, i_{1}\, a_{1}\, a_{2}}_{\ {\bf q},\, 
{\bf q}_{1},\; {\bf k}_{1},\, {\bf k}_{2}}
\label{eq:11y}
\end{equation}
is the scattering probability for the elastic collision of colorless plasminos and plasmons, and the integration measure is defined as
\begin{equation}
d\hspace{0.02cm}{\cal T}_{qg\hspace{0.02cm}\rightarrow\hspace{0.02cm} qg} \equiv
(2\pi)^{4}\,
\delta(\omega^{-}_{{\bf q}} + \omega^{l}_{{\bf k}_{1}} - \omega^{-}_{{\bf q}_{1}} - \omega^{l}_{{\bf k}_{2}})\ \!
\delta({\bf q} + {\bf k}_{1} - {\bf q}_{1} - {\bf k}_{2})\ \!
\frac{d{\bf q}_{1}\hspace{0.02cm} d{\bf k}_{1}\hspace{0.03cm} d{\bf k}_{2}}{(2\pi)^{9}}.
\label{eq:11u}
\end{equation}
\indent The same reasoning leads us, instead of (\ref{eq:10i}), to the following kinetic equation for the plasmons number density $N^{\hspace{0.03cm}l}_{\bf k}$:
\begin{equation}
\frac{d N_{\bf k}^{\hspace{0.03cm}l}}{d\hspace{0.03cm} t} \equiv \frac{\partial N_{\bf k}^{\hspace{0.03cm}l}}{\partial\hspace{0.03cm} t} +
{\bf v}_{\bf k}^{\hspace{0.02cm}l}\cdot\frac{\partial N_{\bf k}^{\hspace{0.03cm}l}}
{\partial\hspace{0.03cm} {\bf x}}
=
-\,\hat{\gamma}_{\cal B}[\hspace{0.03cm} n_{\bf q}^{-}] \, N_{\bf k}^{\hspace{0.03cm}l} 
- 
N_{\bf k}^{\hspace{0.03cm}l}\hspace{0.03cm} \Gamma^{(b)}_{\rm d} [\hspace{0.03cm}n_{\bf q}^{-}, N_{\bf k}^{\hspace{0.03cm}l}] 
+ 
( 1 + N_{\bf k}^{\hspace{0.03cm}l} )\hspace{0.03cm} \Gamma^{(b)}_{\rm i}[\hspace{0.03cm}n_{\bf q}^{-}, N_{\bf k}^{\hspace{0.03cm}l}],
\label{eq:11i}
\end{equation}
where
\[
{\bf v}_{\bf k}^{\hspace{0.02cm}l} = \frac{\partial\hspace{0.04cm} \omega_{\bf k}^{\hspace{0.03cm}l}}
{\partial\hspace{0.03cm} {\bf k}} = - \Biggl[\left(\frac{\partial\hspace{0.02cm} {\rm Re} \,
\varepsilon^{\hspace{0.02cm}l}(k)}{\partial\hspace{0.02cm} {\bf k}} \right)\!
\left( \frac{\partial\hspace{0.02cm} {\rm Re}\,\varepsilon^{\hspace{0.02cm}l}(k)}
{\partial\hspace{0.02cm} \omega} \right)^{\!\!-1}\,\Biggr]
\Bigg\vert_{\omega\hspace{0.03cm} =\hspace{0.03cm} \omega_{\bf k}^{l}}
\]
is the group velocity of longitudinal oscillations, $\varepsilon^{\hspace{0.02cm}l}(k) = (k^{2}\,^{\ast}\!\Delta^{\,l}(k))^{-1}$ is the longitudinal permittivity, $\hat{\gamma}_{\cal F}[\hspace{0.03cm} n_{\bf q}^{-}]$ is the decrement of nonlinear Landau damping of collective Bose-excitations, which represents a linear functional of the plasmino number density $n^{-}_{\bf q}$:
\[
	\hat{\gamma}_{\cal B}[\hspace{0.03cm} n_{\bf q}^{-}]  \equiv \gamma^{l}({\bf k})
	=
	\frac{2}{d_{A}}\hspace{0.03cm}\!\int\!\frac{d\hspace{0.03cm}{\bf q}}{(2\pi)^{3}}\ n^{-}_{{\bf q}}\ \!
	{\rm Im} \Bigl[\hspace{0.03cm}\widetilde{T}^{\,(2)\, i\, i\, a\, a}_{{\bf q},\, {\bf q},\, {\bf k},\, {\bf k}}\Bigr],
\]
and the generalized decay rate $\Gamma^{(b)}_{\rm d}$ and the inverse regeneration rate $\Gamma^{(b)}_{\rm i}$ are nonlinear functionals of the plasmino and plasmon number densities:
\begin{align}
&\Gamma^{(b)}_{\rm d}[\hspace{0.03cm}n_{\bf q}^{-}, N_{\bf k}^{\hspace{0.03cm}l}] =
\int\!d{\cal T}_{\hspace{0.03cm}gq\hspace{0.02cm}\rightarrow\hspace{0.02cm} gq}
{\it w}_{\hspace{0.03cm}gq\hspace{0.02cm}\rightarrow\hspace{0.02cm} gq}
({\bf k}, {\bf q}_{1}; {\bf k}_{1}, {\bf q}_{2})
(1 + N_{{\bf k}_{1}}^{\hspace{0.03cm}l})\hspace{0.03cm} n^{-}_{{\bf q}_{1}}(1 - n_{{\bf q}_{2}}^{-}),
\notag\\[1ex]
&\Gamma^{(b)}_{\rm i} [\hspace{0.03cm}n_{\bf q}^{-}, N_{\bf k}^{\hspace{0.03cm}l}] =
\int\!d{\cal T}_{\hspace{0.03cm}gq\hspace{0.02cm}\rightarrow\hspace{0.02cm} gq}
{\it w}_{\hspace{0.03cm}gq\hspace{0.02cm}\rightarrow\hspace{0.02cm} gq}
({\bf k}, {\bf q}_{1}; {\bf k}_{1}, {\bf q}_{2})
N_{{\bf k}_{1}}^{\hspace{0.03cm}l}(1 - n^{-}_{{\bf q}_{1}})\hspace{0.03cm}n_{{\bf q}_{2}}^{-}.
\notag
\end{align}
Here,
\begin{equation}
{\it w}_{\hspace{0.03cm}gq\hspace{0.02cm}\rightarrow\hspace{0.02cm} gq}
({\bf k}, {\bf q}_{1}; {\bf k}_{1}, {\bf q}_{2}) =
\frac{1}{d_{A}}\,\!\sum\limits_{\lambda_{1},\, \lambda_{2}\, =\,\pm}\!\!
\widetilde{T}^{\,(2)\, i_{1}\, i_{2}\, a\, a_{1}}_{\; {\bf q}_{1},\, {\bf q}_{2},\, {\bf k},\, {\bf k}_{1}}\ \!
\widetilde{T}^{\hspace{0.03cm}\ast\hspace{0.03cm} (2)\hspace{0.03cm} i_{1}\, i_{2}\, a\, a_{1}}_{\; {\bf q}_{1},\, 
{\bf q}_{2},\, {\bf k},\, {\bf k}_{1}}
\label{eq:11o}
\end{equation}
is the scattering probability for an elastic collision of colorless plasmons and plasminos, and the integration measure is defined as
\[
d{\cal T}_{\hspace{0.03cm}gq\hspace{0.02cm}\rightarrow\hspace{0.02cm} gq} \equiv 
(2\pi)^{4}\,
\delta(\omega^{\hspace{0.02cm}l}_{{\bf k}} + \omega^{-}_{{\bf q}_{1}} - \omega^{\hspace{0.02cm}l}_{{\bf k}_{1}} - \omega^{-}_{{\bf q}_{2}})\ \!
\delta({\bf k} + {\bf q}_{1} - {\bf k}_{1} - {\bf q}_{2})\ \!
\frac{d{\bf q}_{1}\hspace{0.02cm} d{\bf q}_{2}\hspace{0.03cm} d{\bf k}_{1}}{(2\pi)^{9}}\, .
\]
In Eq.\,(\ref{eq:11o}) $d_{A} = N^{2}_{c} - 1$ is another invariant for the group $SU(N_{c})$. The equations (\ref{eq:11t}) and (\ref{eq:11i}) constitute a self-consistent system, which defines the evolution of the plasmino and plasmon number densities. In a general case of course, we have to add two more equations connected with the elastic scattering plasminos off plasminos, Eq.\,(\ref{eq:8u}), and the elastic scattering plasmons off plasmons, Eq.\,(4.4) from the paper \cite{markov_2020}.


\section{Explicit form of the vertex functions}
\label{section_12}
\setcounter{equation}{0}

We need to compare the effective amplitude (\ref{eq:5t}) obtained within the framework of (pseudo)clas\-sical approach with the matrix element of corresponding scattering process derived in the context of the hard thermal loop approximation. In \cite{markov_2006}, within the HTL-approximation, the probability of plasmino-plasmon scattering was obtained
\begin{equation}
{\it w}_{qg\rightarrow qg}^{\!(-\hspace{0.03cm}l;-\hspace{0.03cm} l)}
({\bf q}, {\bf k}_{1}; {\bf q}_{1}, {\bf k}_{2}) = \frac{1}{C_{A}}
\sum\limits_{\lambda,\,\lambda_{1}\hspace{0.02cm} =\hspace{0.02cm} \pm \hspace{0.01cm}1}\!\!
{\rm T}^{\, a_{1}\hspace{0.02cm} a_{2},\, i\hspace{0.035cm} i_{1}}_{\lambda\hspace{0.03cm}\lambda_{1}}(-{\bf k}_{1},{\bf k}_{2};{\bf q}_{1},-{\bf q})
\bigl({\rm T}^{\, a_{1}\hspace{0.02cm} a_{2},\, i\hspace{0.035cm} i_{1}}_{\lambda\hspace{0.03cm}\lambda_{1}}(-{\bf k}_{1},{\bf k}_{2};{\bf q}_{1},-{\bf q})\bigr)^{\!\ast}.
\label{eq:12q}
\end{equation}
Here the matrix element ${\rm T}^{\, a_{1}\, a_{2},\, i\hspace{0.035cm} i_{1}}_{\lambda\hspace{0.03cm}\lambda_{1}}$ of elastic plasmino-plasmon scattering has the following structure:
\begin{equation}
{\rm T}^{\, a_{1}\hspace{0.02cm} a_{2},\,i\, i_1}_{\lambda\hspace{0.03cm}\lambda_{1}}(-{\bf k}_1,{\bf k}_2;{\bf q}_1,-{\bf q})
= g^2
\left(\frac{Z_{-}({\bf q})}{2}\right)^{\!\!1/2}\!
\left(\frac{Z_{-}({\bf q}_1)}{2}\right)^{\!\!1/2}\!
\left(\frac{Z_l({\bf k}_1)}{2\hspace{0.03cm} \omega_{{\bf k}_1}^{\hspace{0.02cm} l}}\right)^{\!\!1/2}\!
\left(\frac{Z_l({\bf k}_2)}{2\hspace{0.03cm} \omega_{{\bf k}_2}^{\hspace{0.02cm} l}}\right)^{\!\!1/2}
\label{eq:12w}
\end{equation}
\[
\times\,
\Biggl(\frac{\tilde{u}^{\mu_{1}}(k_{1})}{\sqrt{\bar{u}^2(k_{1})}}\Biggr)
\Biggl(\frac{\tilde{u}^{\mu_{2}}(k_{2})}{\sqrt{\bar{u}^2(k_{2})}}\Biggr)
\Bigl[\,\bar{v}^{(-)}_{\beta}(\hat{\bf q}_1,\lambda_{1})\!\hspace{0.02cm} 
\,^{\ast}\tilde{\bar{\Gamma}}^{(Q)a_1a_2,\,i\, i_1}_{\mu_1\mu_2,\,\alpha\beta}\!
(-k_1,k_2;q_1,-q) v^{(-)}_{\alpha}(\hat{\bf q},\lambda)\Bigr]_{\,{\rm on-shell}},
\]
where, in turn, the effective amplitude $\!\,^{\ast}\tilde{\bar{\Gamma}}^{(Q)\hspace{0.025cm}a_{1}\hspace{0.02cm} a_{2},\,i\hspace{0.035cm} i_{1}}_{\mu_{1}\mu_{2},\,\alpha\beta}(-k_{1}, k_{2}; q_{1}, -q)$ is defined as
\begin{equation}
\!\,^{\ast}\tilde{\bar{\Gamma}}^{(Q)\hspace{0.025cm} a_{1}a_{2},\,i\hspace{0.035cm} i_{1}}_{\mu_{1}\mu_{2},\,\alpha\beta}
(-k_{1},k_{2}; q_{1},-q) 
= -\,\Bigl\{
\delta\hspace{0.02cm}{\Gamma}^{(Q)\hspace{0.025cm} a_{2}\hspace{0.02cm}a_{1},\,i_{1}i}_{\mu_{1}\mu_{2},\,\beta\alpha}
(k_{1},-k_{2};-q_{1},q)
\label{eq:12e}
\end{equation}
\[
+\,[t^{a_{2}},t^{a_{1}}]^{i_{1}i}
\,^{\ast}\Gamma^{(Q)}_{\nu,\,\beta\alpha}(q - q_{1}; -q, q_{1})
\,^{\ast}{\cal D}^{\nu\nu^{\prime}}(k_{1} - k_{2})
\,^\ast\Gamma_{\nu^{\prime}\mu_1\mu_2}(k_{1} - k_{2},-k_{1},k_{2})
\]
\[
-\,(t^{a_{2}}t^{a_{1}})^{i_{1}i}
\,^{\ast}\Gamma^{(Q)}_{\mu_2,\,\beta\gamma}(k_2;-q_1-k_2,q_1)
\,^{\ast}\!S_{\gamma\gamma^{\prime}}(-q_1-k_2)
\,^{\ast}\Gamma^{(Q)}_{\mu_1,\,\gamma^{\prime}\alpha}(-k_{1};-q,q + k_{1})
\hspace{0.1cm}
\]
\[
\hspace{0.25cm}
-\,(t^{a_1}t^{a_2})^{i_1i}
\,^{\ast}\Gamma^{(Q)}_{\mu_1,\,\beta\gamma}(-k_{1};-q_{1} + k_{1},q_{1})
\,^{\ast}\!S_{\gamma\gamma^{\prime}}(-q_{1} + k_{1})
\,^{\ast}\Gamma^{(Q)}_{\mu_2,\,\gamma^{\prime}\alpha}(k_{2};-q, q - k_{2})
\Bigr\}.
\]
The HTL-resumed vertex between a quark pair and a gluon $\!\,^{\ast}\Gamma^{(Q)\hspace{0.02cm}\mu}_{\alpha\beta}(q-q_1;q_1,-q)$ and the HTL-induced vertex between a quark pair and two gluons $\delta\hspace{0.02cm}{\Gamma}^{(Q)\hspace{0.025cm} a_{2}\hspace{0.02cm}a_{1},\,i_{1}i}_{\mu_{1}\mu_{2},\,\beta\alpha}(k_{1},-k_{2};-q_{1},q)$ on the right-hand side of expression (\ref{eq:12e}) are defined by Eqs.\,(\ref{ap:B4}) and (\ref{ap:B7}). The explicit form of a medium modified quark propagator $\!\,^{\ast}\!S_{\alpha\alpha^{\prime}}(q)$, is given by the formulae (\ref{ap:B11})\,--\,(\ref{ap:B13}). The effective amplitude 
(\ref{eq:12e}) possesses the following property:
\[
\!\,^{\ast}\tilde{\bar{\Gamma}}^{(Q)\hspace{0.025cm} a_{1}a_{2},\,i\hspace{0.035cm}i_{1}}_{\mu_{1}\mu_{2},\,\alpha\beta}
(-k_{1},k_{2};q_{1},-q)=
\!\,^{\ast}\tilde{\bar{\Gamma}}^{(Q)\hspace{0.025cm} a_{2}\hspace{0.02cm}a_{1},\,i\hspace{0.035cm}i_{1}}_{\mu_{2}\mu_{1},\,\alpha\beta}
(k_{2},-k_{1};q_{1},-q).
\]
\indent Comparing two expressions (\ref{eq:11y}) and (\ref{eq:12q}) for the plasmino-plasmon scattering probability, we see that the effective amplitude $\widetilde{T}^{(2)\, i\hspace{0.035cm} i_{1}\, a_{1}\hspace{0.02cm} a_{2}}_{\, {\bf q},\, {\bf q}_{1},\, {\bf k}_{1},\, {\bf k}_{2}}$ defined by expression (\ref{eq:5t}) should be identified with the matrix element ${\rm T}^{\, a_{1}\hspace{0.02cm} a_{2},\, i\hspace{0.035cm} i_{1}}_{\lambda\hspace{0.03cm}\lambda_{1}}(-{\bf k}_{1},{\bf k}_{2};{\bf q}_{1},-{\bf q})$ (more exactly with its complex conjugation) calculated by means of high-temperature quantum field theory:
\begin{equation}
\widetilde{T}^{\,(2)\, i\hspace{0.035cm} i_{1}\, a_{1}\hspace{0.02cm} a_{2}}_{\, {\bf q},\, {\bf q}_{1},\, {\bf k}_{1},\, 
{\bf k}_{2}} (\lambda,\lambda_{1})
=
\bigl(\hspace{0.03cm}{\rm T}^{\, a_{1}\hspace{0.02cm} a_{2},\, i\hspace{0.035cm} i_{1}}_{\lambda\hspace{0.03cm}\lambda_{1}}
(-{\bf k}_{1},{\bf k}_{2}; {\bf q}_{1},-{\bf q})\bigr)^{\!\ast}
\label{eq:12r}
\end{equation}
From the expressions for the effective amplitude (\ref{eq:5t}) and for the matrix element (\ref{eq:12w}), (\ref{eq:12e}), we can immediately obtain the explicit form of the amplitude $T^{\,(2)\hspace{0.03cm} i\hspace{0.035cm} i_{1}\, a_{1}\hspace{0.02cm} a_{2}}_{\, {\bf q},\, {\bf q}_{1},\, {\bf k}_{1},\, {\bf k}_{2}}$, which is the vertex function in the definition of the fourth-order Hamiltonian $H^{(4)}$ (\ref{eq:2g}):
\begin{equation}
T^{\,(2)\hspace{0.03cm} i\hspace{0.035cm} i_{1}\, a_{1}\hspace{0.02cm} a_{2}}_{\, {\bf q},\, {\bf q}_{1},\, {\bf k}_{1},\, {\bf k}_{2}}
(\lambda,\lambda_{1})
\equiv
g^{2}
\left(\frac{Z_{-}({\bf q})}{2}\right)^{\!\!1/2}\!
\left(\frac{Z_{-}({\bf q}_1)}{2}\right)^{\!\!1/2}
\times
\label{eq:12t}
\end{equation}
\[
\times\,
\Biggl(
\frac{\epsilon^{\hspace{0.02cm}l}_{{\mu}_{1}}({\bf k}_{1})}{\sqrt{2\hspace{0.03cm}\omega^{\hspace{0.02cm}l}_{{\bf k}_{1}}}}\Biggr)
\Biggl(\frac{\epsilon^{l}_{\mu_{2}}({\bf k}_{2})}{\sqrt{2\hspace{0.03cm}\omega^{l}_{{\bf k}_{2}}}}\Biggr)
\Bigl[\,\bar{v}^{(-)}_{\beta}(\hat{\bf q},\lambda)\hspace{0.02cm}
\delta\hspace{0.02cm}{\Gamma}^{(Q)\hspace{0.025cm} a_{1}\hspace{0.02cm}a_{2},\,i\hspace{0.02cm} i_{1}}_{\mu_{1}\mu_{2},\,\beta\alpha}
(-k_{1}, k_{2}; q_{1}, -q) v^{(-)}_{\alpha}(\hat{\bf q}_1,\lambda_{1})\Bigr]_{\,{\rm on-shell}}.
\]
Here, we have taken into account the relationship (\ref{eq:9j}) between the longitudinal projector and the polarization vector $\epsilon^{\hspace{0.02cm}l}_{\mu}({\bf k})$, and the conjugation property for the HTL-induced vertex
$\delta\hspace{0.02cm}{\Gamma}^{(Q)\hspace{0.025cm} a_{1}\hspace{0.02cm}a_{2}}_{\mu_{1}\mu_{2}}$, Eq.\,(\ref{ap:B10}).\\
\indent Let us now determine an explicit form of the three-point vertex functions ${\mathcal G}^{\; a\, i\,  i_{1}}_{\,{\bf k},\, {\bf q},\, {\bf q}_{1}}$, ${\mathcal K}^{\; a_{1}\, i\,  i_{1}}_{\,{\bf k}_1,\, {\bf q},\, {\bf q}_{1}}$ and ${\mathcal P}^{\; a_{1}\, i\,  i_{1}}_{\,{\bf k}_1,\, {\bf q},\, {\bf q}_{1}}$. For this purpose, in the effective amplitude (\ref{eq:5t}) we keep only the terms with these vertex functions, and in the matrix element (\ref{eq:12w}), (\ref{eq:12e}) we do the terms with HTL-resumed vertices between a quark pair and a gluon. As a result, from the equality (\ref{eq:12r}) follows 
\begin{align}
-\;
4\, &\frac{{\mathcal G}^{\; a_{2}\,  i\, i^{\hspace{0.02cm}\prime}}_{{\bf k}_{2},\, {\bf q},\, {\bf k}_{2} - {\bf q}}\ 
{\mathcal G}^{\hspace{0.03cm}\ast\, a_{1}\, i_{1}\hspace{0.03cm}  i^{\hspace{0.02cm}\prime}}_{{\bf k}_{1},\, {\bf q}_{1},\, {\bf k}_{1} - {\bf q}_{1}}}{\omega^{l}_{{\bf k}_{2}} - \omega^{-}_{{\bf q}}- \omega^{-}_{{\bf k}_{2} - {\bf q}}} 
\;+\;
\frac{{\mathcal P}^{\; a_{2}\, i^{\hspace{0.02cm}\prime}\hspace{0.03cm}  i_{1}}_{{\bf k}_{2},\, {\bf k}_{2} + {\bf q}_{1},\, {\bf q}_{1}}\, 
{\mathcal P}^{\hspace{0.03cm}\ast\, a_{1}\, i^{\hspace{0.02cm}\prime}\,  i}_{{\bf k}_{1},\, {\bf k}_{1} + {\bf q},\,  {\bf q}}}
{\omega^{l}_{{\bf k}_{2}} - \omega^{-}_{ {\bf k}_{2} + {\bf q}_{1}} + \omega^{-}_{{\bf q}_{1}}}
\;-\; 
\frac{{\mathcal P}^{\; a_{2}\, i\,  i^{\hspace{0.02cm}\prime}}_{{\bf k}_{2},\, {\bf q},\, {\bf q} - {\bf k}_{2}}\, 
{\mathcal P}^{\hspace{0.03cm}\ast\, a_{1}\, i_{1}\,  i^{\hspace{0.02cm}\prime}}_{{\bf k}_{1},\, {\bf q}_{1},\,  {\bf q}_{1} - {\bf k}_{1}}}
{\omega^{l}_{{\bf k}_{2}} - \omega^{-}_{{\bf q}} + \omega^{-}_{{\bf q} - {\bf k}_{2}}}
\notag\\[1.5ex]
+\;
4\, &\frac{{\mathcal K}^{\; a_{2}\, i_{1}\,  i^{\hspace{0.02cm}\prime}}_{{\bf k}_{2},\, {\bf q}_{1},\, -{\bf k}_{2} - {\bf q}_{1}}\ 
{\mathcal K}^{\hspace{0.03cm}\ast\, a_{1}\,  i\, i^{\hspace{0.02cm}\prime}}_{{\bf k}_{1},\, {\bf q},\, -{\bf k}_{1} - {\bf q}}}
{\omega^{l}_{{\bf k}_{2}} + \omega^{-}_{{\bf q}_{1}} + \omega^{-}_{-{\bf k}_{2} - {\bf q}_{1}}}
-
2\, \frac{{\mathcal V}^{\; a_{1}\, a_{2}\,  a^{\prime}}_{{\bf k}_{1}, {\bf k}_{2},\, {\bf k}_{1} - {{\bf k}_{2}}}\, 
{\mathcal P}^{\hspace{0.03cm}\ast\, a^{\prime}\, i_{1}\, i}_{\;{\bf q}_{1} - {\bf q},\, {\bf q}_{1},\, {\bf q}}}
{\omega^{l}_{{\bf q}_{1} - {\bf q}} - \omega^{-}_{{\bf q}_{1}} + \omega^{-}_{{\bf q}}}
-
2\, \frac{{\mathcal P}^{\; a^{\prime}\, i\,  i_{1}}_{\;{\bf q} - {\bf q}_{1},\, {\bf q},\, {\bf q}_{1}}\, 
{\mathcal V}^{\hspace{0.03cm}\ast\, a_{2}\, a_{1}\,  a^{\prime}}_{{\bf k}_{2}, {\bf k}_{1},\, {\bf k}_{2} - {\bf k}_{1}}}
{\omega^{l}_{{\bf q} - {\bf q}_{1}} - \omega^{-}_{{\bf q}} + \omega^{-}_{{\bf q}_{1}}}
\notag
\end{align}
\begin{equation}
= g^{2}
\left(\frac{Z_{-}({\bf q})}{2}\right)^{\!\!1/2}\!
\left(\frac{Z_{-}({\bf q}_1)}{2}\right)^{\!\!1/2}\!
\Biggl(
\frac{\epsilon^{\hspace{0.02cm}l}_{{\mu}_{1}}({\bf k}_{1})}{\sqrt{2\hspace{0.03cm}\omega^{\hspace{0.02cm}l}_{{\bf k}_{1}}}}\Biggr)
\Biggl(\frac{\epsilon^{\hspace{0.02cm}l}_{\mu_{2}}({\bf k}_{2})}{\sqrt{2\hspace{0.03cm}\omega^{\hspace{0.02cm}l}_{{\bf k}_{2}}}}\Biggr)
\hspace{0.02cm}
\bar{v}^{(-)}_{\beta}(\hat{\bf q},\lambda)\hspace{0.02cm}v^{(-)}_{\alpha}(\hat{\bf q}_{1},\lambda_{1})\
\label{eq:12y}
\vspace{-0.4cm}
\end{equation}
\begin{align}
\hspace{1cm}\times\hspace{0.035cm}
\Bigl\{&[\,t^{a_{1}},t^{a_{2}}]^{i\hspace{0.03cm} i_{1}}
\,^{\ast}\Gamma^{(Q)\hspace{0.03cm}\nu}_{\beta\alpha}(- q + q_{1}; q, -q_{1})
\,^{\ast}{\cal D}_{\nu\nu^{\prime}}(-k_{1} + k_{2})
\,^\ast\Gamma^{\hspace{0.03cm}\nu^{\prime}\mu_1\mu_2}(-k_{1} + k_{2}, k_{1}, -k_{2})
\notag\\[1.5ex]
-\,&(t^{a_{1}}t^{a_{2}})^{i\hspace{0.035cm} i_{1}}
\,^{\ast}\Gamma^{(Q)\hspace{0.02cm} \mu_{1}}_{\beta\gamma}(k_{1}; q, -q - k_{1})
\,^{\ast}\!S_{\gamma\gamma^{\prime}}(q_{1} + k_{2})
\,^{\ast}\Gamma^{(Q)\hspace{0.02cm} \mu_{2}}_{\gamma^{\prime}\alpha}(-k_{2}; q_{1} + k_{2}, -q_{1})
\notag\\[1.5ex]
-\,&(t^{a_{2}}t^{a_{1}})^{i\hspace{0.035cm} i_{1}}
\,^{\ast}\Gamma^{(Q)\hspace{0.02cm} \mu_{2}}_{\beta\gamma}(-k_{2}; q, -q + k_{2})
\,^{\ast}\!S_{\gamma\gamma^{\prime}}(q_{1} - k_{1})
\,^{\ast}\Gamma^{(Q)\hspace{0.02cm} \mu_{1}}_{\gamma^{\prime}\alpha}(k_{1}; q_{1} - k_{1}, -q_{1})
\Bigr\}\Bigr\vert_{\,{\rm on-shell}}.
\notag
\end{align}
Here, on the right-hand side we have used the conjugation rules for the HTL-induced vertex functions
$\!\,^{\ast}\Gamma_{\mu\mu_{1} \mu_{2}}$ and $\,^{\ast}\Gamma^{(Q)}_{\mu}$, Eqs.\,(\ref{ap:A4}) and (\ref{ap:B8}), and for the effective gluon and quark propaga\-tors, Eqs.\,(\ref{ap:A12}) and (\ref{ap:B14}), correspondingly.\\
\indent  As we have done in section \ref{section_9}, in the effective gluon propagator $\!\,^{\ast}{\cal D}_{\nu\nu^{\prime}}$ on the right-hand side of relation (\ref{eq:12y}) we retain only the terms with the longitudinal projector $\widetilde{Q}_{\nu\nu^{\prime}}$ and perform the substitution
\begin{equation}
\,^{\ast}{\cal D}_{\nu\nu^{\prime}}(-k_{1} + k_{2})
\Rightarrow
-\,\widetilde{Q}_{\nu\nu^{\prime}}(-k_{1} + k_{2}) \,^{\ast}\!\Delta^{l}(-k_{1} + k_{2})
\label{eq:12u}
\end{equation}
\[
\simeq
\Biggl(\frac{\epsilon^{\hspace{0.02cm}l}_{\nu}({\bf q} - {\bf q}_{1})}{\sqrt{2\hspace{0.03cm}\omega^{\hspace{0.02cm}l}_{{\bf q}\hspace{0.02cm} -\hspace{0.02cm} {\bf q}_{1}}}}\Biggr)
\Biggl(\frac{ \epsilon^{\hspace{0.02cm}l}_{\nu^{\prime}}(-{\bf k}_{1} + {\bf k}_{2})}{\sqrt{2\hspace{0.03cm}\omega^{\hspace{0.02cm}l}_{-{\bf k}_{1}\hspace{0.02cm} +\hspace{0.02cm} {\bf k}_{2}}}}\Biggr)
\biggl[\,\frac{1}{\omega^{l}_{{\bf q}_{1} - {\bf q}}\! - \omega^{-}_{{\bf q}_{1}} + \omega^{-}_{{\bf q}}}
\,+\,
\frac{1}{\omega^{l}_{{\bf q} - {\bf q}_{1}}\! - \omega^{-}_{{\bf q}} + \omega^{-}_{{\bf q}_{1}}}\,\biggr],
\]
where on the right-hand side we have taken into account, due to Eq.\,(\ref{eq:11u}), the conservation laws of momentum and energy in an elementary act of scattering plasmino off plasmon 
\begin{equation}
\left\{
\begin{array}{ll}
\omega^{-}_{{\bf q}} + \omega^{l}_{{\bf k}_{1}} = \omega^{-}_{{\bf q}_{1}} + \omega^{l}_{{\bf k}_{2}}, \\[1.5ex]
{\bf q} + {\bf k}_{1} = {\bf q}_{1} + {\bf k}_{2}
\end{array}
\right.
\label{eq:12i}
\end{equation}
and the evenness of the dispersion relation for the plasmon mode.\\
\indent From the approximation of the longitudinal part of gluon propagator (\ref{eq:12u}) we see that it contains contributions proportional to the factors $1/(\omega^{l}_{{\bf q}_{1} - {\bf q}}\! - \omega^{-}_{{\bf q}_{1}} + \omega^{-}_{{\bf q}})$ and $1/(\omega^{l}_{{\bf q} - {\bf q}_{1}}\! - \omega^{-}_{{\bf q}} + \omega^{-}_{{\bf q}_{1}})$, which take place in the last two terms on the left-hand side of (\ref{eq:12y}). By using the expressions for color structure of the vertex functions ${\mathcal V}^{\ \! a\, a_{1}\, a_{2}}_{{\bf k},\, {\bf k}_{1},\, {\bf k}_{2}}$ and ${\mathcal P}^{\; a\, i\;  i_{1}}_{{\bf k},\, {\bf q},\, {\bf q}_{1}}$, Eqs.\,(\ref{eq:2l}) and (\ref{eq:9s}), and taking into account the conjugation properties 
\[
{\mathcal V}^{\hspace{0.03cm}\ast}_{\, {\bf k}_{2},\, {\bf k}_{1},\, {\bf k}_{2} - {\bf k}_{1}}
=
{\mathcal V}_{\, {\bf k}_{1},\, {\bf k}_{2},\, {\bf k}_{1}-{\bf k}_{2}},
\quad
{\mathcal P}^{\hspace{0.03cm}\ast}_{{\bf q}_{1} - {\bf q},\, {\bf q}_{1},\, {\bf q}}
=
-\hspace{0.03cm}{\mathcal P}_{{\bf q} - {\bf q}_{1},\, {\bf q},\, {\bf q}_{1}},
\] 
we can represent these two terms in the following form: 
\[
-\hspace{0.02cm}2\hspace{0.025cm}i\hspace{0.025cm}  [\,t^{a_{1}},t^{a_{2}}]^{i\hspace{0.035cm} i_{1}}
\biggl(\frac{1}{\omega^{l}_{{\bf q}_{1} - {\bf q}}\! - \omega^{-}_{{\bf q}_{1}} + \omega^{-}_{{\bf q}}}
\,+\,
\frac{1}{\omega^{l}_{{\bf q} - {\bf q}_{1}}\! - \omega^{-}_{{\bf q}} + \omega^{-}_{{\bf q}_{1}}}\biggr)
{\mathcal V}_{\,{\bf k}_{1}, {\bf k}_{2},\, {\bf k}_{1} - {{\bf k}_{2}}}
\hspace{0.025cm} {\mathcal P}_{{\bf q} - {\bf q}_{1},\, {\bf q},\,  {\bf q}_{1}}.
\]
This expression should be compared to
\[
[\,t^{a_{1}},t^{a_{2}}]^{i\hspace{0.035cm} i_{1}}
\biggl(\,\frac{1}{\omega^{\hspace{0.02cm}l}_{{\bf q}_{1} - {\bf q}}\! - \omega^{-}_{{\bf q}_{1}} + \omega^{-}_{{\bf q}}}
\,+\,
\frac{1}{\omega^{\hspace{0.02cm}l}_{{\bf q} - {\bf q}_{1}}\! - \omega^{-}_{{\bf q}} + \omega^{-}_{{\bf q}_{1}}}\,\biggr)
\hspace{0.8cm}
\]
\vspace{0.1cm}
\[
\times\hspace{0.04cm}
\Biggl\{g\hspace{0.02cm}
\Biggl(\frac{ \epsilon^{\hspace{0.02cm} l}_{\nu^{\prime}}(-{\bf k}_{1} + {\bf k}_{2})}{\sqrt{2\hspace{0.03cm}\omega^{\hspace{0.02cm}l}_{-{\bf k}_{1}\hspace{0.02cm}+\hspace{0.02cm} {\bf k}_{2}}}}\Biggr)
\Biggl(
\frac{\epsilon^{\hspace{0.02cm} l}_{{\mu}_{1}}({\bf k}_{1})}{\sqrt{2\hspace{0.03cm}\omega^{\hspace{0.02cm}l}_{{\bf k}_{1}}}}\Biggr)
\Biggl(\frac{\epsilon^{\hspace{0.02cm}l}_{\mu_{2}}({\bf k}_{2})}{\sqrt{2\hspace{0.03cm}\omega^{\hspace{0.02cm} l}_{{\bf k}_{2}}}}\Biggr)
\,^\ast\Gamma^{\hspace{0.03cm}\nu^{\prime}\mu_1\mu_2}(-k_{1} + k_{2}, k_{1}, -k_{2})
\Biggr\} 
\]
\vspace{0.2cm}
\[
\times\hspace{0.03cm}
\Biggl\{\! g\! \left(\frac{Z_{-}({\bf q})}{2}\right)^{\!1/2}\!\! \left(\frac{Z_{-}({\bf q}_{1})}{2}\right)^{\!1/2}
\!\!\!\bar{v}^{(-)}_{\beta}(\hat{\bf q},\lambda)\hspace{0.02cm} v^{(-)}_{\alpha}(\hat{\bf q}_{1},\lambda_{1})
\Biggl(\frac{ \epsilon^{\hspace{0.02cm}l}_{\mu}({\bf q} - {\bf q}_{1})}{\sqrt{2\hspace{0.03cm}
\omega^{\hspace{0.02cm}l}_{{\bf q} - {\bf q}_{1}}}}\Biggr)
\!\,^{\ast}\Gamma^{(Q)\mu}_{\beta\hspace{0.03cm}\alpha}(-q\hspace{0.02cm} +\hspace{0.02cm} q_{1}; q, -q_{1}) \Biggr\}\hspace{0.01cm}
\Biggr|_{\rm \,on-shell}.
\]
We see that the vertex functions ${\mathcal V}_{\, -{\bf k}_{1} + {\bf k}_{2},\, - {\bf k}_{1},\, {\bf k}_{2}}$ and 
${\mathcal P}_{{\bf q} - {\bf q}_{1},\, {\bf q},\, {\bf q}_{1}}(\lambda, \lambda_{1})$ given by formulas (\ref{eq:2z}) and (\ref{eq:9g}), correspondingly, are exactly reproduced.\\
\indent Let us consider the remaining contributions in (\ref{eq:12y}), which contain the effective (retarded) quark propagator $\!\,^{\ast}\!\hspace{0.03cm}S_{\gamma\gamma^{\prime}}$. In the hard thermal loop approximation the structure of this propagator is defined by Eqs.\,(\ref{ap:B11})\,--\,(\ref{ap:B13}). However, it is more convenient for us to use a somewhat different representation of the quark propagator suggested by Weldon \cite{weldon_2000} for the case of chirally invariant phase of QCD, namely,
\begin{equation}
\,^{\ast}\!\hspace{0.03cm}S(q_{0},{\bf q}) =
\label{eq:12o}
\end{equation}
\[
h_{+}(\hat{\mathbf q}) \biggl(\frac{Z_{+}({\bf q})}{q_{0} - {\mathcal E}_{+}({\bf q})} + 
\frac{Z^{\ast}_{-}({\bf q})}{q_{0} + {\mathcal E}^{\ast}_{-}({\bf q})} - f(q_{0},|{\bf q}|)\!\biggr)
+\hspace{0.03cm}
h_{-}(\hat{\mathbf q}) \biggl(\frac{Z_{-}({\bf q})}{q_{0} - {\mathcal E}_{-}({\bf q})} + 
\frac{Z^{\ast}_{+}({\bf q})}{q_{0} + {\mathcal E}^{\ast}_{+}({\bf q})} + f^{\ast}(-q^{\ast}_{0},|{\bf q}|)\!\biggr).
\]
The $q_{0}$ poles at ${\mathcal E}_{+}$ and at ${\mathcal E}_{-}$ are due to the particle excitation and hole excitation and the $q_{0}$ poles at $-\hspace{0.02cm}{\mathcal E}^{\ast}_{-}$ and at $-\hspace{0.02cm}{\mathcal E}^{\ast}_{+}$ are due to the antihole excitation and antiparticle excitation, respectively. The propagator in Eq.\,(\ref{eq:12o}) is invariant under chirality as well as under parity, charge conjugation, and time reversal. The function $f(q_{0},|{\bf q}|)$ is an unknown function except for the requirement that it has no singularities in the upper half of the complex $q_{0}$ plane and must satisfy the reflection property
\[
f(q_{0},|{\bf q}|) = -f^{\ast}(-q^{\ast}_{0},-|{\bf q}|).
\]
At the plasmino excitation pole when $q_{0}\rightarrow {\mathcal E}_{-}({\bf q})\simeq \omega_{\bf q}^{-}$ the retarded propagator (\ref{eq:12o}) behaves as
\begin{equation}
\,^{\ast}\!S_{\alpha\alpha^{\prime}}(q_{0},{\bf q}) \sim 
\sum\limits_{\lambda\hspace{0.02cm} = \hspace{0.02cm}\pm\hspace{0.01cm} 1}\!v^{(-)}_{\alpha}(\hat{\bf q},\lambda)
\bar{v}^{(-)}_{\alpha^{\prime}}(\hat{\bf q},\lambda)\, \frac{Z_{-}({\bf q})}{q_{0} - \omega_{\bf q}^{-}}.
\label{eq:12_1o}
\end{equation}  
Here, we have taken into account that by virtue of the second relation in (\ref{ap:B3}), the representation $h_{-}(\hat{\bf q}) =
\sum_{\lambda}\! v^{(-)}(\hat{\bf q},\lambda) \hspace{0.02cm}\bar{v}^{(-)}(\hat{\bf q},\lambda)$ is true. Using this approximation, we obtain the following expres\-sions for the first effective quark propagator on the right-hand side of Eq.\,(\ref{eq:12y}):
\begin{equation}
\hspace{0.8cm}\,^{\ast}\!S_{\gamma\gamma^{\prime}}(q_{1} + k_{2})
\,\simeq\,
\sum\limits_{\lambda^{\prime}\hspace{0.02cm} = \hspace{0.02cm}\pm\hspace{0.01cm} 1}\!
v^{(-)}_{\gamma}(\widehat{{\bf q} + {\bf k}_{1}},\lambda^{\prime}\hspace{0.03cm})
\hspace{0.03cm}\bar{v}^{(-)}_{\gamma^{\prime}}(\widehat{{\bf q}_{1} + {\bf k}_{2}},\lambda^{\prime}\hspace{0.03cm})\,
\frac{Z^{1/2}_{-}({\bf q} + {\bf k}_{1})\hspace{0.02cm}Z^{1/2}_{-}({\bf q}_{1} + {\bf k}_{2})}
{\omega^{-}_{{\bf q}_{1}} + \omega^{l}_{{\bf k}_{2}}\! - \omega^{-}_{{\bf q}_{1} + {\bf k}_{2}}}
\label{eq:12p}
\end{equation}
and for the second propagator
\begin{equation}
\hspace{0.9cm}
\,^{\ast}\!S_{\gamma\gamma^{\prime}}(q_{1} - k_{1})
\,\simeq\,
\sum\limits_{\lambda^{\prime}\hspace{0.02cm} = \hspace{0.02cm}\pm\hspace{0.01cm} 1}\!
v^{(-)}_{\gamma}(\widehat{{\bf q} - {\bf k}_{2}},\lambda^{\prime}\hspace{0.03cm})
\hspace{0.03cm}\bar{v}^{(-)}_{\gamma^{\prime}}(\widehat{{\bf q}_{1} - {\bf k}_{1}},\lambda^{\prime}\hspace{0.03cm})\,
\frac{Z^{1/2}_{-}({\bf q} - {\bf k}_{2})\hspace{0.02cm}Z^{1/2}_{-}({\bf q}_{1} - {\bf k}_{1})}
{\omega^{-}_{{\bf q}_{1}} - \omega^{l}_{{\bf k}_{1}}\! - \omega^{-}_{{\bf q}_{1} - {\bf k}_{1}}}\hspace{0.02cm}.
\label{eq:12a}
\end{equation}
In deriving (\ref{eq:12p}) and (\ref{eq:12a}) we have considered the conservation laws of momentum and energy for elastic scattering plasmino off plasmon, Eq.\,(\ref{eq:12i}).\\ 
\indent The substitution of the approximation (\ref{eq:12p}) into the second term in braces on the right-hand side of (\ref{eq:12y}) gives us:
\[
-\hspace{0.02cm}(t^{a_{1}}t^{a_{2}})^{i\hspace{0.03cm} i_{1}}\hspace{0.02cm}\times
\]
\[
\sum\limits_{\lambda^{\prime}\hspace{0.02cm} =\hspace{0.02cm}\pm}
\Biggl[\!\hspace{0.02cm}\sqrt{2}\hspace{0.02cm}g\!\left(\frac{Z_{-}({\bf q})}{2}\right)^{\!\!1/2}\!\! \left(\frac{Z_{-}({\bf q}\! +\! {\bf k}_{1})}{2}\right)^{\!\!1/2}\!
\!\!\!
\bar{v}^{(-)}_{\gamma}(\widehat{{\bf q}\!\hspace{0.02cm} +\!\hspace{0.02cm} {\bf k}_{1}},\lambda^{\prime})\hspace{0.03cm} 
v^{(-)}_{\beta}(\hat{\bf q},\lambda)
\hspace{0.02cm} 
\Biggl(\frac{ \epsilon^{\hspace{0.02cm}l}_{\mu_{1}}({\bf k}_{1})}{\sqrt{2\hspace{0.02cm}\omega^{l}_{{\bf k}_{1}}}}\Biggr)
\!\,^{\ast}\Gamma^{(Q)\hspace{0.02cm} \mu_{1}}_{\gamma\beta}(k_{1}; -q - k_{1}, q)\Biggr]^{\!\ast}\!\!\!\hspace{0.03cm}\times
\]
\[
\Biggl[\!\hspace{0.02cm}\sqrt{2}\hspace{0.02cm}g\!\left(\frac{Z_{-}({\bf q}_{1})}{2}\right)^{\!\!\!1/2}\!\!\! 
\left(\frac{Z_{-}({\bf q}_{1}\! +\! {\bf k}_{2})}{2}\right)^{\!\!\!1/2}
\!\!\!\bar{v}^{(-)}_{\gamma^{\prime}}(\widehat{{\bf q}_{1}\! +\!\hspace{0.02cm} {\bf k}_{2}},\lambda^{\prime}\hspace{0.03cm}) v^{(-)}_{\alpha}(\hat{\bf q}_{1},\lambda_{1})\hspace{0.03cm} 
\Biggl(\!\frac{ \epsilon^{\hspace{0.02cm}l}_{\mu_{2}}({\bf k}_{2})}{\sqrt{2\hspace{0.02cm}\omega^{l}_{{\bf k}_{2}}}}\!\hspace{0.03cm} \Biggr)
\!\,^{\ast}\Gamma^{(Q)\hspace{0.02cm} \mu_{2}}_{\gamma^{\prime}\alpha}(-k_{2}; q_{1} + k_{2}, -q_{1})\!\Biggr]
\]
\[
\times\,
\frac{1}{\omega^{l}_{{\bf k}_{2}} -\, \omega^{-}_{{\bf k}_{2} + {\bf q}_{1}} +\, \omega^{-}_{{\bf q}_{1}}}\; 
\Biggr|_{\rm \,on-shell}.
\]
From the structure of the last factor here, we see that this contribution should be compared with the second term on the left-hand side of (\ref{eq:12y}). Remembering our agreement on summation over helicity, Eq.(\ref{eq:5y}), and taking into account color structure of the three-point amplitude ${\mathcal P}^{\; a\, i\,  i_{1}}_{{\bf k},\, {\bf q},\, {\bf q}_{1}}$, Eq.\,(\ref{eq:9s}),
we can write this term in the following form: 
\[
(t^{a_{1}}t^{a_{2}})^{ii_{1}}\,
\frac{\sum\limits_{\lambda^{\prime}\hspace{0.02cm} = \hspace{0.02cm}\pm\hspace{0.01cm} 1}
{\mathcal P}_{{\bf k}_{2},\, {\bf q}_{1} + {\bf k}_{2},\, {\bf q}_{1}}(\lambda^{\prime},\lambda_{1})\, 
{\mathcal P}^{\hspace{0.03cm}\ast}_{{\bf k}_{1},\, {\bf q} + {\bf k}_{1},\,  {\bf q}}(\lambda^{\prime},\lambda)}
{\omega^{l}_{{\bf k}_{2}} -\, \omega^{-}_{{\bf k}_{2} + {\bf q}_{1}} +\, \omega^{-}_{{\bf q}_{1}}}.
\]
Comparing this expression with the previous one we see that the vertex function ${\mathcal P}_{{\bf q},\, {\bf q}_{1},\, {\bf q}_{2}}(\lambda,\lambda_{1})$, given by the formula (\ref{eq:9g}), is again correctly reproduced.  
We result in a similar conclusion by substituting the second approximated propagator (\ref{eq:12a}) into the third term on the right-hand side (\ref{eq:12y}) and by comparing the obtained expression with
\[
- \,(t^{a_{2}}t^{a_{1}})^{ii_{1}}\,
\frac{\sum\limits_{\lambda^{\prime}\hspace{0.02cm} =\hspace{0.02cm}\pm\hspace{0.01cm} 1}\,
{\mathcal P}_{{\bf k}_{2},\, {\bf q},\, {\bf q} - {\bf k}_{2}}(\lambda, \lambda^{\prime}\hspace{0.03cm})\,
{\mathcal P}^{\hspace{0.03cm}\ast}_{{\bf k}_{1},\, {\bf q}_{1},\,  {\bf q}_{1} - {\bf k}_{1}}(\lambda_{1}, \lambda^{\prime}\hspace{0.03cm})}
{\omega^{l}_{{\bf k}_{2}} - \omega^{-}_{{\bf q}} + \omega^{-}_{{\bf q} - {\bf k}_{2}}}.
\]
Finally, the coefficient functions ${\mathcal G}^{\; a\, i\,  i_{1}}_{\,{\bf k},\, {\bf q},\, {\bf q}_{1}}$ and ${\mathcal K}^{\; a\, i\,  i_{1}}_{\,{\bf k},\, {\bf q},\, {\bf q}_{1}}$ in the first and in the fourth terms on the left-hand side of (\ref{eq:12y}) should be set equal zero, in according with (\ref{eq:9a}).


\section{Higher coefficient functions $S^{\,(n)\, i\; i_{1}\, i_{2}\, i_{3}}_{\, {\bf q},\, {\bf q}_{1},\, {\bf q}_{2},\, 
{\bf q}_{3}}$ for the canonical transformation (\ref{eq:3y})}
\label{section_13}
\setcounter{equation}{0}

In this section we consider the problem of defining an explicit form of the third-order coefficient functions 
$S^{(n)\, i\, i_{1}\, i_{2}\, i_{3}}_{\ {\bf q},\, {\bf q}_{1},\, {\bf q}_{2},\, {\bf q}_{3}},\, n = 1,\ldots,4$, entering into the canonical transformation of the fermionic variable $b^{\; i}_{{\bf q}}$, Eq.\,(\ref{eq:3y}). Here, we follow the approach  proposed in the paper by Krasitskii \cite{krasitskii_1990}. Let us return once again to the fourth-order interaction Hamiltonian (\ref{eq:2g}), more exactly to its second term. Generally speaking, this term is part of a more general expression:  
\begin{equation}
\begin{split}
&\int\frac{d{\bf q}\hspace{0.03cm} d{\bf q}_{1}\hspace{0.02cm} d{\bf q}_{2}\hspace{0.03cm} d{\bf q}_{3}}{(2\pi)^{12}}\, 
\Bigl(T^{\,(1)\, i\, i_{1}\, i_{2}\, i_{3}}_{{\bf q},\, {\bf q}_{1},\, {\bf q}_{2},\, {\bf q}_{3}}\; 
b^{\,\ast\, i}_{{\bf q}}\, b^{\; i_{1}}_{{\bf q}_{1}}\, b^{\;i_{2}}_{{\bf q}_{2}}\, b^{\;i_{3}}_{{\bf q}_{3}}\,
\,+\,
T^{\hspace{0.03cm}\ast\hspace{0.03cm}(1)\hspace{0.03cm} i\, i_{1}\, i_{2}\, i_{3}}_{{\bf q},\, {\bf q}_{1},\, {\bf q}_{2},\, {\bf q}_{3}}\; 
b^{\; i}_{{\bf q}}\ b^{\,\ast\, i_{1}}_{{\bf q}_{1}}\hspace{0.03cm} b^{\,\ast\;i_{2}}_{{\bf q}_{2}}\hspace{0.03cm} b^{\,\ast\;i_{3}}_{{\bf q}_{3}}
\Bigr)\\[1ex]
&\hspace{5cm}\times(2\pi)^{3}\hspace{0.03cm}\delta({\bf q} - {\bf q}_{1} - {\bf q}_{2} - {\bf q}_{3})\\[1ex]
+\;
\frac{1}{2}&\int\frac{d{\bf q}\hspace{0.03cm} d{\bf q}_{1}\hspace{0.02cm} d{\bf q}_{2}\hspace{0.03cm} d{\bf q}_{3}}{(2\pi)^{12}}\; 
T^{\,(2)\, i\; i_{1}\, i_{2}\, i_{3}}_{{\bf q},\, {\bf q}_{1},\, {\bf q}_{2},\, {\bf q}_{3}}\; 
b^{\,\ast\, i}_{{\bf q}}\hspace{0.03cm} b^{\,\ast\, i_{1}}_{{\bf q}_{1}}\hspace{0.03cm} b^{\;i_{2}}_{{\bf q}_{2}}\, b^{\;i_{3}}_{{\bf q}_{3}}\,
(2\pi)^{3}\hspace{0.03cm}\delta({\bf q} + {\bf q}_{1} - {\bf q}_{2} - {\bf q}_{3})\\[1ex]
+\;
\frac{1}{4}&\int\frac{d{\bf q}\hspace{0.03cm} d{\bf q}_{1}\hspace{0.02cm} d{\bf q}_{2}\hspace{0.03cm} d{\bf q}_{3}}{(2\pi)^{12}}\, 
\Bigl(T^{\,(4)\, i\, i_{1}\, i_{2}\, i_{3}}_{{\bf q},\, {\bf q}_{1},\, {\bf q}_{2},\, {\bf q}_{3}}\; 
b^{\; i}_{{\bf q}}\ b^{\; i_{1}}_{{\bf q}_{1}}\, b^{\;i_{2}}_{{\bf q}_{2}}\, b^{\;i_{3}}_{{\bf q}_{3}}\,
\,+\,
T^{\hspace{0.03cm}\ast\hspace{0.03cm}(4)\hspace{0.03cm} i\; i_{1}\, i_{2}\, i_{3}}_{{\bf q},\, {\bf q}_{1},\, {\bf q}_{2},\, {\bf q}_{3}}\; 
b^{\,\ast\, i}_{{\bf q}}\ b^{\,\ast\, i_{1}}_{{\bf q}_{1}}\, b^{\,\ast\;i_{2}}_{{\bf q}_{2}}\, b^{\,\ast\;i_{3}}_{{\bf q}_{3}}
\Bigr)\\[1ex]
&\hspace{5cm}\times(2\pi)^{3}\hspace{0.03cm}\delta({\bf q} + {\bf q}_{1} + {\bf q}_{2} + {\bf q}_{3}).
\end{split}
\label{eq:13q}
\end{equation}
It is obvious that even if the nonresonant fourth-order contributions with the vertex functions $T^{\,(1)\, i\, i_{1}\, i_{2}\, i_{3}}_{{\bf q},\, {\bf q}_{1},\, {\bf q}_{2},\, {\bf q}_{3}}$ and $T^{\,(4)\, i\, i_{1}\, i_{2}\, i_{3}}_{{\bf q},\, {\bf q}_{1},\, {\bf q}_{2},\, {\bf q}_{3}}$ by virtue of the properties of the system under consideration vanish, still they will inevitably be generated by the canonical transformations (\ref{eq:3t}), (\ref{eq:3y}) from the free-field and third-order Hamiltonians, Eqs.\,(\ref{eq:2d}) and (\ref{eq:2f}), correspondingly. We can determine the higher coefficient functions $S^{\,(n)\, i\, i_{1}\, i_{2}\, i_{3}}_{\, {\bf q},\, {\bf q}_{1},\, {\bf q}_{2},\, {\bf q}_{3}}$ in (\ref{eq:3y}) from the requirement of making these ``induced'' contributions vanish. At least it can be unambiguously done for the functions with $n = 1,\,3$ and $4$.\\
\indent The first step is to find all the contributions proportional to the products $f^{\,\ast\, i}_{{\bf q}} f^{\, i_{1}}_{{\bf q}_{1}} f^{\;i_{2}}_{{\bf q}_{2}} f^{\;i_{3}}_{{\bf q}_{3}}$ and $f^{\, i}_{{\bf q}} f^{\, i_{1}}_{{\bf q}_{1}} f^{\;i_{2}}_{{\bf q}_{2}} f^{\;i_{3}}_{{\bf q}_{3}}$  from the free-field Hamiltonian $H^{(0)}$ given by Eq.\,(\ref{eq:2d}) under the canonical transformation (\ref{eq:3y}). With allowance made for the relations (\ref{eq:3p}) in the case of the product  $f^{\,\ast\, i}_{{\bf q}} f^{\, i_{1}}_{{\bf q}_{1}} f^{\;i_{2}}_{{\bf q}_{2}} f^{\;i_{3}}_{{\bf q}_{3}}$ we obtain the following contributions from $H^{(0)}$
\[
\int\!\frac{d{\bf q}\hspace{0.03cm} d{\bf q}_{1}\hspace{0.02cm} d{\bf q}_{2}\hspace{0.03cm} d{\bf q}_{3}}{(2\pi)^{12}}\, 
\biggl\{\omega^{-}_{{\bf q}}S^{\,(1)\, i\; i_{1}\, i_{2}\, i_{3}}_{\ {\bf q},\, {\bf q}_{1},\, {\bf q}_{2},\, {\bf q}_{3}}
-
\frac{1}{3}\,\Bigl(\omega^{-}_{{\bf q}_{1}}S^{\,\ast\hspace{0.03cm}(3)\, i_{1}\, i_{3}\, i_{2}\, i}_{\ {\bf q}_{1},\, {\bf q}_{3},\, {\bf q}_{2},\, {\bf q}}
-
\omega^{-}_{{\bf q}_{2}}S^{\,\ast\hspace{0.03cm}(3)\, i_{2}\, i_{3}\, i_{1}\, i}_{\ {\bf q}_{2},\, {\bf q}_{3},\, {\bf q}_{1},\, {\bf q}}
-
\omega^{-}_{{\bf q}_{3}}S^{\,\ast\hspace{0.03cm}(3)\, i_{3}\, i_{1}\, i_{2}\, i}_{\ {\bf q}_{3},\, {\bf q}_{1},\, {\bf q}_{2},\, {\bf q}}
\Bigr)
\]
\vspace{-0.3cm}
\begin{align}
+\,
\frac{1}{3}\int\!\frac{d\hspace{0.02cm}{\bf k}^{\prime}}{(2\pi)^{3}}\;\omega^{l}_{{\bf k}^{\prime}}
\Bigl[\hspace{0.03cm}
&\Bigl(F^{(1)\, a^{\prime}\, i_{2}\, i_{3}}_{\ {\bf k}^{\prime},\, {\bf q}_{2},\, {\bf q}_{3}}\, 
F^{\hspace{0.03cm}\ast\hspace{0.03cm}(2)\, a^{\prime}\, i_{1}\, i}_{\ {\bf k}^{\prime},\,{\bf q}_{1},\, {\bf q}}
\,-\,
F^{(2)\, a^{\prime}\, i\, i_{1}}_{\ {\bf k}^{\prime},\, {\bf q},\, {\bf q}_{1}}\, 
F^{\hspace{0.03cm}\ast\hspace{0.03cm}(3)\, a^{\prime}\, i_{2}\, i_{3}}_{\ {\bf k}^{\prime},\, {\bf q}_{2},\, {\bf q}_{3}}\Bigr)
\label{eq:13w}\\[1ex]
-\,
&\Bigl(F^{(1)\, a^{\prime}\, i_{1}\, i_{3}}_{\ {\bf k}^{\prime},\, {\bf q}_{1},\, {\bf q}_{3}}\, 
F^{\hspace{0.03cm}\ast\hspace{0.03cm}(2)\, a^{\prime}\, i_{2}\, i}_{\ {\bf k}^{\prime},\,{\bf q}_{2},\, {\bf q}}
\,-\,
F^{(2)\, a^{\prime}\, i\, i_{2}}_{\ {\bf k}^{\prime},\, {\bf q},\, {\bf q}_{2}}\, 
F^{\hspace{0.03cm}\ast\hspace{0.03cm}(3)\, a^{\prime}\, i_{1}\, i_{3}}_{\ {\bf k}^{\prime},\, {\bf q}_{1},\, {\bf q}_{3}}\Bigr)
\notag\\[1ex]
+\,
&\Bigl(F^{(1)\, a^{\prime}\, i_{1}\, i_{2}}_{\ {\bf k}^{\prime},\, {\bf q}_{1},\, {\bf q}_{2}}\, 
F^{\hspace{0.03cm}\ast\hspace{0.03cm}(2)\, a^{\prime}\, i_{3}\, i}_{\ {\bf k}^{\prime},\,{\bf q}_{3},\, {\bf q}}
\,-\,
F^{(2)\, a^{\prime}\, i\, i_{3}}_{\ {\bf k}^{\prime},\, {\bf q},\, {\bf q}_{3}}\, 
F^{\hspace{0.03cm}\ast\hspace{0.03cm}(3)\, a^{\prime}\, i_{1}\, i_{2}}_{\ {\bf k}^{\prime},\, {\bf q}_{1},\, {\bf q}_{2}}\Bigr)
\hspace{0.03cm}\Bigr]
\biggr\}\hspace{0.03cm}
f^{\,\ast\, i}_{{\bf q}} f^{\; i_{1}}_{{\bf q}_{1}} f^{\;i_{2}}_{{\bf q}_{2}} f^{\;i_{3}}_{{\bf q}_{3}}
\notag
\hspace{3cm}
\end{align}
and in the case of the product $f^{\, i}_{{\bf q}} f^{\, i_{1}}_{{\bf q}_{1}} f^{\;i_{2}}_{{\bf q}_{2}} f^{\;i_{3}}_{{\bf q}_{3}}$ we have
\[
\int\!\frac{d{\bf q}\hspace{0.03cm} d{\bf q}_{1}\hspace{0.02cm} d{\bf q}_{2}\hspace{0.03cm} d{\bf q}_{3}}{(2\pi)^{12}}\, 
\biggl\{\frac{1}{4}\,\Bigl(
\omega^{-}_{{\bf q}}S^{\,\ast\hspace{0.03cm}(4)\, i\, i_{1}\, i_{2}\, i_{3}}_{\ {\bf q},\, {\bf q}_{1},\, {\bf q}_{2},\, {\bf q}_{3}}
-
\omega^{-}_{{\bf q}_{1}}S^{\,\ast\hspace{0.03cm}(4)\, i_{1}\, i\, i_{2}\, i_{3}}_{\ {\bf q}_{1},\, {\bf q},\, {\bf q}_{2},\, {\bf q}_{3}}
+
\omega^{-}_{{\bf q}_{2}}S^{\,\ast\hspace{0.03cm}(4)\, i_{2}\, i\, i_{1}\, i_{3}}_{\ {\bf q}_{2},\, {\bf q},\, {\bf q}_{1},\, {\bf q}_{3}}
-
\omega^{-}_{{\bf q}_{3}}S^{\,\ast\hspace{0.03cm}(4)\, i_{3}\, i\, i_{1}\, i_{2}}_{\ {\bf q}_{3},\, {\bf q},\, {\bf q}_{1},\, {\bf q}_{2}}
\Bigr)
\]
\begin{align}
-\,
\frac{1}{6}\int\!\frac{d\hspace{0.02cm}{\bf k}^{\prime}}{(2\pi)^{3}}\;\omega^{l}_{{\bf k}^{\prime}}\Bigl[
&F^{(1)\, a^{\prime}\, i\, i_{1}}_{\ {\bf k}^{\prime},\, {\bf q},\, {\bf q}_{1}}\, 
F^{\hspace{0.02cm}\ast\hspace{0.03cm}(3)\, a^{\prime}\, i_{2}\, i_{3}}_{\ {\bf k}^{\prime},\,{\bf q}_{2},\, {\bf q}_{3}}
\!-
F^{(1)\, a^{\prime}\, i\, i_{2}}_{\ {\bf k}^{\prime},\, {\bf q},\, {\bf q}_{2}}\, 
F^{\hspace{0.02cm}\ast\hspace{0.03cm}(3)\, a^{\prime}\, i_{1}\, i_{3}}_{\ {\bf k}^{\prime},\,{\bf q}_{1},\, {\bf q}_{3}}
\!+
F^{(1)\, a^{\prime}\, i\, i_{3}}_{\ {\bf k}^{\prime},\, {\bf q},\, {\bf q}_{3}}\, 
F^{\hspace{0.02cm}\ast\hspace{0.03cm}(3)\, a^{\prime}\, i_{1}\, i_{2}}_{\ {\bf k}^{\prime},\,{\bf q}_{1},\, {\bf q}_{2}}
\label{eq:13e}\\[1ex]
+\,
&F^{(1)\, a^{\prime}\, i_{1}\, i_{2}}_{\ {\bf k}^{\prime},\, {\bf q}_{1},\, {\bf q}_{2}}\, 
F^{\hspace{0.02cm}\ast\hspace{0.03cm}(3)\, a^{\prime}\, i\, i_{3}}_{\ {\bf k}^{\prime},\,{\bf q},\, {\bf q}_{3}}
\!-
F^{(1)\, a^{\prime}\, i_{1}\, i_{3}}_{\ {\bf k}^{\prime},\, {\bf q}_{1},\, {\bf q}_{3}}\, 
F^{\hspace{0.02cm}\ast\hspace{0.03cm}(3)\, a^{\prime}\, i\, i_{2}}_{\ {\bf k}^{\prime},\,{\bf q},\, {\bf q}_{2}}
\!+
F^{(1)\, a^{\prime}\, i_{2}\, i_{3}}_{\ {\bf k}^{\prime},\, {\bf q}_{2},\, {\bf q}_{3}}\, 
F^{\hspace{0.02cm}\ast\hspace{0.03cm}(3)\, a^{\prime}\, i\, i_{1}}_{\ {\bf k}^{\prime},\,{\bf q},\, {\bf q}_{1}}\Bigr]\!
\biggr\}
f^{\; i}_{{\bf q}} f^{\; i_{1}}_{{\bf q}_{1}} f^{\;i_{2}}_{{\bf q}_{2}} f^{\;i_{3}}_{{\bf q}_{3}}.
\notag
\end{align}
Here, the integrand in (\ref{eq:13w}) has been antisymmetrized under the interchange of color indices and momentum arguments with the index 1 and  the color indices, and momentum arguments with the indices 2 and 3, whereas the integrand in (\ref{eq:13e}) has been antisymmetrized under the interchange of color indices and momentum arguments with pairs of indices (1,2) and (3,4).\\
\indent Further in the obtained expression (\ref{eq:13w}) we eliminate three coefficient functions $S^{\,\ast\hspace{0.03cm}(1,3,4)}$. For this purpose, we make use of the integral relation (\ref{ap:D2a}) connecting the functions $S^{\hspace{0.03cm}\ast\hspace{0.03cm} (3)\hspace{0.03cm} i_{3}\; i_{2}\; i_{1}\, i}_{\ {\bf q}_{3},\, {\bf q}_{2},\, {\bf q}_{1},\, {\bf q}}$ and $S^{\,(1)\; i\; i_{1}\; i_{2}\; i_{3}}_{\ {\bf q},\, {\bf q}_{1},\, {\bf q}_{2},\, {\bf q}_{3}}$ among themselves and also the symmetry conditions for the function 
$S^{\,(1)\; i\; i_{1}\; i_{2}\; i_{3}}_{\ {\bf q},\, {\bf q}_{1},\, {\bf q}_{2},\, {\bf q}_{3}}$, Eq.\,(\ref{eq:3i}). As a result, instead of the expression in the first line of (\ref{eq:13w}), we get
\[
\int\!\frac{d{\bf q}\hspace{0.03cm} d{\bf q}_{1}\hspace{0.02cm} d{\bf q}_{2}\hspace{0.03cm} d{\bf q}_{3}}{(2\pi)^{12}}\, 
\biggl\{\omega^{-}_{{\bf q}}S^{\,(1)\, i\; i_{1}\, i_{2}\, i_{3}}_{\ {\bf q},\, {\bf q}_{1},\, {\bf q}_{2},\, {\bf q}_{3}}
\!-
\frac{1}{3}\,\Bigl(\omega^{-}_{{\bf q}_{1}}S^{\,\ast\hspace{0.03cm}(3)\, i_{1}\, i_{3}\, i_{2}\, i}_{\ {\bf q}_{1},\, {\bf q}_{3},\, {\bf q}_{2},\, {\bf q}}
-
\omega^{-}_{{\bf q}_{2}}S^{\,\ast\hspace{0.03cm}(3)\, i_{2}\, i_{3}\, i_{1}\, i}_{\ {\bf q}_{2},\, {\bf q}_{3},\, {\bf q}_{1},\, {\bf q}}
-
\omega^{-}_{{\bf q}_{3}}S^{\,\ast\hspace{0.03cm}(3)\, i_{3}\, i_{1}\, i_{2}\, i}_{\ {\bf q}_{3},\, {\bf q}_{1},\, {\bf q}_{2},\, {\bf q}}
\Bigr)\!\biggr\}
\]
\begin{equation}
=\!
\int\!\frac{d{\bf q}\hspace{0.03cm} d{\bf q}_{1}\hspace{0.02cm} d{\bf q}_{2}\hspace{0.03cm} d{\bf q}_{3}}{(2\pi)^{12}}\, 
f^{\,\ast\, i}_{{\bf q}} f^{\; i_{1}}_{{\bf q}_{1}} f^{\;i_{2}}_{{\bf q}_{2}} f^{\;i_{3}}_{{\bf q}_{3}}\,
\biggl\{\bigl(\omega^{-}_{{\bf q}} - \omega^{-}_{{\bf q}_{1}} - \omega^{-}_{{\bf q}_{2}} - \omega^{-}_{{\bf q}_{3}}\bigr)
S^{\,(1)\, i\; i_{1}\, i_{2}\, i_{3}}_{\ {\bf q},\, {\bf q}_{1},\, {\bf q}_{2},\, {\bf q}_{3}}
+
\frac{1}{3}\int\!\frac{d\hspace{0.02cm}{\bf k}^{\prime}}{(2\pi)^{3}}\,\times
\label{eq:13r}
\end{equation}
\vspace{-0.3cm}
\begin{align}
\biggl[\hspace{0.03cm}\omega^{-}_{{\bf q}_{1}}\Bigl(\hspace{0.03cm}
&F^{(1)\, a^{\prime}\, i_{1}\, i_{3}}_{\ {\bf k}^{\prime},\, {\bf q}_{1},\, {\bf q}_{3}}\, 
F^{\hspace{0.03cm}\ast\hspace{0.03cm}(2)\, a^{\prime}\, i_{2}\, i}_{\ {\bf k}^{\prime},\,{\bf q}_{2},\, {\bf q}}
+
F^{(2)\hspace{0.03cm} a^{\prime}\, i\; i_{2}}_{\ {\bf k}^{\prime},\, {\bf q},\, {\bf q}_{2}}\, 
F^{\hspace{0.03cm}\ast\hspace{0.03cm}(3)\, a^{\prime}\, i_{1}\, i_{3}}_{\ {\bf k}^{\prime},\, {\bf q}_{1},\, {\bf q}_{3}}
-
F^{(1)\, a^{\prime}\, i_{1}\, i_{2}}_{\ {\bf k}^{\prime},\, {\bf q}_{1},\, {\bf q}_{2}}\, 
F^{\hspace{0.03cm}\ast\hspace{0.03cm}(2)\, a^{\prime}\, i_{3}\, i}_{\ {\bf k}^{\prime},\,{\bf q}_{3},\, {\bf q}}
-
F^{(2)\, a^{\prime}\, i\; i_{3}}_{\ {\bf k}^{\prime},\, {\bf q},\, {\bf q}_{3}}\, 
F^{\hspace{0.03cm}\ast\hspace{0.03cm}(3)\, a^{\prime}\, i_{1}\, i_{2}}_{\ {\bf k}^{\prime},\, {\bf q}_{1},\, {\bf q}_{2}}\Bigr)
\notag\\[1ex]
-\;
\omega^{-}_{{\bf q}_{2}}\Bigl(&F^{(1)\, a^{\prime}\, i_{2}\, i_{3}}_{\ {\bf k}^{\prime},\, {\bf q}_{2},\, {\bf q}_{3}}\, 
F^{\hspace{0.03cm}\ast\hspace{0.03cm}(2)\, a^{\prime}\, i_{1}\, i}_{\ {\bf k}^{\prime},\,{\bf q}_{1},\, {\bf q}}
+
F^{(2)\hspace{0.03cm} a^{\prime}\, i\, i_{1}}_{\ {\bf k}^{\prime},\, {\bf q},\, {\bf q}_{1}}\, 
F^{\hspace{0.03cm}\ast\hspace{0.03cm}(3)\, a^{\prime}\, i_{2}\, i_{3}}_{\ {\bf k}^{\prime},\, {\bf q}_{2},\, {\bf q}_{3}}
+
F^{(1)\, a^{\prime}\, i_{1}\, i_{2}}_{\ {\bf k}^{\prime},\, {\bf q}_{1},\, {\bf q}_{2}}\, 
F^{\hspace{0.03cm}\ast\hspace{0.03cm}(2)\, a^{\prime}\, i_{3}\, i}_{\ {\bf k}^{\prime},\,{\bf q}_{3},\, {\bf q}}
+
F^{(2)\, a^{\prime}\, i\; i_{3}}_{\ {\bf k}^{\prime},\, {\bf q},\, {\bf q}_{3}}\, 
F^{\hspace{0.03cm}\ast\hspace{0.03cm}(3)\, a^{\prime}\, i_{1}\, i_{2}}_{\ {\bf k}^{\prime},\, {\bf q}_{1},\, {\bf q}_{2}}
\Bigr)
\notag\\[1ex]
+\;
\omega^{-}_{{\bf q}_{3}}\Bigl(&F^{(1)\, a^{\prime}\, i_{1}\, i_{3}}_{\ {\bf k}^{\prime},\, {\bf q}_{1},\, {\bf q}_{3}}\, 
F^{\hspace{0.03cm}\ast\hspace{0.03cm}(2)\, a^{\prime}\, i_{2}\, i}_{\ {\bf k}^{\prime},\,{\bf q}_{2},\, {\bf q}}
+
F^{(2)\, a^{\prime}\, i\; i_{2}}_{\ {\bf k}^{\prime},\, {\bf q},\, {\bf q}_{2}}\, 
F^{\hspace{0.03cm}\ast\hspace{0.03cm}(3)\, a^{\prime}\, i_{1}\, i_{3}}_{\ {\bf k}^{\prime},\, {\bf q}_{1},\, {\bf q}_{3}}
-
F^{(1)\, a^{\prime}\, i_{2}\, i_{3}}_{\ {\bf k}^{\prime},\, {\bf q}_{2},\, {\bf q}_{3}}\, 
F^{\hspace{0.03cm}\ast\hspace{0.03cm}(2)\, a^{\prime}\, i_{1}\, i}_{\ {\bf k}^{\prime},\,{\bf q}_{1},\, {\bf q}}
-
F^{(2)\hspace{0.03cm} a^{\prime}\, i\; i_{1}}_{\ {\bf k}^{\prime},\, {\bf q},\, {\bf q}_{1}}\, 
F^{\hspace{0.03cm}\ast\hspace{0.03cm}(3)\, a^{\prime}\, i_{2}\, i_{3}}_{\ {\bf k}^{\prime},\, {\bf q}_{2},\, {\bf q}_{3}}\Bigr)\! \hspace{0.02cm}\biggr]\! \hspace{0.02cm}\biggr\}.
\notag 
\end{align}
\indent Let us consider the expression in parentheses of (\ref{eq:13e}). Here it is necessary to use the canonicity condition (\ref{ap:D4c}) and the symmetry relation for the function $S^{\,(4)\, i\, i_{1}\, i_{2}\, i_{3}}_{\ {\bf q},\, {\bf q}_{1},\, {\bf q}_{2},\, {\bf q}_{3}}$, Eq.\,(\ref{eq:3o}). As a consequence, we obtain 
\[
\int\!\frac{d{\bf q}\hspace{0.03cm} d{\bf q}_{1}\hspace{0.02cm} d{\bf q}_{2}\hspace{0.03cm} d{\bf q}_{3}}{(2\pi)^{12}}\, 
\biggl\{\frac{1}{4}\,\Bigl(
\omega^{-}_{{\bf q}}S^{\,\ast\hspace{0.03cm}(4)\, i\, i_{1}\, i_{2}\, i_{3}}_{\ {\bf q},\, {\bf q}_{1},\, {\bf q}_{2},\, {\bf q}_{3}}
-
\omega^{-}_{{\bf q}_{1}}S^{\,\ast\hspace{0.03cm}(4)\, i_{1}\, i\, i_{2}\, i_{3}}_{\ {\bf q}_{1},\, {\bf q},\, {\bf q}_{2},\, {\bf q}_{3}}
+
\omega^{-}_{{\bf q}_{2}}S^{\,\ast\hspace{0.03cm}(4)\, i_{2}\, i\, i_{1}\, i_{3}}_{\ {\bf q}_{2},\, {\bf q},\, {\bf q}_{1},\, {\bf q}_{3}}
-
\omega^{-}_{{\bf q}_{3}}S^{\,\ast\hspace{0.03cm}(4)\, i_{3}\, i\, i_{1}\, i_{2}}_{\ {\bf q}_{3},\, {\bf q},\, {\bf q}_{1},\, {\bf q}_{2}}
\Bigr)\!\biggr\}
\]
\begin{equation}
=\!	
\int\!\frac{d{\bf q}\hspace{0.03cm} d{\bf q}_{1}\hspace{0.02cm} d{\bf q}_{2}\hspace{0.03cm} d{\bf q}_{3}}{(2\pi)^{12}}\, 
f^{\; i}_{{\bf q}} f^{\; i_{1}}_{{\bf q}_{1}} f^{\;i_{2}}_{{\bf q}_{2}} f^{\;i_{3}}_{{\bf q}_{3}}\,
\biggl\{\frac{1}{4}\,\bigl(\omega^{-}_{{\bf q}} + \omega^{-}_{{\bf q}_{1}} + \omega^{-}_{{\bf q}_{2}} + \omega^{-}_{{\bf q}_{3}}\bigr)
S^{\,\ast\hspace{0.03cm}(4)\, i\, i_{1}\, i_{2}\, i_{3}}_{\ {\bf q},\, {\bf q}_{1},\, {\bf q}_{2},\, {\bf q}_{3}}
+
\frac{1}{6}\int\!\frac{d\hspace{0.02cm}{\bf k}^{\prime}}{(2\pi)^{3}}\,\times
\label{eq:13t}
\end{equation}
\vspace{-0.3cm}
\begin{align}
\biggl[\hspace{0.03cm}\omega^{-}_{{\bf q}_{1}}\Bigl(\hspace{0.03cm}
&F^{(1)\, a^{\prime}\, i\, i_{2}}_{\ {\bf k}^{\prime},\, {\bf q},\, {\bf q}_{2}}\, 
F^{\hspace{0.03cm}\ast\hspace{0.03cm}(3)\, a^{\prime}\, i_{1}\, i_{3}}_{\ {\bf k}^{\prime},\, {\bf q}_{1},\, {\bf q}_{3}}
-
F^{(1)\, a^{\prime}\, i_{1}\, i_{3}}_{\ {\bf k}^{\prime},\, {\bf q}_{1},\, {\bf q}_{3}}\, 
F^{\hspace{0.03cm}\ast\hspace{0.03cm}(3)\, a^{\prime}\, i\; i_{2}}_{\ {\bf k}^{\prime},\,{\bf q},\, {\bf q}_{2}}
-
F^{(1)\, a^{\prime}\, i\; i_{3}}_{\ {\bf k}^{\prime},\, {\bf q},\, {\bf q}_{3}}\, 
F^{\hspace{0.03cm}\ast\hspace{0.03cm}(3)\, a^{\prime}\, i_{1}\, i_{2}}_{\ {\bf k}^{\prime},\, {\bf q}_{1},\, {\bf q}_{2}}
+
F^{(1)\, a^{\prime}\, i_{1}\, i_{2}}_{\ {\bf k}^{\prime},\, {\bf q}_{1},\, {\bf q}_{2}}\, 
F^{\hspace{0.03cm}\ast\hspace{0.03cm}(3)\, a^{\prime}\, i\; i_{3}}_{\ {\bf k}^{\prime},\,{\bf q},\, {\bf q}_{3}}
\Bigr)
\notag\\[1ex]
-\;
\omega^{-}_{{\bf q}_{2}}\Bigl(&F^{(1)\, a^{\prime}\, i\; i_{1}}_{\ {\bf k}^{\prime},\, {\bf q},\, {\bf q}_{1}}\, 
F^{\hspace{0.03cm}\ast\hspace{0.03cm}(3)\, a^{\prime}\, i_{2}\, i_{3}}_{\ {\bf k}^{\prime},\, {\bf q}_{2},\, {\bf q}_{3}}
-
F^{(1)\, a^{\prime}\, i_{2}\, i_{3}}_{\ {\bf k}^{\prime},\, {\bf q}_{2},\, {\bf q}_{3}}\, 
F^{\hspace{0.03cm}\ast\hspace{0.03cm}(3)\, a^{\prime}\, i\, i_{1}}_{\ {\bf k}^{\prime},\,{\bf q},\, {\bf q}_{1}}
+
F^{(1)\hspace{0.03cm} a^{\prime}\, i\; i_{3}}_{\ {\bf k}^{\prime},\, {\bf q},\, {\bf q}_{3}}\, 
F^{\hspace{0.03cm}\ast\hspace{0.03cm}(3)\, a^{\prime}\, i_{1}\, i_{2}}_{\ {\bf k}^{\prime},\, {\bf q}_{1},\, {\bf q}_{2}}
-
F^{(1)\, a^{\prime}\, i_{1}\, i_{2}}_{\ {\bf k}^{\prime},\, {\bf q}_{1},\, {\bf q}_{2}}\, 
F^{\hspace{0.03cm}\ast\hspace{0.03cm}(3)\, a^{\prime}\, i\; i_{3}}_{\ {\bf k}^{\prime},\,{\bf q},\, {\bf q}_{3}}
\Bigr)
\notag\\[1ex]
+\;
\omega^{-}_{{\bf q}_{3}}\Bigl(&F^{(1)\, a^{\prime}\, i\, i_{2}}_{\ {\bf k}^{\prime},\, {\bf q},\, {\bf q}_{2}}\, 
F^{\hspace{0.03cm}\ast\hspace{0.03cm}(3)\, a^{\prime}\, i_{1}\, i_{3}}_{\ {\bf k}^{\prime},\, {\bf q}_{1},\, {\bf q}_{3}}
-
F^{(1)\, a^{\prime}\, i_{1}\, i_{3}}_{\ {\bf k}^{\prime},\, {\bf q}_{1},\, {\bf q}_{3}}\, 
F^{\hspace{0.03cm}\ast\hspace{0.03cm}(3)\, a^{\prime}\, i\; i_{2}}_{\ {\bf k}^{\prime},\,{\bf q},\, {\bf q}_{2}}
-
F^{(1)\, a^{\prime}\, i\, i_{1}}_{\ {\bf k}^{\prime},\, {\bf q},\, {\bf q}_{1}}\, 
F^{\hspace{0.03cm}\ast\hspace{0.03cm}(3)\, a^{\prime}\, i_{2}\, i_{3}}_{\ {\bf k}^{\prime},\, {\bf q}_{2},\, {\bf q}_{3}}
+
F^{(1)\, a^{\prime}\, i_{2}\, i_{3}}_{\ {\bf k}^{\prime},\, {\bf q}_{2},\, {\bf q}_{3}}\, 
F^{\hspace{0.03cm}\ast\hspace{0.03cm}(3)\, a^{\prime}\, i\; i_{1}}_{\ {\bf k}^{\prime},\,{\bf q},\, {\bf q}_{1}}
\Bigr)\biggr]\!
\biggr\}.
\notag
\end{align}
\indent Further, we will analyze the fourth-order contributions from the Hamiltonian $H^{(3)}$, Eq.\,(\ref{eq:2f}). Passing from the functions ${\mathcal G},\, {\mathcal P}$ and ${\mathcal K}$ to those $F^{(1,2,3)}$ by the rules (\ref{eq:4e}) 
and performing, when required, the relevant antisymmetrization, in the case of the product $f^{\, \ast\, i}_{{\bf q}} f^{\; i_{1}}_{{\bf q}_{1}} f^{\;i_{2}}_{{\bf q}_{2}} f^{\;i_{3}}_{{\bf q}_{3}}$, we obtain 
\begin{equation}
\int\!\frac{d{\bf q}\hspace{0.03cm} d{\bf q}_{1}\hspace{0.02cm} d{\bf q}_{2}\hspace{0.03cm} d{\bf q}_{3}}{(2\pi)^{12}}\, 
f^{\, \ast\, i}_{{\bf q}} f^{\; i_{1}}_{{\bf q}_{1}} f^{\;i_{2}}_{{\bf q}_{2}} f^{\;i_{3}}_{{\bf q}_{3}}
\label{eq:13y}
\end{equation}
\vspace{-0.6cm}
\begin{align}
\times\,\biggl\{
\frac{1}{3}\int\!\frac{d\hspace{0.02cm}{\bf k}^{\prime}}{(2\pi)^{3}}\,
\Bigl[\hspace{0.03cm}
&\bigl(\omega^{l}_{{\bf k}^{\prime}} - \omega^{-}_{{\bf q}_{1}} - \omega^{-}_{{\bf q}_{3}}\bigr)
F^{(1)\, a^{\prime}\, i_{1}\, i_{3}}_{\ {\bf k}^{\prime},\, {\bf q}_{1},\, {\bf q}_{3}}\, 
F^{\hspace{0.03cm}\ast\hspace{0.03cm}(2)\, a^{\prime}\, i_{2}\, i}_{\ {\bf k}^{\prime},\,{\bf q}_{2},\, {\bf q}}
-
\bigl(\omega^{l}_{{\bf k}^{\prime}} - \omega^{-}_{{\bf q}_{1}} - \omega^{-}_{{\bf q}_{2}}\bigr)
F^{(1)\, a^{\prime}\, i_{1}\, i_{2}}_{\ {\bf k}^{\prime},\, {\bf q}_{1},\, {\bf q}_{2}}\, 
F^{\hspace{0.03cm}\ast\hspace{0.03cm}(2)\, a^{\prime}\, i_{3}\, i}_{\ {\bf k}^{\prime},\,{\bf q}_{3},\, {\bf q}}
\notag\\[1ex]
-\,
&\bigl(\omega^{l}_{{\bf k}^{\prime}} - \omega^{-}_{{\bf q}_{2}} - \omega^{-}_{{\bf q}_{3}}\bigr)
F^{(1)\, a^{\prime}\, i_{2}\, i_{3}}_{\ {\bf k}^{\prime},\, {\bf q}_{2},\, {\bf q}_{3}}\, 
F^{\hspace{0.03cm}\ast\hspace{0.03cm}(2)\, a^{\prime}\, i_{1}\, i}_{\ {\bf k}^{\prime},\,{\bf q}_{1},\, {\bf q}}
\Bigr]
\notag\\[1ex]
+\,
\frac{1}{3}\int\!\frac{d\hspace{0.02cm}{\bf k}^{\prime}}{(2\pi)^{3}}\,
\Bigl[\hspace{0.03cm}
&\bigl(\omega^{l}_{{\bf k}^{\prime}} - \omega^{-}_{\bf q} + \omega^{-}_{{\bf q}_{2}}\bigr)
F^{(1)\, a^{\prime}\, i_{1} \, i_{3}}_{\ {\bf k}^{\prime},\, {\bf q}_{1},\, {\bf q}_{3}}\, 
F^{\hspace{0.03cm}\ast\hspace{0.03cm}(2)\, a^{\prime}\, i_{2}\, i}_{\ {\bf k}^{\prime},\,{\bf q}_{2},\, {\bf q}}
-
\bigl(\omega^{l}_{{\bf k}^{\prime}} - \omega^{-}_{\bf q} + \omega^{-}_{{\bf q}_{1}}\bigr)
F^{(1)\, a^{\prime}\, i_{2}\, i_{3}}_{\ {\bf k}^{\prime},\, {\bf q}_{2},\, {\bf q}_{3}}\, 
F^{\hspace{0.03cm}\ast\hspace{0.03cm}(2)\, a^{\prime}\, i_{1}\, i}_{\ {\bf k}^{\prime},\,{\bf q}_{1},\, {\bf q}}
\notag\\[1ex]
+\;
&\bigl(\omega^{l}_{{\bf k}^{\prime}} - \omega^{-}_{{\bf q}_{1}} + \omega^{-}_{{\bf q}}\bigr)
F^{(2)\, a^{\prime}\, i\; i_{1}}_{\ {\bf k}^{\prime},\, {\bf q},\, {\bf q}_{1}}\, 
F^{\hspace{0.03cm}\ast\hspace{0.03cm}(3)\, a^{\prime}\, i_{2}\, i_{3}}_{\ {\bf k}^{\prime},\,{\bf q}_{2},\, {\bf q}_{3}}
\;-
\bigl(\omega^{l}_{{\bf k}^{\prime}} - \omega^{-}_{{\bf q}} + \omega^{-}_{{\bf q}_{3}}\bigr)
F^{(1)\, a^{\prime}\, i_{1}\, i_{2}}_{\ {\bf k}^{\prime},\, {\bf q}_{1},\, {\bf q}_{2}}\, 
F^{\hspace{0.03cm}\ast\hspace{0.03cm}(2)\, a^{\prime}\, i_{3}\, i}_{\ {\bf k}^{\prime},\,{\bf q}_{3},\, {\bf q}}
\notag\\[1.5ex]
+\;
&\bigl(\omega^{l}_{{\bf k}^{\prime}} - \omega^{-}_{{\bf q}_{3}} + \omega^{-}_{{\bf q}}\bigr)
F^{(2)\, a^{\prime}\, i\; i_{3}}_{\ {\bf k}^{\prime},\, {\bf q},\, {\bf q}_{3}}\, 
F^{\hspace{0.03cm}\ast\hspace{0.03cm}(3)\, a^{\prime}\, i_{1}\, i_{2}}_{\ {\bf k}^{\prime},\,{\bf q}_{1},\, {\bf q}_{2}}
-
\bigl(\omega^{l}_{{\bf k}^{\prime}} - \omega^{-}_{{\bf q}_{2}} + \omega^{-}_{{\bf q}}\bigr)
F^{(2)\, a^{\prime}\, i\; i_{2}}_{\ {\bf k}^{\prime},\, {\bf q},\, {\bf q}_{2}}\, 
F^{\hspace{0.03cm}\ast\hspace{0.03cm}(3)\, a^{\prime}\, i_{1}\, i_{3}}_{\ {\bf k}^{\prime},\,{\bf q}_{1},\, {\bf q}_{3}}
\,\Bigr]
\notag\\[1ex]
+\,
\frac{1}{3}\int\!\frac{d\hspace{0.02cm}{\bf k}^{\prime}}{(2\pi)^{3}}\,
\Bigl[\hspace{0.03cm}
&\bigl(\omega^{l}_{{\bf k}^{\prime}} + \omega^{-}_{{\bf q}_{2}} + \omega^{-}_{{\bf q}_{3}}\bigr)
F^{(2)\, a^{\prime}\, i\; i_{1}}_{\ {\bf k}^{\prime},\, {\bf q},\, {\bf q}_{1}}\, 
F^{\hspace{0.03cm}\ast\hspace{0.03cm}(3)\, a^{\prime}\, i_{2}\, i_{3}}_{\ {\bf k}^{\prime},\,{\bf q}_{2},\, {\bf q}_{3}}
-
\bigl(\omega^{l}_{{\bf k}^{\prime}} + \omega^{-}_{{\bf q}_{1}} + \omega^{-}_{{\bf q}_{3}}\bigr)
F^{(2)\, a^{\prime}\, i\; i_{2}}_{\ {\bf k}^{\prime},\, {\bf q},\, {\bf q}_{2}}\, 
F^{\hspace{0.03cm}\ast\hspace{0.03cm}(3)\, a^{\prime}\, i_{1}\, i_{3}}_{\ {\bf k}^{\prime},\,{\bf q}_{1},\, {\bf q}_{3}}
\notag\\[1ex]
+\,
&\bigl(\omega^{l}_{{\bf k}^{\prime}} + \omega^{-}_{{\bf q}_{1}} + \omega^{-}_{{\bf q}_{2}}\bigr)
F^{(2)\, a^{\prime}\, i\; i_{3}}_{\ {\bf k}^{\prime},\, {\bf q},\, {\bf q}_{3}}\, 
F^{\hspace{0.03cm}\ast\hspace{0.03cm}(3)\, a^{\prime}\, i_{1}\, i_{2}}_{\ {\bf k}^{\prime},\,{\bf q}_{1},\, {\bf q}_{2}}
\Bigr]\biggr\}
\notag
\end{align}
and in the case of the product $f^{\, i}_{{\bf q}} f^{\; i_{1}}_{{\bf q}_{1}} f^{\;i_{2}}_{{\bf q}_{2}} f^{\;i_{3}}_{{\bf q}_{3}}$ we have, correspondingly, 
\begin{equation}
\int\!\frac{d{\bf q}\hspace{0.03cm} d{\bf q}_{1}\hspace{0.02cm} d{\bf q}_{2}\hspace{0.03cm} d{\bf q}_{3}}{(2\pi)^{12}}\, 
f^{\, i}_{{\bf q}} f^{\; i_{1}}_{{\bf q}_{1}} f^{\;i_{2}}_{{\bf q}_{2}} f^{\;i_{3}}_{{\bf q}_{3}}
\label{eq:13u}
\end{equation}
\vspace{-0.6cm}
\begin{align}
\times\,\biggl\{
\frac{1}{6}\int\!\frac{d\hspace{0.02cm}{\bf k}^{\prime}}{(2\pi)^{3}}\,
\Bigl[\hspace{0.03cm}
&\bigl(\omega^{l}_{{\bf k}^{\prime}} - \omega^{-}_{\bf q} - \omega^{-}_{{\bf q}_{1}}\bigr)
F^{(1)\, a^{\prime}\, i\; i_{1}}_{\ {\bf k}^{\prime},\, {\bf q},\; {\bf q}_{1}}\, 
F^{\hspace{0.03cm}\ast\hspace{0.03cm}(3)\, a^{\prime}\, i_{2}\, i_{3}}_{\ {\bf k}^{\prime},\,{\bf q}_{2},\, {\bf q}_{3}}
-
\bigl(\omega^{l}_{{\bf k}^{\prime}} - \omega^{-}_{\bf q} - \omega^{-}_{{\bf q}_{2}}\bigr)
F^{(1)\, a^{\prime}\, i\; i_{2}}_{\ {\bf k}^{\prime},\, {\bf q},\, {\bf q}_{2}}\, 
F^{\hspace{0.03cm}\ast\hspace{0.03cm}(3)\, a^{\prime}\, i_{1}\, i_{3}}_{\ {\bf k}^{\prime},\,{\bf q}_{1},\, {\bf q}_{3}}
\notag\\[1ex]
+\;
&\bigl(\omega^{l}_{{\bf k}^{\prime}} - \omega^{-}_{{\bf q}} - \omega^{-}_{{\bf q}_{3}}\bigr)
F^{(1)\, a^{\prime}\, i\; i_{3}}_{\ {\bf k}^{\prime},\, {\bf q},\; {\bf q}_{3}}\, 
F^{\hspace{0.03cm}\ast\hspace{0.03cm}(3)\, a^{\prime}\, i_{1}\, i_{2}}_{\ {\bf k}^{\prime},\,{\bf q}_{1},\, {\bf q}_{2}}
\;+
\bigl(\omega^{l}_{{\bf k}^{\prime}} - \omega^{-}_{{\bf q}_{1}} - \omega^{-}_{{\bf q}_{2}}\bigr)
F^{(1)\, a^{\prime}\, i_{1}\, i_{2}}_{\ {\bf k}^{\prime},\, {\bf q}_{1},\, {\bf q}_{2}}\, 
F^{\hspace{0.03cm}\ast\hspace{0.03cm}(3)\, a^{\prime}\, i\; i_{3}}_{\ {\bf k}^{\prime},\,{\bf q},\; {\bf q}_{3}}
\notag\\[1.5ex]
-\;
&\bigl(\omega^{l}_{{\bf k}^{\prime}} - \omega^{-}_{{\bf q}_{1}} - \omega^{-}_{{\bf q}_{3}}\bigr)
F^{(1)\, a^{\prime}\, i_{1}\, i_{3}}_{\ {\bf k}^{\prime},\, {\bf q}_{1},\, {\bf q}_{3}}\, 
F^{\hspace{0.03cm}\ast\hspace{0.03cm}(3)\, a^{\prime}\, i\; i_{2}}_{\ {\bf k}^{\prime},\,{\bf q},\; {\bf q}_{2}}
+
\bigl(\omega^{l}_{{\bf k}^{\prime}} - \omega^{-}_{{\bf q}_{2}} - \omega^{-}_{{\bf q}_{3}}\bigr)
F^{(1)\, a^{\prime}\, i_{2}\, i_{3}}_{\ {\bf k}^{\prime},\, {\bf q}_{2},\, {\bf q}_{3}}\, 
F^{\hspace{0.03cm}\ast\hspace{0.03cm}(3)\, a^{\prime}\, i\; i_{1}}_{\ {\bf k}^{\prime},\,{\bf q},\, {\bf q}_{1}}
\,\Bigr]
\notag\\[1ex]
+\,
\frac{1}{6}\int\!\frac{d\hspace{0.02cm}{\bf k}^{\prime}}{(2\pi)^{3}}\,
\Bigl[\hspace{0.03cm}
&\bigl(\omega^{l}_{{\bf k}^{\prime}} + \omega^{-}_{{\bf q}_{2}} + \omega^{-}_{{\bf q}_{3}}\bigr)
F^{(1)\, a^{\prime}\, i\; i_{1}}_{\ {\bf k}^{\prime},\, {\bf q},\; {\bf q}_{1}}\, 
F^{\hspace{0.03cm}\ast\hspace{0.03cm}(3)\, a^{\prime}\, i_{2}\, i_{3}}_{\ {\bf k}^{\prime},\,{\bf q}_{2},\, {\bf q}_{3}}
-
\bigl(\omega^{l}_{{\bf k}^{\prime}} + \omega^{-}_{{\bf q}_{1}} + \omega^{-}_{{\bf q}_{3}}\bigr)
F^{(1)\, a^{\prime}\, i\; i_{2}}_{\ {\bf k}^{\prime},\, {\bf q},\; {\bf q}_{2}}\, 
F^{\hspace{0.03cm}\ast\hspace{0.03cm}(3)\, a^{\prime}\, i_{1}\, i_{3}}_{\ {\bf k}^{\prime},\,{\bf q}_{1},\, {\bf q}_{3}}
\notag\\[1ex]
+\;
&\bigl(\omega^{l}_{{\bf k}^{\prime}} + \omega^{-}_{{\bf q}_{1}} + \omega^{-}_{{\bf q}_{2}}\bigr)
F^{(1)\, a^{\prime}\, i\; i_{3}}_{\ {\bf k}^{\prime},\, {\bf q},\; {\bf q}_{3}}\, 
F^{\hspace{0.03cm}\ast\hspace{0.03cm}(3)\, a^{\prime}\, i_{1}\, i_{2}}_{\ {\bf k}^{\prime},\,{\bf q}_{1},\, {\bf q}_{2}}
\;+
\bigl(\omega^{l}_{{\bf k}^{\prime}} + \omega^{-}_{{\bf q}} + \omega^{-}_{{\bf q}_{3}}\bigr)
F^{(1)\, a^{\prime}\, i_{1}\, i_{2}}_{\ {\bf k}^{\prime},\, {\bf q}_{1},\, {\bf q}_{2}}\, 
F^{\hspace{0.03cm}\ast\hspace{0.03cm}(3)\, a^{\prime}\, i\; i_{3}}_{\ {\bf k}^{\prime},\,{\bf q},\; {\bf q}_{3}}
\notag\\[1.5ex]
-\;
&\bigl(\omega^{l}_{{\bf k}^{\prime}} + \omega^{-}_{{\bf q}} + \omega^{-}_{{\bf q}_{2}}\bigr)
F^{(1)\, a^{\prime}\, i_{1}\, i_{3}}_{\ {\bf k}^{\prime},\, {\bf q}_{1},\, {\bf q}_{3}}\, 
F^{\hspace{0.03cm}\ast\hspace{0.03cm}(3)\, a^{\prime}\, i\; i_{2}}_{\ {\bf k}^{\prime},\,{\bf q},\; {\bf q}_{2}}
+
\bigl(\omega^{l}_{{\bf k}^{\prime}} + \omega^{-}_{{\bf q}} + \omega^{-}_{{\bf q}_{1}}\bigr)
F^{(1)\, a^{\prime}\, i_{2}\, i_{3}}_{\ {\bf k}^{\prime},\, {\bf q}_{2},\, {\bf q}_{3}}\, 
F^{\hspace{0.03cm}\ast\hspace{0.03cm}(3)\, a^{\prime}\, i\; i_{1}}_{\ {\bf k}^{\prime},\,{\bf q},\; {\bf q}_{1}}
\,\Bigr]\biggr\}.
\notag
\end{align}
Here, we did not collect similar terms since the expressions (\ref{eq:13y}) and (\ref{eq:13u}) are more convenient for further consideration. In the first and in the third terms on the right-hand side of (\ref{eq:13q}) we make the substitutions: $b^{\,i}_{\bf q} \rightarrow f^{\,i}_{\bf q}$ and  $b^{\,\ast\, i}_{{\bf q}} \rightarrow f^{\,\ast\, i}_{\bf q}$. Next, we put together (\ref{eq:13q}), (\ref{eq:13w}) and (\ref{eq:13e}) and also the expressions (\ref{eq:13y}) and (\ref{eq:13u}) taking into account (\ref{eq:13r}) and (\ref{eq:13t}) and collect similar terms. As in the previous section, the most part of the terms are mutually canceled. The last step is passing from the coefficient functions $F^{(1,2,3)}$ to the ``physical'' functions ${\mathcal G},\, {\mathcal P}$ and ${\mathcal K}$ by the rules (\ref{eq:4e}) and carrying out the  integration over ${\bf k}^{\prime}$. Thus we lead to the following  fourth-order expressions in the new variables $f^{\,i}_{\bf q}$ and $f^{\,\ast\, i}_{{\bf q}}$, which must be added to the effective Hamiltonian (\ref{eq:6r}) 
\begin{equation}
\int\!\frac{d{\bf q}\hspace{0.03cm} d{\bf q}_{1}\hspace{0.02cm} d{\bf q}_{2}\hspace{0.03cm} d{\bf q}_{3}}{(2\pi)^{12}}\, 
f^{\,\ast\, i}_{{\bf q}} f^{\; i_{1}}_{{\bf q}_{1}} f^{\;i_{2}}_{{\bf q}_{2}} f^{\;i_{3}}_{{\bf q}_{3}}
\label{eq:13i}
\end{equation}
\[
\times\,\biggl\{
\bigl(\hspace{0.02cm}\omega^{-}_{{\bf q}} - \omega^{-}_{{\bf q}_{1}} - \omega^{-}_{{\bf q}_{2}} - \omega^{-}_{{\bf q}_{3}}\bigr)
S^{\,(1)\, i\; i_{1}\, i_{2}\, i_{3}}_{\ {\bf q},\, {\bf q}_{1},\, {\bf q}_{2},\, {\bf q}_{3}}
+
\widetilde{T}^{\,(1)\, i\; i_{1}\, i_{2}\, i_{3}}_{{\bf q},\, {\bf q}_{1},\, {\bf q}_{2},\, {\bf q}_{3}}\, 
(2\pi)^{3}\hspace{0.03cm}\delta({\bf q} - {\bf q}_{1} - {\bf q}_{2} - {\bf q}_{3})
\biggr\}
\]
\[
+
\int\!\frac{d{\bf q}\hspace{0.03cm} d{\bf q}_{1}\hspace{0.02cm} d{\bf q}_{2}\hspace{0.03cm} d{\bf q}_{3}}{(2\pi)^{12}}\, 
f^{\, i}_{{\bf q}}\, f^{\; i_{1}}_{{\bf q}_{1}} f^{\;i_{2}}_{{\bf q}_{2}} f^{\;i_{3}}_{{\bf q}_{3}}
\hspace{0.5cm}
\]
\[
\times\,\frac{1}{4}\,\biggl\{
\bigl(\hspace{0.02cm}\omega^{-}_{{\bf q}} + \omega^{-}_{{\bf q}_{1}} + \omega^{-}_{{\bf q}_{2}} + \omega^{-}_{{\bf q}_{3}}\bigr)
S^{\hspace{0.03cm}\ast\hspace{0.03cm}(4)\hspace{0.04cm} i\; i_{1}\, i_{2}\, i_{3}}_{\ {\bf q},\, {\bf q}_{1},\, {\bf q}_{2},\, {\bf q}_{3}}
+
\widetilde{T}^{\,(4)\, i\; i_{1}\, i_{2}\, i_{3}}_{{\bf q},\, {\bf q}_{1},\, {\bf q}_{2},\, {\bf q}_{3}}\, 
(2\pi)^{3}\hspace{0.03cm}\delta({\bf q} + {\bf q}_{1} + {\bf q}_{2} + {\bf q}_{3})
\biggr\},
\]
where the effective amplitudes $\widetilde{T}^{(1)}$ and $\widetilde{T}^{(4)}$ have, correspondingly, the following structures:
\begin{align}
\widetilde{T}^{\,(1)\, i\, i_{1}\, i_{2}\, i_{3}}_{{\bf q},\, {\bf q}_{1},\, {\bf q}_{2},\, {\bf q}_{3}}
=
T^{\,(1)\, i\, i_{1}\, i_{2}\, i_{3}}_{{\bf q},\, {\bf q}_{1},\, {\bf q}_{2},\, {\bf q}_{3}}
\;+
\frac{1}{3}\,\biggl[\,&\frac{\,{\mathcal P}^{\; a\, i\,  i_{1}}_{{\bf q} - {\bf q}_{1},\, {\bf q},\, {\bf q}_{1}}\,
{\mathcal G}^{\hspace{0.03cm}\ast\, a\, i_{2}\, i_{3}}_{{\bf q}_{2} + {\bf q}_{3},\, {\bf q}_{2},\, {\bf q}_{3}}}
{\omega^{l}_{{\bf q}_{2} + {\bf q}_{3}} - \omega^{-}_{{\bf q}_{2}} - \omega^{-}_{{\bf q}_{3}}} 
\;-\;
\frac{{\mathcal P}^{\; a\, i\,  i_{2}}_{{\bf q} - {\bf q}_{2},\, {\bf q},\, {\bf q}_{2}}\,
{\mathcal G}^{\hspace{0.03cm}\ast\, a\, i_{1}\, i_{3}}_{{\bf q}_{1} + {\bf q}_{3},\, {\bf q}_{1},\, {\bf q}_{3}}}
{\omega^{l}_{{\bf q}_{1} + {\bf q}_{3}} - \omega^{-}_{{\bf q}_{1}} - \omega^{-}_{{\bf q}_{3}}} 
\hspace{0.85cm}
\label{eq:13o}\\[1.5ex]
+\;
&\frac{{\mathcal P}^{\; a\, i\, i_{3}}_{{\bf q} - {\bf q}_{3},\, {\bf q},\, {\bf q}_{3}}\,
{\mathcal G}^{\hspace{0.03cm}\ast\, a\, i_{1}\,  i_{2}}_{{\bf q}_{1} + {\bf q}_{2},\, {\bf q}_{1},\, {\bf q}_{2}}}
{\omega^{l}_{{\bf q}_{1} + {\bf q}_{2}} - \omega^{-}_{{\bf q}_{1}} - \omega^{-}_{{\bf q}_{2}}} 
\;-\;
\frac{{\mathcal K}^{\; a\, i_{2}\, i_{3}}_{-{\bf q}_{2} - {\bf q}_{3},\, {\bf q}_{2},\, {\bf q}_{3}}\,
{\mathcal P}^{\hspace{0.03cm}\ast\, a\, i_{1}\,  i}_{-{\bf q} + {\bf q}_{1},\, {\bf q}_{1},\, {\bf q}}}
{\omega^{l}_{-{\bf q}_{2} - {\bf q}_{3}} + \omega^{-}_{{\bf q}_{2}} + \omega^{-}_{{\bf q}_{3}}} 
\notag\\[1.5ex]
+\;
&\frac{{\mathcal K}^{\; a\, i_{1}\, i_{3}}_{-{\bf q}_{1} - {\bf q}_{3},\, {\bf q}_{1},\, {\bf q}_{3}}\,
{\mathcal P}^{\hspace{0.03cm}\ast\, a\, i_{2}\,  i}_{-{\bf q} + {\bf q}_{2},\, {\bf q}_{2},\, {\bf q}}}
{\omega^{l}_{-{\bf q}_{1} - {\bf q}_{3}} + \omega^{-}_{{\bf q}_{1}} + \omega^{-}_{{\bf q}_{3}}} 
\;-\;
\frac{{\mathcal K}^{\; a\, i_{1}\, i_{2}}_{-{\bf q}_{1} - {\bf q}_{2},\, {\bf q}_{1},\, {\bf q}_{2}}\,
{\mathcal P}^{\hspace{0.03cm}\ast\, a\, i_{3}\,  i}_{-{\bf q} + {\bf q}_{3},\, {\bf q}_{3},\, {\bf q}}}
{\omega^{l}_{-{\bf q}_{1} - {\bf q}_{2}} + \omega^{-}_{{\bf q}_{1}} + \omega^{-}_{{\bf q}_{2}}} 
\,\biggr],
\notag
\end{align}
\begin{align}
\widetilde{T}^{\,(4)\, i\, i_{1}\, i_{2}\, i_{3}}_{{\bf q},\, {\bf q}_{1},\, {\bf q}_{2},\, {\bf q}_{3}}
=
T^{\,(4)\, i\, i_{1}\, i_{2}\, i_{3}}_{{\bf q},\, {\bf q}_{1},\, {\bf q}_{2},\, {\bf q}_{3}}
\;+
\frac{2}{3}\,\biggl[\,&\frac{{\mathcal K}^{\; a\, i_{2}\, i_{3}}_{-{\bf q}_{2} - {\bf q}_{3},\, {\bf q}_{2},\, {\bf q}_{3}}\,
{\mathcal G}^{\hspace{0.03cm}\ast\, a\, i\; i_{1}}_{{\bf q} + {\bf q}_{1},\, {\bf q},\, {\bf q}_{1}}}
{\omega^{l}_{-{\bf q}_{2} - {\bf q}_{3}} + \omega^{-}_{{\bf q}_{2}} + \omega^{-}_{{\bf q}_{3}}} 
\;-\;
\frac{{\mathcal K}^{\; a\, i_{1}\, i_{3}}_{-{\bf q}_{1} - {\bf q}_{3},\, {\bf q}_{1},\, {\bf q}_{3}}\,
{\mathcal G}^{\hspace{0.03cm}\ast\, a\, i\; i_{2}}_{{\bf q} + {\bf q}_{2},\, {\bf q},\, {\bf q}_{2}}}
{\omega^{l}_{-{\bf q}_{1} - {\bf q}_{3}} + \omega^{-}_{{\bf q}_{1}} + \omega^{-}_{{\bf q}_{3}}} 
\label{eq:13p}\\[1.5ex]
\;+\;
&\frac{{\mathcal K}^{\; a\, i_{1}\, i_{2}}_{-{\bf q}_{1} - {\bf q}_{2},\, {\bf q}_{1},\, {\bf q}_{2}}\,
{\mathcal G}^{\hspace{0.03cm}\ast\, a\, i\;  i_{3}}_{{\bf q} + {\bf q}_{3},\, {\bf q},\, {\bf q}_{3}}}
{\omega^{l}_{-{\bf q}_{1} - {\bf q}_{2}} + \omega^{-}_{{\bf q}_{1}} + \omega^{-}_{{\bf q}_{2}}} 
\;+\;
\frac{{\mathcal K}^{\; a\, i\; i_{3}}_{-{\bf q} - {\bf q}_{3},\, {\bf q},\, {\bf q}_{3}}\,
{\mathcal G}^{\hspace{0.03cm}\ast\, a\, i_{1}\,  i_{2}}_{{\bf q}_{1} + {\bf q}_{2},\, {\bf q}_{1},\, {\bf q}_{2}}}
{\omega^{l}_{{\bf q}_{1} + {\bf q}_{2}} - \omega^{-}_{{\bf q}_{1}} - \omega^{-}_{{\bf q}_{2}}} 
\notag\\[1.5ex]
-\;
&\frac{{\mathcal K}^{\; a\, i\; i_{2}}_{-{\bf q} - {\bf q}_{2},\, {\bf q},\, {\bf q}_{2}}\,
{\mathcal G}^{\hspace{0.03cm}\ast\, a\, i_{1}\, i_{3}}_{{\bf q}_{1} + {\bf q}_{3},\, {\bf q}_{1},\, {\bf q}_{3}}}
{\omega^{l}_{{\bf q}_{1} + {\bf q}_{3}} - \omega^{-}_{{\bf q}_{1}} - \omega^{-}_{{\bf q}_{3}}} 
\;+\;
\frac{\,{\mathcal K}^{\; a\, i\; i_{1}}_{-{\bf q} - {\bf q}_{1},\, {\bf q},\, {\bf q}_{1}}\,
{\mathcal G}^{\hspace{0.03cm}\ast\, a\, i_{2}\, i_{3}}_{{\bf q}_{2} + {\bf q}_{3},\, {\bf q}_{2},\, {\bf q}_{3}}}
{\omega^{l}_{{\bf q}_{2} + {\bf q}_{3}} - \omega^{-}_{{\bf q}_{2}} - \omega^{-}_{{\bf q}_{3}}} 
\,\biggr].
\notag
\end{align}
The requirement of making the expression (\ref{eq:13i}) vanish uniquely defines the explicit form of the desired higher coefficient functions $S^{(1)}$ and $S^{(4)}$ in the canonical transformation (\ref{eq:3y}):
\begin{equation}
\begin{split}
&S^{\,(1)\, i\; i_{1}\, i_{2}\, i_{3}}_{\ {\bf q},\, {\bf q}_{1},\, {\bf q}_{2},\, {\bf q}_{3}}
=
-\hspace{0.03cm}
\frac{1}{\omega^{-}_{{\bf q}} - \omega^{-}_{{\bf q}_{1}} - \omega^{-}_{{\bf q}_{2}} - \omega^{-}_{{\bf q}_{3}}}\,
\widetilde{T}^{\,(1)\, i\; i_{1}\, i_{2}\, i_{3}}_{{\bf q},\, {\bf q}_{1},\, {\bf q}_{2},\, {\bf q}_{3}}\, 
(2\pi)^{3}\hspace{0.03cm}\delta({\bf q} - {\bf q}_1 - {\bf q}_{2} - {\bf q}_{3}),
\\[1.5ex]
&S^{\,(4)\, i\; i_{1}\, i_{2}\, i_{3}}_{\ {\bf q},\, {\bf q}_{1},\, {\bf q}_{2},\, {\bf q}_{3}}
=
-\hspace{0.03cm}
\frac{1}{\omega^{-}_{{\bf q}} + \omega^{-}_{{\bf q}_{1}} + \omega^{-}_{{\bf q}_{2}} + \omega^{-}_{{\bf q}_{3}}}\,
\widetilde{T}^{\hspace{0.03cm}\ast\hspace{0.03cm}(4)\hspace{0.03cm} i\; i_{1}\, i_{2}\, i_{3}}_{{\bf q},\, {\bf q}_{1},\, {\bf q}_{2},\, {\bf q}_{3}}\, 
(2\pi)^{3}\hspace{0.03cm}\delta({\bf q} + {\bf q}_{1} + {\bf q}_{2} + {\bf q}_{3}).
\end{split}
\label{eq:13a}
\end{equation}
The coefficient function $S^{\,(3)\, i\; i_{1}\, i_{2}\, i_{3}}_{\ {\bf q},\, {\bf q}_{1},\, {\bf q}_{2},\, {\bf q}_{3}}$ is defined through $S^{\,(1)\, i\; i_{1}\, i_{2}\, i_{3}}_{\ {\bf q},\, {\bf q}_{1},\, {\bf q}_{2},\, {\bf q}_{3}}$ with the help of the canonicity condition (\ref{ap:D2a}). In the hard thermal loop approximation it is necessary to set 
\[
T^{\,(1)\, i\; i_{1}\, i_{2}\, i_{3}}_{{\bf q},\, {\bf q}_{1},\, {\bf q}_{2},\, {\bf q}_{3}}
=
T^{\,(4)\, i\; i_{1}\, i_{2}\, i_{3}}_{{\bf q},\, {\bf q}_{1},\, {\bf q}_{2},\, {\bf q}_{3}} = 0.
\]
Furthermore, if one takes into account early obtained expressions (\ref{eq:9a}) in the same approximation, then from the expressions (\ref{eq:13o})\,--\,(\ref{eq:13a}) follows
\begin{equation}
S^{\,(1)\, i\; i_{1}\, i_{2}\, i_{3}}_{\, {\bf q},\, {\bf q}_{1},\, {\bf q}_{2},\, {\bf q}_{3}}
=
S^{\,(3)\, i\; i_{1}\, i_{2}\, i_{3}}_{\, {\bf q},\, {\bf q}_{1},\, {\bf q}_{2},\, {\bf q}_{3}}
=
S^{\,(4)\, i\; i_{1}\, i_{2}\, i_{3}}_{\, {\bf q},\, {\bf q}_{1},\, {\bf q}_{2},\, {\bf q}_{3}}
= 0.
\label{eq:13s}
\end{equation}
Thus, we obtain the vanishing of the third-order coefficient functions $S^{\hspace{0.03cm}(1)}$, $S^{\hspace{0.03cm}(3)}$ and $S^{\hspace{0.03cm}(4)}$ within the HTL-approximation.\\
\indent Now we turn to the determination of the coefficient function $S^{\,(2)\; i\; i_{1}\, i_{2}\, i_{3}}_{\ {\bf q},\, {\bf q}_{1},\, {\bf q}_{2},\, {\bf q}_{3}}$. We recall that this coefficient function enters into the expression for the effective fourth-order Hamiltonian describing the elastic scattering process of plasmino off plasmino:  
\begin{equation}
{\mathcal H}^{(4)}_{qq\rightarrow qq} 
\,=
\frac{1}{2}\int\frac{d{\bf q}\hspace{0.03cm} d{\bf q}_{1}\hspace{0.02cm} d{\bf q}_{2}\hspace{0.03cm} d{\bf q}_{3}}{(2\pi)^{12}}
\label{eq:13d}
\end{equation}
\[
\times\,\biggl\{
\bigl(\hspace{0.02cm}\omega^{-}_{{\bf q}} + \omega^{-}_{{\bf q}_{1}} - \omega^{-}_{{\bf q}_{2}} - \omega^{-}_{{\bf q}_{3}}\bigr)
S^{\,(2)\, i\; i_{1}\, i_{2}\, i_{3}}_{\, {\bf q},\, {\bf q}_{1},\, {\bf q}_{2},\, {\bf q}_{3}}
+
\widetilde{T}^{\, (2)\, i\; i_{1}\, i_{2}\, i_{3}}_{\, {\bf q},\, {\bf q}_{1},\, {\bf q}_{2},\, {\bf q}_{3}}\, 
(2\pi)^{3}\hspace{0.03cm}\delta({\bf q} + {\bf q}_1 - {\bf q}_{2} - {\bf q}_{3})
\biggr\}
f^{\,\ast\, i}_{{\bf q}} f^{\,\ast\, i_{1}}_{{\bf q}_{1}} f^{\;i_{2}}_{{\bf q}_{2}} f^{\;i_{3}}_{{\bf q}_{3}},
\]
where the effective amplitude $\widetilde{T}^{\,(2)}$ is given by the formula (\ref{eq:6t}). However, an explicit form of this coefficient function as opposed to the previous ones, will no longer be so unambiguous. This is due to the fact that the canonical transformation admits a certain freedom (in the case of the coefficients $S^{\hspace{0.03cm}(1)}, S^{\hspace{0.03cm}(3)}$, and $S^{\hspace{0.03cm}(4)}$ this freedom is limited by the condition of exclusion of the nonresonant terms from ${\mathcal H}^{(4)}$). For the coefficient function $S^{\hspace{0.03cm}(2)}$ we have two canonicity conditions (\ref{ap:D2b}) and (\ref{ap:D4a}). As was mentioned above, these conditions are insufficient to uniquely define $S^{\hspace{0.03cm}(2)}$. Let us write out these conditions once again in the form of functional equations:
\begin{equation}
\begin{split}
&S^{\hspace{0.03cm}(2)\, i\; i_{1}\; i_{2}\; i_{3}}_{\, {\bf q},\, {\bf q}_{1},\, {\bf q}_{2},\, {\bf q}_{3}}
+
S^{\hspace{0.035cm}(2)\, i_{1}\, i\; i_{2}\; i_{3}}_{\, {\bf q}_{1},\, {\bf q},\, {\bf q}_{2},\, {\bf q}_{3}}
=
\Phi^{\hspace{0.03cm}(2)\, i\; i_{1}\; i_{2}\; i_{3}}_{\, {\bf q},\, {\bf q}_{1},\, {\bf q}_{2},\, {\bf q}_{3}},
\\[2.5ex]
&S^{\hspace{0.03cm}(2)\; i\; i_{1}\; i_{2}\; i_{3}}_{\, {\bf q},\, {\bf q}_{1},\, {\bf q}_{2},\, {\bf q}_{3}}
+
S^{\hspace{0.03cm} \ast\hspace{0.03cm}(2)\hspace{0.03cm} i_{3}\; i_{2}\; i_{1}\, i}_{\ {\bf q}_{3},\, {\bf q}_{2},\, {\bf q}_{1},\, {\bf q}}
=
\Phi^{(22)\, i\; i_{1}\; i_{2}\; i_{3}}_{\ {\bf q},\, {\bf q}_{1},\, {\bf q}_{2},\, {\bf q}_{3}},
\label{eq:13f}
\end{split}
\end{equation}
where we set by definition
\begin{equation}
\Phi^{\,(2)\, i\; i_{1}\, i_{2}\, i_{3}}_{\, {\bf q},\, {\bf q}_{1},\, {\bf q}_{2},\, {\bf q}_{3}}
\equiv
\label{eq:13g}
\end{equation}
\[
\frac{1}{2}
\int\!\!\frac{d\hspace{0.02cm}{\bf k}^{\prime}}{(2\pi)^{3}}\,\Bigl[
F^{(2)\, a^{\prime}\, i_{1}\, i_{3}}_{\ {\bf k}^{\prime},\, {\bf q}_{1},\, {\bf q}_{3}}\, F^{\,\ast\hspace{0.03cm}(2)\, a^{\prime}\, i_{2}\, i}_{\ {\bf k}^{\prime},\, {\bf q}_{2},\, {\bf q}}
-\,
F^{(2)\, a^{\prime}\, i\, i_{2}}_{\ {\bf k}^{\prime},\, {\bf q},\, {\bf q}_{2}}\, F^{\,\ast\hspace{0.03cm}(2)\, a^{\prime}\, i_{3}\, i_{1}}_{\ {\bf k}^{\prime},\, {\bf q}_{3},\, {\bf q}_{1}}
+\,
F^{(2)\, a^{\prime}\, i\, i_{3}}_{\ {\bf k}^{\prime},\, {\bf q},\, {\bf q}_{3}}\, F^{\,\ast\hspace{0.03cm}(2)\, a^{\prime}\, i_{2}\, i_{1}}_{\ {\bf k}^{\prime},\, {\bf q}_{2},\, {\bf q}_{1}}
-\,
F^{(2)\, a^{\prime}\, i_{1}\, i_{2}}_{\ {\bf k}^{\prime},\, {\bf q}_{1},\, {\bf q}_{2}}\, F^{\,\ast\hspace{0.03cm}(2)\, a^{\prime}\, i_{3}\, i}_{\ {\bf k}^{\prime},\, {\bf q}_{3},\, {\bf q}}\hspace{0.03cm}\Bigr]
\vspace{0.3cm}
\]
and
\begin{equation}
\Phi^{\hspace{0.03cm}(22)\; i\; i_{1}\, i_{2}\, i_{3}}_{\ {\bf q},\, {\bf q}_{1},\, {\bf q}_{2},\, {\bf q}_{3}}
\equiv
\label{eq:13h}
\end{equation}
\[
\frac{1}{2}\int\!\!\frac{d\hspace{0.02cm}{\bf k}^{\prime}}{(2\pi)^{3}}\Bigl[
F^{(2)\, a^{\prime}\, i_{1}\, i_{3}}_{\ {\bf k}^{\prime},\, {\bf q}_{1},\, {\bf q}_{3}}\hspace{0.02cm} F^{\,\ast\hspace{0.03cm}(2)\, a^{\prime}\, i_{2}\, i}_{\ {\bf k}^{\prime},\, {\bf q}_{2},\, {\bf q}}
-
F^{(2)\, a^{\prime}\, i\, i_{2}}_{\ {\bf k}^{\prime},\, {\bf q},\, {\bf q}_{2}}\hspace{0.02cm} F^{\,\ast\hspace{0.03cm}(2)\, a^{\prime}\, i_{3}\, i_{1}}_{\ {\bf k}^{\prime},\, {\bf q}_{3},\, {\bf q}_{1}}
-\hspace{0.02cm}
4 F^{(3)\, a^{\prime}\, i\, i_{1}}_{\ {\bf k}^{\prime},\, {\bf q},\, {\bf q}_{1}}\hspace{0.02cm} F^{\,\ast\hspace{0.03cm}(3)\, a^{\prime}\, i_{2}\, i_{3}}_{\ {\bf k}^{\prime},\, {\bf q}_{2},\, {\bf q}_{3}}
+\hspace{0.02cm}
4 F^{(1)\, a^{\prime}\, i_{2}\, i_{3}}_{\ {\bf k}^{\prime},\, {\bf q}_{2},\, {\bf q}_{3}}\hspace{0.02cm} F^{\,\ast\hspace{0.03cm}(1)\, a^{\prime}\, i\; i_{1}}_{\ {\bf k}^{\prime},\, {\bf q},\, {\bf q}_{1}}\Bigr] .
\]
The first equation in (\ref{eq:13f}) defines the rearrangement rule of the first two indices and correspon\-ding momentum arguments of the function $S^{\,(2)}$, and the second one defines the conjugation rule. A direct check shows that the functions (\ref{eq:13g}) and (\ref{eq:13h}) have the following properties:
\begin{align}
&\Phi^{\hspace{0.03cm}(2)\, i\; i_{1}\, i_{2}\, i_{3}}_{\, {\bf q},\, {\bf q}_{1},\, {\bf q}_{2},\, {\bf q}_{3}}
=\,
\Phi^{\hspace{0.03cm}(2)\, i_{1}\, i\; i_{2}\, i_{3}}_{\, {\bf q}_{1},\, {\bf q},\, {\bf q}_{2},\, {\bf q}_{3}}
=
-\hspace{0.03cm}\Phi^{\hspace{0.03cm}(2)\, i\; i_{1}\, i_{3}\, i_{2}}_{\, {\bf q},\, {\bf q}_{1},\, {\bf q}_{3},\, {\bf q}_{2}},
\notag\\[1.7ex]
&\Phi^{(22)\, i\; i_{1}\, i_{2}\, i_{3}}_{\, {\bf q},\, {\bf q}_{1},\, {\bf q}_{2},\, {\bf q}_{3}}
\,+
\Phi^{(22)\, i_{1}\, i\; i_{2}\, i_{3}}_{\, {\bf q}_{1},\, {\bf q},\, {\bf q}_{2},\, {\bf q}_{3}}
=
\Phi^{\hspace{0.03cm}(2)\, i\; i_{1}\, i_{2}\, i_{3}}_{\, {\bf q},\, {\bf q}_{1},\, {\bf q}_{2},\, {\bf q}_{3}},
\label{eq:13j}\\[1.7ex]
&\Phi^{(22)\, i\; i_{1}\, i_{2}\, i_{3}}_{\, {\bf q},\, {\bf q}_{1},\, {\bf q}_{2},\, {\bf q}_{3}}
\,=
\Phi^{\ast\hspace{0.03cm}(22)\, i_{3}\, i_{2}\, i_{1}\, i}_{\ {\bf q}_{3},\, {\bf q}_{2},\, {\bf q}_{1},\, {\bf q}}.
\notag
\end{align}
\indent We seek a solution of the system of the functional equations (\ref{eq:13f}) in the following form:
\begin{equation}
S^{\hspace{0.03cm}(2)\; i\; i_{1}\, i_{2}\, i_{3}}_{\, {\bf q},\, {\bf q}_{1},\, {\bf q}_{2},\, {\bf q}_{3}}
=
\alpha\hspace{0.04cm}\Phi^{(22)\, i\; i_{1}\, i_{2}\, i_{3}}_{\, {\bf q},\, {\bf q}_{1},\, {\bf q}_{2},\, {\bf q}_{3}}
+
\beta\hspace{0.04cm}\Phi^{\hspace{0.03cm}(2)\, i\; i_{1}\, i_{2}\, i_{3}}_{\, {\bf q},\, {\bf q}_{1},\, {\bf q}_{2},\, {\bf q}_{3}}
+
\gamma\hspace{0.04cm}\Phi^{\hspace{0.02cm}\ast\hspace{0.02cm}(2)\hspace{0.03cm} i_{3}\; i_{2}\; i_{1}\, i}_{\ {\bf q}_{3},\, {\bf q}_{2},\, {\bf q}_{1},\, {\bf q}},   
\label{eq:13k}
\end{equation}
where $\alpha,\,\beta$ and $\gamma$ are unknown, generally speaking, complex coefficients. Substituting this expression into (\ref{eq:13f}) and making use of the properties (\ref{eq:13j}), we get an algebraic system for the unknown coefficients:
\[
\alpha + \alpha^{\ast} = 1, \quad \beta + \gamma^{\ast} = 0, \quad \alpha + 2\hspace{0.03cm}\beta = 1.
\]
A solution of this algebraic system has the form
\[
\alpha = \frac{1}{2} + i\hspace{0.04cm}{\rm Im}\,\alpha,\quad \beta = \frac{1}{2}\,\alpha^{\ast}, 
\quad \gamma = -\hspace{0.02cm}\frac{1}{2}\,\alpha,
\]
where ${\rm Im}\,\alpha$ is an arbitrary numerical parameter. In particular, we can put ${\rm Im}\,\alpha \equiv  0$ and thus, instead of (\ref{eq:13k}), we have 
\begin{equation}
S^{\hspace{0.03cm}(2)\; i\; i_{1}\, i_{2}\, i_{3}}_{\, {\bf q},\, {\bf q}_{1},\, {\bf q}_{2},\, {\bf q}_{3}}
=
\frac{1}{2}\,\Phi^{(22)\, i\; i_{1}\, i_{2}\, i_{3}}_{\, {\bf q},\, {\bf q}_{1},\, {\bf q}_{2},\, {\bf q}_{3}}
\,+\,
\frac{1}{4}\hspace{0.03cm}\bigl(\hspace{0.02cm}\Phi^{\hspace{0.03cm}(2)\, i\; i_{1}\, i_{2}\, i_{3}}_{\, {\bf q},\, {\bf q}_{1},\, {\bf q}_{2},\, {\bf q}_{3}}
-\,
\Phi^{\hspace{0.02cm}\ast\hspace{0.03cm}(2)\, i_{3}\, i_{2}\, i_{1}\, i}_{\ {\bf q}_{3},\, {\bf q}_{2},\, {\bf q}_{1},\, {\bf q}}
\hspace{0.02cm}\bigr).   
\label{eq:13l}
\end{equation}
It only remains to substitute the explicit expressions for the functions 
$\Phi^{(22)\, i\; i_{1}\; i_{2}\; i_{3}}_{\ {\bf q},\, {\bf q}_{1},\, {\bf q}_{2},\, {\bf q}_{3}}$ and $\Phi^{(2)\; i\; i_{1}\; i_{2}\; i_{3}}_{\ {\bf q},\, {\bf q}_{1},\, {\bf q}_{2},\, {\bf q}_{3}}$, Eqs.\,(\ref{eq:13g}) and (\ref{eq:13h}) into (\ref{eq:13l}), to pass from the coefficient functions $F^{(n)}$ to the ``physical'' functions ${\mathcal G},\, {\mathcal P}$ and ${\mathcal K}$ by the rules (\ref{eq:4e}) and to perform the integration over $d\hspace{0.03cm}{\bf k}^{\prime}$. This gives us the required coefficient function $S^{\hspace{0.03cm}(2)}$. However, it is necessary to note that this function will be defined with an accuracy of an arbitrary function $\Lambda^{(2)}$ satisfying the conditions
\[
\Lambda^{(2)\; i\; i_{1}\, i_{2}\, i_{3}}_{\, {\bf q},\, {\bf q}_{1},\, {\bf q}_{2},\, {\bf q}_{3}}
=
-\Lambda^{(2)\; i_{1}\, i\; i_{2}\, i_{3}}_{\, {\bf q}_{1},\, {\bf q},\, {\bf q}_{2},\, {\bf q}_{3}}
=
-\Lambda^{(2)\; i\; i_{1}\, i_{3}\, i_{2}}_{\, {\bf q},\, {\bf q}_{1},\, {\bf q}_{3},\, {\bf q}_{2}}
=
-\Lambda^{\ast\hspace{0.025cm}(2)\, i_{3}\, i_{2}\, i_{1}\, i}_{\, {\bf q}_{3},\, {\bf q}_{2},\, {\bf q}_{1},\, 
{\bf q}}.
\]
Taking into account all above-mentioned we can write out the most general form of the desired coefficient function
\begin{equation}
S^{\hspace{0.03cm}(2)\; i\; i_{1}\, i_{2}\, i_{3}}_{\, {\bf q},\, {\bf q}_{1},\, {\bf q}_{2},\, {\bf q}_{3}}
=
\Lambda^{(2)\; i\; i_{1}\, i_{2}\, i_{3}}_{\, {\bf q},\, {\bf q}_{1},\, {\bf q}_{2},\, {\bf q}_{3}}
\label{eq:13z}
\end{equation}
\begin{align}
+\,
\Biggl\{\,&\frac{{\mathcal G}^{\; a\, i\; i_{1}}_{{\bf q} + {\bf q}_{1},\, {\bf q},\, {\bf q}_{1}}\ 
{\mathcal G}^{\hspace{0.03cm}\ast\, a\, i_{2}\, i_{3}}_{{\bf q}_{2} + {\bf q}_{3},\, {\bf q}_{2},\, {\bf q}_{3}}}
{\bigl(\omega^{l}_{{\bf q} + {\bf q}_{1}} - \omega^{-}_{{\bf q}} - \omega^{-}_{{\bf q}_{1}}\bigr)
\bigl(\omega^{l}_{{\bf q}_{2} + {\bf q}_{3}} - \omega^{-}_{{\bf q}_{2}} - \omega^{-}_{{\bf q}_{3}}\bigr)
} 
-
\frac{{\mathcal K}^{\; a\, i_{2}\, i_{3}}_{-{\bf q}_{2} - {\bf q}_{3},\, {\bf q}_{2},\, {\bf q}_{3}}\
{\mathcal K}^{\hspace{0.03cm}\ast\, a\, i\; i_{1}}_{-{\bf q} - {\bf q}_{1},\, {\bf q},\, {\bf q}_{1}}}
{\bigl(\omega^{l}_{-{\bf q}_{2} - {\bf q}_{3}} + \omega^{-}_{{\bf q}_{2}} + \omega^{-}_{{\bf q}_{3}}\bigr)
\bigl(\omega^{l}_{-{\bf q} - {\bf q}_{1}} + \omega^{-}_{{\bf q}} + \omega^{-}_{{\bf q}_{1}}\bigr)
}
\notag\\[1.5ex]
+\,
\frac{1}{4}\Biggl(
&\frac{{\mathcal P}^{\; a\, i\,  i_{2}}_{{\bf q} - {\bf q}_{2},\, {\bf q},\, {\bf q}_{2}}\ 
{\mathcal P}^{\hspace{0.03cm}\ast\, a\, i_{3}\, i_{1}}_{{\bf q}_{3} - {\bf q}_{1},\, {\bf q}_{3},\,  {\bf q}_{1}}}
{\bigl(\omega^{l}_{{\bf q} - {\bf q}_{2}} - \omega^{-}_{{\bf q}} + \omega^{-}_{{\bf q}_{2}}\bigr)
\bigl(\omega^{l}_{{\bf q}_{3} - {\bf q}_{1}} - \omega^{-}_{{\bf q}_{3}} + \omega^{-}_{{\bf q}_{1}}\bigr)
}
-
\frac{{\mathcal P}^{\; a\, i_{1}\, i_{3}}_{{\bf q}_{1} - {\bf q}_{3},\, {\bf q}_{1},\, {\bf q}_{3}}\ 
{\mathcal P}^{\hspace{0.03cm}\ast\, a\, i_{2}\, i}_{{\bf q}_{2} - {\bf q},\, {\bf q}_{2},\, {\bf q}}}
{\bigl(\omega^{l}_{{\bf q}_{1} - {\bf q}_{3}} - \omega^{-}_{{\bf q}_{1}} + \omega^{-}_{{\bf q}_{3}}\bigr)
\bigl(\omega^{l}_{{\bf q}_{2} - {\bf q}} - \omega^{-}_{{\bf q}_{2}} + \omega^{-}_{{\bf q}}\bigr)
}
\Biggr)
\notag\\[1.5ex]
+\;
\frac{1}{4}\Biggl(
&\frac{{\mathcal P}^{\; a\, i_{1}\, i_{2}}_{{\bf q}_{1} - {\bf q}_{2},\, {\bf q}_{1},\, {\bf q}_{2}}\ 
{\mathcal P}^{\hspace{0.03cm}\ast\, a\, i_{3}\, i}_{{\bf q}_{3} - {\bf q},\, {\bf q}_{3},\, {\bf q}}}
{\bigl(\omega^{l}_{{\bf q}_{1} - {\bf q}_{2}} - \omega^{-}_{{\bf q}_{1}} + \omega^{-}_{{\bf q}_{2}}\bigr)
\bigl(\omega^{l}_{{\bf q}_{3} - {\bf q}} - \omega^{-}_{{\bf q}_{3}} + \omega^{-}_{{\bf q}}\bigr)
}
- 
\frac{{\mathcal P}^{\; a\, i\,  i_{3}}_{{\bf q} - {\bf q}_{3},\, {\bf q},\, {\bf q}_{3}}\ 
{\mathcal P}^{\hspace{0.03cm}\ast\, a\, i_{2}\,  i_{1}}_{{\bf q}_{2} - {\bf q}_{1},\, {\bf q}_{2},\, {\bf q}_{1}}}
{\bigl(\omega^{l}_{{\bf q} - {\bf q}_{3}} - \omega^{-}_{{\bf q}} + \omega^{-}_{{\bf q}_{3}}\bigr)
\bigl(\omega^{l}_{{\bf q}_{2} - {\bf q}_{1}} - \omega^{-}_{{\bf q}_{2}} + \omega^{-}_{{\bf q}_{1}}\bigr)
}
\Biggr)\!\Biggr\}
\notag
\end{align}
\[
\times\hspace{0.03cm} (2\pi)^{3}\hspace{0.03cm}\delta({\bf q} + {\bf q}_1 - {\bf q}_{2} - {\bf q}_{3}).
\]
Here, the right-hand part is written in such a form that it is convenient to compare the structure of the coefficient function $S^{\hspace{0.03cm}(2)\; i\; i_{1}\; i_{2}\; i_{3}}_{\ {\bf q},\, {\bf q}_{1},\, {\bf q}_{2},\, {\bf q}_{3}}$ with the structure of the effective amplitude $\widetilde{T}^{\,(2)\, i\, i_{1}\, i_{2}\, i_{3}}_{{\bf q},\, {\bf q}_{1},\, {\bf q}_{2},\, {\bf q}_{3}}$, Eq.\,(\ref{eq:6t}). By virtue of the arbitrariness of the function $\Lambda^{(2)}$ in (\ref{eq:13z}) the canonical transformation (\ref{eq:3y}) admits a certain freedom. The function $\Lambda^{(2)}$ can be chosen in the form which is convenient (variation of $S^{\hspace{0.03cm}(2)}$ alters simultaneously both the integrand in (\ref{eq:13d}) and $f^{\hspace{0.03cm}i}_{{\bf q}}$, but leaves unchanged $b^{\,i}_{{\bf q}}$ in the canonical transformation (\ref{eq:3y})). In particular, we can put $\Lambda^{(2)}\! \equiv\! 0$, although it is possible that this function will be explicitly defined in constructing the effective Hamiltonian of higher (sixth) order. Finally, in the hard thermal loops approximation the first two terms in braces on the right-hand side of (\ref{eq:13z}) with the functions ${\mathcal G}$ and ${\mathcal K}$ should be set equal to zero.\\
\indent We need to consider in more detail the practical implication of the coefficient function $S^{\hspace{0.03cm}(2)}$ for the Hamilton formalism under consideration when the resonance frequency difference (\ref{eq:6y}) is different from zero. We recall for this purpose that the function $T^{(2)\, i\; i_{1}\; i_{2}\; i_{3}}_{{\bf q},\, {\bf q}_{1},\, {\bf q}_{2},\, {\bf q}_{3}}$ in the initial fourth-order interaction Hamiltonian $H^{(4)}$, Eq.\,(\ref{eq:2g}), satisfies the ``conditions of natural symmetry'' 
\begin{equation}
T^{\,(2)\, i\; i_{1}\, i_{2}\, i_{3}}_{\, {\bf q},\, {\bf q}_{1},\, {\bf q}_{2},\, {\bf q}_{3}} 
= 
 -\hspace{0.03cm}T^{\,(2)\, i_{1}\,  i\; i_{2}\, i_{3}}_{\, {\bf q}_{1},\, {\bf q},\, {\bf q}_{2},\, {\bf q}_{3}}
=
 -\hspace{0.03cm}T^{\,(2)\, i\; i_{1}\, i_{3}\, i_{2}}_{\, {\bf q},\, {\bf q}_{1},\, {\bf q}_{3},\, {\bf q}_{2}}.
\label{eq:13x}
\end{equation}
The requirement of reality of this Hamiltonian leads to another condition
\begin{equation}
T^{\hspace{0.03cm}\ast\hspace{0.03cm} (2)\, i\; i_{1}\, i_{2}\, i_{3}}_{\ {\bf q},\, {\bf q}_{1},\, {\bf q}_{2},\, {\bf q}_{3}}
=
T^{\,(2)\, i_{2}\, i_{3}\, i\; i_{1}}_{\; {\bf q}_{2},\, {\bf q}_{3},\, {\bf q},\, {\bf q}_{1}}. 
\label{eq:13c}
\end{equation}
\indent Let us consider now the effective amplitude $\widetilde{T}^{\,(2)\, i\, i_{1}\, i_{2}\, i_{3}}_{{\bf q},\, {\bf q}_{1},\, {\bf q}_{2},\, {\bf q}_{3}}$, which is defined by the expression (\ref{eq:6t}). If one does not use the four-wave resonance condition  
\begin{equation}
\omega^{-}_{{\bf q}} + \omega^{-}_{{\bf q}_{1}} = \omega^{-}_{{\bf q}_{2}} + \omega^{-}_{{\bf q}_{3}},
\label{eq:13v}
\end{equation}
then by virtue of the symmetry properties (\ref{eq:2h}) and (\ref{eq:2j}), it is not difficult to verify the validity of the following equality:
\[
\widetilde{T}^{\,(2)\, i\; i_{1}\, i_{2}\, i_{3}}_{\, {\bf q},\, {\bf q}_{1},\, {\bf q}_{2},\, {\bf q}_{3}} 
= 
-\hspace{0.03cm}\widetilde{T}^{\,(2)\, i\; i_{1}\, i_{3}\, i_{2}}_{\, {\bf q},\, {\bf q}_{1},\, {\bf q}_{3},\, {\bf q}_{2}}
\]
and at the same time we have
\[
\widetilde{T}^{\,(2)\, i\; i_{1}\, i_{2}\, i_{3}}_{\, {\bf q},\, {\bf q}_{1},\, {\bf q}_{2},\, {\bf q}_{3}} 
\neq 
 -\hspace{0.03cm}\widetilde{T}^{\,(2)\, i_{1}\, i\; i_{2}\, i_{3}}_{\, {\bf q}_{1},\, {\bf q},\, {\bf q}_{2},\, {\bf q}_{3}}
\quad \mbox{and} \quad
\widetilde{T}^{\hspace{0.03cm}\ast\hspace{0.03cm} (2)\, i\; i_{1}\, i_{2}\, i_{3}}_{\ {\bf q},\, {\bf q}_{1},\, {\bf q}_{2},\, {\bf q}_{3}}
\neq
\widetilde{T}^{\,(2)\, i_{2}\, i_{3}\, i\; i_{1}}_{\; {\bf q}_{2},\, {\bf q}_{3},\, {\bf q},\, {\bf q}_{1}}.
\]
In the last two expressions we can put the equal sign only under the condition (\ref{eq:13v}).\\
\indent We set $\Lambda^{(2)} \equiv 0$ in the expression (\ref{eq:13z}) and introduce the following notation:
\begin{equation}
\mathscr{T}^{\,(2)\; i\; i_{1}\, i_{2}\, i_{3}}_{\, {\bf q},\, {\bf q}_{1},\, {\bf q}_{2},\, {\bf q}_{3}}\, 
(2\pi)^{3}\hspace{0.02cm}\delta({\bf q} + {\bf q}_1 - {\bf q}_{2} - {\bf q}_{3})
\equiv
\label{eq:13b}
\end{equation}
\[
\equiv
\bigl(\hspace{0.02cm}\omega^{-}_{{\bf q}} + \omega^{-}_{{\bf q}_{1}} - \omega^{-}_{{\bf q}_{2}} - \omega^{-}_{{\bf q}_{3}}\bigr)
S^{\,(2)\; i\; i_{1}\, i_{2}\, i_{3}}_{\, {\bf q},\, {\bf q}_{1},\, {\bf q}_{2},\, {\bf q}_{3}}
+
\widetilde{T}^{\,(2)\, i\; i_{1}\, i_{2}\, i_{3}}_{\, {\bf q},\, {\bf q}_{1},\, {\bf q}_{2},\, {\bf q}_{3}}\, 
(2\pi)^{3}\hspace{0.03cm}\delta({\bf q} + {\bf q}_1 - {\bf q}_{2} - {\bf q}_{3}).
\]
The function $\mathscr{T}^{\,(2)\, i\; i_{1}\, i_{2}\, i_{3}}_{\, {\bf q},\, {\bf q}_{1},\, {\bf q}_{2},\, {\bf q}_{3}}$ will be called a {\it complete effective amplitude}. Substituting the explicit expressions of the functions $S^{\,(2)}$ and $\widetilde{T}^{\,(2)}$, Eqs.\,(\ref{eq:13z}) and (\ref{eq:6t}), performing simple algebraic transformations, we define an explicit form of the complete effective amplitude $\mathscr{T}^{\,(2)}$:
\begin{equation}
\mathscr{T}^{\,(2)\, i\; i_{1}\, i_{2}\, i_{3}}_{\, {\bf q},\, {\bf q}_{1},\, {\bf q}_{2},\, {\bf q}_{3}}
=
T^{\,(2)\, i\; i_{1}\, i_{2}\, i_{3}}_{\, {\bf q},\, {\bf q}_{1},\, {\bf q}_{2},\, {\bf q}_{3}}
\label{eq:13n}
\vspace{-0.3cm}
\end{equation}
\begin{align}
+\,
\biggl(&\frac{1}{\omega^{l}_{{\bf q} + {\bf q}_{1}} - \omega^{-}_{{\bf q}} - \omega^{-}_{{\bf q}_{1}}} 
\,+\,
\frac{1}{\omega^{l}_{{\bf q}_{2} + {\bf q}_{3}} - \omega^{-}_{{\bf q}_{2}} - \omega^{-}_{{\bf q}_{3}}} 
\biggr)\hspace{0.02cm}
{\mathcal G}^{\; a\, i\; i_{1}}_{{\bf q} + {\bf q}_{1},\, {\bf q},\, {\bf q}_{1}}\, 
{\mathcal G}^{\hspace{0.03cm}\ast\, a\, i_{2}\, i_{3}}_{{\bf q}_{2} + {\bf q}_{3},\, {\bf q}_{2},\, {\bf q}_{3}}
\notag\\[1.5ex]
+\,
\biggl(&\frac{1}{\omega^{l}_{-{\bf q} - {\bf q}_{1}} + \omega^{-}_{{\bf q}} + \omega^{-}_{{\bf q}_{1}}} 
+
\frac{1}{\omega^{l}_{-{\bf q}_{2} - {\bf q}_{3}} + \omega^{-}_{{\bf q}_{2}} + \omega^{-}_{{\bf q}_{3}}} 
\biggr)\hspace{0.02cm}
{\mathcal K}^{\; a\, i_{2}\, i_{3}}_{-{\bf q}_{2} - {\bf q}_{3},\, {\bf q}_{2},\, {\bf q}_{3}}\,
{\mathcal K}^{\hspace{0.03cm}\ast\, a\, i\; i_{1}}_{-{\bf q} - {\bf q}_{1},\, {\bf q},\, {\bf q}_{1}}
\notag\\[1.5ex]
+\,
\frac{1}{4}\,\biggl[\biggl(
&\frac{1}{\omega^{l}_{{\bf q}_{3} - {\bf q}_{1}} - \omega^{-}_{{\bf q}_{3}} + \omega^{-}_{{\bf q}_{1}}}
\,+\,
\frac{1}{\omega^{l}_{{\bf q} - {\bf q}_{2}} - \omega^{-}_{{\bf q}} + \omega^{-}_{{\bf q}_{2}}}
\biggr)\hspace{0.02cm}
{\mathcal P}^{\; a\, i\, i_{2}}_{{\bf q} - {\bf q}_{2},\, {\bf q},\, {\bf q}_{2}}\, 
{\mathcal P}^{\hspace{0.03cm}\ast\, a\, i_{3}\, i_{1}}_{{\bf q}_{3} - {\bf q}_{1},\, {\bf q}_{3},\,  {\bf q}_{1}}
\notag\\[1.5ex]
+\,
\biggl(
&\frac{1}{\omega^{l}_{{\bf q}_{1} - {\bf q}_{3}} - \omega^{-}_{{\bf q}_{1}} + \omega^{-}_{{\bf q}_{3}}}
\,+\,
\frac{1}{\omega^{l}_{{\bf q}_{2} - {\bf q}} - \omega^{-}_{{\bf q}_{2}} + \omega^{-}_{{\bf q}}}
\biggr)\hspace{0.02cm}
{\mathcal P}^{\; a\, i_{1}\, i_{3}}_{{\bf q}_{1} - {\bf q}_{3},\, {\bf q}_{1},\, {\bf q}_{3}}\, 
{\mathcal P}^{\hspace{0.03cm}\ast\, a\, i_{2}\, i}_{{\bf q}_{2} - {\bf q},\, {\bf q}_{2},\, {\bf q}}
\notag\\[1.5ex]
-\,
\biggl(
&\frac{1}{\omega^{l}_{{\bf q}_{1} - {\bf q}_{2}} - \omega^{-}_{{\bf q}_{1}} + \omega^{-}_{{\bf q}_{2}}}
\,+\,
\frac{1}{\omega^{l}_{{\bf q}_{3} - {\bf q}} - \omega^{-}_{{\bf q}_{3}} + \omega^{-}_{{\bf q}}}
\biggr)\hspace{0.02cm}
{\mathcal P}^{\; a\, i_{1}\, i_{2}}_{{\bf q}_{1} - {\bf q}_{2},\, {\bf q}_{1},\, {\bf q}_{2}}\, 
{\mathcal P}^{\hspace{0.03cm}\ast\, a\, i_{3}\, i}_{{\bf q}_{3} - {\bf q},\, {\bf q}_{3},\, {\bf q}}
\notag\\[1.5ex]
-\, 
\biggl(
&\frac{1}{\omega^{l}_{{\bf q}_{2} - {\bf q}_{1}} - \omega^{-}_{{\bf q}_{2}} + \omega^{-}_{{\bf q}_{1}}}
\,+\,
\frac{1}{\omega^{l}_{{\bf q} - {\bf q}_{3}} - \omega^{-}_{{\bf q}} + \omega^{-}_{{\bf q}_{3}}}
\biggr)\hspace{0.02cm}
{\mathcal P}^{\; a\, i\; i_{3}}_{{\bf q} - {\bf q}_{3},\, {\bf q},\, {\bf q}_{3}}\,
{\mathcal P}^{\hspace{0.03cm}\ast\, a\, i_{2}\, i_{1}}_{{\bf q}_{2} - {\bf q}_{1},\, {\bf q}_{2},\, {\bf q}_{1}}
\biggr].
\notag
\end{align}
In deriving this expression we have used only the momentum conservation law in an elementary act of elastic scattering of two plasminos
\begin{equation}
{\bf q} + {\bf q}_1 = {\bf q}_{2} + {\bf q}_{3},
\label{eq:13m}
\end{equation}
which is true by virtue of availability of the corresponding $\delta$-function in (\ref{eq:13b}). The presented form (\ref{eq:13n}) of the complete effective amplitude $\mathscr{T}^{\,(2)}$ makes  practically obvious the validity of symmetry conditions
\[
\mathscr{T}^{\,(2)\, i\; i_{1}\; i_{2}\, i_{3}}_{\, {\bf q},\, {\bf q}_{1},\, {\bf q}_{2},\, {\bf q}_{3}} 
= 
 -\hspace{0.03cm}\mathscr{T}^{\,(2)\, i_{1}\, i\; i_{2}\, i_{3}}_{\, {\bf q}_{1},\, {\bf q},\, {\bf q}_{2},\, {\bf q}_{3}}
=
-\hspace{0.03cm}\mathscr{T}^{\,(2)\, i\; i_{1}\, i_{3}\, i_{2}}_{\, {\bf q},\, {\bf q}_{1},\, {\bf q}_{3},\, {\bf q}_{2}},
\quad\;
\mathscr{T}^{\hspace{0.03cm}\ast\hspace{0.03cm} (2)\, i\; i_{1}\, i_{2}\, i_{3}}_{\ {\bf q},\, {\bf q}_{1},\, {\bf q}_{2},\, {\bf q}_{3}}
=
\mathscr{T}^{\,(2)\, i_{2}\, i_{3}\, i\; i_{1}}_{\, {\bf q}_{2},\, {\bf q}_{3},\, {\bf q},\, {\bf q}_{1}}. 
\]
Thus, a role of the coefficient function $S^{\hspace{0.03cm}(2)}$ in fact is reduced to the total symmetrization of the effective amplitude $\widetilde{T}^{\,(2)}$. This involves the fulfillment of all necessary symmetry conditions without any using the resonance condition (\ref{eq:13v}). In other words, the effective amplitude $\widetilde{T}^{\,(2)}$ satisfies the symmetry conditions of the (\ref{eq:13x}) and (\ref{eq:13c}) type only on the resonance surface described by Eqs.\,(\ref{eq:13v}) and (\ref{eq:13m}) (that generally speaking can result in energy nonconservation), while accounting for the contribution with the function  $S^{\hspace{0.03cm}(2)}$ makes it possible to extend the symmetry conditions throughout the space of the vectors ${\bf q},\, {\bf q}_{1},\, {\bf q}_{2}$ and ${\bf q}_{3}$. In this case the resonance frequency difference $\Delta\hspace{0.02cm}\omega = \omega^{-}_{{\bf q}} + \omega^{-}_{{\bf q}_{1}} - \omega^{-}_{{\bf q}_{2}} - \omega^{-}_{{\bf q}_{3}}$ can be arbitrary and need not be small.


\section{Higher coefficient functions $J^{(n)\, a\, a_{1}\, i_{1}\, i_{2}}_{\, {\bf k},\, {\bf k}_{1},\, {\bf q}_{1},\, {\bf q}_{2}}$ and $R^{\,(n)\, i\, a_{1}\, a_{2}\, i_{1}}_{\; {\bf q},\, {\bf k}_{1},\, {\bf k}_{2},\, {\bf q}_{1}}$ for the canonical transformations (\ref{eq:3t}) and (\ref{eq:3y})}
\label{section_14}
\setcounter{equation}{0}

We are coming now to the problem of defining an explicit form of the third-order coefficient functions  $J^{(n)\, a\, a_{1}\, i_{1}\, i_{2}}_{\ {\bf k},\, {\bf k}_{1},\, {\bf q}_{1},\, {\bf q}_{2}}$ and $R^{\,(n)\, i\, a_{1}\, a_{2}\, i_{1}}_{\ {\bf q},\, {\bf k}_{1},\, {\bf k}_{2},\, {\bf q}_{1}},\, n = 1,\ldots,6$ entering into the canonical transformations of the bosonic (\ref{eq:3t}) and fermionic variables (\ref{eq:3y}). For this purpose, we return again to the fourth-order interaction Hamiltonian (\ref{eq:2g}), namely to its first term. This term is part of a more general expression
\[
\int\frac{d{\bf q}\hspace{0.03cm} d{\bf q}_{1}\hspace{0.02cm} d{\bf k}_{1}\hspace{0.02cm} d{\bf k}_{2}}{(2\pi)^{12}}
\]
\vspace{-0.5cm}
\begin{align}
\times\,
\biggl\{\frac{1}{2}\,\Bigl(&T^{\, (1)\, i\; i_{1}\, a_{1}\, a_{2}}_{\ {\bf q},\, {\bf q}_{1},\, {\bf k}_{1},\, {\bf k}_{2}}\, 
b^{\,\ast\, i}_{{\bf q}}\, b^{\;i_{1}}_{{\bf q}_{1}}\hspace{0.03cm} {a}^{\!\phantom{\ast} a_{1}}_{{\bf k}_{1}}\hspace{0.02cm}  
{a}^{\!\phantom{\ast} a_{2}}_{{\bf k}_{2}}
\,-\,
T^{\,\ast\, (1)\, i\, i_{1}\, a_{1}\, a_{2}}_{\ {\bf q},\, {\bf q}_{1},\, {\bf k}_{1},\, {\bf k}_{2}}\, 
b^{\, i}_{{\bf q}}\, b^{\,\ast\, i_{1}}_{{\bf q}_{1}}\, {a}^{\ast\, a_{1}}_{{\bf k}_{1}}\hspace{0.03cm}  
{a}^{\ast\, a_{2}}_{{\bf k}_{2}}\Bigr)\, 
(2\pi)^{3}\hspace{0.03cm}\delta({\bf q} - {\bf q}_{1} - {\bf k}_{1} - {\bf k}_{2})
\notag\\[1.5ex]
+\,
\Bigl(&T^{\, (11)\, i\, i_{1}\, a_{1}\, a_{2}}_{\ {\bf q},\, {\bf q}_{1},\, {\bf k}_{1},\, {\bf k}_{2}}\,
b^{\, i}_{{\bf q}}\, b^{\;i_{1}}_{{\bf q}_{1}}\,  {a}^{\ast\ \!\!a_{1}}_{{\bf k}_{1}}
{a}^{\!\phantom{\ast} a_{2}}_{{\bf k}_{2}}
\,-
T^{\,\ast\, (11)\, i\, i_{1}\, a_{1}\, a_{2}}_{\ {\bf q},\, {\bf q}_{1},\, {\bf k}_{1},\, {\bf k}_{2}}
b^{\,\ast\, i}_{{\bf q}}\, b^{\,\ast\, i_{1}}_{{\bf q}_{1}}\,  
{a}^{\!\phantom{\ast} a_{1}}_{{\bf k}_{1}}  {a}^{\ast\ \!\!a_{2}}_{{\bf k}_{2}}\Bigr)\, 
(2\pi)^{3}\hspace{0.03cm}\delta({\bf q} + {\bf q}_{1} - {\bf k}_{1} + {\bf k}_{2})
\notag\\[1.5ex]
+\;
&T^{\, (2)\, i\, i_{1}\, a_{1}\, a_{2}}_{\ {\bf q},\, {\bf q}_{1},\, {\bf k}_{1},\, {\bf k}_{2}}\, 
b^{\,\ast\, i}_{{\bf q}}\hspace{0.03cm} b^{\;i_{1}}_{{\bf q}_{1}}\, {a}^{\ast\ \!\!a_{1}}_{{\bf k}_{1}}\hspace{0.03cm}  
{a}^{\!\phantom{\ast} a_{2}}_{{\bf k}_{2}}\ 
(2\pi)^{3}\hspace{0.03cm}\delta({\bf q} - {\bf q}_{1} + {\bf k}_{1} - {\bf k}_{2})
\label{eq:14q}\\[1.5ex]
+\,
\frac{1}{2}\,\Bigl(&T^{\, (22)\, i\, i_{1}\, a_{1}\, a_{2}}_{\ {\bf q},\, {\bf q}_{1},\, {\bf k}_{1},\, {\bf k}_{2}}\, 
b^{\, i}_{{\bf q}}\, b^{\;i_{1}}_{{\bf q}_{1}}\hspace{0.03cm} {a}^{\ast\ \!\!a_{1}}_{{\bf k}_{1}}\hspace{0.03cm}  
{a}^{\ast\ \!\! a_{2}}_{{\bf k}_{2}}
\,-\,
T^{\,\ast\, (22)\, i\, i_{1}\, a_{1}\, a_{2}}_{\ {\bf q},\, {\bf q}_{1},\, {\bf k}_{1},\, {\bf k}_{2}}\, 
b^{\,\ast\, i}_{{\bf q}}\, b^{\,\ast\, i_{1}}_{{\bf q}_{1}}\hspace{0.03cm} {a}^{\!\phantom{\ast} a_{1}}_{{\bf k}_{1}}\hspace{0.03cm}  
{a}^{\!\phantom{\ast} a_{2}}_{{\bf k}_{2}}\Bigr)\,
(2\pi)^{3}\hspace{0.03cm}\delta({\bf q} + {\bf q}_{1} - {\bf k}_{1} - {\bf k}_{2})
\notag\\[1.5ex]
+\,
\frac{1}{2}\,\Bigl(&T^{\, (4)\, i\; i_{1}\, a_{1}\, a_{2}}_{\ {\bf q},\, {\bf q}_{1},\, {\bf k}_{1},\, {\bf k}_{2}}\,
b^{\, i}_{{\bf q}}\, b^{\; i_{1}}_{{\bf q}_{1}}\hspace{0.03cm}  {a}^{\!\phantom{\ast} a_{1}}_{{\bf k}_{1}}
{a}^{\!\phantom{\ast} a_{2}}_{{\bf k}_{2}}
\,-
T^{\,\ast\, (4)\, i\, i_{1}\, a_{1}\, a_{2}}_{\ {\bf q},\, {\bf q}_{1},\, {\bf k}_{1},\, {\bf k}_{2}}
b^{\,\ast\, i}_{{\bf q}}\, b^{\,\ast\, i_{1}}_{{\bf q}_{1}}\,  
{a}^{\ast\ \!\!a_{1}}_{{\bf k}_{1}}  {a}^{\ast\ \!\!a_{2}}_{{\bf k}_{2}}\Bigr)\, 
(2\pi)^{3}\hspace{0.03cm}\delta({\bf q} + {\bf q}_{1} + {\bf k}_{1} + {\bf k}_{2})\biggr\}.
\notag
\end{align}
It is evident that, even though the contributions with the functions $T^{\,(1)\, i\, i_{1}\, i_{2}\, i_{3}}_{{\bf q},\, {\bf q}_{1},\, {\bf q}_{2},\, {\bf q}_{3}}, T^{\, (11)\, i\, i_{1}\, a_{1}\, a_{2}}_{\ {\bf q},\, {\bf q}_{1},\, {\bf k}_{1},\, {\bf k}_{2}}$, $T^{\, (22)\, i\, i_{1}\, a_{1}\, a_{2}}_{\ {\bf q},\, {\bf q}_{1},\, {\bf k}_{1},\, {\bf k}_{2}}$ and $T^{\,(4)\, i\, i_{1}\, i_{2}\, i_{3}}_{{\bf q},\, {\bf q}_{1},\, {\bf q}_{2},\, {\bf q}_{3}}$ by virtue of the properties of the system under study equal zero, they are still generated by the canonical transformations (\ref{eq:3t}) and (\ref{eq:3y}). We determine the coefficient functions  $J^{(n)\, a\, a_{1}\, i_{1}\, i_{2}}_{\ {\bf k},\, {\bf k}_{1},\, {\bf q}_{1},\, {\bf q}_{2}}$ and $R^{\,(n)\, i\, a_{1}\, a_{2}\, i_{1}}_{\ {\bf q},\, {\bf k}_{1},\, {\bf k}_{2},\, {\bf q}_{1}}$ from the requirement of making these ``induced'' contributions vanish. As in the case considered in the previous section, this can be done unambiguously for the functions with $n = 1,\,3,\ldots,6$. The coefficient functions $J^{(2)\, a\, a_{1}\, i_{1}\, i_{2}}_{\ {\bf k},\, {\bf k}_{1},\, {\bf q}_{1},\, {\bf q}_{2}}$ and $R^{\,(2)\, i\, a_{1}\, a_{2}\, i_{1}}_{\ {\bf q},\, {\bf k}_{1},\, {\bf k}_{2},\, {\bf q}_{1}}$ require special consideration, which will be given just below.\\ 
\indent We will not give a detailed derivation of the coefficient functions 
$J^{(n)\, a\, a_{1}\, i_{1}\, i_{2}}_{\ {\bf k},\, {\bf k}_{1},\, {\bf q}_{1},\, {\bf q}_{2}}$ and $R^{\,(n)\, i\, a_{1}\, a_{2}\, i_{1}}_{\ {\bf q},\, {\bf k}_{1},\, {\bf k}_{2},\, {\bf q}_{1}}$ with $n = 1,\,3,\ldots,6$ by virtue of cumbersome calculations. They are similar to those in section \ref{section_13}. The complete expressions of these functions are presented in Appendix \ref{appendix_E}. Here we give their form in the hard thermal loop approximation. In this case we must set 
\[
T^{\, (1)\, i\, i_{1}\, a_{1}\, a_{2}}_{\ {\bf q},\, {\bf q}_{1},\, {\bf k}_{1},\, {\bf k}_{2}}
=
T^{\, (11)\, i\, i_{1}\, a_{1}\, a_{2}}_{\ {\bf q},\, {\bf q}_{1},\, {\bf k}_{1},\, {\bf k}_{2}}
=
T^{\, (22)\, i\, i_{1}\, a_{1}\, a_{2}}_{\ {\bf q},\, {\bf q}_{1},\, {\bf k}_{1},\, {\bf k}_{2}}
=
T^{\, (4)\, i\, i_{1}\, a_{1}\, a_{2}}_{\ {\bf q},\, {\bf q}_{1},\, {\bf k}_{1},\, {\bf k}_{2}} = 0
\]
and take into account the equalities (\ref{eq:9a}). In contrast to (\ref{eq:13s}), here, the situation is somewhat nontrivial. The most part of the coefficient functions in the HTL-approximation in view of the structure of the effective amplitudes (\ref{ap:E2})\,--\,(\ref{ap:E5}) vanishes, i.e.
\[
J^{\,(1)\, a_{1}\, a_{2}\, i\; i_{1}}_{\ {\bf k}_{1},\, {\bf k}_{2},\, {\bf q},\, {\bf q}_{1}}
=
J^{\,(3)\, a_{1}\, a_{2}\, i\; i_{1}}_{\ {\bf k}_{1},\, {\bf k}_{2},\, {\bf q},\, {\bf q}_{1}}
=
J^{\,(4)\, a_{1}\, a_{2}\, i\; i_{1}}_{\ {\bf k}_{1},\, {\bf k}_{2},\, {\bf q},\, {\bf q}_{1}}
=
J^{\,(6)\, a_{1}\, a_{2}\, i\; i_{1}}_{\ {\bf k}_{1},\, {\bf k}_{2},\, {\bf q},\, {\bf q}_{1}}
= 0,
\]
and as a consequence of the canonicity conditions in Appendices \ref{appendix_C} and \ref{appendix_D}, we also have
\[
R^{\,(4)\, i\, a_{1}\, a_{2}\, i_{1}}_{\ {\bf q},\, {\bf k}_{1},\, {\bf k}_{2},\, {\bf q}_{1}}
=
R^{\,(5)\, i\, a_{1}\, a_{2}\, i_{1}}_{\ {\bf q},\, {\bf k}_{1},\, {\bf k}_{2},\, {\bf q}_{1}}
=
R^{\,(6)\, i\, a_{1}\, a_{2}\, i_{1}}_{\ {\bf q},\, {\bf k}_{1},\, {\bf k}_{2},\, {\bf q}_{1}}
= 0.
\]
However there are three nontrivial coefficient functions 
\begin{align}
J^{\,(5)\, a_{1}\, a_{2}\, i\; i_{1}}_{\ {\bf k}_{1},\, {\bf k}_{2},\, {\bf q},\, {\bf q}_{1}}
\,=\;
-\,
\frac{1}{\omega^{-}_{{\bf q}}  - \omega^{-}_{{\bf q}_{1}} + \omega^{\hspace{0.02cm} l}_{{\bf k}_{1}} + \omega^{\hspace{0.02cm} l}_{{\bf k}_{2}}}\,
&\Biggl\{\frac{{\mathcal P}^{\hspace{0.03cm}\ast\, a_{2}\, j\,  i}_{{\bf k}_{2},\,{\bf k}_{2} + {\bf q},\, {\bf q}}\, 
{\mathcal P}^{\hspace{0.03cm}\ast\, a_{1}\, i_{1}\hspace{0.03cm} j}_{{\bf k}_{1},\, {\bf q}_{1},\, {\bf q}_{1} - {\bf k}_{1}}}
{\omega^{\hspace{0.02cm} l}_{{\bf k}_{2}} - \omega^{-}_{{\bf k}_{2} + {\bf q}}\! + \omega^{-}_{{\bf q}}}
\;-\; 
\frac{{\mathcal P}^{\hspace{0.03cm}\ast\, a_{1}\, j\,  i}_{{\bf k}_{1},\,{\bf k}_{1} + {\bf q},\, {\bf q}}\, 
{\mathcal P}^{\hspace{0.03cm}\ast\, a_{2}\, i_{1}\hspace{0.03cm} j}_{{\bf k}_{2},\, {\bf q}_{1},\, {\bf q}_{1} - {\bf k}_{2}}}
{\omega^{\hspace{0.02cm} l}_{{\bf k}_{2}} - \omega^{-}_{{\bf q}_{1}} + \omega^{-}_{{\bf q}_{1} - {\bf k}_{2}}}
\notag\\[1ex]
-\,2\,\Biggl(\hspace{0.02cm}
\frac{{\mathcal U}^{\hspace{0.03cm}\ast\, a_{1}\, a_{2}\,  a}_{\hspace{0.03cm} {\bf k}_{1}, {\bf k}_{2},\, -{\bf k}_{1} - 
{{\bf k}_{2}}}\, 
{\mathcal P}^{\; a\, i\; i_{1}}_{{\bf q} - {\bf q}_{1},\, {\bf q},\, {\bf q}_{1}}}
{\omega^{\hspace{0.02cm} l}_{{\bf q} - {\bf q}_{1}} - \omega^{-}_{{\bf q}} + \omega^{-}_{{\bf q}_{1}}}
\;+\;
&\frac{{\mathcal V}^{\hspace{0.03cm}\ast\, a\, a_{1}\, a_{2}}_{{\bf k}_{1} + {{\bf k}_{2}},\, {\bf k}_{1},\, {\bf k}_{2}}
{\mathcal P}^{\hspace{0.03cm}\ast\, a\, i_{1}\,  i}_{{\bf q}_{1} - {\bf q},\, {\bf q}_{1},\, {\bf q}}\,}
{\omega^{\hspace{0.02cm} l}_{{\bf q}_{1} - {\bf q}} - \omega^{-}_{{\bf q}_{1}} + \omega^{-}_{{\bf q}}}
\Biggr)\!\Biggr\}\, 
(2\pi)^{3}\hspace{0.03cm}\delta({\bf q} - {\bf q}_{1} + {\bf k}_{1} + {\bf k}_{2}),
\notag\\[3ex]
R^{(1)\; i\, a_{1}\, a_{2}\, i_{1}}_{\, {\bf q},\, {\bf k}_{1},\, {\bf k}_{2},\, {\bf q}_{1}}
\!=
-\hspace{0.03cm}\frac{1}{2}\,
\frac{1}{\omega^{-}_{{\bf q}}  - \omega^{-}_{{\bf q}_{1}} - \omega^{\hspace{0.02cm} l}_{{\bf k}_{1}} - \omega^{\hspace{0.02cm} l}_{{\bf k}_{2}}}\,
&\Biggl\{\frac{{\mathcal P}^{\; a_{2}\, j\,  i_{1}}_{{\bf k}_{2},\,{\bf k}_{2} + {\bf q}_{1},\, {\bf q}_{1}}\, 
{\mathcal P}^{\; a_{1}\, i\, j}_{{\bf k}_{1},\, {\bf q},\, {\bf q} - {\bf k}_{1}}}
{\omega^{\hspace{0.02cm} l}_{{\bf k}_{2}} - \omega^{-}_{{\bf k}_{2} + {\bf q}_{1}}\! + \omega^{-}_{{\bf q}_{1}}}
\;+\; 
\frac{{\mathcal P}^{\; a_{1}\, j\,  i_{1}}_{{\bf k}_{1},\,{\bf k}_{1} + {\bf q}_{1},\, {\bf q}_{1}}\, 
{\mathcal P}^{\; a_{2}\, i\, j}_{{\bf k}_{2},\, {\bf q},\, {\bf q} - {\bf k}_{2}}}
{\omega^{\hspace{0.02cm} l}_{{\bf k}_{1}} -  \omega^{-}_{{\bf k}_{1} + {\bf q}_{1}}\! + \omega^{-}_{{\bf q}_{1}}}
\notag\\[1ex]
-\,2\,\Biggl(\hspace{0.02cm}
\frac{{\mathcal U}^{\; a_{1}\, a_{2}\,  a}_{\hspace{0.03cm} {\bf k}_{1}, {\bf k}_{2},\, -{\bf k}_{1} - {{\bf k}_{2}}}\, 
{\mathcal P}^{\hspace{0.03cm}\ast\, a\, i_{1}\, i}_{{\bf q}_{1} - {\bf q},\, {\bf q}_{1},\, {\bf q}}}
{\omega^{\hspace{0.02cm} l}_{{\bf k}_{1}} + \omega^{\hspace{0.02cm} l}_{{\bf k}_{2}} + \omega^{\hspace{0.02cm} l}_{-{\bf k}_{1} - {\bf k}_{2}}}
\;+\;
&\frac{{\mathcal V}^{\; a\, a_{1}\, a_{2}}_{{\bf k}_{1} + {{\bf k}_{2}},\, {\bf k}_{1},\, {\bf k}_{2}}
{\mathcal P}^{\; a\, i\; i_{1}}_{{\bf q} - {\bf q}_{1},\, {\bf q},\, {\bf q}_{1}}\,}
{\omega^{\hspace{0.02cm} l}_{{\bf k}_{1} + {\bf k}_{2}} - \omega^{\hspace{0.02cm} l}_{{\bf k}_{1}} - \omega^{\hspace{0.02cm} l}_{{\bf k}_{2}}}
\Biggr)\!\Biggr\}\, 
(2\pi)^{3}\hspace{0.03cm}\delta({\bf q} - {\bf q}_{1} - {\bf k}_{1} - {\bf k}_{2}),
\notag\\[3ex]
R^{(3)\; i\, a_{1}\, a_{2}\, i_{1}}_{\, {\bf q},\, {\bf k}_{1},\, {\bf k}_{2},\, {\bf q}_{1}}
=\,
\frac{1}{2}\,
\frac{1}{\omega^{-}_{{\bf q}}  - \omega^{-}_{{\bf q}_{1}} + \omega^{\hspace{0.02cm} l}_{{\bf k}_{1}} + \omega^{\hspace{0.02cm} l}_{{\bf k}_{2}}}\,
&\Biggl\{\frac{{\mathcal P}^{\hspace{0.03cm}\ast\, a_{2}\, j\,  i}_{{\bf k}_{2},\,{\bf k}_{2} + {\bf q},\, {\bf q}}\, 
{\mathcal P}^{\hspace{0.03cm}\ast\, a_{1}\, i_{1}\, j}_{{\bf k}_{1},\, {\bf q}_{1},\, {\bf q}_{1} - {\bf k}_{1}}}
{\omega^{\hspace{0.02cm} l}_{{\bf k}_{1}} - \omega^{-}_{{\bf q}_{1}} + \omega^{-}_{ {\bf q}_{1} - {\bf k}_{1}}}
\;+\; 
\frac{{\mathcal P}^{\hspace{0.03cm}\ast\, a_{1}\, j\,  i}_{{\bf k}_{1},\,{\bf k}_{1} + {\bf q},\, {\bf q}}\, 
{\mathcal P}^{\hspace{0.03cm}\ast\, a_{2}\, i_{1}\, j}_{{\bf k}_{2},\, {\bf q}_{1},\, {\bf q}_{1} - {\bf k}_{2}}}
{\omega^{\hspace{0.02cm} l}_{{\bf k}_{2}} - \omega^{-}_{{\bf q}_{1}} + \omega^{-}_{{\bf q}_{1} - {\bf k}_{2}}}
\notag\\[1ex]
+\,2\,\Biggl(\hspace{0.02cm}
\frac{{\mathcal U}^{\hspace{0.03cm}\ast\, a_{1}\, a_{2}\,  a}_{\hspace{0.03cm} {\bf k}_{1}, {\bf k}_{2},\, -{\bf k}_{1} - 
{{\bf k}_{2}}}\, 
{\mathcal P}^{\; a\, i\; i_{1}}_{{\bf q} - {\bf q}_{1},\, {\bf q},\, {\bf q}_{1}}}
{\omega^{\hspace{0.02cm} l}_{{\bf k}_{1}} + \omega^{\hspace{0.02cm} l}_{{\bf k}_{2}} + \omega^{\hspace{0.02cm} l}_{-{\bf k}_{1} - {\bf k}_{2}}}
\;+\;
&\frac{{\mathcal V}^{\hspace{0.03cm}\ast\, a\, a_{1}\, a_{2}}_{{\bf k}_{1} + {{\bf k}_{2}},\, {\bf k}_{1},\, {\bf k}_{2}}
{\mathcal P}^{\hspace{0.03cm}\ast\, a\, i_{1}\,  i}_{{\bf q}_{1} - {\bf q},\, {\bf q}_{1},\, {\bf q}}\, 
}
{\omega^{\hspace{0.02cm} l}_{{\bf k}_{1} + {\bf k}_{2}}\! - \omega^{\hspace{0.02cm} l}_{{\bf k}_{1}} - \omega^{\hspace{0.02cm} l}_{{\bf k}_{2}}}
\Biggr)\!\Biggr\}\, 
(2\pi)^{3}\hspace{0.03cm}\delta({\bf q} - {\bf q}_{1} + {\bf k}_{1} + {\bf k}_{2}).
\notag
\end{align}
The fact that these functions are non-vanishing suggests their possible connection with the processes of plasmino-antiplasmino annihilation into two plasmons and production of plasmino-antiplasmino pair by merging two plasmons. As was mentioned, the processes with antiplasmino branch of fermionic oscillations in the present paper are not considered (see discussion in Conclusion).\\
\indent Now we pass to defining the coefficient functions $J^{(2)\, a_{1}\, a_{2}\, i\, i_{1}}_{\, {\bf k}_{1},\, {\bf k}_{2},\, {\bf q},\, {\bf q}_{1}}$ and $R^{\,(2)\, i\, a_{1}\, a_{2}\, i_{1}}_{\ {\bf q},\, {\bf k}_{1},\, {\bf k}_{2},\, {\bf q}_{1}}$. Recall that the former function enter into the fourth-order effective Hamiltonian describing the elastic scattering of plasmino off plasmon:  
\begin{equation}
{\mathcal H}^{(4)}_{qg\rightarrow qg} 
=
\int\frac{d{\bf q}\hspace{0.03cm} d{\bf q}_{1}\hspace{0.02cm} d{\bf k}_{1}\hspace{0.02cm} d{\bf k}_{2}}{(2\pi)^{12}}
\label{eq:14w}
\end{equation}
\[
\times\,\Bigl\{\!\hspace{0.02cm}
\bigl(\omega^{-}_{{\bf q}} + \omega^{\hspace{0.02cm} l}_{{\bf k}_{1}} - \omega^{-}_{{\bf q}_{1}} - \omega^{\hspace{0.02cm} l}_{{\bf k}_{2}}\bigr)
J^{\,(2)\, a_{1}\, a_{2}\, i\, i_{1}}_{\; {\bf k}_{1},\, {\bf k}_{2},\, {\bf q},\, {\bf q}_{1}}
\hspace{0.03cm}+\hspace{0.03cm}
\widetilde{T}^{\,(2)\, i\, i_{1}\, a_{1}\, a_{2}}_{\; {\bf q},\, {\bf q}_{1},\, {\bf k}_{1},\, {\bf k}_{2}}\, 
(2\pi)^{3}\hspace{0.03cm}\delta({\bf q} + {\bf k}_1 - {\bf q}_{1} - {\bf k}_{2})
\Bigr\}
f^{\,\ast\, i}_{{\bf q}} f^{\;i_{1}}_{{\bf q}_{1}}\hspace{0.03cm} c^{\ast\ \!\!a_{1}}_{{\bf k}_{1}} c^{\hspace{0.03cm}a_{2}}_{{\bf k}_{2}},
\]
where the effective amplitude $\widetilde{T}^{(2)}$ is defined by the expression (\ref{eq:5t}). As stated above an explicit form of this function in contrast to the previous ones is not unambiguous. Unlike the coefficient function $S^{\,(2)}$ from section \ref{section_13}, for the functions $J^{\,(2)}$ and $R^{\,(2)}$ we have self-consistent system of four canonicity conditions (\ref{ap:C1b}), (\ref{ap:C3a}), (\ref{ap:C4b}) and (\ref{ap:D1b}) connecting these two functions $J^{\,(2)}$ and $R^{\,(2)}$ among themselves. These conditions are insufficient for unequivocal determination of $J^{\,(2)}$ and $R^{\,(2)}$. Let us write out these relations as a system of four functional equations:
\begin{equation}
\begin{split}
&J^{\,(2)\, a_{1}\, a_{2}\, i\; i_{1}}_{\ {\bf k}_{1},\, {\bf k}_{2},\, {\bf q},\, {\bf q}_{1}} 
+
J^{\hspace{0.025cm} \ast\hspace{0.025cm} (2)\hspace{0.03cm} a_{2}\, a_{1}\, i_{1}\, i}_{\ {\bf k}_{2},\, {\bf k}_{1},\, {\bf q}_{1},\, {\bf q}} 
=
\Psi^{\hspace{0.03cm} (22)\hspace{0.03cm} a_{1}\, a_{2}\, i\, i_{1}}_{\ {\bf k}_{1},\, {\bf k}_{2},\, {\bf q},\, {\bf q}_{1}},
\\[2.5ex]
&J^{\,(2)\, a_{1}\, a_{2}\, i\; i_{1}}_{\ {\bf k}_{1},\, {\bf k}_{2},\, {\bf q},\, {\bf q}_{1}} 
-
R^{\,(2)\, i\, a_{1}\, a_{2}\, i_{1}}_{\ {\bf q},\, {\bf k}_{1},\, {\bf k}_{2},\, {\bf q}_{1}}
=
\Psi^{\,(2)\,  a_{1}\, a_{2}\, i\; i_{1}}_{\ {\bf k}_{1},\, {\bf k}_{2},\, {\bf q},\, {\bf q}_{1}},
\\[2.5ex]
&J^{\,(2)\, a_{1}\, a_{2}\, i\; i_{1}}_{\ {\bf k}_{1},\, {\bf k}_{2},\, {\bf q},\, {\bf q}_{1}} 
+
R^{\hspace{0.03cm}\ast\hspace{0.03cm} (2)\, i_{1}\, a_{2}\, a_{1}\, i}_{\ {\bf q}_{1},\, {\bf k}_{2},\, {\bf k}_{1},\, {\bf q}_{1}}
=
\Upsilon^{\,(2)\,  a_{1}\, a_{2}\, i\; i_{1}}_{\ {\bf k}_{1},\, {\bf k}_{2},\, {\bf q},\, {\bf q}_{1}},
\\[2.5ex]
&R^{\,(2)\, i\, a_{1}\, a_{2}\, i_{1}}_{\ {\bf q},\, {\bf k}_{1},\, {\bf k}_{2},\, {\bf q}_{1}}
+
R^{\hspace{0.025cm}\ast\hspace{0.025cm} (2)\hspace{0.03cm} i_{1}\, a_{2}\, a_{1}\, i}_{\ {\bf q}_{1},\, {\bf k}_{2},\, {\bf k}_{1},\, {\bf q}}
=
\Upsilon^{\,(22)\,  a_{1}\, a_{2}\, i\, i_{1}}_{\ {\bf k}_{1},\, {\bf k}_{2},\, {\bf q},\, {\bf q}_{1}},
\end{split}
\label{eq:14e}
\end{equation}
where we set by definition
\begin{subequations} 
\label{eq:14r}
\begin{align}
\Psi^{\,(2)\, a_{1}\, a_{2}\, i\, i_{1}}_{\ {\bf k}_{1},\, {\bf k}_{2},\, {\bf q},\, {\bf q}_{1}}
\equiv
2\!\int\!\frac{d\hspace{0.02cm}{\bf k}^{\hspace{0.02cm} \prime}}{(2\pi)^{3}}\,\Bigl[\hspace{0.03cm}
V^{(1)\,a_{1}\, a_{2}\, a^{\prime}}_{\ {\bf k}_{1},\, {\bf k}_{2},\, {\bf k}^{\prime}}\, &F^{(2)\, a^{\prime}\, i\, i_{1}}_{\ {\bf k}^{\prime},\, {\bf q},\, {\bf q}_{1}}
-
V^{\hspace{0.03cm}\ast\hspace{0.03cm} (1)\,a_{2}\, a_{1}\, a^{\prime}}_{\ {\bf k}_{2},\, {\bf k}_{1},\, {\bf k}^{\prime}} F^{\hspace{0.03cm}\ast\hspace{0.03cm} (2)\, a^{\prime}\, i_{1}\, i}_{\ {\bf k}^{\prime},\, {\bf q}_{1},\, {\bf q}}\hspace{0.03cm}\Bigr]
\label{eq:14ra}\\[1ex]
+\!
\int&\!\frac{d{\bf q}^{\prime}}{(2\pi)^{3}}\,\Bigl[\hspace{0.03cm}4\hspace{0.02cm}
F^{(1)\, a_{1}\, i_{1}\, i^{\prime}}_{\ {\bf k}_{1},\, {\bf q}_{1},\, {\bf q}^{\prime}}\, F^{\hspace{0.03cm}\ast\hspace{0.03cm} (1)\, a_{2}\, i\, i^{\prime}}_{\ {\bf k}_{2},\, {\bf q},\, {\bf q}^{\prime}}
+
F^{(2)\, a_{1}\, i^{\prime}\, i_{1}}_{\ {\bf k}_{1},\, {\bf q}^{\prime},\, {\bf q}_{1}}\, F^{\hspace{0.03cm}\ast\hspace{0.03cm} (2)\, a_{2}\, i^{\prime}\, i}_{\ {\bf k}_{2},\,  {\bf q}^{\prime},\, {\bf q}}\hspace{0.03cm}\Bigr],
\notag\\[1ex]
\Psi^{\,(22)\, a_{1}\, a_{2}\, i\, i_{1}}_{\ {\bf k}_{1},\, {\bf k}_{2},\, {\bf q},\, {\bf q}_{1}}
\equiv
\int\!\frac{d{\bf q}^{\prime}}{(2\pi)^{3}}\,\Bigl[\hspace{0.02cm}
4\hspace{0.01cm}F^{(1)\, a_{1}\, i_{1}\, i^{\prime}}_{\ {\bf k}_{1},\, {\bf q}_{1},\, {\bf q}^{\prime}}\, 
&F^{\hspace{0.03cm}\ast\hspace{0.03cm} (1)\, a_{2}\, i\, i^{\prime}}_{\ {\bf k}_{2},\, {\bf q},\, {\bf q}^{\prime}}
\!-
F^{(2)\, a_{1}\, i\, i^{\prime}}_{\ {\bf k}_{1},\, {\bf q},\, {\bf q}^{\prime}}\, 
F^{\hspace{0.03cm}\ast\hspace{0.03cm} (2)\, a_{2}\, i_{1}\, i^{\prime}}_{\ {\bf k}_{2},\, {\bf q}_{1},\, {\bf q}^{\prime}}
\label{eq:14rb}\\[1ex]
&+
F^{(2)\, a_{1}\, i^{\prime}\, i_{1}}_{\ {\bf k}_{1},\, {\bf q}^{\prime},\, {\bf q}_{1}}\, F^{\hspace{0.03cm}\ast\hspace{0.03cm} (2)\, a_{2}\, i^{\prime}\, i}_{\ {\bf k}_{2}, \, {\bf q}^{\prime},\, {\bf q}}
\!-
4\hspace{0.01cm}F^{(3)\, a_{1}\, i\, i^{\prime}}_{\ {\bf k}_{1},\, {\bf q},\, {\bf q}^{\prime}}\, F^{\hspace{0.03cm}\ast\hspace{0.03cm} (3)\, a_{2}\, i_{1}\, i^{\prime}}_{\ {\bf k}_{2},\, {\bf q}_{1},\, {\bf q}^{\prime}}\hspace{0.02cm}\Bigr],
\notag\\[1ex]
\Upsilon^{\,(2)\, a_{1}\, a_{2}\, i\, i_{1}}_{\ {\bf k}_{1},\, {\bf k}_{2},\, {\bf q},\, {\bf q}_{1}}
\equiv
2\!\int\!\frac{d\hspace{0.02cm}{\bf k}^{\hspace{0.02cm} \prime}}{(2\pi)^{3}}\,\Bigl[\hspace{0.03cm}
V^{(1)\,a_{1}\, a_{2}\, a^{\prime}}_{\ {\bf k}_{1},\, {\bf k}_{2},\, {\bf k}^{\prime}}
&F^{(2)\, a^{\prime}\, i\, i_{1}}_{\ {\bf k}^{\prime},\, {\bf q},\, {\bf q}_{1}}
-
V^{\hspace{0.03cm}\ast\hspace{0.03cm} (1)\,a_{2}\, a_{1}\, a^{\prime}}_{\ {\bf k}_{2},\, {\bf k}_{1},\, {\bf k}^{\prime}} 
F^{\hspace{0.03cm}\ast\hspace{0.03cm} (2)\, a^{\prime}\, i_{1}\, i}_{\ {\bf k}^{\prime},\, {\bf q}_{1},\, {\bf q}}\hspace{0.03cm}\Bigr]
\label{eq:14rc}\\[1ex]
-\!
\int&\!\frac{d{\bf q}^{\prime}}{(2\pi)^{3}}\,\Bigl[\hspace{0.03cm}
4\hspace{0.02cm}F^{(3)\, a_{1}\, i\, i^{\prime}}_{\ {\bf k}_{1},\, {\bf q},\, {\bf q}^{\prime}}\, 
F^{\hspace{0.03cm}\ast\hspace{0.03cm} (3)\, a_{2}\, i_{1}\, i^{\prime}}_{\ {\bf k}_{2},\,  {\bf q}_{1},\, {\bf q}^{\prime}}
\!+
F^{(2)\, a_{1}\, i\, i^{\prime}}_{\ {\bf k}_{1},\, {\bf q},\, {\bf q}^{\prime}}\, 
F^{\hspace{0.03cm}\ast\hspace{0.03cm} (2)\, a_{2}\, i_{1}\, i^{\prime}}_{\ {\bf k}_{2},\, {\bf q}_{1},\, {\bf q}^{\prime}}\hspace{0.03cm}\Bigr],
\notag\\[1ex]
\Upsilon^{\hspace{0.03cm}(22)\hspace{0.03cm} a_{1}\, a_{2}\, i\, i_{1}}_{\ {\bf k}_{1},\, {\bf k}_{2},\, {\bf q},\, {\bf q}_{1}}\!
\equiv\!
-\!\!\int\!\frac{d{\bf q}^{\prime}}{(2\pi)^{3}}\,\Bigl[\hspace{0.02cm}
4\hspace{0.01cm}F^{(1)\, a_{1}\, i_{1}\, i^{\prime}}_{\ {\bf k}_{1},\, {\bf q}_{1},\, {\bf q}^{\prime}}\, 
&F^{\hspace{0.03cm}\ast\hspace{0.03cm} (1)\, a_{2}\, i\, i^{\prime}}_{\ {\bf k}_{2},\, {\bf q},\, {\bf q}^{\prime}}
\!+
F^{(2)\, a_{1}\, i\, i^{\prime}}_{\ {\bf k}_{1},\, {\bf q},\, {\bf q}^{\prime}}\, 
F^{\hspace{0.03cm}\ast\hspace{0.03cm} (2)\, a_{2}\, i_{1}\hspace{0.03cm} i^{\prime}}_{\ {\bf k}_{2},\, {\bf q}_{1},\, {\bf q}^{\prime}}\!\!
\label{eq:14rd}\\[1ex]
&+
F^{(2)\, a_{1}\, i^{\prime}\, i_{1}}_{\ {\bf k}_{1},\, {\bf q}^{\prime},\, {\bf q}_{1}}\, F^{\hspace{0.03cm}\ast\hspace{0.03cm} (2)\, a_{2}\, i^{\prime}\, i}_{\ {\bf k}_{2}, \, {\bf q}^{\prime},\, {\bf q}}
+
4\hspace{0.01cm}F^{(3)\, a_{1}\, i\, i^{\prime}}_{\ {\bf k}_{1},\, {\bf q},\, {\bf q}^{\prime}}\, F^{\hspace{0.03cm}\ast\hspace{0.03cm} (3)\, a_{2}\, i_{1}\, i^{\prime}}_{\ {\bf k}_{2},\, {\bf q}_{1},\, {\bf q}^{\prime}}\hspace{0.02cm}\Bigr].
\notag
\end{align}
\end{subequations} 
In the system of four equations (\ref{eq:14e}) only three ones are independent. A direct check shows that the functions (\ref{eq:14ra})\,--\,(\ref{eq:14rd}) possess the following properties:
\begin{align}
&\Upsilon^{\,(2)\,  a_{1}\, a_{2}\, i\; i_{1}}_{\ {\bf k}_{1},\, {\bf k}_{2},\, {\bf q},\, {\bf q}_{1}}
-
\Psi^{\,(2)\,  a_{1}\, a_{2}\, i\; i_{1}}_{\ {\bf k}_{1},\, {\bf k}_{2},\, {\bf q},\, {\bf q}_{1}}
=\,
\Upsilon^{\hspace{0.03cm}(22)\, a_{1}\, a_{2}\, i\, i_{1}}_{\ {\bf k}_{1},\, {\bf k}_{2},\, {\bf q},\, {\bf q}_{1}},
\notag\\[1.7ex]
&\Upsilon^{\,(2)\, a_{1}\, a_{2}\, i\; i_{1}}_{\ {\bf k}_{1},\, {\bf k}_{2},\, {\bf q},\, {\bf q}_{1}}
\,+
\Psi^{\hspace{0.02cm}\ast\hspace{0.03cm} (2)\hspace{0.03cm} a_{2}\, a_{1}\, i_{1}\, i}_{\ {\bf k}_{2},\, {\bf k}_{1},\, {\bf q}_{1},\, {\bf q}}
=
\Psi^{\hspace{0.03cm}(22)\hspace{0.03cm} a_{1}\, a_{2}\, i\, i_{1}}_{\ {\bf k}_{1},\, {\bf k}_{2},\, {\bf q},\, {\bf q}_{1}},
\notag\\[1.7ex]
&\Upsilon^{\hspace{0.02cm}\ast\hspace{0.03cm} (2)\hspace{0.03cm} a_{2}\, a_{1}\, i_{1}\, i}_{\ {\bf k}_{2},\, {\bf k}_{1},\, {\bf q}_{1},\, {\bf q}}
\,+
\Psi^{\, (2)\,  a_{1}\, a_{2}\, i\; i_{1}}_{\ {\bf k}_{1},\, {\bf k}_{2},\, {\bf q},\, {\bf q}_{1}}
=
\Psi^{\,(22)\, a_{1}\, a_{2}\, i\, i_{1}}_{\ {\bf k}_{1},\, {\bf k}_{2},\, {\bf q},\, {\bf q}_{1}},
\label{eq:14t}\\[1.7ex]
&\Upsilon^{\hspace{0.02cm}\ast\hspace{0.03cm} (2)\hspace{0.03cm} a_{2}\, a_{1}\, i_{1}\, i}_{\ {\bf k}_{2},\, {\bf k}_{1},\, {\bf q}_{1},\, {\bf q}}
\,-
\Psi^{\hspace{0.02cm}\ast\hspace{0.03cm} (2)\hspace{0.03cm}  a_{2}\, a_{1}\, i_{1}\, i}_{\ {\bf k}_{2},\, {\bf k}_{1},\, {\bf q}_{1},\, {\bf q}}
=
\Upsilon^{\hspace{0.03cm} (22)\, a_{1}\, a_{2}\, i\, i_{1}}_{\ {\bf k}_{1},\, {\bf k}_{2},\, {\bf q},\, {\bf q}_{1}},
\notag\\[1.7ex]
&\Psi^{\hspace{0.03cm} (22)\, a_{1}\, a_{2}\, i\, i_{1}}_{\ {\bf k}_{1},\, {\bf k}_{2},\, {\bf q},\, {\bf q}_{1}}
\,=
\Psi^{\hspace{0.02cm}\ast\hspace{0.03cm} (22)\, a_{2}\, a_{1}\, i_{1}\, i}_{\ {\bf k}_{2},\, {\bf k}_{1},\, {\bf q}_{1},\, 
{\bf q}},
\notag\\[1.7ex]
&\Upsilon^{\hspace{0.03cm} (22)\, a_{1}\, a_{2}\, i\, i_{1}}_{\ {\bf k}_{1},\, {\bf k}_{2},\, {\bf q},\, {\bf q}_{1}}
\,=
\Upsilon^{\hspace{0.02cm}\ast\hspace{0.03cm} (22)\, a_{2}\, a_{1}\, i_{1}\, i}_{\ {\bf k}_{2},\, {\bf k}_{1},\, {\bf q}_{1},\, 
{\bf q}}.
\notag
\end{align}
Useful consequences of the relations (\ref{eq:14t}) are the ones
\begin{align}
&\Psi^{\,(2)\,  a_{1}\, a_{2}\, i\; i_{1}}_{\ {\bf k}_{1},\, {\bf k}_{2},\, {\bf q},\, {\bf q}_{1}}
\,-\,
\Psi^{\hspace{0.02cm}\ast\hspace{0.03cm} (2)\hspace{0.03cm} a_{2}\, a_{1}\, i_{1}\, i}_{\ {\bf k}_{2},\, {\bf k}_{1},\, {\bf q}_{1},\, {\bf q}}
\,=
\Upsilon^{\,(2)\, a_{1}\, a_{2}\, i\; i_{1}}_{\ {\bf k}_{1},\, {\bf k}_{2},\, {\bf q},\, {\bf q}_{1}}
\,-
\Upsilon^{\hspace{0.02cm}\ast\hspace{0.02cm} (2)\hspace{0.03cm} a_{2}\, a_{1}\, i_{1}\, i}_{\ {\bf k}_{2},\, {\bf k}_{1},\, {\bf q}_{1},\, {\bf q}},
\notag\\[1.7ex]
&\Upsilon^{\,(2)\, a_{1}\, a_{2}\, i\; i_{1}}_{\ {\bf k}_{1},\, {\bf k}_{2},\, {\bf q},\, {\bf q}_{1}}
\,+\,
\Upsilon^{\hspace{0.02cm}\ast\hspace{0.03cm} (2)\hspace{0.03cm} a_{2}\, a_{1}\, i_{1}\, i}_{\ {\bf k}_{2},\, {\bf k}_{1},\, {\bf q}_{1},\, {\bf q}}
\,=
\Psi^{\hspace{0.02cm} (22)\hspace{0.03cm} a_{1}\, a_{2}\, i\, i_{1}}_{\ {\bf k}_{1},\, {\bf k}_{2},\, {\bf q},\, {\bf q}_{1}}
\,+\,
\Upsilon^{\hspace{0.03cm} (22)\hspace{0.03cm} a_{1}\, a_{2}\, i\, i_{1}}_{\ {\bf k}_{1},\, {\bf k}_{2},\, {\bf q},\, {\bf q}_{1}},
\label{eq:14y}\\[1.7ex]
&\Psi^{\,(2)\, a_{1}\, a_{2}\, i\, i_{1}}_{\ {\bf k}_{1},\, {\bf k}_{2},\, {\bf q},\, {\bf q}_{1}}
\,+\,
\Psi^{\hspace{0.02cm}\ast\hspace{0.03cm} (2)\, a_{2}\, a_{1}\, i_{1}\, i}_{\ {\bf k}_{2},\, {\bf k}_{1},\, {\bf q}_{1},\, {\bf q}}
\,=
\Psi^{\hspace{0.03cm} (22)\hspace{0.03cm} a_{1}\, a_{2}\, i\, i_{1}}_{\ {\bf k}_{1},\, {\bf k}_{2},\, {\bf q},\, {\bf q}_{1}}
\,-\,
\Upsilon^{\hspace{0.03cm} (22)\hspace{0.03cm} a_{1}\, a_{2}\, i\, i_{1}}_{\ {\bf k}_{1},\, {\bf k}_{2},\, {\bf q},\, {\bf q}_{1}}.
\notag
\end{align}
\indent We will seek the coefficient functions $J^{\,(2)}$ and $R^{\,(2)}$ in the most general form:
\begin{equation}
J^{\,(2)\, a_{1}\, a_{2}\, i\; i_{1}}_{\ {\bf k}_{1},\, {\bf k}_{2},\, {\bf q},\, {\bf q}_{1}} =
\vspace{0.1cm}
\label{eq:14u}
\end{equation}
\[ 
x\hspace{0.03cm} \Psi^{\hspace{0.03cm} (22)\, a_{1}\, a_{2}\, i\, i_{1}}_{\ {\bf k}_{1},\, {\bf k}_{2},\, {\bf q},\, {\bf q}_{1}}
+
y\hspace{0.03cm} \Upsilon^{\hspace{0.03cm} (22)\, a_{1}\, a_{2}\, i\, i_{1}}_{\ {\bf k}_{1},\, {\bf k}_{2},\, {\bf q},\, {\bf q}_{1}}
+
\alpha\hspace{0.03cm}\Psi^{\,(2)\, a_{1}\, a_{2}\, i\; i_{1}}_{\ {\bf k}_{1},\, {\bf k}_{2},\, {\bf q},\, {\bf q}_{1}}
+
\beta\hspace{0.03cm}\Upsilon^{\,(2)\, a_{1}\, a_{2}\, i\; i_{1}}_{\ {\bf k}_{1},\, {\bf k}_{2},\, {\bf q},\, {\bf q}_{1}}
+
\gamma\hspace{0.03cm}\Psi^{\hspace{0.02cm}\ast\hspace{0.02cm} (2)\hspace{0.03cm} a_{2}\, a_{1}\, i_{1}\, i}_{\ {\bf k}_{2},\, {\bf k}_{1},\, {\bf q}_{1},\, {\bf q}}
+
\delta\hspace{0.03cm}\Upsilon^{\hspace{0.02cm}\ast\hspace{0.02cm} (2)\hspace{0.03cm} a_{2}\, a_{1}\, i_{1}\, i}_{\ {\bf k}_{2},\, {\bf k}_{1},\, {\bf q}_{1},\, {\bf q}},  
\vspace{0.3cm}
\]
\begin{equation}
R^{\,(2)\, i\, a_{1}\, a_{2}\, i_{1}}_{\ {\bf q},\, {\bf k}_{1},\, {\bf k}_{2},\, {\bf q}_{1}} =
\label{eq:14i}
\vspace{0.1cm}
\end{equation}
\[ 
\hat{x}\hspace{0.03cm} \Psi^{\hspace{0.03cm} (22)\hspace{0.03cm} a_{1}\, a_{2}\, i\, i_{1}}_{\ {\bf k}_{1},\, {\bf k}_{2},\, {\bf q},\, {\bf q}_{1}}
+
\hat{y}\hspace{0.03cm} \Upsilon^{\hspace{0.03cm} (22)\hspace{0.03cm} a_{1}\, a_{2}\, i\, i_{1}}_{\ {\bf k}_{1},\, {\bf k}_{2},\, {\bf q},\, {\bf q}_{1}}
+
\hat{\alpha}\hspace{0.03cm}\Psi^{\,(2)\, a_{1}\, a_{2}\, i\; i_{1}}_{\; {\bf k}_{1},\, {\bf k}_{2},\, {\bf q},\, {\bf q}_{1}}
+
\hat{\beta}\hspace{0.03cm}\Upsilon^{\,(2)\, a_{1}\, a_{2}\, i\; i_{1}}_{\; {\bf k}_{1},\, {\bf k}_{2},\, {\bf q},\, {\bf q}_{1}}
+
\hat{\gamma}\hspace{0.03cm}\Psi^{\hspace{0.02cm}\ast\hspace{0.02cm} (2)\hspace{0.03cm} a_{2}\, a_{1}\, i_{1}\, i}_{\ {\bf k}_{2},\, {\bf k}_{1},\, {\bf q}_{1},\, {\bf q}}
+
\hat{\delta}\hspace{0.03cm}\Upsilon^{\hspace{0.02cm}\ast\hspace{0.02cm} (2)\hspace{0.03cm} a_{2}\, a_{1}\, i_{1}\, i}_{\ {\bf k}_{2},\, {\bf k}_{1},\, {\bf q}_{1},\, {\bf q}},  
\]
where $x,\,y,\,\alpha,\,\ldots,\,\hat{x},\,\hat{y},\,\hat{\alpha},\,\ldots$ are unknown coefficients. These coefficients in general case are complex, however for simplicity they will be considered as real. In principle, with the help of the relations (\ref{eq:14t}) and (\ref{eq:14y}), it is possible to get rid of any three functions on the right-hand side of 
(\ref{eq:14u}) and (\ref{eq:14i}), for example of $\Psi^{\,(2)},\,\Psi^{\hspace{0.02cm}\ast\hspace{0.03cm} (2)}$ and 
$\Upsilon^{\hspace{0.02cm}\ast\hspace{0.03cm} (2)}$. However, it is more convenient to do this at the stage of deriving a system of algebraic equations for the coefficients $x,\,y,\,\alpha,\,\ldots,\,\hat{x},\,\hat{y},\,\hat{\alpha},\,\ldots$ .\\
\indent Substituting the presentations (\ref{eq:14u}) and (\ref{eq:14i}) into (\ref{eq:14e}) and then excluding the functions $\Psi^{\,(2)},\,\Psi^{\hspace{0.02cm}\ast\hspace{0.03cm} (2)}$ and $\Upsilon^{\hspace{0.02cm}\ast\hspace{0.03cm} (2)}$, we get the system of algebraic equations for the unknown coefficients:
\vspace{-0.3cm}
\begin{flushleft}
1. the first system
\end{flushleft}
\vspace{-0.4cm}
\begin{equation}
\begin{split}
&2\hspace{0.02cm}x = 1 - \alpha - \beta - \gamma - \delta,
\\[0.8ex]
&2\hspace{0.02cm}y = \alpha - \beta + \gamma - \delta,
\\[0.8ex]
&2\hspace{0.02cm}\hat{x} = - \hat{\alpha} - \hat{\beta} - \hat{\gamma} - \hat{\delta},
\\[0.8ex]
&2\hspace{0.02cm}\hat{y} = 1 + \hat{\alpha} - \hat{\beta} + \hat{\gamma} - \hat{\delta},
\end{split}
\hspace{1.4cm}
\label{eq:14o}
\end{equation}
\vspace{-0.3cm}
\begin{flushleft}
2. the second system
\end{flushleft}
\vspace{-0.4cm}
\begin{equation}
\begin{split}
&x - \hat{x} = - (\gamma - \hat{\gamma}) -  (\delta - \hat{\delta}),
\\[0.8ex]
&y - \hat{y} =   (\alpha - \hat{\alpha}) -  (\delta - \hat{\delta}) - 1,
\\[0.8ex]
&x + \hat{x} = - \gamma - \delta - \hat{\alpha} - \hat{\beta},
\\[0.8ex]
&y + \hat{y} = \alpha - \delta - \hat{\beta} + \hat{\gamma},
\end{split}
\label{eq:14p}
\end{equation}
\vspace{-0.3cm}
\begin{flushleft}
3. an additional relation
\end{flushleft}
\vspace{-0.3cm}
\begin{equation}
(\alpha - \hat{\alpha}) + (\beta - \hat{\beta}) - (\gamma - \hat{\gamma}) - (\delta - \hat{\delta}) = 1.
\label{eq:14a}
\end{equation}
The second system (\ref{eq:14p}) is easily examined to be a trivial consequence of (\ref{eq:14o}) and (\ref{eq:14a}), therefore it can be omitted. Any set of the coefficients $x,\,y,\,\alpha,\,\ldots,\,\hat{x},\,\hat{y},\,\hat{\alpha},\,\ldots$ satisfying the relations (\ref{eq:14o}) and (\ref{eq:14a}) under the substitution into (\ref{eq:14u}) and (\ref{eq:14i}) gives rise to a solution of the system of functional equations (\ref{eq:14e}). Thus, for example, the most simple solutions can be obtained if we set
\[
x= \hat{x} = \frac{1}{2}, \quad \hat{\alpha} = -1,
\]
the remaining coefficients vanish, or 
\[
y = \hat{y} = \frac{1}{2}, \quad \alpha = 1,
\]
and the remaining coefficients vanish. As a result we have two sets solutions   
\[
\left\{
\begin{array}{l}
J^{\,(2)\, a_{1}\, a_{2}\, i\; i_{1}}_{\ {\bf k}_{1},\, {\bf k}_{2},\, {\bf q},\, {\bf q}_{1}} =
\displaystyle\frac{1}{2}\, 
\Psi^{\hspace{0.03cm} (22)\hspace{0.03cm} a_{1}\, a_{2}\, i\, i_{1}}_{\ {\bf k}_{1},\, {\bf k}_{2},\, {\bf q},\, {\bf q}_{1}}, 
\\[3ex]
R^{\,(2)\, a_{1}\, a_{2}\, i\; i_{1}}_{\ {\bf k}_{1},\, {\bf k}_{2},\, {\bf q},\, {\bf q}_{1}} =
\displaystyle\frac{1}{2}\, 
\Psi^{\hspace{0.03cm} (22)\hspace{0.03cm} a_{1}\, a_{2}\, i\, i_{1}}_{\ {\bf k}_{1},\, {\bf k}_{2},\, {\bf q},\, {\bf q}_{1}}
-
\Psi^{\,(2)\, a_{1}\, a_{2}\, i\; i_{1}}_{\ {\bf k}_{1},\, {\bf k}_{2},\, {\bf q},\, {\bf q}_{1}},
\end{array}
\right.
\;
\left\{
\begin{array}{l}
J^{\,(2)\, a_{1}\, a_{2}\, i\; i_{1}}_{\ {\bf k}_{1},\, {\bf k}_{2},\, {\bf q},\, {\bf q}_{1}} =
\displaystyle\frac{1}{2}\, \Upsilon^{\hspace{0.03cm} (22)\hspace{0.03cm} a_{1}\, a_{2}\, i\, i_{1}}_{\ {\bf k}_{1},\, {\bf k}_{2},\, {\bf q},\, 
{\bf q}_{1}} 
+
\Psi^{\,(2)\, a_{1}\, a_{2}\, i\; i_{1}}_{\ {\bf k}_{1},\, {\bf k}_{2},\, {\bf q},\, {\bf q}_{1}},   \\[3ex]
R^{\,(2)\, a_{1}\, a_{2}\, i\; i_{1}}_{\ {\bf k}_{1},\, {\bf k}_{2},\, {\bf q},\, {\bf q}_{1}} =
\displaystyle\frac{1}{2}\, \Upsilon^{\hspace{0.03cm} (22)\hspace{0.03cm} a_{1}\, a_{2}\, i\, i_{1}}_{\ {\bf k}_{1},\, {\bf k}_{2},\, {\bf q},\, 
{\bf q}_{1}}.
\end{array}
\right.
\]
By a direct substitution with the use of (\ref{eq:14t}) and (\ref{eq:14y}), we see that they satisfy all functional equations in the system (\ref{eq:14e}). Note that these two sets of solutions cannot be reduced to one another, i.e. they are independent.\\
\indent We choice the parameters $x,\,y,\,\alpha,\,\ldots,\,\hat{x},\,\hat{y},\,\hat{\alpha},\,\ldots$ so that the form of the coefficient functions $J^{\,(2)}$ and $R^{\,(2)}$ be as close as possible to the form of the coefficient function  
$S^{\,(2)}$, Eq.\,(\ref{eq:13l}). This is achieved by choosing the following values for the parameters:
\[
x = \frac{1}{2},\quad \hat{y} = \frac{1}{2}, \quad \alpha = \frac{1}{4}, \quad \hat{\beta} = -\frac{1}{4}
\]
and the other parameters equal zero. As a result, we obtain
 \begin{equation}
\left\{
\begin{array}{l}
J^{\,(2)\, a_{1}\, a_{2}\, i\; i_{1}}_{\ {\bf k}_{1},\, {\bf k}_{2},\, {\bf q},\, {\bf q}_{1}} =
\displaystyle\frac{1}{2}\, 
\Psi^{\hspace{0.03cm} (22)\hspace{0.03cm} a_{1}\, a_{2}\, i\, i_{1}}_{\ {\bf k}_{1},\, {\bf k}_{2},\, {\bf q},\, {\bf q}_{1}}
+
\frac{1}{4}\,\Bigl(\Psi^{\,(2)\, a_{1}\, a_{2}\, i\; i_{1}}_{\ {\bf k}_{1},\, {\bf k}_{2},\, {\bf q},\, {\bf q}_{1}}
-
\Psi^{\hspace{0.02cm}\ast\hspace{0.02cm} (2)\hspace{0.03cm} a_{2}\, a_{1}\, i_{1}\, i}_{\ {\bf k}_{2},\, {\bf k}_{1},\, 
{\bf q}_{1},\, {\bf q}} \hspace{0.02cm}\Bigr), 
\\[3ex]
R^{\,(2)\, a_{1}\, a_{2}\, i\; i_{1}}_{\ {\bf k}_{1},\, {\bf k}_{2},\, {\bf q},\, {\bf q}_{1}} =
\displaystyle\frac{1}{2}\, 
\Upsilon^{\hspace{0.03cm} (22)\hspace{0.03cm} a_{1}\, a_{2}\, i\, i_{1}}_{\ {\bf k}_{1},\, {\bf k}_{2},\, {\bf q},\, {\bf q}_{1}}
-
\frac{1}{4}\,\Bigl(\Upsilon^{\,(2)\, a_{1}\, a_{2}\, i\; i_{1}}_{\ {\bf k}_{1},\, {\bf k}_{2},\, {\bf q},\, {\bf q}_{1}}
-
\Upsilon^{\hspace{0.02cm}\ast\hspace{0.02cm} (2)\hspace{0.03cm} a_{2}\, a_{1}\, i_{1}\, i}_{\ {\bf k}_{2},\, {\bf k}_{1},\, {\bf q}_{1},\, {\bf q}} \hspace{0.02cm}\Bigr).
\end{array}
\right.
\label{eq:14s}
\end{equation}
We can verify the validity of the obtained solution by direct substitution (\ref{eq:14s}) into (\ref{eq:14e}). Let us insert  the explicit expressions for the functions $\Psi^{\,(22)},\,\Upsilon^{\,(22)},\,\Psi^{\,(2)}$ and $\Upsilon^{\,(2)}$, 
Eqs.\,(\ref{eq:14ra})\,--\,(\ref{eq:14rd}), into (\ref{eq:14s}) and pass from the coefficient functions $V^{(1)}$ and $F^{(n)},\,n = 1,2,3$ to the vertex functions ${\mathcal V},\,{\mathcal G},\, {\mathcal P}$ and ${\mathcal K}$ by the rules (\ref{eq:4w}), (\ref{eq:4e}). Further, performing the integration over $d\hspace{0.03cm}{\bf k}^{\prime}$ and $d\hspace{0.03cm}{\bf q}^{\prime}$, we define the form of the desired coefficient functions $J^{\,(2)}$ and $R^{\,(2)}$. Here, we give the explicit form of the first of them
\begin{equation}
J^{\,(2)\, a_{1}\, a_{2}\, i\; i_{1}}_{\ {\bf k}_{1},\, {\bf k}_{2},\, {\bf q},\, {\bf q}_{1}}
=
\Lambda^{(2)\, a_{1}\, a_{2}\, i\; i_{1}}_{\ {\bf k}_{1},\, {\bf k}_{2},\, {\bf q},\, {\bf q}_{1}}
\label{eq:14d}
\end{equation}
\begin{align}
+\,
\Biggl[\hspace{0.03cm}\frac{1}{2}\hspace{0.03cm}\Biggl(&\frac{
{\mathcal P}^{\; a_{2}\, i\,  j}_{{\bf k}_{2},\, {\bf q},\, {\bf q} - {\bf k}_{2}}\, 
{\mathcal P}^{\hspace{0.03cm}\ast\, a_{1}\, i_{1}\hspace{0.03cm} j}_{{\bf k}_{1},\, {\bf q}_{1},\, {\bf q}_{1} - {\bf k}_{1}}}
{\bigl(\omega^{\hspace{0.02cm} l}_{{\bf k}_{2}} - \omega^{-}_{{\bf q}} + \omega^{-}_{{\bf q} - {\bf k}_{2}}\bigr)
\bigl(\omega^{\hspace{0.02cm} l}_{{\bf k}_{1}} - \omega^{-}_{{\bf q}_{1}} + \omega^{-}_{{\bf q}_{1} - {\bf k}_{1}}\bigr)} 
\,-\,
\frac{{\mathcal P}^{\; a_{2}\, j\, i_{1}}_{{\bf k}_{2},\, {\bf k}_{2} + {\bf q}_{1},\, {\bf q}_{1}}\, 
{\mathcal P}^{\hspace{0.03cm}\ast\, a_{1}\, j\, i}_{{\bf k}_{1},\, {\bf k}_{1} +\hspace{0.03cm}{\bf q},\, {\bf q}}}
{\bigl(\omega^{\hspace{0.02cm} l}_{{\bf k}_{2}} - \omega^{-}_{{\bf k}_{2} + {\bf q}_{1}} + \omega^{-}_{{\bf q}_{1}}\bigr)
\bigl(\omega^{\hspace{0.02cm} l}_{{\bf k}_{1}} - \omega^{-}_{{\bf k}_{1} +\hspace{0.03cm} {\bf q}} + \omega^{-}_{{\bf q}}\bigr)}
\Biggr) 
\notag\\[1.5ex]
+\,
2\hspace{0.02cm}\Biggl(&\frac{{\mathcal G}^{\; a_{2}\, i\,  j}_{{\bf k}_{2},\, {\bf q},\, {\bf k}_{2} - {\bf q}}\; 
{\mathcal G}^{\hspace{0.03cm}\ast\, a_{1}\, i_{1}\hspace{0.03cm} j}_{{\bf k}_{1},\, {\bf q}_{1},\, {\bf k}_{1} - {\bf q}_{1}}}
{\bigl(\omega^{\hspace{0.02cm} l}_{{\bf k}_{2}} - \omega^{-}_{{\bf q}} - \omega^{-}_{{\bf k}_{2} - {\bf q}}\bigr)
\bigl(\omega^{\hspace{0.02cm} l}_{{\bf k}_{1}} - \omega^{-}_{{\bf q}_{1}} - \omega^{-}_{{\bf k}_{1} - {\bf q}_{1}}\bigr)} 
-
\frac{{\mathcal K}^{\; a_{2}\, j\, i_{1}}_{{\bf k}_{2},\, - {\bf k}_{2} - {\bf q}_{1},\, {\bf q}_{1}}\hspace{0.03cm} 
{\mathcal K}^{\hspace{0.03cm}\ast\, a_{1}\, j\, i}_{{\bf k}_{1},\,- {\bf k}_{1} - {\bf q},\, {\bf q}}}
{\bigl(\omega^{\hspace{0.02cm} l}_{{\bf k}_{2}} + \omega^{-}_{-{\bf k}_{2} - {\bf q}_{1}} + \omega^{-}_{{\bf q}_{1}}\bigr)
\bigl(\omega^{\hspace{0.02cm} l}_{{\bf k}_{1}} + \omega^{-}_{-{\bf k}_{1} - {\bf q}} + \omega^{-}_{{\bf q}}\bigr)}
\!\Biggr) 
\notag\\[1.5ex]
+\,
\Biggl(
&\frac{{\mathcal V}^{\; a_{1}\, a_{2}\,  a}_{{\bf k}_{1}, {\bf k}_{2},\, {\bf k}_{1} - {{\bf k}_{2}}}\, 
{\mathcal P}^{\hspace{0.03cm}\ast\, a\, i_{1}\hspace{0.03cm} i}_{{\bf q}_{1} - {\bf q},\, {\bf q}_{1},\, {\bf q}}}
{\bigl(\omega^{\hspace{0.02cm} l}_{{\bf k}_{1}} - \omega^{\hspace{0.02cm} l}_{{\bf k}_{2}} - \omega^{\hspace{0.02cm} l}_{{\bf k}_{1} - {\bf k}_{2}}\bigr)
\bigl(\omega^{\hspace{0.02cm} l}_{{\bf q}_{1} - {\bf q}} - \omega^{-}_{{\bf q}_{1}} + \omega^{-}_{{\bf q}}\bigr)
}
- 
\frac{{\mathcal P}^{\; a\, i\,  i_{1}}_{{\bf q} - {\bf q}_{1},\, {\bf q},\, {\bf q}_{1}}\, 
{\mathcal V}^{\hspace{0.03cm}\ast\, a_{2}\, a_{1}\,  a}_{{\bf k}_{2}, {\bf k}_{1},\, {\bf k}_{2} - {\bf k}_{1}}}
{\bigl(\omega^{\hspace{0.02cm} l}_{{\bf q} - {\bf q}_{1}} - \omega^{-}_{{\bf q}} + \omega^{-}_{{\bf q}_{1}}\bigr)
\bigl(\omega^{\hspace{0.02cm} l}_{{\bf k}_{2}} - \omega^{\hspace{0.02cm} l}_{{\bf k}_{1}} - \omega^{\hspace{0.02cm} l}_{{\bf k}_{2} - {\bf k}_{1}}\bigr)
}
\Biggr)\!\Biggr]
\notag
\end{align}
\[
\times\hspace{0.03cm} (2\pi)^{3}\hspace{0.03cm}\delta({\bf q} + {\bf k}_{1} - {\bf q}_{1} - {\bf k}_{2}).
\]
On the right-hand side we have taken into account the fact that the coefficient functions $J^{\,(2)}$ and $R^{\,(2)}$ are defined up to an additive arbitrary function $\Lambda^{(2)}$ satisfying the only condition
\[
\Lambda^{(2)\, a_{1}\, a_{2}\, i\; i_{1}}_{\ {\bf k}_{1},\, {\bf k}_{2},\, {\bf q},\, {\bf q}_{1}}
=
-\Lambda^{\!\ast\, (2)\, a_{2}\, a_{1}\, i_{1}\, i}_{\ {\bf k}_{2},\, {\bf k}_{1},\, {\bf q}_{1},\, {\bf q}}.
\]
In the hard thermal loop approximation the terms on right-hand side of (\ref{eq:14d}) with the functions ${\mathcal G}$ and ${\mathcal K}$ should be set equal to zero.\\
\indent Let us return to the effective Hamiltonian (\ref{eq:14w}). In the expression (\ref{eq:14d}) we put $\Lambda^{(2)} \equiv 0$. We define a {\it complete effective amplitude} $\mathscr{T}^{\,(2)\, i\, i_{1}\, a_{1}\, a_{2}}_{\; {\bf q},\, {\bf q}_{1},\, {\bf k}_{1},\, {\bf k}_{2}}$ for the scattering process of plasmino off plasmon setting by definition
\begin{equation}
\mathscr{T}^{\,(2)\, i\, i_{1}\, a_{1}\, a_{2}}_{\; {\bf q},\, {\bf q}_{1},\, {\bf k}_{1},\, {\bf k}_{2}}\,
(2\pi)^{3}\hspace{0.03cm}\delta({\bf q} + {\bf k}_1 - {\bf q}_{1} - {\bf k}_{2})
\equiv
\label{eq:14f}
\end{equation}
\[
\equiv
\bigl(\omega^{-}_{{\bf q}} + \omega^{\hspace{0.02cm} l}_{{\bf k}_{1}} - \omega^{-}_{{\bf q}_{1}} - \omega^{\hspace{0.02cm} l}_{{\bf k}_{2}}\bigr)
J^{\hspace{0.03cm} (2)\hspace{0.03cm} a_{1}\, a_{2}\, i\; i_{1}}_{\; {\bf k}_{1},\, {\bf k}_{2},\, {\bf q},\, {\bf q}_{1}}
\hspace{0.03cm}+\hspace{0.03cm}
\widetilde{T}^{\hspace{0.03cm} (2)\hspace{0.03cm} i\; i_{1}\, a_{1}\, a_{2}}_{\; {\bf q},\, {\bf q}_{1},\, {\bf k}_{1},\, {\bf k}_{2}}\, 
(2\pi)^{3}\hspace{0.03cm}\delta({\bf q} + {\bf k}_1 - {\bf q}_{1} - {\bf k}_{2}).
\]
Substituting the explicit expressions for the functions $J^{\,(2)}$ and $\widetilde{T}^{\,(2)}$, Eqs.\,(\ref{eq:14d}) and 
(\ref{eq:5t}), perfor\-ming simple algebraic transformations, we define the complete effective amplitude $\mathscr{T}^{\,(2)}$:
\begin{equation}
\mathscr{T}^{\hspace{0.03cm} (2)\hspace{0.03cm} i\; i_{1}\, a_{1}\, a_{2}}_{\; {\bf q},\, {\bf q}_{1},\, {\bf k}_{1},\, {\bf k}_{2}}
=
T^{\hspace{0.03cm} (2)\hspace{0.03cm} i\; i_{1}\, a_{1}\, a_{2}}_{\; {\bf q},\, {\bf q}_{1},\, {\bf k}_{1},\, {\bf k}_{2}}
\label{eq:14g}
\vspace{-0.3cm}
\end{equation}
\begin{align}
-\,
\frac{1}{2}\,\Biggl[
\Biggl(&\frac{1}{\omega^{\hspace{0.02cm} l}_{{\bf k}_{2}} - \omega^{-}_{{\bf q}} + \omega^{-}_{{\bf q} - {\bf k}_{2}}} 
\,+\,
\frac{1}{\omega^{\hspace{0.02cm} l}_{{\bf k}_{1}} - \omega^{-}_{{\bf q}_{1}} + \omega^{-}_{{\bf q}_{1} - {\bf k}_{1}}} 
\Biggr)\hspace{0.02cm}
{\mathcal P}^{\; a_{2}\, i\,  j}_{{\bf k}_{2},\, {\bf q},\, {\bf q} - {\bf k}_{2}}\, 
{\mathcal P}^{\hspace{0.03cm}\ast\, a_{1}\, i_{1}\hspace{0.03cm} j}_{{\bf k}_{1},\, {\bf q}_{1},\, {\bf q}_{1} - {\bf k}_{1}}
\notag\\[1.5ex]
-\,
\Biggl(&\frac{1}{\omega^{\hspace{0.02cm} l}_{{\bf k}_{2}} - \omega^{-}_{{\bf k}_{2} +\hspace{0.03cm} {\bf q}_{1}} + 
\omega^{-}_{{\bf q}_{1}}} 
+
\frac{1}{\omega^{\hspace{0.02cm} l}_{{\bf k}_{1}} - \omega^{-}_{{\bf k}_{1} +\hspace{0.03cm} {\bf q}} + \omega^{-}_{{\bf q}}} 
\Biggr)\hspace{0.02cm}
{\mathcal P}^{\; a_{2}\, j\, i_{1}}_{{\bf k}_{2},\, {\bf k}_{2} + {\bf q}_{1},\, {\bf q}_{1}}\, 
{\mathcal P}^{\hspace{0.03cm}\ast\, a_{1}\, j\, i}_{{\bf k}_{1},\, {\bf k}_{1} + {\bf q},\, {\bf q}}
\Biggr]
\notag
\end{align}
\begin{align}
-\,
2\,\Biggl[\Biggl(
&\frac{1}{\omega^{\hspace{0.02cm} l}_{{\bf k}_{2}} - \omega^{-}_{{\bf q}} - \omega^{-}_{{\bf k}_{2} - {\bf q}}}
\,+\,
\frac{1}{\omega^{\hspace{0.02cm} l}_{{\bf k}_{1}} - \omega^{-}_{{\bf q}_{1}} - \omega^{-}_{{\bf k}_{1} - {\bf q}_{1}}}
\Biggr)\hspace{0.02cm}
{\mathcal G}^{\; a_{2}\, i\hspace{0.03cm} j}_{{\bf k}_{2},\, {\bf q},\, {\bf k}_{2} - {\bf q}}\; 
{\mathcal G}^{\hspace{0.03cm}\ast\, a_{1}\, i_{1}\,  j}_{{\bf k}_{1},\, {\bf q}_{1},\, {\bf k}_{1} - {\bf q}_{1}}
\notag\\[1.5ex]
-\,
\Biggl(
&\frac{1}{\omega^{\hspace{0.02cm} l}_{{\bf k}_{2}} + \omega^{-}_{-{\bf k}_{2} - {\bf q}_{1}} + \omega^{-}_{{\bf q}_{1}}}
\,+\,
\frac{1}{\omega^{\hspace{0.02cm} l}_{{\bf k}_{1}} + \omega^{-}_{-{\bf k}_{1} - {\bf q}} + \omega^{-}_{{\bf q}}}
\Biggr)\hspace{0.02cm}
{\mathcal K}^{\; a_{2}\, j\, i_{1}}_{{\bf k}_{2},\, - {\bf k}_{2} - {\bf q}_{1},\, {\bf q}_{1}}\, 
{\mathcal K}^{\hspace{0.03cm}\ast\, a_{1}\, j\, i}_{{\bf k}_{1},\,- {\bf k}_{1} - {\bf q},\, {\bf q}}
\Biggr]
\notag\\[1.5ex]
+\,
\Biggl(
&\frac{1}{\omega^{\hspace{0.02cm} l}_{{\bf k}_{1}} - \omega^{\hspace{0.02cm} l}_{{\bf k}_{2}} - \omega^{\hspace{0.02cm} l}_{{\bf k}_{1} - {\bf k}_{2}}}
\,-\,
\frac{1}{\omega^{\hspace{0.02cm} l}_{{\bf q}_{1} - {\bf q}} - \omega^{-}_{{\bf q}_{1}} + \omega^{-}_{{\bf q}}}
\Biggr)\hspace{0.02cm}
{\mathcal V}^{\; a_{1}\, a_{2}\,  a}_{{\bf k}_{1}, {\bf k}_{2},\, {\bf k}_{1} - {{\bf k}_{2}}}\, 
{\mathcal P}^{\hspace{0.03cm}\ast\, a\, i_{1}\, i}_{{\bf q}_{1} - {\bf q},\, {\bf q}_{1},\, {\bf q}}
\notag\\[1.5ex]
+\, 
\Biggl(
&\frac{1}{\omega^{\hspace{0.02cm} l}_{{\bf k}_{2}} - \omega^{\hspace{0.02cm} l}_{{\bf k}_{1}} - \omega^{\hspace{0.02cm} l}_{{\bf k}_{2} - {\bf k}_{1}}}
\,-\,
\frac{1}{\omega^{\hspace{0.02cm} l}_{{\bf q} - {\bf q}_{1}} - \omega^{-}_{{\bf q}} + \omega^{-}_{{\bf q}_{1}}}
\Biggr)\hspace{0.02cm}
{\mathcal P}^{\; a\, i\;  i_{1}}_{{\bf q} - {\bf q}_{1},\, {\bf q},\, {\bf q}_{1}}\, 
{\mathcal V}^{\hspace{0.03cm}\ast\, a_{2}\, a_{1}\,  a}_{{\bf k}_{2}, {\bf k}_{1},\, {\bf k}_{2} - {\bf k}_{1}}.
\notag
\end{align}
In deriving this expression we have used only the momentum conservation law in an elementary act of elastic scattering of plasmino off plasmon
\begin{equation}
{\bf q} + {\bf k}_{1} = {\bf q}_{1} + {\bf k}_{2},
\label{eq:14h}
\end{equation}
that is valid on strength of the corresponding $\delta$-function in (\ref{eq:14f}). The coefficient function 
$J^{\,(2)\, a_{1}\, a_{2}\, i\, i_{1}}_{\; {\bf k}_{1},\, {\bf k}_{2},\, {\bf q},\, {\bf q}_{1}}$ in the definition (\ref{eq:14f}) plays the same role as the coefficient function 
$S^{\,(2)\, i\; i_{1}\, i_{2}\, i_{3}}_{\, {\bf q},\, {\bf q}_{1},\, {\bf q}_{2},\, {\bf q}_{3}}$ in the definition (\ref{eq:13b}), namely, it leads to the total symmetrization of original effective amplitude $\widetilde{T}^{\,(2)\, i\, i_{1}\, a_{1}\, a_{2}}_{\; {\bf q},\, {\bf q}_{1},\, {\bf k}_{1},\, {\bf k}_{2}}$. As we have shown in section \ref{section_2}, the first term on the right-hand side of (\ref{eq:14g}) satisfies the condition that follows from the requirement of the reality of the Hamiltonian $H^{(4)}$:
\[
T^{\,(2)\, i\; i_{1}\, a_{1}\, a_{2}}_{\ {\bf q},\, {\bf q}_{1},\, {\bf k}_{1},\, {\bf k}_{2}}
=
T^{\hspace{0.03cm}\ast\hspace{0.02cm}(2)\hspace{0.03cm} i_{1}\, i\; a_{2}\; a_{1}}_{\ {\bf q}_{1},\, {\bf q},\, {\bf k}_{2},\, {\bf k}_{1}}.
\]
For the effective amplitude $\widetilde{T}^{\,(2)\, i\, i_{1}\, a_{1}\, a_{2}}_{\; {\bf q},\, {\bf q}_{1},\, {\bf k}_{1},\, {\bf k}_{2}}$, as it was defined by the expression (\ref{eq:5t}), this property will hold only on the resonant manifold given by the wavenumber condition (\ref{eq:14h}) and by the frequency one
\begin{equation}
\omega^{-}_{{\bf q}} + \omega^{\hspace{0.02cm} l}_{{\bf k}_{1}} = \omega^{-}_{{\bf q}_{1}} + \omega^{\hspace{0.02cm} l}_{{\bf k}_{2}}.
\label{eq:14j}
\end{equation}
It is easily to verify by means of direct calculation that the complete effective amplitude $\mathscr{T}^{\hspace{0.03cm} (2)\hspace{0.03cm} i\; i_{1}\, a_{1}\, a_{2}}_{\; {\bf q},\, {\bf q}_{1},\, {\bf k}_{1},\, {\bf k}_{2}}$ satisfies the condition of the reality
\[
\mathscr{T}^{\,(2)\, i\; i_{1}\, a_{1}\, a_{2}}_{\ {\bf q},\, {\bf q}_{1},\, {\bf k}_{1},\, {\bf k}_{2}}
=
\mathscr{T}^{\hspace{0.03cm}\ast\hspace{0.02cm}(2)\hspace{0.03cm} i_{1}\, i\; a_{2}\; a_{1}}_{\ {\bf q}_{1},\, {\bf q},\, {\bf k}_{2},\, {\bf k}_{1}}
\]
without using the four-wave resonance equation (\ref{eq:14j}), in other words, for arbitrary values of the resonance frequency difference $\Delta\hspace{0.02cm}\omega = \omega^{-}_{{\bf q}} + \omega^{\hspace{0.02cm} l}_{{\bf k}_{1}} - \omega^{-}_{{\bf q}_{1}} - \omega^{\hspace{0.02cm} l}_{{\bf k}_{2}}$.


\section{Conclusion}
\label{section_15}
\setcounter{equation}{0}

In this paper we have set up the (pseudo)classical Hamiltonian formalism needed to describe the processes of nonlinear interaction for soft gluon and soft quark-antiquark excitations in a high-temperature plasma with  non-Abelian type of interaction. In solving the problem posed, the effective Braaten\hspace{0.02cm}-\hspace{0.02cm}Pisarski theory (the hard thermal loop approximation) was used. It must be emphasized that the most of the results in Sections 2\hspace{0.02cm}-\hspace{0.02cm}8 do not strictly rely on the HTL approximation, but were obtained by employing the methods of kinetic theory and nonequilibrium statistical physics, which are sufficiently general for a wide class of nonequilibrium media.	
The canonical transformations of the bosonic $a^{a}_{{\bf k}}$ and fermionic $b^{\,i}_{{\bf q}}$ normal variables, Eqs.\,(\ref{eq:3t}) and (\ref{eq:3y}), are constructed in the explicit form, which enabled us to exclude the third-order interaction Hamiltonian $H^{(3)}$ (\ref{eq:2f}). This in turn allowed to define new effective interaction Hamiltonians ${\mathcal H}^{(4)}_{qg\rightarrow qg}$ and ${\mathcal H}^{(4)}_{qq\rightarrow qq}$, Eqs.\,(\ref{eq:5r}) and (\ref{eq:6r}), with the gauge-covariant scattering amplitudes of plasmino off plasmons $\widetilde{T}^{(1)\, i\, i_{1}\, a_{1}\, a_{2}}_{\, {\bf q},\, {\bf q}_{1},\, {\bf k}_{1},\, {\bf k}_{2}}$ and plasmino off plasmino $\widetilde{T}^{(2)\, i\, i_{1}\, i_{2}\, i_{3}}_{{\bf q},\, {\bf q}_{1},\, {\bf q}_{2},\, {\bf q}_{3}}$, Eqs.\,(\ref{eq:5t}) and  (\ref{eq:6t}), correspondingly. On the basis of the canonicity conditions (\ref{eq:3e}) and (\ref{eq:3r}) we have determined a complete system of independent relations of the integral type connecting the lowest and highest coefficient functions in the integrands of various terms in the canonical transfor\-mations. We sought for the latter in the form of an integer-degree series in the new normal variables $c^{a}_{{\bf k}}$ and $f^{\,i}_{{\bf q}}$. Further, we note that the corresponding construction of the kinetic equations for soft plasma modes taking into account at the same time boson and fermion degrees of freedom is in general a direct extension of the purely bosonic case \cite{markov_2020} at the conceptual level, but very cumbersome and tangled in practical terms. Therefore, here we have restricted ourselves to the detailed consideration of only the simplest processes of nonlinear interaction of soft purely collective excitations in a quark-gluon plasma: elastic scattering of plasmino off plasmino and of plasmino off plasmon.\\ 
\indent It should also be mentioned that the very formulation of the problem, taking into account only plasmons and plasminos, is hard to see any applicable to real physics of the quark-gluon plasma due to the fact that the residues $Z_{l}({\bf k})$ and $Z_{-}({\bf q})$ of these QGP oscillation modes are exponentially suppressed for large momenta and, in addition, these modes are not stable relative to the high order corrections to the one-loop approximation for the HTL propagators and should disappear at all at some $|{\bf k}|, |{\bf q}| \sim g\hspace{0.02cm}T$, as already discussed in the Introduction. This is just done for sake of simplicity, since the number of branches for the various oscillation modes in a hot QCD plasma is sufficiently large. Since the consideration was carried out at a quite general level, it is not difficult in principle to rewrite the obtained kinetic equations for the soft transverse gluon and normal quark excitations. In fact, the matter here is reduced to some replacements. For example, instead of separating the longitudinal part of the gluon propagator $\,^{\ast}\widetilde{\cal D}_{\mu\mu^{\prime}} (k)$, as was done in Eq.\,(\ref{eq:9u}), we now need to leave the transverse part 
\[
\,^{\ast}\widetilde{\cal D}^{\mu\mu^{\prime}}(k)
\Rightarrow
- P^{\mu\hspace{0.02cm}\mu^{\prime}}(k) \,^{\ast}\!\Delta^t(k),
\]
where, in turn, the transverse projector $P^{\mu\hspace{0.02cm}\mu^{\prime}}(k)$ must be represented in the form of expansion in eigenvectors of transverse polarization \cite{kalashnikov_1980}
	\[
	P^{\mu\hspace{0.02cm}\mu^{\prime}}(k)
	=
	\sum\limits_{\zeta\hspace{0.02cm}=\hspace{0.02cm}1,\,2}
	\left(\frac{\epsilon^{\,\mu}(k,\zeta)}
	{\sqrt{\epsilon^2(k,\zeta)}}\right)\!
	\left(\frac{\epsilon^{\,\mu^{\prime}}(k,\zeta)}
	{\sqrt{\epsilon^2(k,\zeta)}}\right)\!,
	\]
	\[
	\epsilon^{\,\mu}(k,\zeta)\equiv {\rm e}^{(\zeta)\mu}k^2
	- k^{\mu}({\rm e}^{(\zeta)}\cdot k), \quad (k\cdot \epsilon\hspace{0.02cm}(k,\zeta))=0.
	\]
Here, ${\rm e}^{(\zeta)\hspace{0.02cm}\mu}$ are some four-vectors such that $\epsilon^{\,\mu}(k,\zeta)$ are linearly independent. This in turn will lead to the corresponding replacement of the gluon wave function (\ref{eq:9j})
	\[
	\left(\frac{Z_l({\bf k})}{2\hspace{0.03cm}\omega_{{\bf k}}^l}\right)^{\!1/2}\!
	\Biggl(\frac{\tilde{u}^{\mu}(k)}{\sqrt{\bar{u}^{\hspace{0.03cm}2}(k)}}
	\Biggr)
	\Rightarrow
	\left(\frac{Z_t({\bf k})}{2\hspace{0.03cm}\omega_{{\bf k}}^t}\right)^{\!1/2}\!
	\left(\frac{\epsilon^{\,\mu}(k,\zeta)}
	{\sqrt{\epsilon^{\hspace{0.03cm}2}(k,\zeta)}}\right),
	\]
and to proper choice of mass-shell conditions. Similar substitutions need to be performed in the fermionic sector of excitations. From the quark propagator $\,^{\ast}\!S_{\alpha\alpha^{\prime}}(q)$ it is necessary to separate the part related to the normal quark mode of oscillations. Instead of the representation (\ref{eq:12_1o}) at the plasmino excitation pole, we must now use in the case when $q_{0}\rightarrow {\mathcal E}_{+}({\bf q})\simeq \omega_{\bf q}^{+}$
	\[
	\,^{\ast}\!S_{\alpha\alpha^{\prime}}(q_{0},{\bf q}) \sim 
	\sum\limits_{\lambda\hspace{0.02cm} = \hspace{0.02cm}\pm\hspace{0.01cm} 1}\!u^{(+)}_{\alpha}(\hat{\bf q},\lambda)
	\bar{u}^{(+)}_{\alpha^{\prime}}(\hat{\bf q},\lambda)\, \frac{Z_{+}({\bf q})}{q_{0} - \omega_{\bf q}^{+}}
	\]
and carry out, where appropriate, the simple replacements of quark wave functions
	\[
	\left(\!\frac{Z_{-}({\bf q})}{2}\!\right)^{\!\!1/2}\!\!\!\!
	v^{(-)}_{\alpha}(\hat{\bf q},\lambda)\!
	\rightarrow\!\!
	\left(\!\frac{Z_{+}({\bf q})}{2}\!\right)^{\!\!1/2}\!\!\!\!
	u^{(+)}_{\alpha}(\hat{\bf q},\lambda),
	\quad
	\left(\!\frac{Z_{-}({\bf q})}{2}\!\right)^{\!\!1/2}\!\!\!\!
	\bar{v}^{(-)}_{\alpha}(\hat{\bf q},\lambda)\!
	\rightarrow\!\!
	\left(\!\frac{Z_{+}({\bf q})}{2}\!\right)^{\!\!1/2}\!\!\!\!
	\bar{u}^{(+)}_{\alpha}(\hat{\bf q},\lambda)
	\]
and corresponding choice of mass-shell conditions. Such replacements will allow us to obtain the kinetic equations for the colorless number density of soft transverse gluon excitations $N^{ab\,t}_{{\bf k}\!\!}\! \equiv \delta^{a\hspace{0.02cm}b} N^{t}_{{\bf k}}$ and for the number density of colorless quark normal excitations $n^{ij\,+}_{{\bf q}\!\!}\! \equiv \delta^{\hspace{0.02cm}ij} n^{+}_{{\bf q}}$ instead of (\ref{eq:8u}), (\ref{eq:11t}) and (\ref{eq:11i}).\\
\indent A qualitatively new difficulties arise here if we want to consider the 'big picture', i.e., when we take into account {\it simultaneously} all branches of oscillation modes in the quark-gluon plasma. Thus, for example, the introduction of transverse oscillations results in appearing additional channels of interaction (this circumstance has been studied in detail at the time for the ordinary electron-ion plasma, see e.g. \cite{tsytovich_1970, tsytovich_1977}). So three--soft-gluon decays can occur in the isotropic QCD plasma. Any process is possible in which a high-frequency wave emits (absorbs) a low-frequency wave, changing to a wave of the same type as the primary. In particular, this allows the following processes:
\[
t \rightleftarrows t + l,
\]
where $t$ is transverse branch and $l$ is longitudinal branch of weakly damped waves in an isotropic quark-gluon plasma. The arrows indicate the directions that the process can take. It is apparent from the law of energy conservation that the processes $t \rightleftarrows t + l$ are possible not only for waves with $\omega^{\hspace{0.03cm}t} \gg \omega^{\hspace{0.03cm}l}$, but also for waves with $\omega^{\hspace{0.03cm}t}$ of the same order as $\omega^{\hspace{0.03cm}l}$. In addition the following process of the fusion of two plasmons into one transverse eigenwave and the inverse decay process are also permitted:
\[
l + l \rightleftarrows t.
\]
The processes are possible only for transverse waves whose frequencies are near $2\hspace{0.03cm}\omega_{pl}$. We can also expect the appearance of new interaction channels in the quark-antiquark sector of soft excitations of QGP. All these processes need to be taken into account for the total balance of changes in the number densities of soft QGP excitations, which will lead to the appearance of qualitatively new terms in the collision integrals in the corresponding kinetic equations.\\  
\indent Here, we have not touched on a number of interesting issues closely related to the subject of our study. The first of them is concerned with consideration of an additional degree of freedom of the system connected with the antiplasmino branch of collective quark-antiquark oscillations. Including the antiplasmino mode in a general scheme of the construction of Hamiltonian formalism requires appropriate generalization of the canonical transformations, the canonicity conditions, the original interaction Hamiltonian etc. The main point here is to construct the effective fourth-order interaction Hamiltonian that would describe the annihilation process of plasmino with antiplasmino into two plasmons and the plasmino-antiplasmino pair production by merging two plasmons and to obtain a self-consistent system of three kinetic equations determining an evolution of the number densities of plasminos, antiplasminos and plasmons by virtue of their interaction among themselves. The introduction of antiplas\-mino mode is straightforward, but this leads to a noticeable complication of the formalism under consideration. For example, instead of the canonical transformations of (\ref{eq:3q}) and (\ref{eq:3w}), it is necessary to consider the transformations from the three original bosonic and fermionic functions $a^{a}_{\bf k}$, $b^{\,i}_{\bf q}$ and $d^{\,i}_{\bf q}$ to three new functions $c^{a}_{\bf k}$, $f^{\,i}_{\bf q}$ and $\bar{f}^{\,i}_{\bf q}$:
\begin{align}
&a^{a}_{\bf k} = a^{a}_{{\bf k}}(c^{a}_{\bf k},\, c^{\ast\ \!\!a}_{{\bf k}}\,f^{\,i}_{\bf q},\,f^{\,\ast\ \!\!i}_{{\bf q}},\,\bar{f}^{\,i}_{\bf q},\,\bar{f}^{\,\ast\ \!\!i}_{{\bf q}}),
\notag\\[0.8ex]
&b^{\,i}_{\bf q} = b^{\,i}_{{\bf q}}(c^{a}_{\bf k},\, c^{\ast\ \!\!a}_{{\bf k}}\,f^{\,i}_{\bf q},\,f^{\,\ast\ \!\!i}_{{\bf q}},\,\bar{f}^{\,i}_{\bf q},\,\bar{f}^{\,\ast\ \!\!i}_{{\bf q}}),
\notag\\[0.8ex]
&d^{\,i}_{\bf q} = d^{\,i}_{{\bf q}}(c^{a}_{\bf k},\, c^{\ast\ \!\!a}_{{\bf k}}\,f^{\,i}_{\bf q},\,f^{\,\ast\ \!\!i}_{{\bf q}},\,\bar{f}^{\,i}_{\bf q},\,\bar{f}^{\,\ast\ \!\!i}_{{\bf q}}).
\notag
\end{align}
A consequence of this will be appearance of new terms in the super-Poisson brackets (\ref{eq:3e}), (\ref{eq:3r}), complication of the canonicity conditions given in Appendices \ref{appendix_C} and \ref{appendix_D}, and so on. For this reason, the construction of Hamilton's formalism for the description of dynamics of collective fermion excitations with antiplasminos  requires independent consideration and it was consciously exclu\-ded from the general analysis in the present paper. This will be the subject of another paper.\\
\indent Other interesting, but somewhat abstract issue not included in this paper is an analysis of a possibility for the construction of so-called odd Poisson bracket (antibracket) with respect to the Grassmann grading and of the corresponding odd Hamiltonian for the system of soft Fermi- and Bose-excitations. A determining condition for this is the fact that the equations of motion reproduced by the odd bracket with Grassmann-odd Hamiltonian have to be identical to the equations obtained in the even Poisson superbracket by means of Grassmann-even Hamiltonian. There exist only few papers in which the authors have analyzed such a possibility for the description of exactly integrable systems \cite{kupershmidt_1985}, of systems with finite degrees of freedom \cite{volkov_1989} and for the description of hydrodynamics \cite{volkov_1995}. In the paper by Volkov {\it et al} \cite{volkov_1989} on the example of  Witten's supersymmetric mechanics it is shown that not only an even but also an odd Poisson bracket is inherently presented in Hamiltonian systems with equal numbers of pairs of Grassmann even and odd canonical variables. Although the treatment was given here for the simplest example of Witten's classical supersymmetric mechanics, which as is well known has a 4-dimensional phase superspace, the authors suggested that the equivalence of the description by means of the brackets of different parities are general in nature and inherent in all of the dynamical systems mentioned above irrespective of the dimension of the phase superspace. Thus, a question arises regarding the extension of this analyses to considered  physical system of soft Fermi- and Bose-excitations in QGP with known even Hamiltonian $H_{even} = H^{(0)} + H^{(3)} + H^{(4)}$, Eqs.\,(\ref{eq:2d})\,--\,(\ref{eq:2g}), and the even Poisson superbracket (\ref{eq:2p}). Thus, it is necessary to develop an algorithm for recovering with respect to the even Hamiltonian and the even (super)bracket relevant odd Hamiltonian $H_{odd}$ and odd bracket and to show how the description of the dynamics in the phase superspace using the brackets of different Grassmann parity will be related to each other.\\
\indent Final interesting point of purely academic interest is the consideration of the most general structure of the free-field and interaction Hamiltonians  $H^{(0)}$, $H^{(3)}$ and $H^{(4)}$, Eqs.\,(\ref{eq:2d})\,--\,(\ref{eq:2g}), and also of the canonical transformations (\ref{eq:3t}) and (\ref{eq:3y}) for which some of the vertex and coefficient functions can take values in a Grassmann algebra. For the first time, this issue within the framework of general quadratic non-stationary  systems with interaction of Bose- and Fermi-operators was discussed by Dodonov and Man\hspace{0.02cm}\!'\hspace{0.02cm}ko \cite{dodonov_1983, dodonov_1987}. In the model proposed by these authors, among the coefficients of the quadratic form there could be both usual commutative and anticommutative (Grassmanian) coefficients naturally arising when dealing with boson-fermion systems. Unfortunately, Dodonov and Man\hspace{0.02cm}\!'\hspace{0.02cm}ko have not given any specific physical system that corresponds to these purely formal constructions.\\
\indent In principle we can formally also extend the structure of (Grassmann-even) interaction Hamiltonians $H^{(3)}$ and $H^{(4)}$ and of the canonical transformations, assumed that some of the vertex and coefficient functions can be anticommuting ones. In other words among the expansion terms the ones with odd power of the Grassmann-valued functions $(b^{\ast\, i}_{{\bf q}}, b^{\; i}_{{\bf q}})$ or $(f^{\ast\ \!\!i}_{{\bf q}}, f^{\,i}_{\bf q})$ may appear. In this case the canonical transformations can be considered in fact as supertransformations as it takes place in supersymmetry. One can extend on the formal level all further reasoning. However, this raises an important question of whether we can give any physical meaning to the vertex and coefficient functions taking values in the Grassmann algebra.


\section*{\bf Acknowledgment}

The authors thank the reviewer for careful reading of the article and for a number of valuable remarks and comments.


\begin{appendices}
\numberwithin{equation}{section}
\section{Effective gluon vertices and gluon propagator}
\numberwithin{equation}{section}
\label{appendix_A}

In this Appendix, we have provided the explicit form of vertex functions and gluon propagator in the hard thermal loop (HTL) approximation \cite{blaizot_2002, ghiglieri_2020, braaten_1990}.\\
\indent Effective three-gluon vertex
\begin{equation}
\,^{\ast} \Gamma^{\mu\hspace{0.02cm} \nu  \rho}(k, k_{1}, k_{2}) \equiv
\Gamma^{\mu\hspace{0.02cm} \nu  \rho}(k, k_{1}, k_{2}) +
\delta\hspace{0.025cm} \Gamma^{\mu\hspace{0.02cm} \nu  \rho}(k, k_{1}, k_{2})
\label{ap:A1}
\end{equation}
is the sum of bare three-gluon vertex
\begin{equation}
\Gamma^{\mu\hspace{0.02cm} \nu  \rho}(k, k_{1}, k_{2}) =
g^{\mu\hspace{0.02cm} \nu } (k - k_{1})^{\rho} + g^{\nu \rho} (k_{1} - k_{2})^{\mu} +
g^{\mu \rho} (k_{2} - k)^{\nu}
\label{ap:A2}
\end{equation}
and the corresponding HTL correction
\begin{equation}
\delta\hspace{0.025cm} \Gamma^{\mu\hspace{0.02cm} \nu  \rho}(k, k_{1}, k_{2}) =
3\hspace{0.045cm}\omega^{2}_{\rm pl}\!\int\!\frac{d\hspace{0.035cm}\Omega}{4 \pi} \,
\frac{v^{\mu}\hspace{0.02cm} v^{\nu} v^{\rho}}{v\cdot k + i\hspace{0.025cm}\epsilon} \,
\Biggl(\frac{\omega_{2}}{v\cdot k_{2} - i\epsilon} -
\frac{\omega_1}{v\cdot k_{1} - i\epsilon}\Biggr),
\quad \epsilon\rightarrow +\hspace{0.02cm}0,
\label{ap:A3}
\end{equation}
where $v^{\mu} = (1,{\bf {\bf v}})$, $k^{\mu} = (\omega, {\bf k})$ is a gluon four-momentum with $k  + k_{1} + k_{2} = 0$, $d\hspace{0.035cm}\Omega$ is a differential solid angle and   $\omega_{\rm pl}^2 = g^2(2N_c+N_f) T^2/18$ is the plasma frequency squared of the gluon sector of plasma excitations. We present below useful properties of the three-gluon HTL-resumed vertex function for complex conjugation and permutation of momenta:
\begin{equation}
\left(\!\,^{\ast}\Gamma_{\mu\hspace{0.02cm} \mu_{1} \mu_{2}}(-k_{1} - k_{2}, k_{1}, k_{2})\right)^{\ast} =
-\!\,^{\ast}\Gamma_{\mu\hspace{0.02cm} \mu_{1} \mu_{2}}(k_{1} + k_{2}, -k_{1}, -k_{2}) 
= \!\,^{\ast}\Gamma_{\mu\hspace{0.02cm} \mu_{2}\mu_{1}}(k_{1} + k_{2}, -k_{2}, -k_{1}).
\label{ap:A4}
\end{equation}
Further, the effective four-gluon vertex
\begin{equation}
^{\ast} \Gamma^{\mu\hspace{0.02cm} \nu  \lambda \sigma}(k,k_1,k_2,k_3) \equiv
\Gamma^{\mu\hspace{0.02cm} \nu  \lambda \sigma}(k,k_1,k_2,k_3) +
\delta\hspace{0.025cm} \Gamma^{\mu\hspace{0.02cm} \nu  \lambda \sigma}(k,k_1,k_2,k_3)
\label{ap:A5}
\end{equation}
is the sum of bare four-gluon vertex
\begin{equation}
\Gamma^{\mu\hspace{0.02cm} \nu  \lambda \sigma} =
2\hspace{0.02cm}g^{\mu\hspace{0.02cm} \nu }g^{\lambda \sigma} - g^{\mu \sigma}g^{\nu \lambda} -
g^{\mu \lambda}g^{\sigma \nu}
\label{ap:A6}
\end{equation}
and the corresponding HTL correction
\begin{equation}
\delta\hspace{0.025cm} \Gamma^{\mu\hspace{0.02cm} \nu  \lambda \sigma}(k, k_1, k_2,k_3) 
= 
3\hspace{0.045cm}\omega^2_{\rm pl}\!\int\!\frac{d\hspace{0.035cm}\Omega}{4 \pi} \, \frac{v^{\mu}\hspace{0.02cm}v^{\nu}\hspace{0.02cm} v^{\lambda}v^{\sigma}}{v\cdot k + i\hspace{0.025cm}\epsilon}
\label{ap:A7}
\end{equation}
\[
\times\Biggl[
\,\frac{1}{v\cdot (k + k_1) + i\epsilon}\, 
\Biggl(\frac{\omega_{2}}{v\cdot k_2 - i \epsilon} - \frac{\omega_3}{v\cdot k_3 - i \epsilon} \Biggr)
- \frac{1}{v\cdot (k + k_3) + i \epsilon}\,
\Biggl(\frac{\omega_{1}}{v\cdot k_1 - i \epsilon} - \frac{\omega_2}{v\cdot k_2 - i \epsilon} \Biggr)\Biggr].
\]
Finally, the expression
\begin{equation}
^{\ast}\widetilde{\cal D}_{\mu\hspace{0.02cm} \nu }(k) = 
- P_{\mu\hspace{0.02cm} \nu }(k) \,^{\ast}\!\Delta^t(k) - \widetilde{Q}_{\mu\hspace{0.02cm} \nu }(k) \,^{\ast}\!\Delta^l(k)
- \xi_{0}\ \!\frac{k^{2}}{(k\cdot u)^{2}}\ \!D_{\mu\hspace{0.02cm} \nu }(k)
\label{ap:A8}
\end{equation}
is a gluon (retarded) propagator in the $A_0$\hspace{0.02cm}-\hspace{0.02cm}gauge, which is modified by effects of the medium. Here, the ``scalar'' transverse and longitudinal propagators have the following form
\begin{equation}
\hspace{-1cm}\,^{\ast}\!\Delta^{t}(k) = \frac{1}{k^2 - \Pi^{t}(k)},
\qquad\quad\;
\,^{\ast}\!\Delta^{l}(k) = \frac{1}{k^2 - \Pi^{l}(k)},
\label{ap:A9}
\end{equation}
where
\[
\Pi^{\hspace{0.025cm} t}(k) = \frac{1}{2}\, \Pi^{\mu\nu}(k) P_{\mu\nu}(k),
\qquad
\Pi^{\hspace{0.025cm} l}(k) = \Pi^{\mu\nu}(k) \widetilde{Q}_{\mu\nu}(k).
\hspace{0.2cm}
\]
The polarization tensor $\Pi_{\mu\hspace{0.02cm} \nu }(k)$ in the HTL approximation takes the form
\[
\Pi^{\mu\hspace{0.02cm} \nu }(k) = 3\hspace{0.045cm}\omega_{\rm pl}^{2}\!\hspace{0.02cm}
\left( u^{\mu}u^{\nu} - \omega\!\int\!\frac{d\hspace{0.035cm}\Omega}{4 \pi}
\,\frac{v^{\mu}\hspace{0.02cm} v^{\nu}}{v\cdot k + i \epsilon} \right)
\]
and the longitudinal and transverse projectors are defined by the expressions
\begin{equation}
\begin{split}
&\widetilde{Q}_{\mu\hspace{0.02cm} \nu }(k) =
\frac{\tilde{u}_{\mu}(k)\hspace{0.02cm} \tilde{u}_{\nu}(k)}{\bar{u}^2(k)}\,,\\[0.7ex]
&P_{\mu\nu}(k) = g_{\mu\nu} - u_{\mu}u_{\nu}
- \widetilde{Q}_{\mu\hspace{0.02cm} \nu }(k)\,\frac{(k\cdot u)^{2}}{k^{2}}\, ,
\end{split}
\label{ap:A10}
\end{equation}
respectively. Two four-vectors 
\begin{equation}
\tilde{u}_{\mu} (k) = \frac{k^2}{(k\cdot u)}\ \! \Bigl(k_{\mu} - u_{\mu}(k\cdot u)\Bigr)
\quad \mbox{and} \quad
\bar{u}_{\mu} (k) = k^2 u_{\mu} - k_{\mu}(k\cdot u)
\label{ap:A11}
\end{equation}
are the projectors onto the longitudinal direction of wavevector ${\bf k}$ written in the Lorentz-invariant form in the Hamilton and Lorentz gauges, respectively. Here, $u^{\mu}$ is the four-velocity of the medium. In the rest frame of the system $u^{\mu}=(1,0,0,0)$. The complex conjugation rule for the gluon propagator (\ref{ap:A8}) has the following form:
\begin{equation}
\bigl(^{\ast}\widetilde{\cal D}_{\mu\hspace{0.02cm} \nu }(k)\bigr)^{\ast} = \!\,^{\ast}\widetilde{\cal D}_{\mu\hspace{0.02cm} \nu }(-k).
\label{ap:A12}
\end{equation}


\numberwithin{equation}{section}
\section{Effective quark-gluon vertices and quark pro\-pagator}
\numberwithin{equation}{section}
\label{appendix_B}

In this Appendix we give an explicit form of the spinors $u^{(\pm)}$ and $v^{(\pm)}$, of the effective quark-gluon vertex functions and quark propagator in the hard-loop framework. The spinor solutions to the free massless Dirac equation (\ref{eq:2ta}) are \cite{weldon_1989(1), weldon_1989(2)}
\begin{equation}
u^{(+)}(\hspace{0.03cm}\hat{\bf q},\lambda) = \frac{1}{\sqrt{2}}\!\hspace{0.02cm}
\left(
\begin{array}{c}
1 \\ \lambda
\end{array}
\right)\!
\chi(\hspace{0.03cm}\hat{\bf q},\lambda), 
\quad
u^{(-)}(\hspace{0.03cm}\hat{\bf q},\lambda) = \frac{1}{\sqrt{2}}\!\hspace{0.02cm}
\left(
\begin{array}{c}
-\lambda \\ 1
\end{array}
\right)\!
\chi(\hspace{0.03cm}\hat{\bf q},\lambda)
\label{ap:B1}
\end{equation}
and similarly, the negative-energy eigenvectors of (\ref{eq:2tb}) are
\begin{equation}
v^{(+)}(-\hspace{0.03cm}\hat{\bf q},\lambda) = \frac{1}{\sqrt{2}}\!\hspace{0.02cm}
\left(
\begin{array}{c}
-\lambda \\ 1
\end{array}
\right)\!
\chi(\hspace{0.03cm}\hat{\bf q},\lambda), 
\quad
v^{(-)}(-\hspace{0.03cm}\hat{\bf q},\lambda) = \frac{1}{\sqrt{2}}\!\hspace{0.02cm}
\left(
\begin{array}{c}
1 \\ \lambda
\end{array}
\right)\!
\chi(\hspace{0.03cm}\hat{\bf q},\lambda),
\label{ap:B2}
\end{equation}
where $\lambda = \pm 1$. The  two-component spinors $\chi(\hspace{0.03cm}\hat{\bf q},\lambda)$ are eigenstates of helicity:
$$
(\boldsymbol{\sigma}\cdot\hat{\bf q})\hspace{0.02cm}\chi(\hspace{0.03cm}\hat{\bf q},\lambda) 
= 
\lambda\hspace{0.03cm} \chi(\hspace{0.02cm}\hat{\bf q},\lambda)
$$
normalized to 
$$
\chi^{\dagger}(\hspace{0.03cm}\hat{\bf q},\lambda)\chi(\hspace{0.03cm}\hat{\bf q},\lambda) = 1,
\quad
\chi(\hspace{0.03cm}\hat{\bf q},\lambda)\chi^{\dagger}(\hspace{0.03cm}\hat{\bf q},\lambda) 
=
\frac{1}{2}\,\bigl({\textrm I} +\! \lambda\hspace{0.05cm} \boldsymbol{\sigma}\cdot\hat{\bf q}\bigr),
$$
such that $\sum_{\lambda} \chi(\hspace{0.03cm}\hat{\bf q},\lambda)\chi^{\dagger}(\hspace{0.03cm}\hat{\bf q},\lambda) = {\textrm I}$. Here, ${\textrm I}$ is the $2\times2$ unit matrix, $\boldsymbol{\sigma} = ( \sigma_{x}, \sigma_{y}, \sigma_{z})$ are the three Pauli matrices. The normalization factors in (\ref{ap:B1}) and (\ref{ap:B2}) are chosen so that $(u^{(\pm)})^{\dagger}\hspace{0.02cm}u^{(\pm)} = (v^{(\pm)})^{\dagger}\hspace{0.02cm}v^{(\pm)} = 1$. It is easy to see that the spinors (\ref{ap:B1}) and (\ref{ap:B2}) satisfy the following relations
\begin{equation}
\begin{split}
&\sum\limits_{\lambda\hspace{0.02cm} = \hspace{0.02cm} \pm\hspace{0.02cm} 1} u^{(\pm)}_{\alpha}(\hspace{0.03cm}\hat{\bf q},\lambda)
\,\bar{u}^{(\pm)}_{\beta}(\hspace{0.03cm}\hat{\bf q},\lambda)
=
\frac{1}{2}\,\bigl(\gamma^0 \mp \hat{{\bf q}}\cdot\boldsymbol{\gamma}\bigr)_{\alpha\beta}
\equiv (h_{\pm}(\hat{\bf q}))_{\alpha\beta},\\[1ex]
&\sum\limits_{\lambda\hspace{0.02cm} = \hspace{0.02cm}\pm\hspace{0.02cm} 1}\! v^{(\pm)}_{\alpha}(-\hat{\bf q},\lambda)
\hspace{0.02cm}\bar{v}^{(\pm)}_{\beta}(-\hat{\bf q},\lambda)
=
\frac{1}{2}\,\bigl(\gamma^0 \pm \hat{{\bf q}}\cdot\boldsymbol{\gamma}\bigr)_{\alpha\beta}
\equiv (h_{\mp}(\hat{\bf q}))_{\alpha\beta}.
\label{ap:B3}
\end{split}
\end{equation}
\indent The effective (i.e., HTL-resummed) vertex between a quark pair and a gluon
\begin{equation}
\,^{\ast}\Gamma^{(Q)\,a}_{\mu}(k; q_{1}, q_{2}) =
t^a\,^{\ast}\Gamma^{(Q)}_{\mu}(k; q_{1}, q_{2}) 
\equiv
t^a(\gamma_{\mu} + \delta\hspace{0.026cm} \Gamma^{(Q)}_{\mu}(k; q_{1}, q_{2}))
\label{ap:B4}
\end{equation}
is the sum of the bare vertex $\gamma_{\mu}$ and the corresponding HTL-correction \cite{braaten_1990, frenkel_1990, blaizot_1994(2)}
\begin{equation}
\delta\hspace{0.025cm} \Gamma^{(Q)}_{\mu}(k;q_{1},q_{2}) =
- \, \omega^{2}_0\!\int\!\frac{{\rm d} \Omega}{4 \pi} \, \frac{v_{\mu}\!\!\hspace{0.02cm} \not\!v}
{(v\cdot q_{1} + i \epsilon)(v\cdot q_{2} - i \epsilon)},
\quad \epsilon\rightarrow + \hspace{0.02cm}0,
\label{ap:B5}
\end{equation}
where $\not\!v = \gamma_{\mu}v^{\mu}$ and $k  + q_{1} + q_{2} = 0$. The superscript $(Q)$ denotes that the vertex $\delta\hspace{0.025cm} \Gamma_{\mu}(x;y_{1},y_{2})$ corresponds to the function (\ref{ap:B5}) in the coordinate representation, where the time arguments satisfy $y_{1}^0 \geq x^{0} \geq y_{2}^{0}$ (boundary conditions).\\ 
\indent Further, the effective two-quarks--one-gluon vertex function, where now (in the coordinate representation) the time arguments satisfy $x^{0} \geq {\rm max}\,(y_{1}^{0},y_{2}^{0})$ and the chronological order of $y_{1}$ and $y_{2}$ is arbitrary, has the following structure \cite{blaizot_1994(2)}:
\begin{equation}
\,^{\ast}\Gamma^{(G)}_{\mu}(k; q_{1}, q_{2}) = \gamma_{\mu} + \delta\hspace{0.025cm} 
\Gamma^{(G)}_{\mu}(k; q_{1}, q_{2}),
\label{ap:B6}
\end{equation}
\[
\delta\hspace{0.025cm} \Gamma^{(G)}_{\mu}(k; q_{1}, q_{2}) =
- \, \omega^{2}_{0}\!\int\!\frac{d \Omega}{4 \pi} \, \frac{v_{\mu}\!\!\hspace{0.02cm} \not\!v}
{(v\cdot q_{1} - i \epsilon)(v\cdot q_{2} - i \epsilon)}.
\]
The time argument of an external gluon leg incoming in the vertex is largest, as indicated by superscript $(G)$.\\
\indent The explicit form of the HTL-induced vertex between quark pair and two gluons is defined by the following expression:
\begin{equation}
\delta\hspace{0.025cm} \Gamma^{(Q)\hspace{0.02cm} ab}_{\mu\nu}(k_{1},k_{2}; q_{1},q_{2}) 
\!=\!
-\,\omega_0^2\!\!\int\!\frac{{\rm d} \Omega}{4 \pi}\, \frac{v_{\mu}v_{\nu}\!\!\hspace{0.02cm} \not\!v}
{(v\cdot q_{1} + i \epsilon)
(v\cdot q_2 - i \epsilon)}\bigg(\frac{t^{a}\hspace{0.02cm}  t^{b}}{v\cdot (q_{1} + k_{1}) + i \epsilon}
+ \frac{t^{b}\hspace{0.02cm}  t^{a}}{v\cdot (q_{1} + k_{2}) + i \epsilon} \bigg).
\label{ap:B7}
\end{equation}
This vertex does not exist at tree level, and at the leading order it arises entirely from the HTL \cite{braaten_1990, frenkel_1990, blaizot_1994(2)}. The superscript $(Q)$ denotes that the vertex $\delta\hspace{0.025cm} \Gamma_{\mu}(x_{1},x_{2};y_{1},y_{2})$ corresponds to the function (\ref{ap:B7}) in the coordinate representation, where the time arguments satisfy $y_{1}^{0} \geq x_{1}^{0} \geq x_{2}^{0} \geq y_{2}^{0}$ (boundary conditions) for the first term in parentheses on the right-hand side of Eq.\,(\ref{ap:B7}) and for the second term we have $y_{1}^0 \geq x_{2}^{0} \geq x_{1}^{0} \geq y_{2}^{0}$, i.e., the time argument of the external quark leg incoming in the vertex function (\ref{ap:B7}) is largest.\\
\indent Below we list the properties of HTL-resumed two-quark\,--\,one-gluon and two-quark\,--\,two-gluon vertex functions, which are used in the text
\begin{align}
&\gamma^{0}
(\!\,^{\ast}\Gamma^{(Q)}_{\mu}(k; q_{1}, q_{2}))^{\dagger}\gamma^{0}
= \!\,^{\ast}\Gamma^{(Q)}_{\mu}(-k; -q_{1}, -q_{2})
= \!\,^{\ast}\Gamma^{(Q)}_{\mu}(k; q_{2}, q_{1}),
\label{ap:B8}\\[1ex]
&\gamma^{0}
(\!\,^{\ast}\Gamma^{(G)}_{\mu}(k;q_1,q_2))^{\dagger}\gamma^0
= \!\,^{\ast}\Gamma^{(G)}_{\mu}(-k;-q_1,-q_2)
= \!\,^{\ast}\Gamma^{(G)}_{\mu}(-k;-q_2,-q_1),
\label{ap:B9}
\end{align}
\begin{equation}
\delta\hspace{0.025cm} \Gamma^{(Q)\hspace{0.025cm}  ab}_{\mu\hspace{0.02cm} \nu }(k_1,k_2;q_1,q_2)=
\delta\hspace{0.025cm} \Gamma^{(Q)\hspace{0.025cm}  ab}_{\mu\hspace{0.02cm} \nu }(-k_2,-k_1;-q_2,-q_1)=
\delta\hspace{0.025cm} \Gamma^{(Q)\hspace{0.025cm}  ba}_{\mu\hspace{0.02cm} \nu }(k_2,k_1;q_1,q_2),
\label{ap:B10}
\end{equation}
$$
\gamma^0
(\delta\hspace{0.025cm} \Gamma^{(Q)\hspace{0.025cm}  ab}_{\mu\hspace{0.02cm} \nu }(k_1,k_2;q_1,q_2))^{\dagger}\gamma^0
= -\,\delta\hspace{0.025cm} \Gamma^{(Q)\hspace{0.025cm}  ab}_{\mu\hspace{0.02cm} \nu }(k_1,k_2;q_2,q_1)
= -\,\delta\hspace{0.025cm} \Gamma^{(Q)\hspace{0.025cm}  ba}_{\mu\hspace{0.02cm} \nu }(-k_1,-k_2;-q_1,-q_2),
$$
$$
\gamma^0
(\delta\hspace{0.025cm} \Gamma^{(G)\hspace{0.025cm}  ab}_{\mu\hspace{0.02cm} \nu }(k_1,k_2;q_1,q_2))^{\dagger}
\gamma^0
= -\,\delta\hspace{0.025cm} \Gamma^{(G)\hspace{0.025cm}  ba}_{\mu\hspace{0.02cm} \nu }(-k_1,-k_2;-q_1,-q_2).
$$
\indent Finally, the expression
\begin{equation}
\,^{\ast}\!S(q) = h_{+}(\hat{\mathbf q}) \,^{\ast}\!\triangle_{+}(q) +
h_{-}(\hat{\mathbf q}) \,^{\ast}\!\triangle_{-}(q)
\label{ap:B11}
\end{equation}
is a quark (retarded) propagator modified by effects of the medium. The matrix functions $h_{\pm}(\hat{\bf q}) = (\gamma^0 \mp \hat{{\bf q}}\cdot\boldsymbol{\gamma})/2$ with $\hat{{\bf q}} \equiv {\bf q}/\vert {\bf q} \vert$ are the spinor projectors onto eigenstates of helicity and
\begin{equation}
\,^{\ast}\!\triangle_{\pm}(q) =
-\,\frac{1}{q^0\mp [\,\vert{\bf q}\vert + \delta\Sigma_{\pm}(q)]},
\label{ap:B12}
\end{equation}
where 
\begin{equation}
\delta\Sigma_{\pm}(q) 
=
\frac{\omega_{0}^{2}}{|{\bf q}|} \biggl[1 - \biggl(1\mp\frac{|{\bf q}|}{q^0}\biggr)
F\biggl(\frac{q^0}{\vert {\bf q}\vert}\bigg)\bigg],
\quad
F (x) \equiv \frac{x}{2} \bigg[ \ln \bigg \vert \frac{1 + x}{1 - x}
\bigg \vert - i \pi\theta (1 - \vert x \vert ) \bigg]
\label{ap:B13}
\end{equation}
are scalar quark self-energies for normal $(+)$ and plasmino $(-)$ modes and $\omega_{0}^{2}= g^2C_FT^2/8$ is plasma frequency squared of the quark sector of plasma excitations. The rule of hermitian conjugation for the effective quark propagator has the following form:
\begin{equation}
\bigl(\!\,^{\ast}\!\hspace{0.02cm}S(q)\bigr)^{\dagger} = -\hspace{0.02cm}\gamma^{0}\,^{\ast}\!\hspace{0.02cm}S(-q)\hspace{0.03cm}\gamma^{0}.
\label{ap:B14}
\end{equation}


\section{First system of the canonicity conditions}
\numberwithin{equation}{section}
\label{appendix_C}

Here, we present a system of the canonicity conditions, which follows from the Poisson superbrackets (\ref{eq:3ea})\,--\,(\ref{eq:3ed}). This system connects the coefficient functions of the second and third orders in the integrands of the canonical transformations (\ref{eq:3t}) and (\ref{eq:3y}) among themselves. Substituting the transformation (\ref{eq:3t}) into equation (\ref{eq:3ea}), equating the coefficients of the same powers of a product of the functions $c^{a}_{\bf k},\, f^{\,i}_{\bf q}$ and their conjugations to zero, and taking into account the relations (\ref{eq:3p}), we obtain two independent systems of the canonicity conditions. The first of them has the form
\begin{align}
3\hspace{0.025cm}W^{(1)\, a\, a_{1}\, a_{2}\, a_{3}}_{\ {\bf k},\, {\bf k}_{1},\, {\bf k}_{2},\, {\bf k}_{3}}
+
W^{\hspace{0.025cm} \ast\hspace{0.025cm} (3)\hspace{0.03cm} a_{3}\, a_{1}\, a_{2}\, a}_{\ {\bf k}_{3},\, {\bf k}_{1},\, {\bf k}_{2},\, {\bf k}}
\,
- 2\!\int\!\frac{d\hspace{0.02cm}{\bf k}^{\hspace{0.015cm}\prime}}{(2\pi)^{3}}\,\Bigl[\hspace{0.02cm}
&V^{(1)\, a\, a_{1}\, a^{\prime}}_{\ \ {\bf k},\, {\bf k}_{1},\, {\bf k}^{\prime}}\,
V^{(1)\, a^{\prime}\, a_{2}\, a_{3}}_{\ \ {\bf k}^{\prime},\, {\bf k}_{2},\, {\bf k}_{3}} 
-
V^{\hspace{0.025cm}\ast\hspace{0.025cm}(1)\, a_{1}\, a^{\prime}\, a}_{\ \  {\bf k}_{1},\, {\bf k}^{\prime},\, {\bf k}}\,
V^{\hspace{0.025cm}\ast\hspace{0.025cm}(3)\, a_{3}\, a_{2}\, a^{\prime}}_{\ \ {\bf k}_{3},\ {\bf k}_{2},\ {\bf k}^{\prime}}
\notag\\[1ex]
&+ 
V^{(1)\, a\, a_{2}\, a^{\prime}}_{\ \ {\bf k},\, {\bf k}_{2},\, {\bf k}^{\prime}}\,
V^{(1)\, a^{\prime}\, a_{1}\, a_{3}}_{\ \ {\bf k}^{\prime},\, {\bf k}_{1},\, {\bf k}_{3}} 
-
V^{\hspace{0.025cm}\ast\hspace{0.025cm}(1)\, a_{2}\, a^{\prime}\, a}_{\ \  {\bf k}_{2},\, {\bf k}^{\prime},\, {\bf k}}\,
V^{\hspace{0.025cm}\ast\hspace{0.025cm}(3)\, a_{3}\, a_{1}\, a^{\prime}}_{\ \ {\bf k}_{3},\, {\bf k}_{1},\, {\bf k}^{\prime}}\Bigr] = 0,
\notag\\[1ex]
W^{(2)\, a\, a_{1}\, a_{2}\, a_{3}}_{\ {\bf k},\, {\bf k}_{1},\, {\bf k}_{2},\, {\bf k}_{3}}
+
W^{\hspace{0.025cm}\ast\hspace{0.025cm}(2)\, a_{3}\, a_{2}\, a_{1}\, a}_{\ {\bf k}_{3},\, {\bf k}_{2},\, {\bf k}_{1},\, {\bf k}}
+
2\!\int\!\frac{d\hspace{0.02cm}{\bf k}^{\hspace{0.015cm}\prime}}{(2\pi)^{3}}\,\Bigl[\hspace{0.02cm}
&V^{(1)\, a^{\prime}\, a_{2}\, a_{3}}_{\ \  {\bf k}^{\prime},\, {\bf k}_{2},\, {\bf k}_{3}}\,
V^{\hspace{0.025cm}\ast\hspace{0.025cm}(1)\,a^{\prime}\, a_{1}\, a}_{\ \ {\bf k}^{\prime},\, {\bf k}_{1},\, {\bf k}}
+
V^{(1)\, a\, a_{2}\, a^{\prime}}_{\ \ {\bf k},\, {\bf k}_{2},\, {\bf k}^{\prime}}\,
V^{\hspace{0.025cm}\ast\hspace{0.025cm}(1)\, a_{3}\, a_{1}\, a^{\prime}}_{\ \ \ {\bf k}_{3},\, {\bf k}_{1},\, {\bf k}^{\prime}}
\notag\\[1ex]
&-
V^{(3)\, a\, a_{1}\, a^{\prime}}_{\ \ {\bf k},\, {\bf k}_{1},\, {\bf k}^{\prime}}\,
V^{\hspace{0.025cm}\ast\hspace{0.025cm}(3)\, a_{3}\, a_{2}\, a^{\prime}}_{\ \ {\bf k}_{3},\, {\bf k}_{2},\, {\bf k}^{\prime}}
-
V^{(1)\, a_{1}\, a^{\prime}\, a_{3}}_{\ \  {\bf k}_{1},\,{\bf k}^{\prime},\, {\bf k}_{3}}\,
V^{\hspace{0.025cm}\ast\hspace{0.025cm}(1)\, a_{2}\, a^{\prime}\, a}_{\ \ {\bf k}_{2},\, {\bf k}^{\prime},\, {\bf k}}
\Bigr] = 0,
\notag
\end{align}
and the second one is, correspondingly,
\begin{subequations} 
\label{ap:C1}
\begin{align}
J^{\,(1)\, a\, a_{1}\, i_{1}\, i_{2}}_{\ {\bf k},\, {\bf k}_{1},\, {\bf q}_{1},\, {\bf q}_{2}} 
-\, 
J^{\hspace{0.025cm}\ast\hspace{0.025cm}(3)\hspace{0.03cm} a_{1}\, a\, i_{1}\, i_{2}}_{\ {\bf k}_{1},\, {\bf k},\, {\bf q}_{1},\, {\bf q}_{2}} 
\,+
\!\int\!\frac{d{\bf q}^{\prime}}{(2\pi)^{3}}\,\Bigl[\hspace{0.02cm}
&F^{(1)\, a\, i_{1}\, i^{\prime}}_{\ {\bf k},\, {\bf q}_{1},\, {\bf q}^{\prime}}\, 
F^{\hspace{0.025cm}\ast\hspace{0.025cm}(2)\hspace{0.03cm} a_{1}\, i_{2}\, i^{\prime}}_{\ {\bf k}_{1},\, {\bf q}_{2},\, {\bf q}^{\prime}}
-
F^{\,(1)\, a\, i_{2}\, i^{\prime}}_{\ {\bf k},\, {\bf q}_{2},\, {\bf q}^{\prime}}\, 
F^{\hspace{0.025cm}\ast\hspace{0.025cm}(2)\hspace{0.03cm} a_{1}\, i_{1}\, i^{\prime}}_{\ {\bf k}_{1},\, {\bf q}_{1},\, {\bf q}^{\prime}}
\label{ap:C1a}\\[1ex]
&+
F^{(2)\, a\, i^{\prime}\, i_{2}}_{\ {\bf k},\, {\bf q}^{\prime},\, {\bf q}_{2}}\, 
F^{\hspace{0.025cm}\ast\hspace{0.025cm}(3)\hspace{0.03cm} a_{1}\, i_{1}\, i^{\prime}}_{\ {\bf k}_1,\, {\bf q}_{1},\, {\bf q}^{\prime}}
-
F^{(2)\, a\, i^{\prime}\, i_{1}}_{\ {\bf k},\, {\bf q}^{\prime},\, {\bf q}_{1}}\, 
F^{\hspace{0.025cm}\ast\hspace{0.025cm}(3)\hspace{0.03cm} a_{1}\, i_{2}\, i^{\prime}}_{\ {\bf k}_1,\, {\bf q}_{2},\, {\bf q}^{\prime}}\Bigr] = 0, \notag\\[1ex]
J^{\,(2)\, a\, a_{1}\, i_{1}\, i_{2}}_{\ {\bf k},\, {\bf k}_{1},\, {\bf q}_{1},\, {\bf q}_{2}} 
+ 
J^{\hspace{0.025cm}\ast\hspace{0.025cm}(2)\hspace{0.03cm} a_{1}\, a\, i_{2}\, i_{1}}_{\ {\bf k}_{1},\, {\bf k},\, {\bf q}_{2},\, {\bf q}_{1}} 
+
\!\int\!\frac{d{\bf q}^{\prime}}{(2\pi)^{3}}\,\Bigl[\hspace{0.02cm}
4\hspace{0.01cm}&F^{(1)\, a\, i_{2}\, i^{\prime}}_{\ {\bf k},\, {\bf q}_{2},\, {\bf q}^{\prime}}\, 
F^{\hspace{0.025cm}\ast\hspace{0.025cm}(1)\hspace{0.03cm} a_{1}\, i^{\prime}\, i_{1}}_{\ {\bf k}_{1},\,{\bf q}^{\prime},\, {\bf q}_{1}}
+
F^{(2)\, a\, i_{1}\, i^{\prime}}_{\ {\bf k},\, {\bf q}_{1},\, {\bf q}^{\prime}}\, 
F^{\hspace{0.025cm}\ast\hspace{0.025cm}(2)\hspace{0.03cm} a_{1}\, i_{2}\, i^{\prime}}_{\ {\bf k}_{1},\, {\bf q}_{2},\, {\bf q}^{\prime}}
\label{ap:C1b}\\[1ex]
&-
F^{(2)\, a\, i^{\prime}\, i_{2}}_{\ {\bf k},\, {\bf q}^{\prime},\, {\bf q}_{2}}\, 
F^{\hspace{0.025cm}\ast\hspace{0.025cm}(2)\hspace{0.03cm} a_{1}\, i^{\prime}\, i_{1}}_{\ {\bf k}_{1}, \, {\bf q}^{\prime},\, {\bf q}_{1}}
-
4\hspace{0.01cm}F^{(3)\, a\, i^{\prime}\, i_{1}}_{\ {\bf k},\, {\bf q}^{\prime},\, {\bf q}_{1}}\, 
F^{\hspace{0.025cm}\ast\hspace{0.025cm}(3)\hspace{0.03cm} a_{1}\, i_{2}\, i^{\prime}}_{\ {\bf k}_1,\, {\bf q}_{2},\, {\bf q}^{\prime}}\hspace{0.02cm}\Bigr] = 0. \notag
\end{align}
\end{subequations} 
We have obtained the first system for purely bosonic case in \cite{markov_2020}. Here, we give it for the completeness of the description. Substituting the transformation (\ref{eq:3t}) into (\ref{eq:3eb}), we find two further independent systems of the canonicity conditions: the first one \cite{markov_2020} is
\begin{align}
W^{(2)\, a\, a_{1}\, a_{2}\, a_{3}}_{\ {\bf k},\, {\bf k}_{1},\, {\bf k}_{2},\, {\bf k}_{3}}
-
W^{(2)\, a_{1}\, a\, a_{2}\, a_{3}}_{\ {\bf k}_{1},\, {\bf k},\, {\bf k}_{2},\, {\bf k}_{3}}
-
2\!\int\!\frac{d\hspace{0.02cm}{\bf k}^{\hspace{0.015cm}\prime}}{(2\pi)^{3}}\,\Bigl[\hspace{0.02cm}
&V^{(1)\, a_{1}\, a_{2}\, a^{\prime}}_{\ \ {\bf k}_{1},\, {\bf k}_{2},\, {\bf k}^{\prime}}\,
V^{\hspace{0.025cm}\ast\hspace{0.025cm}(1)\hspace{0.03cm} a_{3}\, a^{\prime}\, a}_{\ \ {\bf k}_{3},\, {\bf k}^{\prime},\, {\bf k}}
-
V^{(1)\, a\, a_{3}\, a^{\prime}}_{\ \ {\bf k},\, {\bf k}_{3},\, {\bf k}^{\prime}}
\,
V^{\hspace{0.025cm}\ast\hspace{0.025cm}(1)\, a_{2}\,  a^{\prime}\, a_{1}}_{\ \ {\bf k}_{2},\, {\bf k}^{\prime},\, {\bf k}_{1}}
\notag\\[1ex]
&+
V^{(1)\, a_{1}\, a_{3}\, a^{\prime}}_{\ \ {\bf k}_{1},\, {\bf k}_{3},\, {\bf k}^{\prime}}\
V^{\hspace{0.025cm}\ast\hspace{0.025cm}(1)\hspace{0.03cm} a_{2}\, a^{\prime}\, a}_{\ \ {\bf k}_{2},\, {\bf k}^{\prime},\, {\bf k}}
-
V^{(1)\, a\, a_{2}\, a^{\prime}}_{\ \ {\bf k},\, {\bf k}_{2},\, {\bf k}^{\prime}}
\,
V^{\hspace{0.025cm}\ast\hspace{0.025cm}(1)\hspace{0.03cm} a_{3}\,  a^{\prime}\, a_{1}}_{\ \ {\bf k}_{3},\, {\bf k}^{\prime},\, {\bf k}_{1}}
\Bigr] = 0,
\notag\\[1ex]
W^{(3)\, a\, a_{1}\, a_{2}\, a_{3}}_{\ {\bf k},\, {\bf k}_{1},\, {\bf k}_{2},\, {\bf k}_{3}}
-
W^{(3)\, a_{2}\, a_{1}\, a\, a_{3}}_{\ {\bf k}_{2},\, {\bf k}_{1},\, {\bf k},\, {\bf k}_{3}}
+
2\!\int\!\frac{d\hspace{0.02cm}{\bf k}^{\hspace{0.015cm}\prime}}{(2\pi)^{3}}\,\Bigl[\hspace{0.02cm}
&V^{\hspace{0.025cm}\ast\hspace{0.025cm}(1)\hspace{0.03cm} a^{\prime}\, a_{1}\, a_{2}}_{\ \ {\bf k}^{\prime},\, {\bf k}_{1},\, {\bf k}_{2}}\,
V^{\hspace{0.025cm}\ast\hspace{0.025cm}(1)\hspace{0.03cm} a_{3}\, a^{\prime}\, a}_{\ \  {\bf k}_{3},\, {\bf k}^{\prime},\, {\bf k}}\,
+
V^{(1)\, a_{2}\, a_{3}\, a^{\prime}}_{\ \ {\bf k}_{2},\, {\bf k}_{3},\, {\bf k}^{\prime}}\,
V^{(3)\, a\, a_{1}\, a^{\prime}}_{\ \ {\bf k},\, {\bf k}_{1},\, {\bf k}^{\prime}}
\notag\\[1ex]
&-
V^{\hspace{0.025cm}\ast\hspace{0.025cm}(1)\hspace{0.03cm} a^{\prime}\, a_{1}\, a}_{\ \ {\bf k}^{\prime},\, {\bf k}_{1},\, {\bf k}}\,
V^{\hspace{0.025cm}\ast\hspace{0.025cm}(1)\hspace{0.03cm} a_{3}\, a^{\prime}\, a_{2}}_{\ \ {\bf k}_{3},\, {\bf k}^{\prime},\, {\bf k}_{2}}
-
V^{(1)\, a\, a_{3}\, a^{\prime}}_{\ \ {\bf k},\, {\bf k}_{3},\, {\bf k}^{\prime}}\,
V^{(3)\, a_{2}\, a_{1}\, a^{\prime}}_{\ \ {\bf k}_{2},\, {\bf k}_{1},\, {\bf k}^{\prime}}
\Bigl] = 0,
\notag\\[1ex]
3\hspace{0.02cm}W^{(4)\, a\, a_{1}\, a_{2}\, a_{3}}_{\ {\bf k},\, {\bf k}_{1},\, {\bf k}_{2},\, {\bf k}_{3}}
-
3\hspace{0.02cm}W^{(4)\, a_{3}\, a_{1}\, a_{2}\, a}_{\ \hspace{0.02cm} {\bf k}_{3},\, {\bf k}_{1},\, {\bf k}_{2},\, {\bf k}}
-
\,2\!\int\!\frac{d\hspace{0.02cm}{\bf k}^{\hspace{0.015cm}\prime}}{(2\pi)^{3}}\,\Bigl[\hspace{0.02cm}
&V^{\hspace{0.025cm}\ast\hspace{0.025cm}(1)\hspace{0.03cm} a^{\prime}\, a_{1}\, a_{3}}_{\ \ {\bf k}^{\prime},\, {\bf k}_{1},\, {\bf k}_{3}}\,
V^{(3)\, a\, a_{2}\, a^{\prime}}_{\ \ {\bf k},\, {\bf k}_{2},\, {\bf k}^{\prime}}
-
V^{\hspace{0.025cm}\ast\hspace{0.025cm}(1)\hspace{0.03cm} a^{\prime}\, a_{2}\, a}_{\ \ {\bf k}^{\prime},\, {\bf k}_{2},\, {\bf k}}\,
V^{(3)\, a_{3}\, a_{1}\, a^{\prime}}_{\ \ {\bf k_{3}},\, {\bf k}_{1},\, {\bf k}^{\prime}}
\notag\\[1ex]
&+
V^{\hspace{0.025cm}\ast\hspace{0.025cm}(1)\hspace{0.03cm} a^{\prime}\, a_{2}\, a_{3}}_{\ \ {\bf k}^{\prime},\, {\bf k}_{2},\, {\bf k}_{3}}\,
V^{(3)\, a\, a_{1}\, a^{\prime}}_{\ \ {\bf k},\, {\bf k}_{1},\, {\bf k}^{\prime}}
-
V^{\hspace{0.025cm}\ast\hspace{0.025cm}(1)\hspace{0.03cm} a^{\prime}\, a_{1}\, a}_{\ \ {\bf k}^{\prime},\, {\bf k}_{1},\, {\bf k}}\,
V^{(3)\, a_{3}\, a_{2}\, a^{\prime}}_{\ \ {\bf k_{3}},\, {\bf k}_{2},\, {\bf k}^{\prime}}\Bigr] = 0
\notag
\end{align}
and the second one is
\begin{subequations} 
\label{ap:C2}
\begin{align}
J^{\,(4)\, a\, a_{1}\, i_{1}\, i_{2}}_{\; {\bf k},\, {\bf k}_{1},\, {\bf q}_{1},\, {\bf q}_{2}} 
-
J^{\,(4)\, a_{1}\, a\, i_{1}\, i_{2}}_{\; {\bf k_{1}},\, {\bf k},\, {\bf q}_{1},\, {\bf q}_{2}} 
-
\!\int\!\frac{d{\bf q}^{\prime}}{(2\pi)^{3}}\,\Bigl[\hspace{0.03cm}
&F^{(1)\, a\, i_{1}\, i^{\prime}}_{\ {\bf k},\, {\bf q}_{1},\, {\bf q}^{\prime}}\, F^{(2)\, a_{1}\, i^{\prime}\, i_{2}}_{\ {\bf k}_1,\, {\bf q}^{\prime},\, {\bf q}_{2}}
-
F^{(1)\, a_{1}\, i_{1}\, i^{\prime}}_{\ {\bf k}_{1},\, {\bf q}_{1},\, {\bf q}^{\prime}}\, F^{(2)\, a\, i^{\prime}\, i_{2}}_{\ {\bf k},\, {\bf q}^{\prime},\, {\bf q}_{2}}
\label{ap:C2a}\\[1ex]
&+
F^{(2)\, a\, i^{\prime}\, i_{1}}_{\ {\bf k},\, {\bf q}^{\prime},\, {\bf q}_{1}}\,F^{(1)\, a_{1}\, i_{2}\, i^{\prime}}_{\ {\bf k}_1,\, {\bf q_{2}},\, {\bf q}^{\prime}}
-
F^{(2)\, a_{1}\, i^{\prime}\, i_{1}}_{\ {\bf k}_{1},\, {\bf q}^{\prime},\, {\bf q}_{1}}\,F^{(1)\, a\, i_{2}\, i^{\prime}}_{\ {\bf k},\, {\bf q_{2}},\, {\bf q}^{\prime}}\hspace{0.03cm}\Bigr] = 0,
\notag\\[1ex]
J^{\,(5)\, a\, a_{1}\, i_{1}\, i_{2}}_{\; {\bf k},\, {\bf k}_{1},\, {\bf q}_{1},\, {\bf q}_{2}} 
-
J^{\,(5)\, a_{1}\, a\, i_{1}\, i_{2}}_{\; {\bf k_{1}},\, {\bf k},\, {\bf q}_{1},\, {\bf q}_{2}} 
-
\!\int\!\frac{d{\bf q}^{\prime}}{(2\pi)^{3}}\,\Bigl[\hspace{0.03cm}
&F^{(2)\, a\, i_{1}\, i^{\prime}}_{\ {\bf k},\, {\bf q}_{1},\, {\bf q}^{\prime}}\, F^{(2)\, a_{1}\, i^{\prime}\, i_{2}}_{\ {\bf k}_1,\, {\bf q}^{\prime},\, {\bf q}_{2}}
-
F^{(2)\, a_{1}\, i_{1}\, i^{\prime}}_{\ {\bf k}_{1},\, {\bf q}_{1},\, {\bf q}^{\prime}}\, F^{(2)\, a\, i^{\prime}\, i_{2}}_{\ {\bf k},\, {\bf q}^{\prime},\, {\bf q}_{2}}
\label{ap:C2b}\\[1ex]
&+
4\hspace{0.02cm}F^{(3)\, a\, i^{\prime}\, i_{1}}_{\ {\bf k},\, {\bf q}^{\prime},\, {\bf q}_{1}}\,F^{(1)\, a_{1}\, i_{2}\, i^{\prime}}_{\ {\bf k}_1,\, {\bf q_{2}},\, {\bf q}^{\prime}}
-
4\hspace{0.02cm}F^{(3)\, a_{1}\, i^{\prime}\, i_{1}}_{\ {\bf k}_{1},\, {\bf q}^{\prime},\, {\bf q}_{1}}\,F^{(1)\, a\, i_{2}\, i^{\prime}}_{\ {\bf k},\, {\bf q_{2}},\, {\bf q}^{\prime}}\hspace{0.03cm}\Bigr] = 0,
\notag\\[1ex]
J^{\,(6)\, a\, a_{1}\, i_{1}\, i_{2}}_{\; {\bf k},\, {\bf k}_{1},\, {\bf q}_{1},\, {\bf q}_{2}} 
-
J^{\,(6)\, a_{1}\, a\, i_{1}\, i_{2}}_{\; {\bf k_{1}},\, {\bf k},\, {\bf q}_{1},\, {\bf q}_{2}} 
-
\!\int\!\frac{d{\bf q}^{\prime}}{(2\pi)^{3}}\,\Bigl[\hspace{0.03cm}
&F^{(2)\, a\, i_{1}\, i^{\prime}}_{\ {\bf k},\, {\bf q}_{1},\, {\bf q}^{\prime}}\, F^{(3)\, a_{1}\, i^{\prime}\, i_{2}}_{\ {\bf k}_1,\, {\bf q}^{\prime},\, {\bf q}_{2}}
-
F^{(2)\, a_{1}\, i_{1}\, i^{\prime}}_{\ {\bf k}_{1},\, {\bf q}_{1},\, {\bf q}^{\prime}}\, F^{(3)\, a\, i^{\prime}\, i_{2}}_{\ {\bf k},\, {\bf q}^{\prime},\, {\bf q}_{2}}
\label{ap:C2c}\\[1ex]
&+
F^{(3)\, a\, i^{\prime}\, i_{1}}_{\ {\bf k},\, {\bf q}^{\prime},\, {\bf q}_{1}}\,F^{(2)\, a_{1}\, i_{2}\, i^{\prime}}_{\ {\bf k}_1,\, {\bf q_{2}},\, {\bf q}^{\prime}}
-
F^{(3)\, a_{1}\, i^{\prime}\, i_{1}}_{\ {\bf k}_{1},\, {\bf q}^{\prime},\, {\bf q}_{1}}\,F^{(2)\, a\, i_{2}\, i^{\prime}}_{\ {\bf k},\, {\bf q_{2}},\, {\bf q}^{\prime}}\hspace{0.03cm}\Bigr] = 0.
\notag
\end{align}
\end{subequations} 
The system (\ref{ap:C2}) defines the rearrangement rule for the color indices $a$ and $a_{1}$, and for the corresponding momentum arguments ${\bf k}$ and ${\bf k}_{1}$ of the third-order coefficient functions  $J^{(4)}, J^{(5)}$, and $J^{(6)}$.\\
\indent Substituting further (\ref{eq:3t}) and  (\ref{eq:3y}) into the third equation in (\ref{eq:3e}), we obtain more nontrivial relations mixing the fermion and boson degrees of freedom of the system
\begin{subequations} 
\label{ap:C3}
\begin{align}
R^{\,(2)\, i\, a\, a_{1}\, i_{1}}_{\, {\bf q},\, {\bf k},\, {\bf k}_{1},\, {\bf q}_{1}} 
-
J^{\,(2)\, a\, a_{1}\, i\; i_{1}}_{\ {\bf k},\, {\bf k}_{1},\, {\bf q},\, {\bf q}_{1}} 
+
2\!\int\!\frac{d\hspace{0.02cm}{\bf k}^{\hspace{0.015cm}\prime}}{(2\pi)^{3}}\,\Bigl[\hspace{0.03cm}
&V^{(1)\hspace{0.03cm} a\, a_{1}\, a^{\prime}}_{\ {\bf k},\, {\bf k}_{1},\, {\bf k}^{\prime}}\hspace{0.01cm} 
F^{(2)\, a^{\prime}\, i\, i_{1}}_{\ {\bf k}^{\prime},\, {\bf q},\, {\bf q}_{1}}
-
V^{\hspace{0.025cm}\ast\hspace{0.025cm}(1)\hspace{0.03cm} a_{1}\, a^{\prime}\, a}_{\ {\bf k}_{1},\, {\bf k}^{\prime},\, {\bf k}}
\hspace{0.02cm}
F^{\hspace{0.025cm}\ast\hspace{0.025cm}(2)\, a^{\prime}\, i_{1}\, i}_{\ {\bf k}^{\prime},\, {\bf q}_{1},\, {\bf q}}\hspace{0.03cm}\Bigr]
\label{ap:C3a}\\[1ex]
+\!
\int&\!\frac{d{\bf q}^{\prime}}{(2\pi)^{3}}\,\Bigl[\hspace{0.03cm}4\hspace{0.02cm}
F^{(1)\hspace{0.03cm} a\, i_{1}\, i^{\prime}}_{\ {\bf k},\, {\bf q}_{1},\, {\bf q}^{\prime}}\, 
F^{\hspace{0.025cm}\ast\hspace{0.025cm}(1)\hspace{0.03cm} a_{1}\, i\, i^{\prime}}_{\ {\bf k}_1,\, {\bf q},\, {\bf q}^{\prime}}
\!+
F^{(2)\, a\, i^{\prime}\, i_{1}}_{\ {\bf k},\, {\bf q}^{\prime},\, {\bf q}_{1}}\, 
F^{\hspace{0.025cm}\ast\hspace{0.025cm}(2)\hspace{0.03cm} a_{1}\, i^{\prime}\, i}_{\ {\bf k}_{1},\,  {\bf q}^{\prime},\, {\bf q}}\hspace{0.03cm}\Bigr] = 0,
\notag\\[1ex]
2\hspace{0.015cm}R^{\,(3)\, i\, a\, a_{1}\, i_{1}}_{\, {\bf q},\, {\bf k},\, {\bf k}_{1},\, {\bf q}_{1}} 
-
J^{\,(5)\, a\, a_{1}\, i\; i_{1}}_{\ {\bf k},\, {\bf k}_{1},\, {\bf q},\, {\bf q}_{1}}
+
2\!\int\!\frac{d\hspace{0.02cm}{\bf k}^{\hspace{0.015cm}\prime}}{(2\pi)^{3}}\,\Bigl[\hspace{0.03cm}
&V^{(3)\hspace{0.03cm} a\, a_{1}\, a^{\prime}}_{\ {\bf k},\, {\bf k}_{1},\, {\bf k}^{\prime}}
\hspace{0.01cm} 
F^{\hspace{0.025cm}\ast\hspace{0.025cm}(2)\hspace{0.03cm} a^{\prime}\, i_{1}\, i}_{\ {\bf k}^{\prime},\, {\bf q}_{1},\, {\bf q}}
\!-
V^{\hspace{0.025cm}\ast\hspace{0.025cm}(1)\hspace{0.03cm} a^{\prime}\, a_{1}\, a}_{\ {\bf k}^{\prime},\, {\bf k}_{1},\, {\bf k}}
\hspace{0.02cm} 
F^{(2)\hspace{0.03cm} a^{\prime}\, i\; i_{1}}_{\ {\bf k}^{\prime},\, {\bf q},\, {\bf q}_{1}}\hspace{0.03cm}\Bigr]
\label{ap:C3b}\\[1ex]
+\!
\int&\!\frac{d{\bf q}^{\prime}}{(2\pi)^{3}}\,\Bigl[\hspace{0.03cm}4\hspace{0.02cm}
F^{(1)\, a\, i_{1}\, i^{\prime}}_{\ {\bf k},\, {\bf q}_{1},\, {\bf q}^{\prime}}\, 
F^{(3)\hspace{0.03cm} a_{1}\, i\; i^{\prime}}_{\ {\bf k}_1,\, {\bf q},\, {\bf q}^{\prime}}
-
F^{(2)\, a\, i^{\prime}\, i_{1}}_{\ {\bf k},\, {\bf q}^{\prime},\, {\bf q}_{1}}\, 
F^{(2)\hspace{0.03cm} a_{1}\, i\; i^{\prime}}_{\ {\bf k}_{1},\,  {\bf q},\, {\bf q}^{\prime}}\hspace{0.03cm}\Bigr] = 0,
\notag\\[1ex]
R^{\,(5)\, i\, a\, a_{1}\, i_{1}}_{\, {\bf q},\, {\bf k},\, {\bf k}_{1},\, {\bf q}_{1}} 
-
2\hspace{0.02cm}J^{\,(3)\, a\, a_{1}\, i\; i_{1}}_{\ {\bf k},\, {\bf k}_{1},\, {\bf q},\, {\bf q}_{1}}
+
4\!\int\!\frac{d\hspace{0.02cm}{\bf k}^{\hspace{0.015cm}\prime}}{(2\pi)^{3}}\,\Bigl[\hspace{0.03cm}
&V^{(1)\,a\, a_{1}\, a^{\prime}}_{\ {\bf k},\, {\bf k}_{1},\, {\bf k}^{\prime}}
\hspace{0.02cm}
F^{(3)\hspace{0.03cm} a^{\prime}\, i\; i_{1}}_{\ {\bf k}^{\prime},\, {\bf q},\, {\bf q}_{1}}
+
V^{\hspace{0.025cm}\ast\hspace{0.025cm}(1)\hspace{0.03cm} a_{1}\, a^{\prime}\, a}_{\ {\bf k}_{1},\, {\bf k}^{\prime},\, {\bf k}}
\hspace{0.02cm}
F^{\hspace{0.025cm}\ast\hspace{0.025cm}(1)\hspace{0.03cm} a^{\prime}\, i\, i_{1}}_{\ {\bf k}^{\prime},\, {\bf q},\, {\bf q}_{1}}\hspace{0.03cm}\Bigr]
\label{ap:C3c}\\[1ex]
+\,
2\!\int&\!\frac{d{\bf q}^{\prime}}{(2\pi)^{3}}\,\Bigl[\hspace{0.03cm}
F^{(2)\, a\, i_{1}\, i^{\prime}}_{\ {\bf k},\, {\bf q}_{1},\, {\bf q}^{\prime}}\, F^{\hspace{0.025cm}\ast\hspace{0.025cm}(1)\, a_{1}\, i\, i^{\prime}}_{\ {\bf k}_1,\, {\bf q},\, {\bf q}^{\prime}}
+
F^{(3)\, a\, i^{\prime}\, i_{1}}_{\ {\bf k},\, {\bf q}^{\prime},\, {\bf q}_{1}}\, F^{\hspace{0.025cm}\ast\hspace{0.025cm}(2)\, a_{1}\, i^{\prime}\, i}_{\ {\bf k}_{1},\,  {\bf q}^{\prime},\, {\bf q}}\hspace{0.03cm}\Bigr] = 0,
\notag\\[1ex]
R^{\,(6)\, i\, a\, a_{1}\, i_{1}}_{\, {\bf q},\, {\bf k},\, {\bf k}_{1},\, {\bf q}_{1}} 
-
J^{\,(6)\, a\, a_{1}\, i\; i_{1}}_{\ {\bf k},\, {\bf k}_{1},\, {\bf q},\, {\bf q}_{1}}
-
2\!\int\!\frac{d\hspace{0.02cm}{\bf k}^{\hspace{0.015cm}\prime}}{(2\pi)^{3}}\,\Bigl[\hspace{0.03cm}
&V^{\hspace{0.025cm}\ast\hspace{0.025cm}(1)\hspace{0.03cm} a^{\prime}\, a_{1}\, a}_{\ {\bf k}^{\prime},\, {\bf k}_{1},\, {\bf k}}
\hspace{0.02cm}
F^{(3)\hspace{0.03cm} a^{\prime}\, i\; i_{1}}_{\ {\bf k}^{\prime},\, {\bf q},\, {\bf q}_{1}}
+
V^{\,(3)\,a\, a_{1}\, a^{\prime}}_{\ {\bf k},\, {\bf k}_{1},\, {\bf k}^{\prime}}
\hspace{0.02cm}
F^{\hspace{0.025cm}\ast\hspace{0.025cm}(1)\hspace{0.03cm} a^{\prime}\, i\, i_{1}}_{\ {\bf k}^{\prime},\, {\bf q},\, {\bf q}_{1}}\hspace{0.03cm}\Bigr]
\label{ap:C3d}\\[1ex]
+\,
\int&\!\frac{d{\bf q}^{\prime}}{(2\pi)^{3}}\,\Bigl[\hspace{0.03cm}
F^{(2)\, a\, i_{1}\, i^{\prime}}_{\ {\bf k},\, {\bf q}_{1},\, {\bf q}^{\prime}}\, 
F^{(3)\hspace{0.03cm} a_{1}\, i\; i^{\prime}}_{\ {\bf k}_1,\, {\bf q},\, {\bf q}^{\prime}}
-
F^{(3)\, a\, i^{\prime}\, i_{1}}_{\ {\bf k},\, {\bf q}^{\prime},\, {\bf q}_{1}}\, 
F^{(2)\hspace{0.03cm} a_{1}\, i\; i^{\prime}}_{\ {\bf k}_{1},\,  {\bf q},\,{\bf q}^{\prime}}
\hspace{0.03cm}\Bigr] = 0.
\notag
\end{align}
\end{subequations} 
\indent Finally, the substitution of (\ref{eq:3t}) and (\ref{eq:3y}) into the fourth equation (\ref{eq:3ed}) results in
\begin{subequations} 
\label{ap:C4}
\begin{align}
2\hspace{0.01cm}R^{\hspace{0.025cm}\ast\hspace{0.025cm}(1)\hspace{0.03cm} i\, a_{1}\, a\, i_{1}}_{\ \,{\bf q},\, 
{\bf k}_{1},\, {\bf k},\, {\bf q}_{1}} 
+
J^{\,(5)\, a\, a_{1}\, i_{1}\, i}_{\; {\bf k},\, {\bf k}_{1},\, {\bf q}_{1},\, {\bf q}} 
+
2\!\int\!\frac{d\hspace{0.02cm}{\bf k}^{\hspace{0.015cm}\prime}}{(2\pi)^{3}}\,\Bigl[\hspace{0.03cm}
&V^{\hspace{0.025cm}\ast\hspace{0.025cm}(1)\hspace{0.03cm} a^{\prime}\, a_{1}\, a}_{\ {\bf k}^{\prime},\, {\bf k}_{1},\, {\bf k}}
\hspace{0.03cm}
F^{(2)\, a^{\prime}\, i_{1}\, i}_{\ {\bf k}^{\prime},\, {\bf q}_{1},\, {\bf q}}
-
V^{(3)\hspace{0.03cm} a\, a_{1}\, a^{\prime}}_{\ {\bf k},\, {\bf k}_{1},\, {\bf k}^{\prime}}
\hspace{0.02cm} 
F^{\hspace{0.025cm}\ast\hspace{0.025cm}(2)\hspace{0.03cm} a^{\prime}\, i\, i_{1}}_{\ {\bf k}^{\prime},\, {\bf q},\, 
{\bf q}_{1}}\hspace{0.03cm}\Bigr]
\label{ap:C4a}\\[1ex]
-
\!\int&\!\frac{d{\bf q}^{\prime}}{(2\pi)^{3}}\,\Bigl[\hspace{0.03cm}
F^{(2)\, a\, i_{1}\, i^{\prime}}_{\ {\bf k},\, {\bf q}_{1},\, {\bf q}^{\prime}}\, F^{(2)\, a_{1}\, i^{\prime}\, i}_{\ {\bf k}_1,\, 
{\bf q}^{\prime},\, {\bf q}}
-
4\hspace{0.01cm}F^{(3)\, a\, i^{\prime}\, i_{1}}_{\ {\bf k},\, {\bf q}^{\prime},\, {\bf q}_{1}}\, F^{(1)\, a_{1}\, 
i^{\prime}\, i}_{\ {\bf k}_{1},\,  {\bf q}^{\prime},\, {\bf q}}\hspace{0.04cm}\Bigr] = 0,
\notag\\[1ex]
R^{\hspace{0.025cm}\ast\hspace{0.025cm}(2)\hspace{0.03cm} i\, a_{1}\, a\, i_{1}}_{\ \,{\bf q},\, {\bf k}_{1},\, {\bf k},\, 
{\bf q}_{1}} 
+
J^{\,(2)\, a\, a_{1}\, i_{1}\, i}_{\; {\bf k},\, {\bf k}_{1},\, {\bf q}_{1},\, {\bf q}} 
-
2\!\int\!\frac{d\hspace{0.02cm}{\bf k}^{\hspace{0.015cm}\prime}}{(2\pi)^{3}}\,\Bigl[\hspace{0.03cm}
&V^{(1)\hspace{0.03cm} a\, a_{1}\, a^{\prime}}_{\ {\bf k},\, {\bf k}_{1},\, {\bf k}^{\prime}}
\hspace{0.02cm}
F^{(2)\, a^{\prime}\, i_{1}\, i}_{\ {\bf k}^{\prime},\, {\bf q}_{1},\, {\bf q}}
-
V^{\hspace{0.025cm}\ast\hspace{0.025cm}(1)\hspace{0.03cm} a_{1}\, a^{\prime}\, a}_{\ {\bf k}_{1},\, {\bf k}^{\prime},\, {\bf k}}
\hspace{0.02cm}
F^{\hspace{0.025cm}\ast\hspace{0.025cm}(2)\, a^{\prime}\, i\; i_{1}}_{\ {\bf k}^{\prime},\, {\bf q},\, {\bf q}_{1}}\hspace{0.03cm}\Bigr]
\label{ap:C4b}\\[1ex]
+\!
\int&\!\frac{d{\bf q}^{\prime}}{(2\pi)^{3}}\,\Bigl[\hspace{0.03cm}
F^{(2)\, a\, i_{1}\, i^{\prime}}_{\ {\bf k},\, {\bf q}_{1},\, {\bf q}^{\prime}}
\, 
F^{\hspace{0.025cm}\ast\hspace{0.025cm}(2)\hspace{0.03cm} a_{1}\, i\, i^{\prime}}_{\ {\bf k}_1,\, {\bf q},\, 
{\bf q}^{\prime}}
-
4\hspace{0.01cm}F^{(3)\, a\, i^{\prime}\, i_{1}}_{\ {\bf k},\, {\bf q}^{\prime},\, {\bf q}_{1}}
\, 
F^{\hspace{0.025cm}\ast\hspace{0.025cm}(3)\hspace{0.03cm} a_{1}\, i\, i^{\prime}}_{\ {\bf k}_{1},\,  {\bf q},\, 
{\bf q}^{\prime}}\hspace{0.03cm}\Bigr] = 0,
\notag\\[1ex]
R^{\hspace{0.025cm}\ast\hspace{0.025cm}(4)\hspace{0.03cm} i\, a_{1}\, a\, i_{1}}_{\ \,{\bf q},\, {\bf k}_{1},\, {\bf k},\, 
{\bf q}_{1}} 
+
J^{\,(4)\, a\, a_{1}\, i_{1}\, i}_{\; {\bf k},\, {\bf k}_{1},\, {\bf q}_{1},\, {\bf q}} 
+
2\!\int\!\frac{d\hspace{0.02cm}{\bf k}^{\hspace{0.015cm}\prime}}{(2\pi)^{3}}\,\Bigl[\hspace{0.03cm}
&V^{\hspace{0.025cm}\ast\hspace{0.025cm}(1)\hspace{0.03cm} a^{\prime}\, a_{1}\, a}_{\ {\bf k}^{\prime},\, {\bf k}_{1},\, {\bf k}}
\hspace{0.02cm} 
F^{(1)\, a^{\prime}\, i_{1}\, i}_{\ {\bf k}^{\prime},\, {\bf q}_{1},\, {\bf q}}
-
V^{(3)\hspace{0.03cm} a\, a_{1}\, a^{\prime}}_{\ {\bf k},\, {\bf k}_{1},\, {\bf k}^{\prime}}
\hspace{0.02cm}
F^{\hspace{0.025cm}\ast\hspace{0.025cm}(3)\hspace{0.03cm} a^{\prime}\, i\; i_{1}}_{\ {\bf k}^{\prime},\, {\bf q},\, 
{\bf q}_{1}}\hspace{0.03cm}\Bigr]
\label{ap:C4c}\\[1ex]
-
\!\int&\!\frac{d{\bf q}^{\prime}}{(2\pi)^{3}}\,\Bigl[\hspace{0.03cm}
F^{(1)\, a\, i_{1}\, i^{\prime}}_{\ {\bf k},\, {\bf q}_{1},\, {\bf q}^{\prime}}\, F^{(2)\, a_{1}\, i^{\prime}\, i}_{\ {\bf k}_1,\, 
{\bf q}^{\prime},\, {\bf q}}
-
F^{(2)\, a\, i^{\prime}\, i_{1}}_{\ {\bf k},\, {\bf q}^{\prime},\, {\bf q}_{1}}\, F^{(1)\, a_{1}\, i^{\prime}\, i}_{\ {\bf k}_{1},\,  
{\bf q}^{\prime},\, {\bf q}}\hspace{0.04cm}\Bigr] = 0,
\notag\\[1ex]
R^{\hspace{0.025cm}\ast\hspace{0.025cm}(5)\, i\, a_{1}\, a\, i_{1}}_{\ \,{\bf q},\, {\bf k}_{1},\, {\bf k},\, {\bf q}_{1}} 
+
2\hspace{0.02cm}J^{\,(1)\, a\, a_{1}\, i_{1}\, i}_{\; {\bf k},\, {\bf k}_{1},\, {\bf q}_{1},\, {\bf q}} 
-
4\!\int\!\frac{d\hspace{0.02cm}{\bf k}^{\hspace{0.015cm}\prime}}{(2\pi)^{3}}\,\Bigl[\hspace{0.03cm}
&V^{(1)\hspace{0.03cm} a\, a_{1}\, a^{\prime}}_{\ {\bf k},\, {\bf k}_{1},\, {\bf k}^{\prime}}
\hspace{0.02cm} 
F^{(1)\, a^{\prime}\, i_{1}\, i}_{\ {\bf k}^{\prime},\, {\bf q}_{1},\, {\bf q}}
-
V^{\hspace{0.025cm}\ast\hspace{0.025cm}(1)\hspace{0.03cm} a_{1}\, a^{\prime}\, a}_{\ {\bf k}_{1},\, {\bf k}^{\prime},\, {\bf k}}
\hspace{0.02cm}
F^{\hspace{0.025cm}\ast\hspace{0.025cm}(3)\hspace{0.03cm} a^{\prime}\, i\; i_{1}}_{\ {\bf k}^{\prime},\, {\bf q},\, 
{\bf q}_{1}}\hspace{0.03cm}\Bigr]
\label{ap:C4d}\\[1ex]
+\,
2\!\int&\!\frac{d{\bf q}^{\prime}}{(2\pi)^{3}}\,\Bigl[\hspace{0.03cm}
F^{(1)\, a\, i_{1}\, i^{\prime}}_{\ {\bf k},\, {\bf q}_{1},\, {\bf q}^{\prime}}
\, 
F^{\hspace{0.025cm}\ast\hspace{0.025cm}(2)\hspace{0.03cm} a_{1}\, i\; i^{\prime}}_{\ {\bf k}_1,\, {\bf q},\, 
{\bf q}^{\prime}}
-
F^{(2)\, a\, i^{\prime}\, i_{1}}_{\ {\bf k},\, {\bf q}^{\prime},\, {\bf q}_{1}}
\, 
F^{\hspace{0.025cm}\ast\hspace{0.025cm}(3)\hspace{0.03cm} a_{1}\, i\; i^{\prime}}_{\ {\bf k}_{1},\,  {\bf q},\, 
{\bf q}^{\prime}}\hspace{0.03cm}\Bigr] = 0.
\notag
\end{align}
\end{subequations}


\numberwithin{equation}{section}
\section{Second system of the canonicity conditions}
\numberwithin{equation}{section}
\label{appendix_D}

\indent Here we give the second system of the canonicity conditions, which follows from Eqs.\,(\ref{eq:3ra})\,--\,(\ref{eq:3rd}). Substituting (\ref{eq:3t}) and (\ref{eq:3y}) into Eq.\,(\ref{eq:3ra}) and taking into account the relations (\ref{eq:3p}), we lead to four independent relations for the coefficient functions. The first pair of the canonicity conditions has the following form:
\begin{subequations} 
\label{ap:D1}
\begin{align}
R^{\hspace{0.025cm}\ast\hspace{0.025cm}(1)\hspace{0.03cm} i_{1}\, a_{1}\, a_{2}\, i}_{\ {\bf q}_{1},\, {\bf k}_{1},\, {\bf k}_{2},\, {\bf q}} 
+
R^{\,(3)\, i\, a_{1}\, a_{2}\, i_{1}}_{\; {\bf q},\, {\bf k}_{1},\, {\bf k}_{2},\, {\bf q}_{1}} 
-\frac{1}{2}
\int\!\frac{d{\bf q}^{\prime}}{(2\pi)^{3}}\,\Bigl[\hspace{0.03cm}
&F^{(2)\, a_{1}\, i\, i^{\prime}}_{\ {\bf k}_{1},\, {\bf q},\, {\bf q}^{\prime}}\, 
F^{(2)\, a_{2}\, i^{\prime}\, i_{1}}_{\ {\bf k}_{2},\, {\bf q}^{\prime},\, {\bf q}_{1}}
+
F^{(2)\, a_{2}\, i\, i^{\prime}}_{\ {\bf k}_{2},\, {\bf q},\, {\bf q}^{\prime}}\, 
F^{(2)\, a_{1}\, i^{\prime}\, i_{1}}_{\ {\bf k}_{1},\, {\bf q}^{\prime},\, {\bf q}_{1}}
\label{ap:D1a}\\[1ex]
&-
4\hspace{0.02cm}F^{(3)\, a_{1}\, i\, i^{\prime}}_{\ {\bf k}_{1},\, {\bf q},\, {\bf q}^{\prime}}\,F^{(1)\, a_{2}\, i_{1}\, i^{\prime}}_{\ {\bf k}_{2},\, {\bf q_{1}},\, {\bf q}^{\prime}}
-
4\hspace{0.02cm}F^{(3)\, a_{2}\, i\, i^{\prime}}_{\ {\bf k}_{2},\, {\bf q},\, {\bf q}^{\prime}}\,F^{(1)\, a_{1}\, i_{1}\, i^{\prime}}_{\ {\bf k}_{1},\, {\bf q_{1}},\, {\bf q}^{\prime}}\hspace{0.03cm}\Bigr] = 0,
\notag\\[1ex]
R^{\hspace{0.025cm}\ast\hspace{0.025cm}(2)\hspace{0.03cm} i_{1}\, a_{2}\, a_{1}\, i}_{\ {\bf q}_{1},\, {\bf k}_{2},\, {\bf k}_{1},\, {\bf q}} 
\,+\,
R^{\,(2)\, i\, a_{1}\, a_{2}\, i_{1}}_{\; {\bf q},\, {\bf k}_{1},\, {\bf k}_{2},\, {\bf q}_{1}} 
+
\int\!\frac{d{\bf q}^{\prime}}{(2\pi)^{3}}\,\Bigl[\hspace{0.03cm}
&F^{(2)\, a_{1}\, i\, i^{\prime}}_{\ {\bf k}_{1},\, {\bf q},\, {\bf q}^{\prime}}\, 
F^{\hspace{0.025cm}\ast\hspace{0.025cm}(2)\hspace{0.03cm} a_{2}\,  i_{1}\, i^{\prime}}_{\ {\bf k}_{2},\,  {\bf q}_{1},\, {\bf q}^{\prime}}
+\,
F^{(2)\, a_{1}\hspace{0.03cm} i^{\prime}\, i_{1}}_{\ {\bf k}_{1},\, {\bf q}^{\prime},\,  {\bf q}_{1}}\, 
F^{\hspace{0.025cm}\ast\hspace{0.025cm}(2)\hspace{0.03cm} a_{2}\, i^{\prime}\, i}_{\ {\bf k}_{2},\, {\bf q}^{\prime},\, {\bf q}}
\label{ap:D1b}\\[1ex]
&+
4\hspace{0.02cm}F^{(3)\, a_{1}\, i\, i^{\prime}}_{\ {\bf k}_{1},\, {\bf q},\, {\bf q}^{\prime}}\,F^{\hspace{0.025cm}\ast\hspace{0.025cm}(3)\hspace{0.03cm} a_{2}\, i_{1}\, i^{\prime}}_{\ {\bf k}_{2},\, {\bf q}_{1},\, {\bf q}^{\prime}}
+
4\hspace{0.02cm}F^{(1)\, a_{1}\, i_{1}\, i^{\prime}}_{\ {\bf k}_{1},\, {\bf q}_{1},\, {\bf q}^{\prime}}\,F^{\hspace{0.025cm}\ast\hspace{0.025cm}(1)\hspace{0.03cm} a_{2}\, i\, i^{\prime}}_{\ {\bf k}_{2},\, {\bf q},\, {\bf q}^{\prime}}\hspace{0.03cm}\Bigr] = 0
\notag
\end{align}
\end{subequations} 
and, correspondingly, the second one is
\begin{subequations} 
\label{ap:D2}
\begin{align}
3\hspace{0.02cm}S^{\,(1)\; i\; i_{1}\; i_{2}\; i_{3}}_{\; {\bf q},\, {\bf q}_{1},\, {\bf q}_{2},\, {\bf q}_{3}}
+\,
S^{\hspace{0.025cm}\ast\hspace{0.025cm}(3)\hspace{0.03cm} i_{3}\; i_{2}\; i_{1}\, i}_{\ {\bf q}_{3},\, {\bf q}_{2},\, {\bf q}_{1},\, {\bf q}}\,
-
\int\!\frac{d\hspace{0.02cm}{\bf k}^{\hspace{0.015cm}\prime}}{(2\pi)^{3}}\,\Bigl[\hspace{0.03cm}
&F^{(1)\, a^{\prime}\, i_{3}\, i_{2}}_{\ {\bf k}^{\prime},\, {\bf q}_{3},\, {\bf q}_{2}}\, 
F^{\hspace{0.025cm}\ast\hspace{0.025cm}(2)\hspace{0.03cm} a^{\prime}\, i_{1}\, i}_{\ {\bf k}^{\prime},\, {\bf q}_{1},\, {\bf q}}
+
F^{(2)\, a^{\prime}\, i\, i_{1}}_{\ {\bf k}^{\prime},\, {\bf q},\, {\bf q}_{1}}\, 
F^{\hspace{0.025cm}\ast\hspace{0.025cm}(3)\hspace{0.03cm} a^{\prime}\, i_{3}\, i_{2}}_{\ {\bf k}^{\prime},\, {\bf q}_{3},\, {\bf q}_{2}}
\label{ap:D2a}\\[1ex]
&-
F^{(1)\, a^{\prime}\, i_{3}\, i_{1}}_{\ {\bf k}^{\prime},\, {\bf q}_{3},\, {\bf q}_{1}}\, 
F^{\hspace{0.025cm}\ast\hspace{0.025cm}(2)\hspace{0.03cm} a^{\prime}\, i_{2}\, i}_{\ {\bf k}^{\prime},\, {\bf q}_{2},\, {\bf q}}
-
F^{(2)\, a^{\prime}\, i\, i_{2}}_{\ {\bf k}^{\prime},\, {\bf q},\, {\bf q}_{2}}\, 
F^{\hspace{0.025cm}\ast\hspace{0.025cm}(3)\hspace{0.03cm} a^{\prime}\, i_{3}\, i_{1}}_{\ {\bf k}^{\prime},\, {\bf q}_{3},\, {\bf q}_{1}}\hspace{0.03cm}\Bigr] = 0,
\notag\\[1ex]
S^{\,(2)\; i\; i_{1}\; i_{2}\; i_{3}}_{\; {\bf q},\, {\bf q}_{1},\, {\bf q}_{2},\, {\bf q}_{3}}
+
S^{\hspace{0.025cm}\ast\hspace{0.025cm}(2)\hspace{0.03cm} i_{3}\; i_{2}\; i_{1}\, i}_{\ {\bf q}_{3},\, {\bf q}_{2},\, 
{\bf q}_{1},\, {\bf q}}
-
\frac{1}{2}\!\hspace{0.03cm}\int\!\frac{d\hspace{0.02cm}{\bf k}^{\hspace{0.015cm}\prime}}{(2\pi)^{3}}\,\Bigl[\hspace{0.03cm}
&F^{(2)\, a^{\prime}\, i_{1}\, i_{3}}_{\ {\bf k}^{\prime},\, {\bf q}_{1},\, {\bf q}_{3}}\, 
F^{\hspace{0.025cm}\ast\hspace{0.025cm}(2)\hspace{0.03cm} a^{\prime}\, i_{2}\, i}_{\ {\bf k}^{\prime},\, {\bf q}_{2},\, {\bf q}}
-
F^{(2)\, a^{\prime}\, i\, i_{2}}_{\ {\bf k}^{\prime},\, {\bf q},\, {\bf q}_{2}}\, 
F^{\hspace{0.025cm}\ast\hspace{0.025cm}(2)\hspace{0.03cm} a^{\prime}\, i_{3}\, i_{1}}_{\ {\bf k}^{\prime},\, {\bf q}_{3},\, {\bf q}_{1}}
\label{ap:D2b}\\[1ex]
&+
4\hspace{0.02cm}
F^{(3)\, a^{\prime}\, i\, i_{1}}_{\ {\bf k}^{\prime},\, {\bf q},\, {\bf q}_{1}}\, 
F^{\hspace{0.025cm}\ast\hspace{0.025cm}(3)\hspace{0.03cm} a^{\prime}\, i_{3}\, i_{2}}_{\ {\bf k}^{\prime},\, {\bf q}_{3},\, {\bf q}_{2}}
-
4\hspace{0.02cm}
F^{(1)\, a^{\prime}\, i_{3}\, i_{2}}_{\ {\bf k}^{\prime},\, {\bf q}_{3},\, {\bf q}_{2}}\, 
F^{\hspace{0.025cm}\ast\hspace{0.025cm}(1)\hspace{0.03cm} a^{\prime}\, i\, i_{1}}_{\ {\bf k}^{\prime},\, {\bf q},\, 
{\bf q}_{1}}\hspace{0.03cm}\Bigr] = 0.
\notag
\end{align}
\end{subequations} 
Further, from the second equation in (\ref{eq:3rb}) two systems of relations containing the three canonicity conditions each follow. The first system has the form: 
\begin{subequations} 
\label{ap:D3}
\begin{align}
R^{\,(4)\, i\, a_{1}\, a_{2}\, i_{1}}_{\, {\bf q},\, {\bf k}_{1},\, {\bf k}_{2},\, {\bf q}_{1}} 
\,+\,
R^{\,(4)\, i_{1}\, a_{1}\, a_{2}\, i}_{\, {\bf q}_{1},\, {\bf k}_{1},\, {\bf k}_{2},\, {\bf q}} 
-
\int\frac{d{\bf q}^{\prime}}{(2\pi)^{3}}\,\Bigl[\hspace{0.03cm}
&F^{\hspace{0.025cm}\ast\hspace{0.025cm}(2)\hspace{0.03cm} a_{1}\, i^{\prime}\, i}_{\ {\bf k}_{1},\, {\bf q}^{\prime},\, {\bf q}}\, 
F^{\hspace{0.025cm}\ast\hspace{0.025cm}(1)\hspace{0.03cm} a_{2}\,  i_{1}\, i^{\prime}}_{\ {\bf k}_{2},\,  {\bf q}_{1},\, {\bf q}^{\prime}}
+\,
F^{\hspace{0.025cm}\ast\hspace{0.025cm}(2)\hspace{0.03cm} a_{1}\, i^{\prime}\, i_{1}}_{\ {\bf k}_{1},\, {\bf q}^{\prime},\,  {\bf q}_{1}}\, 
F^{\hspace{0.025cm}\ast\hspace{0.025cm}(1)\hspace{0.03cm} a_{2}\, i\, i^{\prime}}_{\ {\bf k}_{2},\, {\bf q},\, {\bf q}^{\prime}}
\label{ap:D3a}\\[1ex]
&+
F^{\hspace{0.025cm}\ast\hspace{0.025cm}(2)\hspace{0.03cm} a_{2}\, i^{\prime}\, i}_{\ {\bf k}_{2},\, {\bf q}^{\prime},\, {\bf q}}\,F^{\hspace{0.025cm}\ast\hspace{0.025cm}(1)\hspace{0.03cm} a_{1}\, i_{1}\, i^{\prime}}_{\ {\bf k}_{1},\, {\bf q}_{1},\, {\bf q}^{\prime}}
+\,
F^{\hspace{0.025cm}\ast\hspace{0.025cm}(2)\hspace{0.03cm} a_{2}\, i^{\prime}\, i_{1}}_{\ {\bf k}_{2},\, {\bf q}^{\prime},\, {\bf q}_{1}}\,F^{\hspace{0.025cm}\ast\hspace{0.025cm}(1)\hspace{0.03cm} a_{1}\, i\, i^{\prime}}_{\ {\bf k}_{1},\, {\bf q},\, {\bf q}^{\prime}}\hspace{0.03cm}\Bigr] = 0,
\notag\\[1ex]
R^{\,(5)\, i\, a_{1}\, a_{2}\, i_{1}}_{\, {\bf q},\, {\bf k}_{1},\, {\bf k}_{2},\, {\bf q}_{1}} 
+
R^{\,(5)\, i_{1}\, a_{1}\, a_{2}\, i}_{\, {\bf q}_{1},\, {\bf k}_{1},\, {\bf k}_{2},\, {\bf q}} 
+2\!
\int\frac{d{\bf q}^{\prime}}{(2\pi)^{3}}\,\Bigl[\hspace{0.03cm}
&F^{(2)\, a_{1}\, i\, i^{\prime}}_{\ {\bf k}_{1},\, {\bf q},\, {\bf q}^{\prime}}\, 
F^{\hspace{0.025cm}\ast\hspace{0.025cm}(1)\hspace{0.03cm} a_{2}\,  i_{1}\, i^{\prime}}_{\ {\bf k}_{2},\,  {\bf q}_{1},\, {\bf q}^{\prime}}
+\,
F^{(2)\, a_{1}\, i_{1}\, i^{\prime}}_{\ {\bf k}_{1},\, {\bf q}_{1},\,  {\bf q}^{\prime}}\, 
F^{\hspace{0.025cm}\ast\hspace{0.025cm}(1)\hspace{0.03cm} a_{2}\, i\, i^{\prime}}_{\ {\bf k}_{2},\, {\bf q},\, {\bf q}^{\prime}}
\label{ap:D3b}\\[1ex]
&-
F^{(3)\, a_{1}\, i_{1}\, i^{\prime}}_{\ {\bf k}_{1},\, {\bf q}_{1},\, {\bf q}^{\prime}}\,
F^{\hspace{0.025cm}\ast\hspace{0.025cm}(2)\hspace{0.03cm} a_{2}\, i^{\prime}\, i}_{\ {\bf k}_{2},\, {\bf q}^{\prime},\, {\bf q}}
-\,
F^{(3)\, a_{1}\, i\, i^{\prime}}_{\ {\bf k}_{1},\, {\bf q},\, {\bf q}^{\prime}}\,
F^{\hspace{0.025cm}\ast\hspace{0.025cm}(2)\hspace{0.03cm} a_{2}\, i^{\prime}\, i_{1}}_{\ {\bf k}_{2},\, {\bf q}^{\prime},\, {\bf q}_{1}}\hspace{0.03cm}\Bigr] = 0,
\notag\\[1ex]
R^{\,(6)\, i\, a_{1}\, a_{2}\, i_{1}}_{\, {\bf q},\, {\bf k}_{1},\, {\bf k}_{2},\, {\bf q}_{1}} 
\,+\,
R^{\,(6)\, i_{1}\, a_{1}\, a_{2}\, i}_{\, {\bf q}_{1},\, {\bf k}_{1},\, {\bf k}_{2},\, {\bf q}} 
+
\int\frac{d{\bf q}^{\prime}}{(2\pi)^{3}}\,\Bigl[\hspace{0.03cm}
&F^{(2)\, a_{1}\, i\, i^{\prime}}_{\ {\bf k}_{1},\, {\bf q},\, {\bf q}^{\prime}}\, F^{(3)\, a_{2}\,  i_{1}\, i^{\prime}}_{\ {\bf k}_{2},\,  {\bf q}_{1},\, {\bf q}^{\prime}}
\,+\,
F^{(2)\, a_{1}\, i_{1}\, i^{\prime}}_{\ {\bf k}_{1},\, {\bf q}_{1},\,  {\bf q}^{\prime}}\, F^{(3)\, a_{2}\, i\, i^{\prime}}_{\ {\bf k}_{2},\, {\bf q},\, {\bf q}^{\prime}}
\label{ap:D3c}\\[1ex]
&+
F^{(2)\, a_{2}\, i\, i^{\prime}}_{\ {\bf k}_{2},\, {\bf q},\, {\bf q}^{\prime}}\, F^{(3)\, a_{1}\,  i_{1}\, i^{\prime}}_{\ {\bf k}_{1},\,  {\bf q}_{1},\, {\bf q}^{\prime}}
\,+\,
F^{(2)\, a_{2}\, i_{1}\, i^{\prime}}_{\ {\bf k}_{2},\, {\bf q}_{1},\,  {\bf q}^{\prime}}\, F^{(3)\, a_{1}\, i\, i^{\prime}}_{\ {\bf k}_{1},\, {\bf q},\, {\bf q}^{\prime}}\hspace{0.03cm}\Bigr] = 0.
\notag
\end{align}
\end{subequations} 
The second system can be written in a similar form:
\begin{subequations} 
\label{ap:D4}
\begin{align}
S^{\,(2)\; i\; i_{1}\; i_{2}\; i_{3}}_{\; {\bf q},\, {\bf q}_{1},\, {\bf q}_{2},\, {\bf q}_{3}}
+\,
S^{\,(2)\; i_{1}\, i\; i_{2}\; i_{3}}_{\; {\bf q}_{1},\, {\bf q},\, {\bf q}_{2},\, {\bf q}_{3}}
- \frac{1}{2}\!\hspace{0.03cm}
\int\!\frac{d\hspace{0.02cm}{\bf k}^{\hspace{0.015cm}\prime}}{(2\pi)^{3}}\,\Bigl[\hspace{0.03cm}
&F^{(2)\, a^{\prime}\, i_{1}\, i_{3}}_{\ {\bf k}^{\prime},\, {\bf q}_{1},\, {\bf q}_{3}}\, 
F^{\hspace{0.025cm}\ast\hspace{0.025cm}(2)\hspace{0.03cm} a^{\prime}\, i_{2}\, i}_{\ {\bf k}^{\prime},\, {\bf q}_{2},\, {\bf q}}
+\,
F^{(2)\hspace{0.03cm}  a^{\prime}\, i\,\hspace{0.02cm} i_{3}}_{\ {\bf k}^{\prime},\, {\bf q},\, {\bf q}_{3}}\, 
F^{\hspace{0.025cm}\ast\hspace{0.025cm}(2)\hspace{0.03cm} a^{\prime}\, i_{2}\, i_{1}}_{\ {\bf k}^{\prime},\, {\bf q}_{2},\, {\bf q}_{1}}
\label{ap:D4a}\\[1ex]
&-
F^{(2)\hspace{0.03cm} a^{\prime}\, i\,\hspace{0.02cm} i_{2}}_{\ {\bf k}^{\prime},\, {\bf q},\, {\bf q}_{2}}\, 
F^{\hspace{0.025cm}\ast\hspace{0.025cm}(2)\hspace{0.03cm} a^{\prime}\, i_{3}\, i_{1}}_{\ {\bf k}^{\prime},\, {\bf q}_{3},\, {\bf q}_{1}}
-
F^{(2)\, a^{\prime}\, i_{1}\, i_{2}}_{\ {\bf k}^{\prime},\, {\bf q}_{1},\, {\bf q}_{2}}\, 
F^{\hspace{0.025cm}\ast\hspace{0.025cm}(2)\hspace{0.03cm} a^{\prime}\, i_{3}\, i}_{\ {\bf k}^{\prime},\, {\bf q}_{3},\, {\bf q}}\hspace{0.03cm}\Bigr] = 0,
\notag\\[1ex]
S^{\,(3)\; i\; i_{1}\; i_{2}\; i_{3}}_{\; {\bf q},\, {\bf q}_{1},\, {\bf q}_{2},\, {\bf q}_{3}}
\,+\,
S^{\,(3)\; i_{1}\, i\; i_{2}\; i_{3}}_{\; {\bf q}_{1},\, {\bf q},\, {\bf q}_{2},\, {\bf q}_{3}}\,
\,+\,
\int\!\frac{d\hspace{0.02cm}{\bf k}^{\hspace{0.015cm}\prime}}{(2\pi)^{3}}\,\Bigl[\hspace{0.03cm}
&F^{(2)\, a^{\prime}\, i_{1}\, i_{3}}_{\ {\bf k}^{\prime},\, {\bf q}_{1},\, {\bf q}_{3}}\, 
F^{\hspace{0.025cm}\ast\hspace{0.025cm}(1)\hspace{0.03cm} a^{\prime}\, i\, i_{2}}_{\ {\bf k}^{\prime},\, {\bf q},\, {\bf q}_{2}}
+\,
F^{(3)\, a^{\prime}\, i_{1}\, i_{2}}_{\ {\bf k}^{\prime},\, {\bf q}_{1},\, {\bf q}_{2}}\, F^{\hspace{0.025cm}\ast\hspace{0.025cm}(2)\, a^{\prime}\, i_{3}\, i}_{\ {\bf k}^{\prime},\, {\bf q}_{3},\, {\bf q}}
\label{ap:D4b}\\[1ex]
&+
F^{(3)\hspace{0.03cm} a^{\prime}\, i\,\hspace{0.02cm} i_{2}}_{\ {\bf k}^{\prime},\, {\bf q},\, {\bf q}_{2}}\, 
F^{\hspace{0.025cm}\ast\hspace{0.025cm}(2)\hspace{0.03cm} a^{\prime}\, i_{3}\, i_{1}}_{\ {\bf k}^{\prime},\, {\bf q}_{3},\, {\bf q}_{1}}
+\,
F^{(2)\hspace{0.03cm} a^{\prime}\, i\,\hspace{0.02cm} i_{3}}_{\ {\bf k}^{\prime},\, {\bf q},\, {\bf q}_{3}}\, 
F^{\hspace{0.025cm}\ast\hspace{0.025cm}(1)\hspace{0.03cm} a^{\prime}\, i_{1}\, i_{2}}_{\ {\bf k}^{\prime},\, {\bf q}_{1},\, {\bf q}_{2}}\hspace{0.03cm}\Bigr] = 0,
\hspace{0.3cm}
\notag\\[1ex]
3\hspace{0.01cm}S^{\,(4)\; i\; i_{1}\; i_{2}\; i_{3}}_{\; {\bf q},\, {\bf q}_{1},\, {\bf q}_{2},\, {\bf q}_{3}}
+
3\hspace{0.01cm}S^{\,(4)\; i_{1}\, i\; i_{2}\; i_{3}}_{\; {\bf q}_{1},\, {\bf q},\, {\bf q}_{2},\, {\bf q}_{3}}
+ 2\!
\int\!\frac{d\hspace{0.02cm}{\bf k}^{\hspace{0.015cm}\prime}}{(2\pi)^{3}}\,\Bigl[\hspace{0.03cm}
&F^{(3)\, a^{\prime}\, i_{1}\, i_{3}}_{\ {\bf k}^{\prime},\, {\bf q}_{1},\, {\bf q}_{3}}\, 
F^{\hspace{0.025cm}\ast\hspace{0.025cm}(1)\hspace{0.03cm} a^{\prime}\, i\, i_{2}}_{\ {\bf k}^{\prime},\, {\bf q},\, {\bf q}_{2}}
+
F^{(3)\hspace{0.03cm} a^{\prime}\, i\,\hspace{0.02cm} i_{3}}_{\ {\bf k}^{\prime},\, {\bf q},\, {\bf q}_{3}}\, 
F^{\hspace{0.025cm}\ast\hspace{0.025cm}(1)\hspace{0.03cm} a^{\prime}\, i_{1}\, i_{2}}_{\ {\bf k}^{\prime},\, {\bf q}_{1},\, {\bf q}_{2}}
\label{ap:D4c}\\[1ex]
&-
F^{(3)\hspace{0.03cm} a^{\prime}\, i\,\hspace{0.02cm} i_{2}}_{\ {\bf k}^{\prime},\, {\bf q},\, {\bf q}_{2}}\, 
F^{\hspace{0.025cm}\ast\hspace{0.025cm}(1)\hspace{0.03cm} a^{\prime}\, i_{1}\, i_{3}}_{\ {\bf k}^{\prime},\, {\bf q}_{1},\, {\bf q}_{3}}
-
F^{(3)\, a^{\prime}\, i_{1}\, i_{2}}_{\ {\bf k}^{\prime},\, {\bf q}_{1},\, {\bf q}_{2}}\, 
F^{\hspace{0.025cm}\ast\hspace{0.025cm}(1)\hspace{0.03cm} a^{\prime}\, i\, i_{3}}_{\ {\bf k}^{\prime},\, {\bf q},\, {\bf q}_{3}}\hspace{0.03cm}\Bigr] = 0.
\notag
\end{align}
\end{subequations} 
The systems (\ref{ap:D3}) and (\ref{ap:D4}) define the rearrangement rule for the color indices $i$ and $i_{1}$, and for the corresponding momentum arguments ${\bf q}$ and ${\bf q}_{1}$ of the third-order coefficient functions  $R^{(4)}, R^{(5)}$,  $R^{(6)}$, $S^{(2)}, S^{(3)}$, and $S^{(4)}$.\\
\indent The substitution of (\ref{eq:3t}) and (\ref{eq:3y}) into the third equation in the system (\ref{eq:3r}) leads us to four relations derived above, the system (\ref{ap:C3}). Finally, the substitution (\ref{eq:3t}) and (\ref{eq:3y}) into (\ref{eq:3rd}) gives us a system of the four canonicity conditions complex-conjugated to the canonicity ones of the system (\ref{ap:C4}).


\numberwithin{equation}{section}
\section{Explicit expressions of the coefficient functions $J^{(n)\, a\, a_{1}\, i_{1}\, i_{2}}_{\ {\bf k},\, {\bf k}_{1},\, {\bf q}_{1},\, {\bf q}_{2}}$ and $R^{\,(n)\, i\, a\, a_{1}\, i_{1}}_{\ {\bf q},\, {\bf k},\, {\bf k}_{1},\, {\bf q}_{1}}$}
\numberwithin{equation}{section}
\label{appendix_E}

In this Appendix we write out an explicit form of the coefficient functions $J^{(n)\, a\, a_{1}\, i_{1}\, i_{2}}_{\ {\bf k},\, {\bf k}_{1},\, {\bf q}_{1},\, {\bf q}_{2}}$ for $n = 3, 4, 5$ and 6, which enter into the integrands of the canonical transformation (\ref{eq:3t}):
\begin{equation}
\begin{split}
&J^{\,(3)\, a_{1}\, a_{2}\, i\; i_{1}}_{\; \hspace{0.03cm}{\bf k}_{1},\, {\bf k}_{2},\, {\bf q},\, {\bf q}_{1}}
=
\frac{1}{\omega^{-}_{{\bf q}} + \omega^{-}_{{\bf q}_{1}} + \omega^{\hspace{0.02cm} l}_{{\bf k}_{1}} - \omega^{\hspace{0.02cm} l}_{{\bf k}_{2}}}\,
\widetilde{T}^{\hspace{0.025cm}\ast\hspace{0.025cm}(11)\hspace{0.03cm} i\, i_{1}\, a_{2}\, a_{1}}_{\ {\bf q},\, {\bf q}_{1},\, {\bf k}_{2},\, {\bf k}_{1}}\, 
(2\pi)^{3}\hspace{0.03cm}\delta({\bf q} + {\bf q}_{1} + {\bf k}_{1} - {\bf k}_{2}),
\\[1.5ex]
&J^{\,(4)\, a_{1}\, a_{2}\, i\; i_{1}}_{\; \hspace{0.04cm} {\bf k}_{1},\, {\bf k}_{2},\, {\bf q},\, {\bf q}_{1}}
=
\frac{1}{\omega^{-}_{{\bf q}} + \omega^{-}_{{\bf q}_{1}} - \omega^{\hspace{0.02cm} l}_{{\bf k}_{1}} - \omega^{\hspace{0.02cm} l}_{{\bf k}_{2}}}\,
\widetilde{T}^{\,(22)\hspace{0.03cm} i\, i_{1}\, a_{2}\, a_{1}}_{\ {\bf q},\, {\bf q}_{1},\, {\bf k}_{2},\, {\bf k}_{1}}\, 
(2\pi)^{3}\hspace{0.03cm}\delta({\bf q} + {\bf q}_{1} - {\bf k}_{1} - {\bf k}_{2}),
\\[1.5ex]
&J^{\,(5)\, a_{1}\, a_{2}\, i\; i_{1}}_{\; \hspace{0.03cm} {\bf k}_{1},\, {\bf k}_{2},\, {\bf q},\, {\bf q}_{1}}
=
-\hspace{0.03cm}
\frac{1}{\omega^{-}_{{\bf q}} - \omega^{-}_{{\bf q}_{1}} + \omega^{\hspace{0.02cm} l}_{{\bf k}_{1}} + \omega^{\hspace{0.02cm} l}_{{\bf k}_{2}}}\,
\widetilde{T}^{\hspace{0.025cm}\ast\hspace{0.025cm}(1)\hspace{0.03cm} i_{1}\, i\, a_{1}\, a_{2}}_{\ {\bf q}_{1},\, {\bf q},\, {\bf k}_{1},\, {\bf k}_{2}}\, 
(2\pi)^{3}\hspace{0.03cm}\delta({\bf q} - {\bf q}_{1} + {\bf k}_{1} + {\bf k}_{2}),
\\[1.5ex]
&J^{\,(6)\, a_{1}\, a_{2}\, i\; i_{1}}_{\; \hspace{0.035cm} {\bf k}_{1},\, {\bf k}_{2},\, {\bf q},\, {\bf q}_{1}}
=
\frac{1}{\omega^{-}_{{\bf q}} + \omega^{-}_{{\bf q}_{1}} + \omega^{\hspace{0.02cm} l}_{{\bf k}_{1}} + \omega^{\hspace{0.02cm} l}_{{\bf k}_{2}}}\,
\widetilde{T}^{\hspace{0.025cm}\ast\hspace{0.025cm}(4)\hspace{0.03cm} i\, i_{1}\, a_{1}\, a_{2}}_{\ {\bf q},\, {\bf q}_{1},\, {\bf k}_{1},\, {\bf k}_{2}}\, 
(2\pi)^{3}\hspace{0.03cm}\delta({\bf q} + {\bf q}_{1} + {\bf k}_{1} + {\bf k}_{2}),
\end{split}
\label{ap:E1}
\end{equation}
where
\begin{align}
\widetilde{T}^{\,(11)\hspace{0.03cm} i\, i_{1}\, a_{1}\, a_{2}}_{\ {\bf q},\, {\bf q}_{1},\, {\bf k}_{1},\, {\bf k}_{2}}
\,=\,
T^{\,(11)\hspace{0.03cm} i\, i_{1}\, a_{1}\, a_{2}}_{\ {\bf q},\, {\bf q}_{1},\, {\bf k}_{1},\, {\bf k}_{2}}
\;\,+\;\,
&\frac{{\mathcal G}^{\hspace{0.03cm}\ast\, a_{1}\,i\, j}_{{\bf k}_{1},\,{\bf q},\, {\bf k}_{1} - {\bf q}}\, 
{\mathcal P}^{\; a_{2}\,j\, i_{1}}_{{\bf k}_{2},\, {\bf k}_{2} + {\bf q}_{1},\,  {\bf q}_{1}}}
{\omega^{\hspace{0.02cm} l}_{{\bf k}_{1}} - \omega^{-}_{{\bf q}} - \omega^{-}_{{\bf k}_{1} - {\bf q}}}
\;-\; 
\frac{{\mathcal G}^{\hspace{0.03cm}\ast\, a_{1}\,i_{1}\, j}_{{\bf k}_{1},\,{\bf q}_{1},\, {\bf k}_{1} - {\bf q}_{1}}\, 
{\mathcal P}^{\; a_{2}\,j\, i}_{{\bf k}_{2},\, {\bf k}_{2} + {\bf q},\,  {\bf q}}}
{\omega^{\hspace{0.02cm} l}_{{\bf k}_{1}} - \omega^{-}_{{\bf q}_{1}} - \omega^{-}_{{\bf k}_{1} - {\bf q}_{1}}}
\label{ap:E2}\\[1.5ex]
\;+\;\,
&\frac{{\mathcal P}^{\hspace{0.03cm}\ast\, a_{1}\, i\, j}_{{\bf k}_{1},\,{\bf q},\, {\bf q} - {\bf k}_{1}}\, 
{\mathcal K}^{\; a_{2}\, i_{1}\, j}_{{\bf k}_{2},\, {\bf q}_{1},\, -{\bf k}_{2} - {\bf q}_{1}}}
{\omega^{\hspace{0.02cm} l}_{{\bf k}_{1}} - \omega^{-}_{{\bf q}} + \omega^{-}_{{\bf q} - {\bf k}_{1}}}
\;-\; 
\frac{{\mathcal P}^{\hspace{0.03cm}\ast\, a_{1}\, i_{1}\,j}_{{\bf k}_{1},\,{\bf q}_{1},\, {\bf q}_{1} - {\bf k}_{1}}\, 
{\mathcal K}^{\; a_{2}\,i\, j}_{{\bf k}_{2},\, {\bf q},\, -{\bf k}_{2} - {\bf q}}}
{\omega^{\hspace{0.02cm} l}_{{\bf k}_{1}} - \omega^{-}_{{\bf q}_{1}} + \omega^{-}_{{\bf q}_{1} - {\bf k}_{1}}}
\notag\\[1.5ex]
+\,
2\hspace{0.03cm}\Biggl(
&\frac{{\mathcal G}^{\hspace{0.03cm}\ast\, a\, i\; i_{1}}_{{\bf q} + {\bf q}_{1},\, {\bf q},\, {\bf q}_{1}} 
{\mathcal V}^{\; a_{1}\, a_{2}\,  a}_{{\bf k}_{1}, {\bf k}_{2},\, {\bf k}_{1} - {{\bf k}_{2}}}}
{\omega^{\hspace{0.02cm} l}_{{\bf q} + {\bf q}_{1}} - \omega^{-}_{{\bf q}} - \omega^{-}_{{\bf q}_{1}}}
\;-\;
\frac{{\mathcal K}^{\; a\, i\; i_{1}}_{-{\bf q} - {\bf q}_{1},\, {\bf q},\, {\bf q}_{1}} 
{\mathcal V}^{\hspace{0.03cm}\ast\, a_{2}\, a_{1}\,  a}_{{\bf k}_{2}, {\bf k}_{1},\, {\bf k}_{2} - {{\bf k}_{1}}}}
{\omega^{\hspace{0.02cm} l}_{-{\bf q} - {\bf q}_{1}} + \omega^{-}_{{\bf q}} + \omega^{-}_{{\bf q}_{1}}}
\Biggr),
\notag
\end{align}
\begin{align}
\widetilde{T}^{\,(22)\hspace{0.03cm} i\, i_{1}\, a_{1}\, a_{2}}_{\ {\bf q},\, {\bf q}_{1},\, {\bf k}_{1},\, {\bf k}_{2}}
\,=\,
T^{\,(22)\hspace{0.03cm} i\, i_{1}\, a_{1}\, a_{2}}_{\ {\bf q},\, {\bf q}_{1},\, {\bf k}_{1},\, {\bf k}_{2}}
\;\,+\;\,
&\frac{{\mathcal G}^{\hspace{0.03cm}\ast\, a_{2}\,i\hspace{0.04cm} j}_{{\bf k}_{2},\,{\bf q},\, {\bf k}_{2} - {\bf q}}\, 
{\mathcal P}^{\hspace{0.03cm}\ast\, a_{1}\, i_{1}\, j}_{{\bf k}_{1},\, {\bf q}_{1},\, {\bf q}_{1} - {\bf k}_{1}}}
{\omega^{\hspace{0.02cm} l}_{{\bf k}_{1}} - \omega^{-}_{{\bf q}_{1}} + \omega^{-}_{{\bf q}_{1} - {\bf k}_{1}}}
\;-\; 
\frac{{\mathcal G}^{\hspace{0.03cm}\ast\, a_{2}\,i_{1}\, j}_{{\bf k}_{2},\,{\bf q}_{1},\, {\bf k}_{2} - {\bf q}_{1}}\, 
{\mathcal P}^{\hspace{0.03cm}\ast\, a_{1}\; i\, j}_{{\bf k}_{1},\, {\bf q},\, {\bf q} - {\bf k}_{1}}}
{\omega^{\hspace{0.02cm} l}_{{\bf k}_{1}} - \omega^{-}_{{\bf q}} + \omega^{-}_{{\bf q} - {\bf k}_{1}}}
\label{ap:E3}\\[1.5ex]
\;+\;\,
&\frac{{\mathcal P}^{\hspace{0.03cm}\ast\, a_{2}\, i\hspace{0.04cm} j}_{{\bf k}_{2},\,{\bf q},\, {\bf q} - {\bf k}_{2}}\, 
{\mathcal G}^{\hspace{0.03cm}\ast\, a_{1}\,i_{1}\,j}_{{\bf k}_{1},\, {\bf q}_{1},\, {\bf k}_{1} - {\bf q}_{1}}}
{\omega^{\hspace{0.02cm} l}_{{\bf k}_{1}} - \omega^{-}_{{\bf q}_{1}} - \omega^{-}_{{\bf k}_{1} - {\bf q}_{1}}}
\;-\; 
\frac{{\mathcal P}^{\hspace{0.03cm}\ast\, a_{2}\, i_{1}\,j}_{{\bf k}_{2},\,{\bf q}_{1},\, {\bf q}_{1} - {\bf k}_{2}}\, 
{\mathcal G}^{\hspace{0.03cm}\ast\, a_{1}\,i\hspace{0.04cm} j}_{{\bf k}_{1},\, {\bf q},\,{\bf k}_{1} - {\bf q}}}
{\omega^{\hspace{0.02cm} l}_{{\bf k}_{1}} - \omega^{-}_{{\bf q}} - \omega^{-}_{{\bf k}_{1} - {\bf q}}}
\notag\\[1.5ex]
+\,
2\hspace{0.03cm}\Biggl(
&\frac{{\mathcal V}^{\hspace{0.03cm}\ast\, a\; a_{1}\, a_{2}}_{{\bf k}_{1} + {{\bf k}_{2}},\, {\bf k}_{1}, {\bf k}_{2}}\,{\mathcal G}^{\hspace{0.03cm}\ast\, a\, i\; i_{1}}_{{\bf q} + {\bf q}_{1},\, {\bf q},\, {\bf q}_{1}}}
{\omega^{\hspace{0.02cm} l}_{{\bf q} + {\bf q}_{1}} - \omega^{-}_{{\bf q}} - \omega^{-}_{{\bf q}_{1}}}
\;-\;
\frac{{\mathcal U}^{\hspace{0.03cm}\ast\, a\; a_{1}\, a_{2}}_{-{\bf k}_{1} - {\bf k}_{2},\, {\bf k}_{1}, {\bf k}_{2}}\,
{\mathcal K}^{\; a\, i\; i_{1}}_{-{\bf q} - {\bf q}_{1},\, {\bf q},\, {\bf q}_{1}}}
{\omega^{\hspace{0.02cm} l}_{-{\bf q} - {\bf q}_{1}} + \omega^{-}_{{\bf q}} + \omega^{-}_{{\bf q}_{1}}}\Biggr),
\notag
\end{align}

\begin{align}
\widetilde{T}^{\,(1)\, i\, i_{1}\, a_{1}\, a_{2}}_{\; {\bf q},\, {\bf q}_{1},\, {\bf k}_{1},\, {\bf k}_{2}}
\,=\,
T^{\,(1)\, i\, i_{1}\, a_{1}\, a_{2}}_{\; {\bf q},\, {\bf q}_{1},\, {\bf k}_{1},\, {\bf k}_{2}}
\;\,+\;\,
&\frac{{\mathcal P}^{\; a_{2}\,j\,  i_{1}}_{{\bf k}_{2},\,{\bf k}_{2} + {\bf q}_{1},\, {\bf q}_{1}}\hspace{0.03cm} 
{\mathcal P}^{\; a_{1}\, i\, j}_{{\bf k}_{1},\, {\bf q},\, {\bf q} - {\bf k}_{1}}}
{\omega^{\hspace{0.02cm} l}_{{\bf k}_{2}} - \omega^{-}_{{\bf k}_{2} +\hspace{0.02cm} {\bf q}_{1}}\! + \omega^{-}_{{\bf q}_{1}}}
\;-\; 
\frac{{\mathcal P}^{\; a_{1}\,j\,  i_{1}}_{{\bf k}_{1},\,{\bf k}_{1} + {\bf q}_{1},\, {\bf q}_{1}}\hspace{0.03cm} 
{\mathcal P}^{\; a_{2}\, i\, j}_{{\bf k}_{2},\, {\bf q},\, {\bf q} - {\bf k}_{2}}}
{\omega^{\hspace{0.02cm} l}_{{\bf k}_{2}} - \omega^{-}_{{\bf q}} + \omega^{-}_{{\bf q} - {\bf k}_{2}}}
\label{ap:E4}\\[1.5ex]
-\,4\hspace{0.03cm}\Biggl(
&\frac{{\mathcal K}^{\; a_{1}\,j\,  i_{1}}_{\hspace{0.03cm}{\bf k}_{1},\, -{\bf k}_{1} - {\bf q}_{1},\, {\bf q}_{1}}\, 
{\mathcal G}^{\; a_{2}\, i\, j}_{{\bf k}_{2},\, {\bf q},\, {\bf k}_{2} - {\bf q}}}
{\omega^{\hspace{0.02cm} l}_{{\bf k}_{2}} - \omega^{-}_{{\bf q}} - \omega^{-}_{{\bf k}_{2} - {\bf q}}} 
\;-\;
\frac{{\mathcal K}^{\; a_{2}\,j\,  i_{1}}_{\hspace{0.03cm}{\bf k}_{2},\, -{\bf k}_{2} - {\bf q}_{1},\, {\bf q}_{1}}\, 
{\mathcal G}^{\; a_{1}\, i\, j}_{{\bf k}_{1},\, {\bf q},\, {\bf k}_{1} - {\bf q}}}
{\omega^{\hspace{0.02cm} l}_{{\bf k}_{2}} + \omega^{-}_{-{\bf k}_{2} - {\bf q}_{1}} + \omega^{-}_{{\bf q}_{1}}} 
\Biggr)
\notag\\[1.5ex]
-\,
2\hspace{0.03cm}\Biggl(\hspace{0.02cm}
&\frac{{\mathcal U}^{\; a_{1}\, a_{2}\,  a}_{\hspace{0.03cm} {\bf k}_{1}, {\bf k}_{2},\, -{\bf k}_{1} - {{\bf k}_{2}}}\, 
{\mathcal P}^{\hspace{0.03cm}\ast\, a\, i_{1}\, i}_{{\bf q}_{1} - {\bf q},\, {\bf q}_{1},\, {\bf q}}}
{\omega^{\hspace{0.02cm} l}_{{\bf q}_{1} - {\bf q}} - \omega^{-}_{{\bf q}_{1}} + \omega^{-}_{{\bf q}}}
\;+\;
\frac{{\mathcal V}^{\; a\; a_{1}\, a_{2}}_{{\bf k}_{1} + {{\bf k}_{2}},\, {\bf k}_{1},\, {\bf k}_{2}}
{\mathcal P}^{\; a\, i\,  i_{1}}_{{\bf q} - {\bf q}_{1},\, {\bf q},\, {\bf q}_{1}}\, 
}
{\omega^{\hspace{0.02cm} l}_{{\bf q} - {\bf q}_{1}} - \omega^{-}_{{\bf q}} + \omega^{-}_{{\bf q}_{1}}}\Biggr),
\notag
\end{align}
\begin{align}
\widetilde{T}^{\,(4)\, i\, i_{1}\, a_{1}\, a_{2}}_{\; {\bf q},\, {\bf q}_{1},\, {\bf k}_{1},\, {\bf k}_{2}}
\,=\,
T^{\,(4)\, i\, i_{1}\, a_{1}\, a_{2}}_{\; {\bf q},\, {\bf q}_{1},\, {\bf k}_{1},\, {\bf k}_{2}}
\;\,+\;\,
&\frac{{\mathcal P}^{\; a_{2}\,j\,  i}_{{\bf k}_{2},\,{\bf k}_{2} + {\bf q},\, {\bf q}}\, 
{\mathcal K}^{\; a_{1}\,j\,  i_{1}}_{\hspace{0.03cm} {\bf k}_{1},\, -{\bf k}_{1} - {\bf q}_{1},\, {\bf q}_{1}}}
{\omega^{\hspace{0.02cm} l}_{{\bf k}_{2}} - \omega^{-}_{{\bf k}_{2} + {\bf q}}\! + \omega^{-}_{{\bf q}}}
\;-\; 
\frac{{\mathcal P}^{\; a_{2}\,j\,  i_{1}}_{{\bf k}_{2},\,{\bf k}_{2} + {\bf q}_{1},\, {\bf q}_{1}}\, 
{\mathcal K}^{\; a_{1}\,j\, i}_{\hspace{0.03cm} {\bf k}_{1},\, -{\bf k}_{1} - {\bf q},\, {\bf q}}}
{\omega^{\hspace{0.02cm} l}_{{\bf k}_{2}} - \omega^{-}_{{\bf k}_{2} + {\bf q}_{1}}\! + \omega^{-}_{{\bf q}_{1}}}
\label{ap:E5}\\[1.5ex]
\;+\;\,
&\frac{{\mathcal K}^{\; a_{2}\,j\, i}_{\hspace{0.03cm} {\bf k}_{2},\, -{\bf k}_{2} - {\bf q},\, {\bf q}}\, 
{\mathcal P}^{\; a_{1}\,j\, i_{1}}_{{\bf k}_{1},\, {\bf k}_{1} + {\bf q}_{1},\ {\bf q}_{1}}}
{\omega^{\hspace{0.02cm} l}_{{\bf k}_{2}} + \omega^{-}_{-{\bf k}_{2} - {\bf q}} + \omega^{-}_{{\bf q}}} 
\;-\;
\frac{{\mathcal K}^{\; a_{2}\,j\, i_{1}}_{\hspace{0.03cm} {\bf k}_{2},\, -{\bf k}_{2} - {\bf q}_{1},\, {\bf q}_{1}}\, 
{\mathcal P}^{\; a_{1}\,j\, i}_{{\bf k}_{1},\, {\bf k}_{1} + {\bf q},\, {\bf q}}}
{\omega^{\hspace{0.02cm} l}_{{\bf k}_{2}} + \omega^{-}_{-{\bf k}_{2} - {\bf q}_{1}}\! + \omega^{-}_{{\bf q}_{1}}} 
\notag\\[1.5ex]
+\,
2\hspace{0.03cm}\Biggl(\hspace{0.02cm}
&\frac{{\mathcal U}^{\; a_{1}\, a_{2}\; a}_{{\bf k}_{1}, {\bf k}_{2},\, -{\bf k}_{1} - {{\bf k}_{2}}}\, 
{\mathcal G}^{\; a\, i\; i_{1}}_{{\bf q} + {\bf q}_{1},\, {\bf q},\, {\bf q}_{1}}}
{\omega^{\hspace{0.02cm} l}_{{\bf q} + {\bf q}_{1}}\! - \omega^{-}_{{\bf q}} - \omega^{-}_{{\bf q}_{1}}}
\;-\;
\frac{{\mathcal V}^{\; a\; a_{1}\, a_{2}}_{{\bf k}_{1} + {{\bf k}_{2}},\, {\bf k}_{1},\, {\bf k}_{2}}
{\mathcal K}^{\; a\, i\;  i_{1}}_{-{\bf q} - {\bf q}_{1},\, {\bf q},\, {\bf q}_{1}}\, 
}
{\omega^{\hspace{0.02cm} l}_{-{\bf q} - {\bf q}_{1}}\! + \omega^{-}_{{\bf q}} + \omega^{-}_{{\bf q}_{1}}}\Biggr).
\notag
\end{align}
The integral relations in Appendix \ref{appendix_C} enable us to restore easy an explicit form of the coefficient functions $J^{(1)\, a\, a_{1}\, i_{1}\, i_{2}}_{\, {\bf k},\, {\bf k}_{1},\, {\bf q}_{1},\, {\bf q}_{2}}$ and $R^{\,(n)\, i\, a\, a_{1}\, i_{1}}_{\; {\bf q},\, {\bf k},\, {\bf k}_{1},\, {\bf q}_{1}} $ for $n = 1, 3, 4, 5, 6$ in the integrands of the canonical transformations (\ref{eq:3t}) and  (\ref{eq:3y}). For instance, the function $J^{\,(3)\, a_{1}\, a_{2}\, i\; i_{1}}_{\ {\bf k}_{1},\, {\bf k}_{2},\, {\bf q},\, {\bf q}_{1}}$ makes it possible to restore an explicit form of two other coefficient functions, namely $J^{\,(1)\, a_{1}\, a_{2}\, i\; i_{1}}_{\ {\bf k}_{1},\, {\bf k}_{2},\, {\bf q},\, {\bf q}_{1}}$ and $R^{(5)\; i\, a_{1}\, a_{2}\, i_{1}}_{\, {\bf q},\, {\bf k}_{1},\, {\bf k}_{2},\, {\bf q}_{1}}$. The former function is defined from the integral relation (\ref{ap:C1a}), whereas the latter one is obtained from the relation (\ref{ap:C3c}).\\
\indent Further, the function $J^{\,(4)\, a_{1}\, a_{2}\, i\; i_{1}}_{\ {\bf k}_{1},\, {\bf k}_{2},\, {\bf q},\, {\bf q}_{1}}$ allows to restore an explicit form of the coefficient function $R^{(4)\; i\, a_{1}\, a_{2}\, i_{1}}_{\, {\bf q},\, {\bf k}_{1},\, {\bf k}_{2},\, {\bf q}_{1}}$. It is defined from the canonicity condition (\ref{ap:C4c}). The function $J^{\,(5)\, a_{1}\, a_{2}\, i\; i_{1}}_{\ {\bf k}_{1},\, {\bf k}_{2},\, {\bf q},\, {\bf q}_{1}}$ enables us to restore an explicit form of another two coefficient functions, namely $R^{(1)\; i\, a_{1}\, a_{2}\, i_{1}}_{\, {\bf q},\, {\bf k}_{1},\, {\bf k}_{2},\, {\bf q}_{1}}$ and $R^{(3)\; i\, a_{1}\, a_{2}\, i_{1}}_{\, {\bf q},\, {\bf k}_{1},\, {\bf k}_{2},\, {\bf q}_{1}}$. The first function is defined from the canonicity condition  (\ref{ap:C4a}), while the second one is derived from the relation (\ref{ap:C3b}).\\
\indent Lastly, the function $J^{\,(6)\, a_{1}\, a_{2}\, i\; i_{1}}_{\ {\bf k}_{1},\, {\bf k}_{2},\, {\bf q},\, {\bf q}_{1}}$ provides us with an explicit form of the coefficient function $R^{(6)\; i\, a_{1}\, a_{2}\, i_{1}}_{\, {\bf q},\, {\bf k}_{1},\, {\bf k}_{2},\, {\bf q}_{1}}$. It is defined from the canonicity condition (\ref{ap:C3d}).

\end{appendices}


\end{document}